%% file: main.tex
\documentclass[aps,
showpacs,preprintnumbers,nofootinbib,superscriptaddress,groupedaddress]{SPhdThesis}

\usepackage[T1]{fontenc}
\usepackage{lmodern}
\usepackage{amsmath}
\usepackage{graphicx,amssymb,amsmath,amsthm,amsfonts,epsfig,epsf}
\usepackage{epstopdf}
\definecolor{darkred}{rgb}{0.5,0,0}
\usepackage{aas_macros}
\usepackage{bm}
\usepackage{dcolumn}
\usepackage[utf8]{inputenc}
\usepackage{latexsym}
\usepackage{rotating}
\usepackage{longtable}

\usepackage{enumerate}
\usepackage{tensor,multirow}
\usepackage{mathtools}
\usepackage{url}
\usepackage{epigraph}
\usepackage{titlesec}
\usepackage[sort&compress,numbers]{natbib}
\usepackage{doi}
\usepackage{colortbl}
\usepackage[toc,page]{appendix}
\usepackage{minitoc}
\usepackage{emptypage}
\usepackage{bbold}
\usepackage{bbm}
\usepackage{flushend}
\usepackage{slashed}
\usepackage{enumitem}
\setdescription{font=\normalfont}
\usepackage{hyperref}
\hypersetup{
  unicode=true,          
  pdftitle={PhD Thesis},
  pdfauthor={},
  pdfsubject={},
  pdfcreator={LaTeX},          
  pdfproducer={LaTeX},         
  pdfkeywords={},
  colorlinks=true,            
  linkcolor  = cyan!70!black,
  citecolor  = cyan!70!black,
  urlcolor   = cyan!70!black,
}

\DeclareSymbolFont{matha}{OML}{txmi}{m}{it}
\DeclareMathSymbol{\varv}{\mathord}{matha}{118}
\def\p{\partial}
\def\Lie{\mathcal{L}}
\DeclareMathAlphabet\mathbfcal{OMS}{cmsy}{b}{n}
\newcommand{\s}{{\mathbbmss{s}}}

\graphicspath{ {./images/} }

\makeatletter
\titleformat{\part}[display]
{\Huge\scshape\filright}
{\partname~\thepart:}
{20pt}
{\thispagestyle{epigraph}}
\makeatother
\SgSetTitle{Universidade de Lisboa ~~~~~~~~~~~~~~~~~~~~~~~~~~~~~~~~~~~
  Instituto Superior Técnico}
\SgSetAuthorDegrees{Athanasios Giannakopoulos}
\SgSetAuthor{Characteristic formulations of general relativity\\ and applications}
\SgSetYear{{\bf 2022}}
\SgSetDegree{
  \begin{tabular}{rl}
    {\bf Supervisor:}& \textbf{Doctor Miguel Rodrigues Zilh\~ao Nogueira}\\
    {\bf Co-Supervisor:}& \textbf{Doctor David Matthew Hilditch}\\
  \end{tabular}}

\SgSetDepartment{
  \textbf{
    Thesis approved in public session to obtain the PhD Degree in Physics
    \\
    \vspace{0.2cm}
    Jury final classification: Pass with Distinction}
}
\SgSetUniversity{$\;$}

\begin{document}
\SgAddTitle
\cleardoublepage
\SgAddTitlesecond
\cleardoublepage

\begin{frontmatter}
  \input{sections/resumo}%
  \input{sections/abstract}%
  \input{sections/citcit2}%
  \cleardoublepage
  \input{sections/acknowledgements}%
  \cleardoublepage
  \dominitoc
  \SgAddToc
  \cleardoublepage
  \SgAddLof
  \cleardoublepage
  \input{sections/preface}%
  \cleardoublepage
  \input{sections/acronyms}%
  \cleardoublepage
\end{frontmatter}

\input{sections/intro} 
\cleardoublepage

\epigraphhead[450]{{\it This part is based on Refs.~\cite{GiaHilZil20, GiaBisHil21}}}
\part{Hyperbolicity and well-posedness}
\cleardoublepage

\input{sections/PDE_theory}
\cleardoublepage
\input{sections/Bondi-like_properties}
\cleardoublepage
\input{sections/gauge_structure}
\cleardoublepage
\input{sections/hyperbolicity}
\cleardoublepage
\input{sections/well-posedness}
\cleardoublepage
\input{sections/numerical_experiments}
\cleardoublepage

\epigraphhead[450]{{\it This part is based on Refs.~\cite{
      BeaCasGia21b, BeaCasGia22}}}
\part{Applications to holography}
\cleardoublepage

\input{sections/jecco}
\cleardoublepage
\input{sections/final_remarks}
\cleardoublepage

\epigraphhead[450]{}
\part{Appendix}
\appendix
 \cleardoublepage
 \input{sections/appendix}
\cleardoublepage

\SgIncludeBib{sections/refs}

\end{document}

%% file: sections/resumo.tex
\begin{resumo}

  A relatividade geral descreve v\'arios sistemas gravitacionais de
  relev\^{a}ncia astrof\'isica, como buracos negros e estrelas de
  neutr\~oes. Pode tamb\'em descrever sistemas fortemente acoplados
  atrav\'es da dualidade hologr\'afica. Al\'em disso, s\~ao t\'opicos
  de investiga\c{c}ao ativa tamb\'em aspectos mais formais da teoria,
  como a estabilidade de espa\c{c}os-tempos e a forma\c{c}\~ao de
  singularidades. Muitas vezes, solu\c{c}\~oes a problemas em aberto
  n\~ao s\~ao conhecidas em forma anal\'itica e quando os m\'etodos
  perturbativos s\~ao inadequados \'e preciso utilizar t\'ecnicas
  num\'ericas.

  O problema valores inicias (e fronteira) característicos tem muitas
  aplica\c{c}\~oes na relatividade geral, envolve geralmente estudos
  num\'ericos e, frequentemente, \'e formulado usando coordenadas de
  tipo Bondi. A boa-formula\c{c}\~ao dos sistemas resultantes das
  equa\c{c}\~oes diferenciais parciais, no entanto, permanece uma
  quest\~ao em aberto. A resposta a esta pergunta afeta a precis\~ao
  e, potencialmente, a confiabilidade das conclus\~oes extra\'idas de
  estudos num\'ericos baseados em tais formula\c{c}\~oes. Uma
  aproxima\c{c}\~ao num\'erica pode convergir para o limite do
  cont\'inuo apenas para sistemas bem-formulados. A no\c{c}\~ao de
  boa-formula\c{c}\~ao est\'a intimamente relacionada \`a da
  hiperbolicidade e inclui a especificação de uma norma.
  
  Na primeira parte desta tese, expandimos a nossa compreens\~ao da
  hiperbolicidade e da boa-formula\c{c}\~ao de sistemas de
  evolu\c{c}\~ao livre de tipo Bondi. Mostramos que v\'arios
  prot\'otipos de formula\c{c}\~oes de tipo Bondi s\~ao apenas
  fracamente hiperb\'olicos e examinamos a causa desse resultado. Numa
  an\'alise linear identificamos a gauge, as restri\c{c}\~oes e os
  blocos f\'isicos na parte principal das equa\c{c}\~oes de campo de
  Einstein nessa gauge, e mostramos que o subsistema relacionado com
  as vari\'aveis de gauge \'e apenas fracamente hiperb\'olico. A
  hiperbolicidade fraca do sistema completo segue como
  consequ\^{e}ncia em v\'arios casos. Demonstramos tamb\'em isso
  explicitamente em atrav\'es de exemplos e, portanto, argumentamos
  que gauge de tipo Bondi resultam em sistemas de evolu\c{c}\~ao livre
  fracamente hiperb\'olicos sob condi\c{c}\~oes bastante
  gerais. Consequentemente, o problema dos valor inicial
  característico em relatividade geral nestas gauges torna-se
  mal-formulado nas normas mais simples que se gostaria de
  utilizar. Discutimos as implica\c{c}\~oes deste resultado em
  m\'etodos de modelagem precisos de sinais de ondas gravitacionais e
  trabalhamos para a constru\c{c}\~ao de normas alternativas que
  possam ser mais apropriadas. Tamb\'em apresentamos testes
  num\'ericos que demonstram a hiperbolicidade fraca na pr\'atica e
  sublinham as caracter\'isticas importantes para realiz\'a-los de
  forma eficaz.

  Na segunda parte, dirigimos a nossa aten\c{c}\~ao para
  aplica\c{c}\~oes dessas formula\c{c}\~oes em sistemas fortemente
  acoplados via holografia. O objectivo principal \'e perceber o
  comportamento qualitativo de plasmas fortemente acoplados, mas
  devido \'a fraca hiperbolicidade, n\~ao podemos realizar estimativas
  de erro rigorosas. Apresentamos ainda o~\texttt{Jecco}, um c\'odigo
  caracter\'istico recentemente desenvolvido que nos permite simular a
  din\^{a}mica de plasmas fortemente acoplados. S\~ao fornecidos
  tamb\'em exemplos representativos das simula\c{c}\~oes que podem ser
  realizadas com este c\'odigo, nomeadamente a din\^{a}mica fora de
  equil\'ibrio de plasmas que sofrem transi\c{c}\~oes de fase. Este
  pode ser um dos cen\'arios do universo primitivo e estas
  simula\c{c}\~oes podem fornecer respostas sobre quest\~oes de
  natureza fundamental.

  \noindent{\bf{Palavras-chave:}} Relatividade geral; Ondas
  gravitacionais; Formula\c{c}\~oes caracter\'isticas;
  Hiperbolicidade; Boa-formula\c{c}\~ao.

\end{resumo}

%% file: sections/abstract.tex
\begin{abstract}

  General relativity can describe various gravitational systems of
  astrophysical relevance, like black holes and neutron stars, or even
  strongly coupled systems through the holographic duality. In
  addition, more formal aspects of the theory like the stability of
  spacetimes and the formation of singularities are still topics of
  active research. In several cases, solutions in closed analytic form
  are not known, and perturbative methods are inadequate, leading to
  the employment of numerical techniques.

  The characteristic initial (boundary) value problem has numerous
  applications in general relativity involving numerical studies and
  is often formulated using Bondi-like coordinates. Well-posedness of
  the resulting systems of partial differential equations, however,
  remains an open question. The answer to this question affects the
  accuracy, and potentially the reliability of conclusions drawn from
  numerical studies based on such formulations. A numerical
  approximation can converge to the continuum limit only for
  well-posed systems. The notion of well-posedness is tightly related
  to that of hyperbolicity and includes the specification of a norm.

  In the first part of this thesis, we expand our understanding of the
  hyperbolicity and well-posedness of Bondi-like free evolution
  systems. We show that several prototype Bondi-like formulations are
  only weakly hyperbolic and examine the root cause of this result. In
  a linear analysis we identify the gauge, constraint and physical
  blocks in the principal part of the Einstein field equations in such
  a gauge, and we show that the subsystem related to the gauge
  variables is only weakly hyperbolic. Weak hyperbolicity of the full
  system follows as a consequence in many cases. We demonstrate this
  explicitly in specific examples, and thus argue that Bondi-like
  gauges result in weakly hyperbolic free evolution systems under
  quite general conditions. Consequently, the characteristic initial
  (boundary) value problem of general relativity in these gauges is
  rendered ill-posed in the simplest norms one would like to
  employ. We discuss the implications of this result in accurate
  gravitational waveform modeling methods and work towards the
  construction of alternative norms that might be more appropriate. We
  also present numerical tests that demonstrate weak hyperbolicity in
  practice and highlight important features to perform them
  effectively.

  In the second part, we turn our attention to applications of these
  formulations to strongly coupled systems via holography. We expect
  these studies to shed more light on the qualitative behavior of
  strongly coupled plasmas, but due to weak hyperbolicity, we cannot
  perform rigorous error estimates to our satisfaction. We
  present~\texttt{Jecco}, a newly developed characteristic code that
  allows us to simulate the dynamics of strongly coupled
  plasmas. Representative examples of the simulations that can be
  achieved with this code are provided, namely the out-of-equilibrium
  dynamics of said plasmas that undergo phase transitions. This is a
  possible 
  scenario of the early universe and such simulations might
  provide insights into questions of fundamental nature.

\end{abstract}

%% file: sections/citcit2.tex
\begin{citcit}

\vspace*{\fill}
\noindent
\makebox[0.8\textwidth]{\it to my parents}

\vfill

\end{citcit}


%% file: sections/acknowledgements.tex
\begin{acknowledgements}

  Reaching the end of my PhD, I realize that I have had an amazing
  experience during this adventure and for this I am grateful to a
  number of people.

  
  Miguel and David, thank you for all the guidance, patience,
  teaching, and understanding all these years. I really appreciate the
  time you took and the effort you made, from the early days sitting
  next to me and helping me fix bugs and errors, to the many
  whiteboard calculations and the numerous online discussions later
  on. I have learned a lot next to you, growing both personally and
  professionally. I believe that the PhD is a unique experience for
  each person, and I definitely feel lucky to have done it under your
  supervision.

  To all my collaborators, thank you for all the work we did together,
  as well as for all the scientific interaction we had. It has been
  very exciting and educational for me. Especially to Nigel, Denis,
  Yago and Mikel, thank you for the hours we spent together discussing
  science.

  I am thankful to all the Jury members for their time and care in
  reading this thesis, as well as their questions, comments and
  suggestions.
  
  I am grateful to all the GRIT and CENTRA members for creating a
  friendly and vibrant atmosphere. It has been a pleasure to be a
  member of this team. Especially to all my fellow PhD colleagues and
  office-mates, thank you for all the time we spent together and all
  the discussions we had. You made my time there more interesting and
  memorable.

  To S\'ergio, Manuel, Jo\~ao, and all the IT team of CENTRA, thank you
  for your constant availability and hard work. You provided crucial
  support, helped me with many technical problems, and allowed me to
  stay focused on my work. The same goes for Rita Sousa. You have
  always been there helping me with all sorts of bureaucratic issues
  that appeared and made my life easier. I really appreciate it.

  Special thanks goes to Miguel Duarte for his help and suggestions
  with the early manuscript.
  

  Lorenzo and Kyriako, I deeply thank you for your friendship, you
  took the Lisbon experience to another level for me and I will always
  remember our times together. Diogo, Vera, and Rebecca, it has been
  amazing living with you. You made me feel at home, I am grateful to
  you. Isa, colega, it has been great sharing this academic path and
  all its challenges and frustrations, thank you. Fernanda, Arianna,
  Rodrigo, Francisco, Krinio, Stefano, thank you all, it has been
  great sharing time and experiences with you.
  
  To my life-long friends from Greece, thank you all for your
  friendship, it means a lot to me. Panteli, the online coffee
  meetings have definitely made the pandemic period better.
  
  To Chrysalena, my long-time partner, you have always been my
  constant companion all these years and in many adventures. I cannot
  thank you enough for all your love, understanding, and patience. I
  love you.

  To my family, my brother and my parents, I cannot imagine how I
  would have done this without you. I am deeply grateful for all your
  support and help, all these years, in pursuing my dreams. I
  appreciate everything you have done for me. You are always in my
  heart.
  
  
  Finally, I am grateful to the Funda\c{c}ao para a Ci\^encia e a
  Tecnologia for funding this work via the grants PD/BD/135425/2017,
  COVID/BD/152483/2022 and BL68/2022-IST-ID and to the GWverse COST
  Action CA16104, ``Black holes, gravitational waves and fundamental
  physics'' for networking and partial financial support. I am also
  grateful to the CENTRA/IST cluster ``Baltasar-Sete-S\'ois'', and the
  MareNostrum supercomputer at the BSC, for providing computational
  resources and technical expertise that helped completing this work.
  
\end{acknowledgements}

%% file: sections/preface.tex
\begin{preface}

  The research presented in this thesis has been carried out at the
  Center for Astrophysics and Gravitation (CENTRA) in the Physics
  department of Instituto Superior T\'ecnico - Universidade de Lisboa.

  I declare that this thesis is not substantially the same as any that
  I have submitted for a degree, diploma or other qualification at any
  other university and that no part of it has already been or is
  concurrently submitted for any such degree, diploma or other
  qualification.

  The following thesis has been the result of several collaborations.

  A complete list of the articles included in this thesis is displayed
  below:

  \begin{center}
    Part I
  \end{center}
  
  \begin{itemize}
  \item[$\diamond$] \cite{GiaHilZil20}:
    {T. Giannakopoulos, D. Hilditch, and M. Zilh\~ao},
    ``Hyperbolicity of General Relativity in Bondi-like gauges'',
    \href{https://journals.aps.org/prd/abstract/10.1103/PhysRevD.102.064035}
    {\it Phys. Rev. D 102, 064035 },
    \href{https://arxiv.org/abs/2007.06419}
    {arXiv:2007.06419 [gr-qc]}

  \item[$\diamond$] \cite{GiaBisHil21}:
    {T. Giannakopoulos, N. T. Bishop, D. Hilditch, D. Pollney, and M. Zilh\~ao},
    ``Gauge structure of the Einstein field equations in Bondi-like coordinates'',
    \href{https://journals.aps.org/prd/abstract/10.1103/PhysRevD.105.084055}
    {\it Phys. Rev. D 105, 084055 },
    \href{https://arxiv.org/abs/2111.14794}
    {arXiv:2111.14794 [gr-qc]}
  \end{itemize}

  \begin{center}
    Part II
  \end{center}
  
  \begin{itemize}
  \item[$\diamond$] \cite{BeaCasGia21b}
    {Y. Bea, J. Casalderrey-Solana, T. Giannakopoulos, A. Jansen,
      S. Krippendorf, D. Mateos, M. Sanchez-Garitaonandia, and
      M. Zilh\~ao},
    ``Spinodal Gravitational Waves'',
    \href{https://arxiv.org/abs/2112.15478}
    {arXiv:2112.15478 [hep-th]}
    
  \item[$\diamond$] \cite{BeaCasGia22}:
    {Y. Bea, J. Casalderrey-Solana, T. Giannakopoulos, A. Jansen,
      D. Mateos, M. Sanchez-Garitaonandia, M. Zilh\~ao},
    ``Holographic Bubbles with Jecco: Expanding, Collapsing and Critical'',
    \href{https://link.springer.com/article/10.1007/JHEP09(2022)008}
    {\it J. High Energ. Phys. 2022, 8 (2022) },
    \href{https://arxiv.org/abs/2202.10503}
    {arXiv:2202.10503 [hep-th]}
        
  \end{itemize}

  The following articles include work of the author of this thesis but
  are not discussed here:

  \begin{itemize}
  \item[$\diamond$] \cite{BeaDiaGia21}:
    {Y. Bea, O. J. C. Dias, T. Giannakopoulos, D. Mateos,
      M. Sanchez-Garitaonandia J. E. Santos, and M. Zilh\~ao},
    ``Crossing a large-N phase transition at finite volume'',
    \href{https://link.springer.com/article/10.1007/JHEP02(2021)061}
    {\it J. High Energ. Phys. 2021, 61 (2021) },
    \href{https://arxiv.org/abs/2007.06467} {arXiv:2007.06467 [hep-th]}

  \item[$\diamond$] \cite{BeaCasGia21}:
    {Y. Bea, J. Casalderrey-Solana, T. Giannakopoulos, D. Mateos,
      M. Sanchez-Garitaonandia, and M. Zilh\~ao},
    ``Bubble Wall Velocity from Holography'',
    \href{https://journals.aps.org/prd/abstract/10.1103/PhysRevD.104.L121903}
    {\it Phys. Rev. D 104, L121903},
    \href{https://arxiv.org/abs/2104.05708} {arXiv:2104.05708 [hep-th]}
    
  \item[$\diamond$] \cite{BeaCasGia21a}:
    {Y. Bea, J. Casalderrey-Solana, T. Giannakopoulos, D. Mateos,
      M. Sanchez-Garitaonandia, and M. Zilh\~ao},
    ``Domain Collisions'',
    \href{https://link.springer.com/article/10.1007/JHEP06(2022)025}
    {\it J. High Energ. Phys. 2022, 25 (2022)},
    \href{https://arxiv.org/abs/2111.03355}
    {arXiv:2111.03355 [hep-th]}

  \end{itemize}

\end{preface}

%% file: sections/acronyms.tex
\begin{acronyms}
  \begin{tabular}{rl}
    {\bf AAdS}& Asymptotically Anti-de Sitter\\
    {\bf ADM}& Arnowitt-Deser-Misner\\
    {\bf AdS}& Anti-de Sitter\\
    {\bf CCE}& Cauchy-Characteristic Extraction\\
    {\bf CCM}& Cauchy-Characteristic Matching\\
    {\bf CIBVP}& Characteristic Initial Boundary Value Problem\\
    {\bf CIVP}& Characteristic Initial Value Problem\\
    {\bf DF}& Dual Frame\\
    {\bf EFE}& Einstein Field Equations\\
    {\bf GR}& General Relativity\\
    {\bf GW}& Gravitational Wave\\
    {\bf IBVP}&  Initial Boundary Value Problem\\
    {\bf IVP}&  Initial Value Problem\\
    {\bf ODE}& Ordinary Differential Equation\\
    {\bf PDE}& Partial Differential Equation\\
    {\bf SH}& Strongly Hyperbolic\\
    {\bf WH}& Weakly Hyperbolic\\
  \end{tabular}
\end{acronyms}

%% file: sections/intro.tex
\chapter{Introduction}

\minitoc

General relativity (GR) is a theory of gravity introduced by Albert
Einstein in 1915~\cite{Ein15, Ein16aen}.
%
%
Since its creation, GR has grant us with several predictions that have
been verified by experimental tests, such as the recent detections of
the gravitational wave (GW) signals produced by binaries of compact
objects like black holes and neutron stars~\cite{AbbAbbAbb16,
  AbbAbbAbb17, PhysRevX6041015, PhysRevX9031040,
  LIGOScientific:2021djp}, and the observation of black hole
shadows~\cite{EHT:2019dse, EHT:2022xnr}.
To make these predictions, one has to solve the equations of motion of
GR, which form a system of coupled, non-linear partial differential
equations (PDEs). Generically, this is a particularly difficult
problem to solve and various techniques may be employed in different
regimes and setups.

Fully analytical solutions to the Einstein field equations (EFE) are
possible in limited cases~\cite{KraHer80}, such as those with high
symmetry e.g. the Minkowski, anti-de Sitter (AdS),
Schwarzschild~\cite{Sch16}, and Kerr spacetimes~\cite{Ker63}. However,
many gravitational systems of interest do not fall into this category
and hence different methods are needed to understand their
behavior. Perturbative schemes can provide good approximations in
various scenarios, for instance in the inspiral and ringdown phases
during the evolution of a binary formed by compact objects. The merger
phase however is highly non-linear, which makes many perturbative
treatments inadequate there. A way to find solutions for gravitational
systems in the highly dynamical, non-linear, strong gravity regime
like the merger is by means of numerical methods. The subfield of
gravitational research that exploits these methods to obtain
approximate solutions to the EFE is often called~\textit{numerical
  relativity}.

A physical process in GR does not depend on the way we choose to
describe it. Motivated by the special features and symmetries of a
specific gravitational setup, we choose appropriate coordinates to
express its spacetime, for example in a setup with spherical symmetry,
spherical polar coordinates can be a good choice.
In numerical relativity, we typically use coordinates to foliate the
spacetime with hypersurfaces of a constant coordinate that we
associate with time.
Quite frequently these hypersurfaces are spacelike, for instance when
modeling the region near the merger of two compact objects. In other
cases, it is more convenient to employ null hypersurfaces. This setup
is called~\textit{characteristic} and the coordinate that labels the
different hypersurfaces is the advanced or retarded time.
In GR the speed of light is the upper bound at which physical signals
can propagate and defines the causal structure of spacetime. This is
also the speed at which GWs travel and together with light rays, they
move along null hypersurfaces. Hence, characteristic formulations are
particularly convenient to describe radiative processes in GR.

\section{Preliminaries}

This section is a collection of some tools that are useful in this
thesis. For a more complete presentation one can consult standard
textbooks of GR, e.g.~\cite{MisThoWhe73c, Wal84a, Car03, Poi04}.

GR adopts a geometric viewpoint for gravity and makes extensive use of
differential geometry as a tool. A central object is the
differentiable manifold, typically denoted by~$\mathcal{M}$. Let us
assume that~$\mathcal{M}$ is covered by a set of
coordinates~$x^\mu$, where~$\mu = 0,1,2,...,n-1$ for a
n-dimensional manifold. Let also~$V_p$ be the space tangent to a
point~$p$ of~$\mathcal{M}$.  Then~$\p_\mu \equiv \p/\p x^\mu$ defines
an element of a coordinate basis for~$V_p$. A vector~$\mathbf{v}$
on~$V_p$ can be expressed in this basis via\
\begin{align*}
  \mathbf{v} = v^\mu \p_\mu\,,
\end{align*}
where~$v^\mu$ are the $n$ components of~$\mathbf{v}$ in the
basis~$\{\p_\mu\}$. We use the Einstein summation convention
throughout, i.e. repeated indices imply summation. If we use a
different basis~$\bar{\p}_{\mu'}$ induced by
coordinates~$\bar{x}^{\mu'}$, then the components of~$\mathbf{v}$
between the two bases are related via
\begin{align*}
  \bar{v}^{\nu'} = v^\mu \frac{\p x^{\nu'}}{\p x^{\mu}}
  \,.
\end{align*}

A vector space~$V_p^*$ dual to~$V_p$ can be defined and the vectors
that live on it are called dual vectors. Given a vector~$\mathbf{v}$
and a dual vector~$\mathbf{w}$ there is a bracket
operation~$\langle \mathbf{v},\mathbf{w} \rangle$ that returns a
number. The elements~$dx^\mu$ can provide a coordinate basis for dual
vectors defined via
\begin{align*}
  \langle dx^\mu , \p_\nu \rangle = \delta^\mu {}_\nu
  \,,
\end{align*}
with~$\delta^\mu{}_\nu=1$ for~$\mu=\nu$ and~$0$ otherwise. A dual
vector~$\mathbf{w}$ can be expressed in this basis via
\begin{align*}
  \mathbf{w} = w_\mu dx^\mu
  \,,
\end{align*}
where~$w_\mu$ are its components in the basis~$\{ dx^\mu \}$ and obey the
transformation rule
\begin{align*}
  \bar{w}_{\nu'} = w_\mu \frac{\p x^\mu}{\p x^{\nu'}}
  \,,
\end{align*}
to a different basis~$\{ \bar{dx}^{\mu'} \}$ induced by the
coordinates~$\bar{x}^{\mu'}$.

A tensor~$T$ of rank~$(k,l)$ is a generalization of vectors and dual
vectors, which takes~$k$ dual vectors and~$l$ vectors and returns a
number. The bases~$\{ \p_\mu \}$ and~$\{ dx^\mu \}$ induced by the
coordinates~$x^\mu$ on~$\mathcal{M}$ provide a basis for tensors as
well, in which the components of~$T$ are written as
\begin{align*}
  T^{\mu_1...\mu_k}{}_{\nu_1...\nu_l}
  \,.
\end{align*}
For brevity we may say that these are the components of the tensor~$T$
on the basis~$x^\mu$, implying the tensor basis induced by the
coordinates. The components of the tensor~$T$ on the
basis~$\bar{x}^{\mu'}$ are related to those on~$x^\mu$ via the
following transformation rule:
\begin{align*}
  \bar{T}^{\mu_1'...\mu_k'}{}_{\nu_1'...\nu_l'}
  =
  T^{\mu_1...\mu_k}{}_{\nu_1...\nu_l}
  \, \frac{\p \bar{x}^{\mu_1'}}{\p x^{\mu_1}}
  ...
  \frac{\p \bar{x}^{\mu_k'}}{\p x^{\mu_k}}
  \frac{\p x^{\nu_1}}{\p \bar{x}^{\nu_1'}}
  ...
  \frac{\p x^{\nu_l}}{\p \bar{x}^{\nu_l'}}
  \,.
\end{align*}

A special tensor is the metric~$g$, a symmetric tensor of
rank~$(0,2)$. It allows us to measure the infinitesimal square
distance~$ds$ between two points on~$\mathcal{M}$ via
\begin{align*}
  ds^2 = g_{\mu\nu} dx^\mu dx^\nu
  \,,
\end{align*}
as well as define the inner product between two
vectors~$\mathbf{u},\mathbf{v}$ through
\begin{align*}
  \mathbf{u} \cdot \mathbf{v} \equiv g_{\mu \nu} u^\mu v^\nu
  \,.
\end{align*}
The metric also allows us to raise and lower indices of any tensor
e.g. given the components~$v^\mu$ of a vector~$\mathbf{v}$---which is
a tensor of rank~$(1,0)$---we can obtain the components~$v_\mu$ of its
dual vector
\begin{align*}
  v_\mu = g_{\mu \nu} v^\nu
  \,.
\end{align*}
Finally, given a metric we can always find a basis of~$V_p^*$ such
that~$g_{\mu\nu}=0$ if~$\mu \neq \nu$ and~$g_{\mu\nu}= \pm 1$
if~$\mu = \nu$. The number of~$+$ and~$-$ signs occurring in this
basis is called the signature of the metric. If there are no~$-$
signs, the metric is positive definite and is called Riemannian.
In GR the spacetime is understood as a manifold equipped with a metric
of signature~$(-,+,...,+)$~\footnote{It can also be~$(+,-,...,-)$ if
  another convention is adopted.}{}, which is called a Lorentzian
manifold.

To understand how tensors change on a manifold, the notion of
tensorial differentiation is needed. A covariant derivative~$\nabla$
takes a tensor or rank~$(k,l)$ to one of rank~$(k,l+1)$ and can be
written as
\begin{align*}
  \nabla_\lambda
  T^{\mu_1...\mu_k}{}_{\nu_1...\nu_l}
  =
  \p_\lambda T^{\mu_1...\mu_k}{}_{\nu_1...\nu_l}
  +
  \Gamma^{\mu_1} {}_{\lambda \sigma} T^{\sigma \cdots \mu_k}{}_{\nu_1...\nu_l}
  +
  \cdots
  -
  \Gamma^{\sigma} {}_{\lambda \nu_1 } T^{\mu_1...\mu_k}{}_{\sigma ...\nu_l}
  -
  \cdots
  \,,
\end{align*}
in the basis~$x^\mu$, where~$\Gamma^{\mu}{}_{\nu \sigma}$ is called
the connection and is symmetric in~$\nu, \sigma$. The connection does
not transform as a tensor and essentially provides a way to relate the
tensor bases of different points on the manifold. There are different
ways to define~$\nabla$, but in GR we often choose a metric compatible
covariant derivative, that is
\begin{align*}
  \nabla_\sigma g_{\mu \nu} = 0
  \,.
\end{align*}
Metric compatibility and the symmetry of the connection allow one to
fully determine it from the metric via
\begin{align*}
  \Gamma^\alpha{}_{\beta \gamma}
  = \frac{1}{2} g^{\alpha \delta}
  \left(
  \p_\gamma g_{\delta \beta}
  + \p_\beta g_{\delta \gamma}
  - \p_\delta g_{\beta \gamma}
  \right)
  \,.
\end{align*}
This connection is called the Christoffel symbol. The metric can also
help us build the Riemann curvature tensor, the components of which in
the~$x^\mu$ basis can be written as
\begin{align*}
  R^\alpha{}_{\beta \gamma \delta}
  =
  \p_\gamma \Gamma^\alpha{}_{\beta \delta}
  - \p_\delta \Gamma^\alpha{}{}_{\beta \gamma}
  + \Gamma^\alpha{}_{\lambda \gamma} \Gamma^\lambda{}_{\beta \delta}
  - \Gamma^\alpha{}_{\lambda \delta} \Gamma^\lambda{}_{\beta \gamma}
  \,.
\end{align*}
The Ricci tensor and scalar can be defined by the relations
\begin{align*}
  R_{\mu \nu} \equiv R^\lambda{}_{\mu \lambda \nu}
  \,, \qquad
  R \equiv g^{\mu \nu} R_{\mu \nu}
  \,,
\end{align*}
respectively. The EFE relate the curvature of the spacetime to its
matter content via
\begin{align*}
  G_{\mu \nu}
  + \Lambda g_{\mu \nu}
  = \kappa T_{\mu \nu}  \,,
\end{align*}
where
\begin{align*}
  G_{\mu \nu} \equiv R_{\mu \nu} - \frac{1}{2} R g_{\mu \nu}
  \,,
\end{align*}
is the Einstein tensor,~$\Lambda$ is the cosmological
constant,~$T_{\mu \nu}$ is the stress-energy tensor of matter,
and~$\kappa \equiv 8 \pi G c^{-4}$, with~$G$ the Newton's constant
and~$c$ the speed of light. The EFE can be derived as the
Euler-Lagrange equations of the action
\begin{align*}
  S = \int
  \left[
  \frac{1}{2 \kappa}
  \left(
  R - 2 \Lambda\right)
  + \mathcal{L}_M
  \right]
  \sqrt{-g} \, d^nx
  \,,
\end{align*}
by varying with respect to the inverse metric, where~$\mathcal{L}_M$
is the Lagrangian density for the matter content that vanishes in
vacuum.

\begin{figure}[!t]
  \begin{center}
    \includegraphics[width=0.77\textwidth]{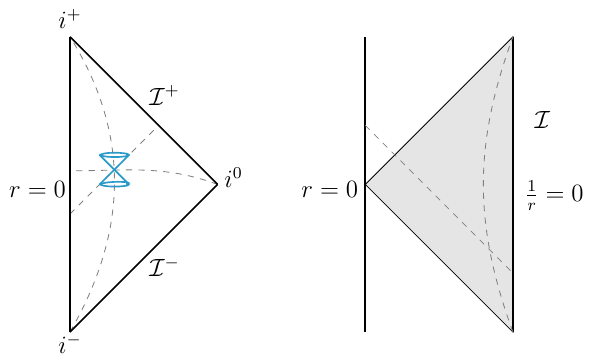}
  \end{center}
  \caption[Examples of conformal diagrams]
  {The conformal (or Carter-Penrose) diagrams for Minkowski (left),
    and AdS (right) spacetimes. Only two dimensions are depicted, the
    time and radius. If the spacetime is 4-dimensional and we use
    spherical polar coordinates, each point represents a two-sphere of
    radius~$r$. In asympotically flat spacetimes, there are the past
    and future temporal infinities, future and past null infinities
    and spatial infinity, denoted by
    ~$i^-$,~$i^+$,~$\mathcal{I}^-$,~$\mathcal{I}^+$ and~$i^0$,
    respectively. The causal structure on the conformal diagram is the
    same as the full spacetime it depicts and is illustrated by a
    light-cone on the left diagram. The asymptotic
    boundary~$\mathcal{I}$ of AdS is timelike. The shaded region is
    the Poincare patch of AdS.}
  \label{Fig:conf_diags_ex}
\end{figure}

The speed of light is constant for every observer in GR and is the
fastest speed at which a physical signal can travel. These properties
make trajectories of light and null hypersurfaces essential in
understanding the causal structure of the spacetime. This is a global
property of the spacetime and is better understood using conformal
diagrams, introduced by Penrose~\cite{Pen63}. Examples of such
diagrams are shown in Fig.~\ref{Fig:conf_diags_ex} for Minkowski
(left) and a region of AdS spacetime, called the Poincare patch (see
e.g.~\cite{Nat20} for more details).

Finally, we should mention that in this work we sometimes use the
abstract index notation, see for instance Sec. 2.4 of~\cite{Wal84a}
for details. Briefly, in this notation, a tensor~$T$ of type~$(k,l)$
is denoted as
\begin{align*}
  T^{a_1 \cdots a_k} {}_{b_a \cdots b_l}
  \,.
\end{align*}
The lowercase Latin indices here are reminders of the number and type
of variables the tensor acts on and not basis components. Using the
abstract index notation one can write true tensor equations that are
valid in any basis.

For the rest of the thesis we adopt the geometric unit convention
i.e.~$G=c=1$.

\section{Motivation}

There are several areas where characteristic formulations of GR have
advantages over more standard spacelike foliations. In the next few
paragraphs we attempt to provide a quick and non-exhaustive overview
of some, which is meant to serve mainly as a motivation for the work
of the thesis.

\subsection*{Precision gravitational wave astronomy}

Arguably, one of the major and most timely areas where characteristic
formulations of GR find use is that of accurate gravitational waveform
modeling. A waveform here typically refers to the GW signal detected
when a binary system of two compact objects inspirals, merges and
relaxes to a final compact object. Given the increasing sensitivity of
GW detectors---ground-based interferometers such as the advanced LIGO,
Virgo and Kagra~\cite{LIGOScientific:2014pky, VIRGO:2014yos,
  KAGRA:2020cvd}, future space-borne detectors like
TianQin~\cite{TianQin:2015yph}, Taiji~\cite{Ruan:2018tsw},
LISA~\cite{lisa17}, and the Einstein
Telescope~\cite{Maggiore:2019uih}---waveforms of high fidelity are
essential to maximize the discovery potential. Collections of these
waveforms form catalogues which are then compared against
observational data and consequently the fundamental properties of the
sources are inferred. These catalogues require high numbers of
different waveforms, and so finding economic but accurate techniques
to produce them is a topic of extensive research. Some of these
methods are the effective-one-body formalism~\cite{BuoDam98},
phenomenological waveforms~\cite{AjiBabChe07}, and numerical
relativity surrogates~\cite{FieGalHes13}. Nevertheless, numerical
relativity approximations are still necessary to a large extent,
e.g. to calibrate some of the aforementioned models.

In the modeling process, a GW detector is typically assumed to be
infinitely far away from the source. After emission, GWs propagate
towards future null infinity at the speed of light, where they can be
detected. Since characteristic formulations are based on null
hypersurfaces, future null infinity can be naturally included in the
computational domain. This is the region where quantities such as the
Bondi news function, that provides a way to determine the energy flux
of gravitational radiation, are unambiguously defined. Different
approaches can be exploited to compute such quantities at infinity
accurately. A common one is to solve the initial boundary value
problem (IBVP) for two compact object in GR using a spacelike
formulation, for a finite region of spacetime.
After solving the same IBVP, but placing the outer boundary of the
computational domain at different radii~$r_{\textrm{out}}$, an
extraction process that utilizes an~$1/r$ expansion can be used to
compute the GW signal at null infinity.
This is an extrapolation technique that allows us to understand the
signal at infinity by data in a finite region, but also introduces
systematic errors which contaminate the accuracy of the
waveform~\cite{BisRez16}.

\begin{figure}[!t]
  \begin{center}
    \includegraphics[width=0.5\textwidth]{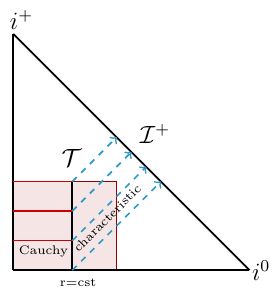}
  \end{center}
  \caption[CCE and CCM depiction]
  {Depiction of the Cauchy-Characteristic extraction and matching
    setups for an asymptotically flat spacetime. The Cauchy setup is
    used to model the region near the GW source and the characteristic
    one the propagation of GW to future null infinity. The
    worldtube~$\mathcal{T}$ is the boundary between the Cauchy and
    characteristic domains. In CCE, information flows only from the
    Cauchy to the characteristic domain, whereas in CCM it is
    communicated both ways, that is the Cauchy and characteristic
    problems are solved simultaneously.}
  \label{Fig:CCE_CCM_conf_diagram}
\end{figure}

To avoid such errors an alternative method has been proposed, which
comes by the name of~\textit{Cauchy-Characteristic Extraction} (CCE)
\cite{BisGomLeh96a, BisGomLeh97a, ZloGomHus03, GomBarFri07,
  ReiBisPol09, ReiBisPol09a, HanSzi14, BarMoxSch19, MoxSchTeu20,
  Ioz21, Mit21, IozBoyDep21, MitMoxSch21, Fou21, MoxSchTeu21}. The
term \textit{Cauchy} \footnote{Note the abuse of the
  term~\textit{Cauchy} here, since the data on this spacelike
  hypersurface cannot provide information for the whole spacetime and
  boundary data are necessary. To avoid possible confusion, we use the
  term~\textit{Cauchy-type} instead for such setups.}{} here refers to
the IBVP constructed using a more standard spacelike formulation.
Once a numerical approximation is obtained for the PDE problem, the
information is communicated to another domain where GR is formulated
using the characteristic approach. The latter is the characteristic
initial boundary value problem (CIBVP) of GR and the boundary is the
outer domain of the IBVP through which information from the spacelike
to the characteristic domain is communicated. In CCE the Cauchy
problem is solved first and the characteristic follows i.e. the
information through the boundary flows only from the Cauchy to the
characteristic domain. Even though CCE improves the waveform accuracy
and has even been used as a benchmark to evaluate the error from
extrapolation techniques, it is still prone to errors that arise from
artificial boundary conditions imposed during the
evolution. The~\textit{Cauchy-Characteristic Matching} (CCM) method
goes a step further and suggests a solution for these errors as
well~\cite{Szi00, Win12, BisRez16}. The way to do this is to solve the
Cauchy-type and the characteristic problems simultaneously and allow
for information to move via their common boundary in both ways, and
not just from the spacelike to the characteristic domain. Then, no
artificial boundary conditions need to be imposed for the IBVP. In
Fig.~\ref{Fig:CCE_CCM_conf_diagram} the CCE and CCM setups are
depicted for an asymptotically flat spacetime.
To the best of our knowledge, CCE has been implemented for GR
successfully in the sense that the relevant codes can run for large
timescales, but CCM has only done so for the wave
equation~\cite{KreWin11, BabKreWin14}.

For completeness, we should mention that characteristic formulations
are not the only strategy to include null infinity in the
computational domain. One alternative is the hyperboloidal approach,
which uses hypersurfaces that are everywhere spacelike but become null
only at infinity~\cite{SteFri82, MonRin08, Zen10, BarSarBuc11,
  VanHusHil14, HilHarBug16, AnsMac16, VanHus17,
  PanossoMacedo:2022fdi}. One advantage is that hyperboloids can be
used both in the near and far zones of the source. On the contrary,
null trajectories are expected to form caustics in the near zone,
which makes the use of the characteristic approach there
challenging. Another alternative is the conformal approach to the
EFE~\cite{Hub99, DouFra16, DouFraSte19}. Both strategies come with
their own advantages and challenges and are topics of active
research. We will not elaborate more on them however since they are
beyond the scope of this thesis.

\subsection*{Strongly coupled systems and numerical holography}

In addition to astrophysically relevant setups, numerical relativity
can provide insights into strongly coupled systems when combined with
holography. Holography roughly states that in certain limits a
classical gravitational theory can be mapped to a non-gravitational,
strongly coupled quantum field theory, that resides in one dimension
less~\cite{tHooft:1993dmi, Susskind:1994vu}. A particular realization
of this duality comes through the AdS/CFT conjecture, proposed by
Maldacena~\cite{Mal97}. This correspondence provides us with the
ability to explore the dynamics of strongly coupled quantum field
theory systems via the equations of motion of classical gravitational
ones.

Typically, the gravitational theory under consideration is GR, with or
without matter content. By solving the equations of motion on the
gravity side one hopes to obtain qualitative results and universal
relations of strongly coupled systems. One such example is the
universality of the viscosity-entropy ratio for strongly coupled
plasmas~\cite{Kovtun:2003wp}, which has been in good agreement with
experimental data~\cite{Casalderrey-Solana:2011dxg}. Characteristic
formulations of GR and numerical relativity tools are often used to
follow the out-of-equilibrium behavior of strongly coupled plasmas
that can be formed in terrestrial heavy ion
collisions~\cite{Casalderrey-Solana:2011dxg}. They are also believed
to exist in the early universe and undergo phase transitions and
turbulent flows that can produce GWs~\cite{Hindmarsh:2020hop}, which
may be detected by future detectors like
LISA~\cite{Caprini:2019egz}. This line of research has also provided
insights into the applicability of hydrodynamics in describing the
near and out-of-equilibrium processes of these systems, which can be
helpful for improving relevant fluid dynamics
codes~\cite{DelZanna:2013eua, Nonaka:2006yn,
  Becattini:2015ska}. Cauchy-type setups are also used in holographic
studies and can model e.g. confinement-deconfinement transitions in
these plasmas~\cite{BanFigMat20, BanFigRos21}, but are not discussed
in this thesis.

\subsection*{Mathematical relativity and more}

Characteristic formulations have found extensive use in mathematical
relativity as well. For example, in~\cite{Mos17} they were used to
show the instability of the AdS spacetime in spherical symmetry with
reflective boundary conditions at infinity, in~\cite{Chr91} to study
the formation conditions for black holes and singularities, and
in~\cite{Dafermos:2021cbw, Aretakis:2011ha, Aretakis:2012ei} to
understand the stability properties of black hole spacetimes. In the
study of gravitational collapse, codes based on null foliations offer
a practical alternative to the standard spacelike foliation
approach. Their advantage lies in the compactness of the system of
PDEs solved~\cite{Gar95, CreOliWin19, GunBauHil19, SieFonMue03,
  AlcBarOli21} as well as the inclusion of null infinity in the
computational domain. The aforementioned setups are usually considered
in asymptotically flat geometries; though see~\cite{OliSop16, Oli17b}
for gravitational collapse in asymptotically AdS (AAdS) spacetimes.

Characteristic setups have also been efficient in studying isolated
gravitational objects, as e.g. relativistic stars~\cite{PapFon99,
  SieFonMul02}. In fact, the first stable evolution of isolated black
holes has been achieved in characteristic
formulations~\cite{GomLehMar98, GomLehMar98a}. They have also been
utilized in the study of the superradiant instability in AAdS
spacetimes~\cite{Che21} and even in explorations of cosmological
scenarios~\cite{vdWBis12}. Finally, they play an important role in the
fluid/gravity correspondence~\cite{Bhattacharyya:2007vjd,
  Rangamani:2009xk, Hubeny:2011hd}.

\section{Thesis outline and main results}

In this thesis we focus on characteristic formulations of GR and their
applications, and adopt the standpoint of numerical relativity. The
thesis is divided in three parts, with the first and second composing
the main body, while the third being appendix material.

In Part I we analyze the hyperbolicity and well-posedness of the
vacuum EFE in Bondi-like characteristic coordinates.
The characteristic problem of GR is commonly formulated using this
type of coordinates (or gauges).
%
%
Since the central interest of this part
is hyperbolicity and well-posedness, these notions are introduced
briefly in Chap.~\ref{chap:pde_theory} together with other necessary
textbook material of PDE analysis. The main properties of Bondi-like
characteristic formulations are presented in
Chap.~\ref{chap:bondi_properties}, along with the mapping of the
associated PDEs to the Arnowitt-Deser-Misner (ADM) formulation of
GR. The latter is a tool we use to understand the pure gauge structure
of Bondi-like PDEs, the notion of which was introduced
in~\cite{HilRic13} and we briefly present in Chap.~\ref{chap:PG}. The
material presented in these three chapters is employed in an extensive
hyperbolicity analysis of some common Bondi-like systems, which is the
topic of Chap.~\ref{chap:bondi-hyp}. More specifically, we analyse the
vacuum EFE in the affine null, the Bondi-Sachs proper and the double
null coordinates and find that all these systems are only weakly
hyperbolic (WH). We identify the pure gauge structure of the angular
sector of the systems as the root cause of weak hyperbolicity. We
further conjecture that PDE systems which result from the EFE and
include up to second order metric derivatives are at most WH beyond
spherical symmetry, if they are constructed in a Bondi-like gauge.

In light of this result we investigate the implications to the
well-posedness of the CIBVP of GR in Bondi-like gauges, in
Chap.~\ref{chap:bondi_well-p}. We start by introducing toy models and
determine the conditions and norms in which their CIBVP is
well-posed. We find that the CIBVP of the strongly hyperbolic (SH) toy
model is well-posed in a version of the common~$L^2$-norm adapted to
the characteristic setup, as expected. The WH model that mimics the
structure of Bondi-like systems is weakly well-posed in a lopsided
norm that is not equivalent to~$L^2$. However, this version of
well-posedness is weaker than that of the SH system, since it is
affected by lower order source terms. We identify the structure of the
source terms that breaks this weak well-posedness. Based on the
aforementioned systems, we examine well-posedness of model CCE and CCM
setups. The former is weakly well-posed for the WH toy model, whereas
the latter is ill-posed when the Cauchy-type setup is formulated with
the SH model and the characteristic with the WH one. The associated
IBVP and CIBVP for this CCM model are separately well-posed in norms
that are incompatible with each other and hence the composite problem
is ill-posed.

Returning our attention to GR in Bondi-like gauges, we know from
textbook results that the CIBVP of GR in these gauges is ill-posed in
the~$L^2$-norm. We thus explore alternative norms for the CIBVP of the
Bondi-Sachs system linearized about flat space. We fail to find such a
norm and pinpoint the structure of the system that leads to this
shortcoming. Nevertheless, motivated by symmetric hyperbolic PDE
systems in Bondi-like gauges which include higher than second order
metric derivatives, we investigate norms for the well-posedness of
their CIBVP. These systems do not fall into the class covered by our
earlier conjecture and we expect a deeper understanding of their CIBVP
to guide us in building appropriate lopsided norms for the CIBVP of
the Bondi-like systems that are covered by our conjecture. This
process is work in progress and we hope to report further results
elsewhere. A possible existence of such a norm would have great
implications for applications of characteristic formulations built
upon Bondi-like gauges, in the case where only the characteristic
setup is solved, i.e. CCE. Regarding precise GW modeling, it could
help us validate the error estimates of waveforms produced via CCE.

Using discrete approximates of the aforementioned norms we demonstrate
the effects of weak hyperbolicity in numerical experiments performed
in the characteristic domain, in Chap.~\ref{chap:numerics}. We adapt
well known robust stability tests to the CIBVP for both toy models and
the Bondi-Sachs proper system. We find that noisy given data are
necessary to identify weak hyperbolicity in practice, as well as
employing norms that are suitable for the specific problem.

In principle, the aforementioned well-posedness result has
implications to any numerical approximation produced with these
characteristic setups, like those related to strongly coupled systems
which is the topic of Part II. We recognize this fact, but due to the
lack of a better alternative at the moment, we exploit these WH
characteristic setups to explore strongly coupled systems via
holography. With this strategy, our main goal is to advance our
understanding of their qualitative behavior, rather than describe
them precisely. To pursue this line of research, we have
developed~\texttt{Jecco} a new modular characteristic code written in
the Julia programming language. In Chap.~\ref{chap:jecco} we provide
details on the models that can be explored with~\texttt{Jecco} and the
PDE systems for which the code currently provides solutions, as well
as information on the numerical implementation and algorithms. We also
present performance, validation and convergence tests of the code, and
provide a brief overview of some physical setups that we have modeled
with~\texttt{Jecco} so far. In Chap.~\ref{chap:final_remarks} we
present our final remarks and suggestions for future work.
  

%% file: sections/PDE_theory.tex
\chapter{A PDE theory toolbox}
\label{chap:pde_theory}

\minitoc

The goal of this chapter is to provide the reader with the necessary
knowledge of PDE theory to understand the results presented in the
thesis. By the end of the chapter the difference between
a~\textit{weakly hyperbolic} and a~\textit{strongly hyperbolic} PDE
system should be clear, as well as the notions of
the~\textit{characteristic initial (boundary) value PDE problem} and
of~\textit{well-posedness} of a PDE problem.
A more complete discussion can be found e.g. in ~\cite{KreLor89,
  GusKreOli95, Val16, Hil13}. The reader familiar with these topics
may prefer to skip the chapter.

\section{PDE classes and causal definitions}
\label{sec:pde_theory:causal_defs}

A rough way to classify PDE systems is into \textit{elliptic},
\textit{parabolic} and \textit{hyperbolic}. From the physical
perspective, elliptic systems have no intrinsic notion of time, with
the prototype elliptic example being the Laplace equation. Both
parabolic and hyperbolic PDEs have an intrinsic notion of
time. However, in parabolic systems the propagation of information has
infinite speed, whereas in hyperbolic ones it is finite and so there
is a clear notion of causality. The heat equation is the prototype
parabolic example and the wave equation the model hyperbolic one. A
system of PDEs may fall into more than one of the aforementioned
classes e.g. mixed hyperbolic-parabolic. There are also examples of
PDEs that can change classification depending on their coefficients as
for example the Tricomi equation
\begin{align*}
  \p_x^2 u(x,y) + x \p_y^2 u(x,y) = 0
  \,,
\end{align*}
which is elliptic for~$x>0$ and hyperbolic for~$x<0$.

In this thesis we focus on hyperbolic PDE systems of the form
\begin{align}
  \mathbfcal{A}^t(\mathbf{u},x^\mu) \, \p_t \mathbf{u}
  + \mathbfcal{A}^p(\mathbf{u},x^\mu)\,
  \p_p \mathbf{u} +  \mathbfcal{S}(\mathbf{u},x^\mu) = 0
  \,,
  \label{eqn:gen_hyper_PDE}
\end{align}
where~$\mathbf{u} = (u_1, u_2, \dots, u_q )^T \, ,$ is the state
vector of the system,
\begin{align*}
  \mathbfcal{A}^\mu =
  \begin{pmatrix}
    a^{\mu}_{11}  & \dots  & a^{\mu}_{1q} \\
    \vdots & \ddots & \vdots \\
    a^{\mu}_{q1}  & \dots  & a^{\mu}_{qq} 
  \end{pmatrix}
\end{align*}
denotes the principal part matrices,~$\mathbfcal{S}(\mathbf{u},x^\mu)$
the source terms and~$x^\mu = (t,x^p)$ some local coordinate system.
The causal nature of a hypersurface~$\Sigma_t$ of constant~$t$ is
\begin{itemize}
\item Spacelike
  if~$\det \left( \mathbfcal{A}^t(\mathbf{u}, x^\mu) \right) \neq 0$
  and~$ \mathbfcal{A}^t(\mathbf{u}, x^\mu)$ is positive-definite
  \footnote{A real valued symmetric matrix is positive-definite if all
    its eigenvalues are real and positive.}{}.
\item Timelike
  if~$\det \left( \mathbfcal{A}^t(\mathbf{u}, x^\mu) \right) \neq 0$
  and~$ \mathbfcal{A}^t(\mathbf{u}, x^\mu)$ is not positive-definite.
\item Null or characteristic
  if~$\det \left( \mathbfcal{A}^t(\mathbf{u}, x^\mu) \right) = 0$.
\end{itemize}
If the PDE system describes GR then the vector normal to a
characteristic hypersurface is also null in the GR notion. In terms of
initial data~$\mathbf{u}_0 \equiv \mathbf{u}(0,x^p)$, the initial
value problem (IVP) and the characteristic initial value problem
(CIVP) have an important difference:
\paragraph*{If~$\Sigma_0$ is spacelike:} All~$q$ elements of the initial
state vector~$\mathbf{u}_0$ have to be provided. Since~$\mathbf{u}_0$
is known, then so is~$\p_p \mathbf{u}_0$ and
\begin{align}
  \mathbfcal{A}^t (0,x^p,\mathbf{u}_0) \, \partial_t \mathbf{u}_0 + 
  \mathbfcal{A}^p (0,x^p,\mathbf{u}_0) \, \partial_p \mathbf{u}_0 
  =
  -\mathbfcal{S} (0,x^p, \mathbf{u}_0)
  \, , 
\end{align}
provides a system of~$q$ PDEs for the~$q$
unknowns~$( \partial_0 \mathbf{u}_0 )$, that lead to a time-dependent
solution $\mathbf{u}$ for the system.

\paragraph*{If~$\Sigma_0$ is characteristic:} Not all~$q$ elements
of~$\mathbf{u}_0$ are freely specifiable on~$\Sigma_0$ and some of
them satisfy constraint equations on it.
For~$m \equiv \textrm{rank}\left(\mathbfcal{A}^t(x^\mu,\mathbf{u})
\right)$ with~$m < q$, there are only~$m$ linearly independent rows
in~$\mathbfcal{A}^t$ and the remaining~$q-m$ rows can be eliminated
resulting in a system of the from
\begin{align}
  \Bar{\mathbfcal{A}}^t (0,x^p,\mathbf{u}_0) \, \partial_t \mathbf{u}_0 + 
  \Bar{\mathbfcal{A}}^p (0,x^p,\mathbf{u}_0) \, \partial_p \mathbf{u}_0 
  =
  -\Bar{\mathbfcal{S}} (0,x^p, \mathbf{u}_0)
  \, ,
  \label{eqn:row_reduced_system}
\end{align}
with
\begin{align*}
  \Bar{\mathbfcal{A}}^t
  &
    = \begin{pmatrix}
      a^t_{11}(0, x^p, \mathbf{u}_0) & \cdots & a^t_{1q}(0, x^p, \mathbf{u}_0) \\
      \vdots & \ddots & \vdots \\
      a^t_{m1}(0, x^p, \mathbf{u}_0) & \cdots & a^t_{mq}(0, x^p, \mathbf{u}_0)\\
      0 & \cdots & 0 \\
      \vdots & \ddots & \vdots \\
      0 & \cdots & 0
    \end{pmatrix}
                   \, ,
                   \quad
  \Bar{\mathbfcal{S}} 
  =
    \begin{pmatrix}             
      S_1(0, x^p, \mathbf{u}_0) \\
      \vdots \\
      S_m(0, x^p, \mathbf{u}_0) \\
      S_{m+1}(0, x^p, \mathbf{u}_0)\\
      \vdots \\
      S_q(0, x^p, \mathbf{u}_0)
    \end{pmatrix}
  \,.
\end{align*}
Since there are~$q-m$ zero rows in~$\bar{\mathbfcal{A}}^t$, then there
are~$q-m$ equations that do not involve derivatives transversal
to~$\Sigma_0$, the~\textit{intrinsic equations}. The state vector can
be split in transversal and intrinsic to~$\Sigma_0$
components~$\mathbf{u}_0 = \left( \mathbf{u}_0^\textrm{tr},
  \mathbf{u}_0^\textrm{int} \right)^T$. The same decomposition holds for the
principal matrices as well as the source terms, namely:
\begin{align*}
  \Bar{\mathbfcal{A}}^{\mu} (0,x^p, \mathbf{u}_0) = 
  \begin{pmatrix}
    \Bar{\mathbfcal{A}}^{\mu}_\textrm{tr} (0,x^p, \mathbf{u}_0)
    \\
    \Bar{\mathbfcal{A}}^{\mu}_\textrm{int} (0,x^p, \mathbf{u}_0)
  \end{pmatrix}
  \, , \qquad
  \Bar{\mathbfcal{S}} (0,x^p, \mathbf{u}_0) = 
  \begin{pmatrix}
    \Bar{\mathbfcal{S}}_\textrm{tr} (0,x^p, \mathbf{u}_0)
    \\
    \Bar{\mathbfcal{S}}_\textrm{int} (0,x^p, \mathbf{u}_0)
  \end{pmatrix} \, , 
\end{align*}
with
\begin{align*}
  \textrm{size}
  \left( \Bar{\mathbfcal{A}}^{\mu}_\textrm{tr} \right)
  &= m \times q \, ,
  \quad \,
  \textrm{size} \left( \Bar{\mathbfcal{A}}^{\mu}_\textrm{int} \right) = (q-m) \times q \, ,
  \nonumber
  \\
  \textrm{length}
  \left( \Bar{\mathbfcal{S}}_\textrm{tr} \right)
  &= m \, ,
  \qquad 
  \textrm{length} \left( \Bar{\mathbfcal{S}}_\textrm{int} \right) = (q-m) \, .
\end{align*}
From the system of intrinsic equations
\begin{align}
  \label{eqn:intrinsic_eqs_formal}
  \Bar{\mathbfcal{A}}^p_\textrm{int} (0,x^p,\mathbf{u}_0) \, \partial_p \mathbf{u}_0 
  =
  -\Bar{\mathbfcal{S}}_\textrm{int} (0,x^p, \mathbf{u}_0)
  \, ,
\end{align}
let us analyze one equation (the analysis of the rest is identical):
\begin{align} 
  & 
    a^p_{q1} \,
    \p_p u_{01}^\textrm{tr} 
    + \dots +
    a^p_{qm} \,
    \p_p u_{0m}^\textrm{tr} 
    + \,
    a^p_{q (m+1)} \,
    \p_p u_{0 m+1}^\textrm{int} 
      + \dots +
    a^p_{qq} \,
    \p_p u_{0 q}^\textrm{int} 
    =
    -S_{q}
      \label{eqn:intrinsic_eq_expand} \, ,
\end{align}
where
\begin{align*}
  a^p_{qj} = a^p_{qj}(0,x^p,\mathbf{u}_0^\textrm{tr},\mathbf{u}_0^\textrm{int}) \,,
  \quad
  S_{q} = S_{q}(0,x^p,\mathbf{u}_0^\textrm{tr},\mathbf{u}_0^\textrm{int}) \,.
\end{align*}
This is a constraint equation that the initial data have to satisfy
on~$\Sigma_0$. In other words, the behavior of the chosen
data~$\mathbf{u}_0$ that is intrinsic to the initial hypersurface has to
comply with the above equation and all the rest of the constraint
(intrinsic) equations. Note that if~$m=0$, then the hypersurface is
a~\textit{total characteristic} of the system and there are $q$
constraint equations i.e. no element of~$\mathbf{u}_0$ is freely
specifiable.

Regarding the transversal part of the
system~\eqref{eqn:row_reduced_system}
\begin{align} \label{eqn:transversal_eqs_formal}
  \Bar{\mathbfcal{A}}^{0}_\textrm{tr} (0,x^p,\mathbf{u}_0) \, \partial_t \mathbf{u}_0 
  +
  \Bar{\mathbfcal{A}}^p_\textrm{tr} (0,x^p,\mathbf{u}_0) \, \partial_p \mathbf{u}_0 
  =
  -\Bar{\mathbfcal{S}}_\textrm{tr} (0,x^p, \mathbf{u}_0)
  \, ,
\end{align}
let us analyze one equation (the rest are identical):
\begin{align} 
  &
    a^t_{11} \,
    \p_t u_{01}^\textrm{tr} 
    + \dots +
    a^t_{1m} \,
    \p_t u_{0m}^\textrm{tr} 
    + \,
    a^t_{1 (m+1)} \,
    \p_t u_{0 m+1}^\textrm{int} 
    + \dots +
    a^t_{1q} \,
    \p_t u_{0 q}^\textrm{int} 
    +
    \nonumber \\
  &
    a^p_{11} \,
    \p_p u_{0 1}^\textrm{tr}
    + \dots +
    a^p_{1m} \,
    \p_p u_{0 m}^\textrm{tr} 
    + \,
    a^p_{1 (m+1)} \,
    \p_p u_{0 m+1}^\textrm{int} 
    + \dots +
    a^p_{1q} \,
    \p_p u_{0 q}^\textrm{int}
    = -S_1
    \label{eqn:transversal_eq_expanded} \, ,
\end{align}
where
\begin{align*}
  a^{\mu}_{1j} = a^{\mu}_{1j}(0,x^p,\mathbf{u}_0^\textrm{tr},\mathbf{u}_0^\textrm{int}) \,,
  \quad
  S_{1} = S_{1}(0,x^p,\mathbf{u}_0^\textrm{tr},\mathbf{u}_0^\textrm{int}) \,.
\end{align*}
The quantities $a^{\mu}_{1j}$, $\mathbf{u}_0$
and~$\partial_p \mathbf{u}_0$ are known for~$t=0$ provided that the
intrinsic equations are satisfied. However, there are the~$q$
unknowns~$(\p_t \mathbf{u}^\textrm{tr}_0, \p_t \mathbf{u}^\textrm{int}_0)$ for
the~$m$ transversal equations (with~$m<q$). One can formally build a
system of~$q$ equations for these~$q$ unknowns from the intrinsic
equations. Acting from the left with~$\partial_t$
on~\eqref{eqn:intrinsic_eq_expand} and permuting the derivatives, one
can obtain:
\begin{align*} 
  &
    \p_t a^p_{q1} \,
    \p_p u_{0 1}^\textrm{tr}
    + \dots +
    \p_t a^p_{qm} \,
    \p_p u_{0 m}^\textrm{tr}
    + \,
    \p_t a^p_{q (m+1)} \,
    \p_p u_{0 m+1}^\textrm{int} 
    + \dots +
    \p_t a^p_{qq} \,
    \p_p u_{0q}^\textrm{int} 
    +
    \nonumber \\
  &
    a^p_{q1} \,
    \p_p \p_t u_{01}^\textrm{tr} 
    + \dots +
    a^p_{qm} \,
    \p_p \p_t u_{0m}^\textrm{tr} 
    + \,
    a^p_{q (m+1)} \,
    \p_p \p_t u_{0 m+1}^\textrm{int} 
    + \dots +
    a^p_{qq} \,
    \p_p \p_t u_{0q}^\textrm{int} 
    =
    - \p_t S_q
    \, ,
\end{align*}
where
\begin{align*}
  \p_t a^p_{q j} = \p_t a^p_{q j}(0,x^p,
  \mathbf{u}_0^\textrm{tr}, \mathbf{u}_0^\textrm{int} ,
  \p_t \mathbf{u}_0^\textrm{tr},\p_t \mathbf{u}_0^\textrm{int})
  \, , \quad
  \p_t S_{q} = \p_t S_{q}(0,x^p, \mathbf{u}_0^\textrm{tr},
  \mathbf{u}_0^\textrm{int} ,\p_t \mathbf{u}_0^\textrm{tr},
  \p_t \mathbf{u}_0^\textrm{int})
  \,.
\end{align*}
By demanding that the constraint equations are satisfied at later
times i.e. they are solved
by~$\mathbf{u}_0 + \partial_t \mathbf{u}_0$, one ends up with~$q-m$
equations of the form:
\begin{align} 
  \p_t a^p_{q1} \,
  \p_p u_{01}^\textrm{tr} 
  + \dots +
  \p_t a^p_{qm} \,
  \p_p u_{0m}^\textrm{tr} 
  + \,
  \p_t a^p_{q (m+1)} \,
  \p_p u_{0 m+1}^\textrm{int} 
  + \dots +
  \p_t a^p_{qq} \,
  \p_p u_{0q}^\textrm{int} 
  = 0
  \, ,
  \label{eqn:intrinsic_partial0_eq_expand_final}
\end{align}
where~$(\p_p \mathbf{u}^\textrm{tr}_0,\p_p \mathbf{u}^\textrm{int}_0 )$ are known
from the given data on~$\Sigma_0$. The~$q$
unknowns~$( \p_t \mathbf{u}_0^\textrm{tr},\p_t \mathbf{u}_0^\textrm{int} )$ still
appear in the terms~$\p_t a^p_{qj}$ and so the union of the~$q-m$
equations of the form~\eqref{eqn:intrinsic_partial0_eq_expand_final} with
the~$m$ transversal equations of the
form~\eqref{eqn:transversal_eq_expanded} provides with~$q$ equations for
these unknowns, while simultaneously assuring that the constraint
equations are satisfied at later times.

\section{Degree of hyperbolicity}
\label{sec:pde_theory:hyp_degree}

Within the hyperbolic class of PDEs there are sub-classes characterized
by their degree of hyperbolicity. The standard way to perform this
characterization is by constructing the principal symbol
\begin{align}
  \mathbf{P}^s = \left(\mathbfcal{A}^t \right)^{-1}
  \mathbfcal{A}^p \, s_p
  \,,
  \label{eqn:principal_symbol}
\end{align}
where~$s^i$ is an arbitrary unit spatial vector. Here, the explicit
dependence of the principal part matrices on~$\mathbf{u}$ and~$x^\mu$
is suppressed. To characterize a PDE system with variable
coefficients~$\mathbf{P}^s$ has to be constructed everywhere in the
domain of interest. To build~$\mathbf{P}^s$ the principal part
matrix~$\mathbfcal{A}^t$ associated with time derivatives has to be
invertible i.e.~$\det(\mathbfcal{A}^t) \neq 0$.

We may refer to a PDE system with~$\det(\mathbfcal{A}^t)=0$ as a
\textit{characteristic PDE system}. For such a system,~$\mathbf{P}^s$
cannot be constructed directly in the chosen coordinates. In this case
a convenient coordinate transformation is utilized to write the system
with an invertible time principal part matrix. Crucially, this
transformation does not alter the degree of hyperbolicity of the
system, but is merely a tool to form~$\mathbf{P}^s$.

If~$\mathbf{P}^s$ has real eigenvalues for all~$s^i$, then the PDE
system is called \textit{weakly hyperbolic}, whereas if in
addition~$\mathbf{P}^s$ is diagonalizable for all~$s^i$, and there
exists a constant~$K$ independent of~$s^i$ such that
\begin{align*}
|\mathbf{T}_s|+|\mathbf{T}_s^{-1}|\leq K,
\end{align*}
with~$\mathbf{T}_s$ the similarity matrix that
diagonalizes~$\mathbf{P}^s$, it is called \textit{strongly
  hyperbolic}. If all eigenvalues of~$\mathbf{P}^s$ are distinct for
all~$s^i$ then the system is called~\textit{strictly
  hyperbolic}. Strict hyperbolicity implies strong
hyperbolicity. Finally, a PDE system is called~\textit{symmetric
  hyperbolic} if all the matrices~$\mathbfcal{A}^\mu$ are Hermitian,
or symmetric for purely real-valued setups. If not all
the~$\mathbfcal{A}^\mu$ matrices are Hermitian (symmetric) in their
original form, but there exists one matrix~$\mathbf{H}$ which can
symmetrize them all via a similarity transformation, then the system
is still symmetric hyperbolic.

\section{Well-posedness and norms}
\label{sec:pde_theory:well-posedness}

\textit{Well-posedness} is a property of a PDE problem which states
that the PDE problem has a unique solution that depends continuously
on the given data, in some appropriate norm. The PDE problem consists
of the PDE system, the domain in which we seek a solution, as well as
the given data (initial and possibly boundary data). The degree of
hyperbolicity of the PDE system is tightly connected to
well-posedness. More specifically, assuming there exists a unique
solution of the problem, the degree of hyperbolicity of the system
affects the existence and the form of the norm in which the problem
can be well-posed.

Consider the Cauchy problem for the linear, constant coefficient
system,
\begin{align}
  \p_t\mathbf{u}
     =
     \mathbfcal{B}^p\p_p\mathbf{u} + \mathbfcal{B} \mathbf{u}
     \,.
    \label{eqn:prototype_1st_order_lin_const_coef_PDE}
\end{align}
The initial value problem (IVP) for a SH system is well-posed in
the~$L^2$-norm
\begin{align}
  ||\mathbf{u}||_{L^2} =
  \left( \int_{\Sigma_t} \mathbf{u}^\dag \mathbf{u} \right)^{1/2}
  \,,
\end{align}
where~$\int_{\Sigma_t}$ denotes the integral over a spacelike
hypersurface~$\Sigma_t$. To be well-posed in the~$L^2$-norm means that
there exist real constants~$K\geq1$ and~$\alpha \in \mathbb{R}$ such
that
\begin{align}
  | e^{\mathbf{P}(i \omega) t}|
  \leq K e^{\alpha t}
  \, ,
  \label{eqn:wp_inequality}
\end{align}
for all~$t\geq0$ and all~$\omega \in \mathbb{R}^n$. Here
\begin{align}
  \mathbf{P}(i \omega) =  i \omega_p \mathbfcal{B}^p + \mathbfcal{B}
  \label{eqn:symbol_P_Fourier}
\end{align}
is the constant-coefficient symbol of the PDE after Fourier
transforming in space, with~$i\omega_p\mathbfcal{B}^p$ the principal
symbol and~$\mathbfcal{B} \, \mathbf{u} = -\mathbfcal{S}$ the lower
order term related to sources. Essentially,
inequality~\eqref{eqn:wp_inequality} states that the solution of the
PDE has to be bounded at each time by an exponential that is
independent of the Fourier mode~$\omega_p$. In this manner one can
obtain an estimate of the solution~$\mathbf{u}$ at all times by the
initial data~$f$
\begin{align*}
  || \mathbf{u}( \cdot \, ,t) ||_{L^2}
  = || e^{\mathbf{P} (i \omega) t } \hat{f}(\omega) ||_{L^2}
  \leq
  K e^{\alpha t} || \hat{f} ||_{L^2}
  = K e^{\alpha t} ||f||_{L^2}\, .
\end{align*}
Crucially, the form of the source terms does not affect well-posedness
for a SH system~\cite{SarTig12, KreLor89}. This result provides the
basis to show well-posedness for the IVP of variable-coefficient SH
systems, as well as non-linear systems with a SH linearization.

In the terminology of~\cite{KreLor89}, if a Cauchy problem instead
satisfies only
\begin{align}
  | e^{\mathbf{P}(i \omega) t}|
  \leq K_1 e^{\alpha t}  \left(1 + | \omega|^q  \right)\, ,
  \label{eqn:weak_wp_inequality}
\end{align}
with~$q$ some natural number, it is called weakly well-posed. This
type of estimate is weaker than~\eqref{eqn:wp_inequality}, because the
explicit appearance of~$\omega$ on the right-hand-side makes it
impossible to bound the solution by an exponential independent
of~$\omega$. If, rather than insisting on~$L^2$ we allow also some
{\it specific} derivative, determined by the system, within the norm,
we can nevertheless obtain the estimate
\begin{align*}
  || \mathbf{u}( \cdot \, ,t) ||_q \leq K_2 \, e ^{\alpha t} ||f||_q \, ,
\end{align*}
for the solution~$\mathbf{u}$. This would not be terrible, except that
if the PDE is only weakly well-posed, then perturbations to the system
by generic lower order terms can lead to frequency dependent
exponential growth of the solution, that
is~$|e^{\mathbf{P}(i \omega) t}|$ grows faster than any polynomial
in~$|\omega|$, and the resulting perturbed problem is ill-posed in any
sense. In Sec.~\ref{sec:bondi_well-p:algebraic_char} we show this
explicitly for our WH models. More examples can be found in subsection
2.2.3 of~\cite{KreLor89} and Example 10
of~\cite{SarTig12}.

Practically, to understand if a weakly well-posed PDE problem becomes
ill-posed due to lower order perturbations, one can focus on the
large~$\omega$ behavior of the eigenvalues of~$\mathbf{P}(i
\omega)$. If there is an eigenvalue with positive real part in this
limit, then it gives rise to solutions that grow exponentially
with~$\omega$, for fixed~$t$. This becomes more clear when considering
how the matrix norm~$|e^{\mathbf{P}(i \omega)t}|$ can be computed. For
completeness, we briefly present a way to perform this computation, as
given in~\cite{SarTig12}. Let us denote as~$\mathbf{M}^*$ the
transposed and complex conjugate of a~$k \times l$
matrix~$\mathbf{M}$. Then, the matrix norm~$|\mathbf{M}|$ can be
computed as
\begin{align*}
  |\mathbf{M}| =
  \sqrt{
  \rho
  \left(\mathbf{M}^* \mathbf{M}\right)
  }
  \,,
\end{align*}
where~$\rho \left(\mathbf{M}^* \mathbf{M} \right)$ is the spectral
radius of the square matrix~$\mathbf{M}^* \mathbf{M}$. The spectrum of
this matrix is the set of all its eigenvalues and the spectral radius
is their greatest absolute value.


%% file: sections/Bondi-like_properties.tex
\chapter{Properties of Bondi-like characteristic
  formulations}
\label{chap:bondi_properties}

\minitoc

In this chapter we review the main features of Bondi-like formulations
and map the corresponding equations and variables to the ADM
language. To the best of our knowledge, such a mapping has been
performed only in spherical symmetry so far~\cite{Fri06}.
To achieve this map, we employ a coordinate transformation between
generalized Bondi-like coordinates and coordinates adapted to the ADM
setup. We may refer to the latter as the ADM coordinates. The gauge,
and thus the PDE character of the system are fixed by the Bondi-like
coordinates. This choice determines for instance which metric
components and/or derivatives thereof vanish. The subsequent
transformation to the ADM coordinates merely results in relabeling
variables and expressing directional derivatives of the Bondi-like
basis in terms of those of the ADM basis. A more geometric description
of the main properties of Bondi-like formulations is provided in terms
of coordinate light speeds in
Sec.~\ref{sec:bondi_properties:coord_lightspeeds}.

\section{Main features of Bondi-like formulations}
\label{sec:bondi_properties:main_features}

To demonstrate relevant features common to all Bondi-like gauges we
work with the generalized Bondi-Sachs formulation of~\cite{CaoHe13}
with line element
\begin{equation}
  \begin{aligned}
    ds^2
    & = g_{uu} du^2 + 2 g_{ur} du\,dr
    + 2 g_{u \theta} du \, d\theta
    + 2 g_{u \phi} du \, d\phi
    + g_{\theta \theta} d\theta^2
    + 2 g_{\theta \phi} d\theta \, d\phi
    + g_{\phi \phi} d \phi^2
    \,.
  \end{aligned}
  \label{eqn:gen_BS_line_element}
\end{equation}
We consider a four dimensional spacetime and identify the
coordinates~$\theta,\phi$ with the usual spherical polar angles on the
two-sphere. All seven nontrivial metric components
of~\eqref{eqn:gen_BS_line_element} are functions of the characteristic
coordinates~$x^{\mu'}=(u,r,\theta,\phi)$, with the hypersurfaces of
constant~$u$ null and henceforth denoted by~$\mathcal{N}_u$. The null
vector~$(\p/\p r)^a$ is both tangent and normal to~$\mathcal{N}_u$ and
hence orthogonal to the spatial vectors~$(\p/\p \theta)^a$
and~$(\p/\p \phi)^a$ that lie within~$\mathcal{N}_u$. This vector
basis guarantees that
\begin{align}
  g^{uu} = g^{u\theta} = g^{u\phi} = 0
  \,,
  \label{eqn:gauge_cond_null}
\end{align}
and every distinct null geodesic in~$\mathcal{N}_u$ can be labeled
by~$\theta,\phi$. The characteristic hypersurface~$\mathcal{N}_u$ can
be either outgoing or ingoing. If the formulation incorporates both
types of null hypersurfaces, then the double null gauge~\cite{Chr08}
is imposed. In this case~$g^{rr}=0$ and the coordinates~$u,r$
correspond to the advanced and retarded time rather than an advanced
(or retarded) time and the radial coordinate.

For convenience here we use the trace-reversed form of the EFE, but
the rest of the analysis is equivalent in the standard form. A free
evolution PDE system for the vacuum EFE in an asymptotically flat
spacetime in a Bondi-like gauge consists of
\begin{align}
  R_{rr} = R_{r\theta} = R_{r\phi} = R_{\theta \theta} = R_{\theta \phi}
  = R_{\phi \phi} = 0
  \,,
  \label{eqn:main_BS_sys}
\end{align}
which is often called the main system. The equation~$R_{ur}=0$ is
commonly referred to as the trivial equation, because solutions to the
main system automatically satisfy it, as shown in~\cite{BonBurMet62,
  CaoHe13} via the contracted Bianchi identities. The supplementary
equations
\begin{align*}
  R_{uu} = R_{u\theta} = R_{u\phi} = 0
  \,,
\end{align*}
are guaranteed to be satisfied in~$\mathcal{N}_{u}$ if they are
satisfied on a cross-section~\cite{BonBurMet62, CaoHe13}.

Regarding terminology, notice that in a standard spacelike foliation,
the constraint equations are intrinsic to the spacelike hypersurfaces
of the foliation.
In a free evolution scheme, the data chosen on the initial
hypersurface should satisfy the constraint equations, in order for the
solution of the PDE problem to be a solution to the EFEs. If the
constraint equations are satisfied initially, they are satisfied also
at later times, due to the Bianchi identities. Therefore, they are not
explicitly solved in a free evolution scheme, but mostly their
violation by a given approximate solution is examined at some stages
of the evolution.
%
%
The supplementary and trivial equations for a characteristic free
evolution scheme are treated as the aforementioned contraints for the
spacelike free evolution scheme. In a characteristic setup however,
the actual constraint equations--in the PDE sense as described in
Sec.~\ref{sec:pde_theory:causal_defs}--are intrinsic to null
hypersurfaces and are part of the evolution scheme, meaning they are
solved at each step of a time evolution. To avoid possible confusion,
we call the characteristic constraint equations \emph{intrinsic}.

The intrinsic equations of the main system can often acquire a nested
structure. Given a certain subset of unknowns as initial data on a
null hypersurface, the nested equations can be integrated in a
specific sequence to obtain a solution to the characteristic PDE
problem. In this case, each nested equation can be integrated
requiring knowledge only of the initial data and the functions
obtained by integrating the previous nested equations. This special
structure becomes apparent in the systems analyzed in the following
chapters, is common in most Bondi-like setups used for numerical
studies, and results in reduced computational cost and time. In
numerical implementations, these nested equations are often treated as
a system of \emph{effectively} ordinary differential equations (ODEs)
in the radial direction. By~``effectively'' here we mean that even
though formally these are partial differential equations, provided
that the necessary functions are given on the computational domain,
each equation can be solved as a standard ODE for one function since
the rest of the elements of the equation are known
quantities. Regarding its hyperbolic properties however the system is
still treated as an actual PDE system. In fact, this viewpoint of a
sequential system of effective ODEs may be misleading about the
well-posedness of the respective PDE problem. In other words, the
notion of sources in the numerical implementation can refer to
quantities that are known during the integration of an
equation. Regarding hyperbolicity however, for first order linear PDEs
with constant coefficients, sources are only terms with no
derivatives. So, a partial derivative of a function obtained from the
previous nested equation can be viewed as a source for the numerical
integration of the next nested equation, but not from the perspective
of the hyperbolicity analysis.

The main system provides six evolution equations for the seven unknown
metric functions. Usually, a definition for the determinant of the
induced metric on the two-spheres is made, namely
\begin{align}
  g_{\theta \theta} \, g_{\phi \phi} - g_{\theta \phi}^2
  = \hat{R}^4 \sin^2\theta
  \,,
  \label{eqn:gauge_cond_det}
\end{align}
where~$\hat{R}$ is taken to be a function of the coordinates, and
reduces to the areal radius of the two-sphere in spherical symmetry.

The aforementioned are common to all Bondi-like gauges. There is a
residual gauge freedom which corresponds to the choice of the
coordinate labeling the position within the null geodesic. This is
done differently in the various Bondi-like gauges. We focus on three
common choices:
\begin{description}  
\item[\textit{Affine null}~\cite{Win13, CreOliWin19}] The final choice
  of equations is achieved by setting~$g^{ur}= - 1$ for
  outgoing~$\mathcal{N}_u$ and~$g^{ur}= 1$ for
  ingoing~$\mathcal{N}_u$.~$\hat{R}$ is then taken to be an unknown of
  the problem.
\item[\textit{Bondi-Sachs proper}~\cite{BonBurMet62}] The radial
  coordinate matches the areal radius~$\hat{R}=r$ and so the
  definition~\eqref{eqn:gauge_cond_det} reduces the number of unknowns
  to six.
\item[\textit{Double null}~\cite{Chr08}] The residual gauge freedom is
  fixed by the condition~$g^{rr}=0$.
\end{description}

\section{From the characteristic to the ADM equations}
\label{sec:bondi_properties:BS_to_ADM_eqs}

We now map from the characteristic to the ADM variables and present
the system equivalent to~\eqref{eqn:main_BS_sys} in ADM formalism. We
assume that~$\mathcal{N}_u$ are outgoing, but an analogous analysis
can be performed for ingoing null hypersurfaces. To begin, we choose
the ADM coordinates~$x^{\mu}=(t,\rho,\theta,\phi)$. They are related
to the characteristic coordinates via
\begin{align}
  u = t - f(\rho) \,, \quad r = \rho
  \,.
  \label{eqn:coord_transf}
\end{align}
As in~\cite{Fri06}, the quantity~$-df/dr$ determines the slope of the
constant~$t$ spacelike hypersurface~$\Sigma_t$ on the~$u,r$ plane. The
angular coordinates~$\theta,\phi$ are unchanged and in this subsection
we may label them with the Latin indices~$A,B$.

The lapse of proper time between~$\Sigma_t$ and~$\Sigma_{t+dt}$ along
their normal observers is~$d \tau = \alpha(t,x^i) dt$, with the lapse
function defined by
\begin{align*}
  \alpha^{-2}(t,x^i)  \equiv -
  g^{\mu \nu} \nabla_\mu t
  \nabla_\nu t
  \,.
\end{align*}
The relative velocity between the trajectory of those observers and
the lines of constant spatial coordinates is given
by~$\beta^i(t,x^j)$, where~$x^i_{t+dt} = x^i_t -
\beta^i(t,x^j)dt$. The quantity~$\beta^i$ is called the shift vector.
The future directed unit normal 4-vector on~$\Sigma_t$ is
\begin{align*}
  n^{\mu} \equiv - \alpha \nabla^{\mu} t
  =
  \alpha^{-1} \left(1, -\beta^i \right)
  \,,
\end{align*}
and its covector form is
\begin{align*}
  n_{\mu} = g_{\mu \nu} n^{\nu}
  =
   \left(-\alpha, 0,0,0 \right)
  \,.
\end{align*}
The metric induced on~$\Sigma_t$ is
\begin{align*}
  \gamma_{\mu\nu} \equiv g_{\mu\nu} + n_\mu n_\nu
  \,.
\end{align*}
The ADM form of the equations is obtained by systematic contraction
with~$n^{\mu}$ and~$\gamma_{\mu \nu}$. This geometric construction is
discussed in most numerical relativity textbooks~\cite{Alc08, Gou07,
  BauSha10}. The spacetime metric takes the form
\begin{align*}
  g_{\mu \nu} =
  \begin{pmatrix}
    -\alpha^{2} + \beta_k \beta^k & \beta_{i} \\
    \beta_{j}  & \gamma_{i j} \\
  \end{pmatrix}\,,
\end{align*}
where lowercase Latin indices denote spatial components. The inverse
of~$g_{\mu \nu}$ is
\begin{align*}
  g^{\mu \nu} =
  \begin{pmatrix}
    -\alpha^{-2}  & \alpha^{-2} \beta^{i} \\
    \alpha^{-2} \beta^{j}  & \gamma^{i j}
    - \alpha^{-2} \beta^{i} \beta^{j} \\
  \end{pmatrix}\,.
\end{align*}
By comparing the~$3+1$ form of the metric and its inverse to the
generalized Bondi version~\eqref{eqn:gen_BS_line_element} we can
interpret the Bondi-like gauges in terms of lapse and shift, and
relate the characteristic variables to the ADM ones. Every Bondi-like
vector basis gives~\eqref{eqn:gauge_cond_null}, which in ADM
coordinates reads
\begin{equation*}
  \begin{aligned}
    g^{uu}
    = \frac{\p u}{\p x^\mu}  \frac{\p u}{\p x^\nu} g^{\mu \nu}
    = g^{tt} -2 f'g^{t \rho} + (f')^2 g^{\rho \rho} = 0
    \,,
    \quad
    g^{uA}
    = \frac{\p u}{\p x^\mu}  \frac{\p x^A}{\p x^\nu} g^{\mu \nu}
    = g^{tA} - f'g^{\rho A} = 0
    \,,
  \end{aligned}
\end{equation*}
and leads to
\begin{equation}
  \begin{aligned}
    \gamma^{\rho \rho}
    &= \left(\frac{1 + f' \beta^\rho}{f' \alpha}\right)^2
    \,,
    \quad
    \gamma^{\rho A}
     = \beta^A \frac{1 + f' \beta^\rho }{f' \alpha^2}
    \,.
  \end{aligned}
  \label{eqn:gauge_cond_null_gamma_uu}
\end{equation}
The Bondi-like metric ansatz \eqref{eqn:gen_BS_line_element} implies
\begin{align*}
  g_{rr} = g_{r A} = 0
  \,,
\end{align*}
which after using~$\beta_i = \gamma_{ij} \beta^j$ yields
\begin{equation}
  \begin{aligned}
    \gamma_{\rho \rho}
    &= \frac{(f')^2 (\alpha^2 + \beta^A \beta^B \gamma_{AB})}
    {(1 + f' \beta^\rho)^2}
    \,,
    \quad
    \gamma_{\rho A}
    = - \frac{f' }{1 + f' \beta^\rho} \beta^B \gamma_{AB} 
    \,.
  \end{aligned}
  \label{eqn:gauge_cond_null_gamma_dd}
\end{equation}
Using the latter
and~$g_{\mu^\prime \nu^\prime} = \frac{\p x^\mu}{\p x^{\mu^\prime}}
\frac{\p x^\nu}{\p x^{\nu^\prime}} g_{\mu \nu}$ provides the following
relations between the characteristic and ADM variables, for all
Bondi-like gauges:
\begin{equation}
 \begin{aligned}
   g_{uu}
   = \frac{\beta^A \beta_A -\alpha^2 (1 + 2 f' \, \beta^\rho)}
   {(1 + f' \, \beta^\rho)^2}
   \,,
   \quad
   g_{ur}
   = \frac{ - f' \, \alpha^2}{1+ f' \beta^\rho}
   \,,
   \quad
   g_{uA}
   = - \gamma_{\rho A}/f'
   \,, \quad
   g_{AB}  = \gamma_{AB} \,.
 \end{aligned}
 \label{eqn:gen_BS_vars_to_ADM}
\end{equation}
The above combined
with~$\gamma^{AB} - \alpha^{-2} \beta^A \beta^B = g^{AB}$ further
yield
\begin{align}
  &\gamma^{\theta \theta}
    = \left( \frac{\beta^\theta}{\alpha}\right)^2
    + \frac{\gamma_{\phi \phi}}{\det(g_{AB})}
    \,, \quad
    \gamma^{\theta \phi}
    = \frac{\beta^\theta \beta^\phi}{\alpha^2}
    - \frac{\gamma_{\theta \phi}}{\det(g_{AB})}
    \,,
    \nonumber
  \\
  &
    \gamma^{\phi \phi}
    = \left( \frac{\beta^\phi}{\alpha}\right)^2
    + \frac{\gamma_{\theta \theta}}{\det(g_{AB})}
    \,.
    \label{eqn:gauge_cond_null_gamma_uu_sphere}
\end{align}
for all Bondi-like gauges.

To proceed with the mapping between characteristic and ADM formalism,
we simply take the standard tensor transformation rule. The main
system~\eqref{eqn:main_BS_sys} written in the ADM coordinates is then
\begin{equation}
  \begin{aligned}
    R_{rr} & = (f')^2 R_{tt} + 2 f' R_{t \rho} + R_{\rho \rho} = 0 
    \,,
    \\
    R_{r A} &= f' R_{t A} + R_{\rho A} = 0 
    \,,
    \\
    R_{AB} & = 0
    \,.
  \end{aligned}
  \label{eqn:main_BS_in_ADM_coords}
\end{equation}
The complete orthogonal projection onto~$\Sigma_t$ is given by
\begin{align}
  \gamma^{\lambda}{}_{\mu} \gamma^{\sigma}{}_{\nu} R_{\lambda \sigma}
  \equiv R^\perp_{\mu \nu}
  &=
    -\mathcal{L}_n K_{\mu \nu}
    - \frac{1}{\alpha} D_{\mu} D_{\nu} \alpha
    + {}^{(3)}R_{\mu \nu}
    + K K_{\mu \nu}
    -2 K_{\mu \lambda} K^{\lambda}{}_{\nu}
    \,,
    \label{eqn:gamma_gamma_Rdd}
\end{align}
with~$R^\perp_{\mu \nu}$ a purely spatial tensor, and
\begin{align}
  \label{eqn:projector_general}
  \gamma^{\mu}{}_{\nu}
  &= \delta^{\mu}{}_{\nu} + n^{\mu} n_{\nu}
    \,,
    \quad
  K_{\mu \nu}
      =
      - \left( \nabla_{\mu} n_{\nu}
      + n_{\mu} n^{\kappa} \nabla_{\kappa} n_{\nu} \right)
      \,,
\end{align}
the orthogonal projector and the extrinsic curvature of~$\Sigma_t$
when embedded in the full spacetime, respectively. The following
purely spatial quantities have been used
\begin{align*}
   D_{\mu} S_{\nu \lambda}
  &=
    \perp \nabla_\mu  S_{\nu \lambda}  
    \,,
  \, \qquad 
  {}^{(3)}\Gamma^\mu{}_{\nu \lambda}
  =
    \perp \Gamma^\mu{}_{\nu \lambda}
    \,,
  \\
  {}^{(3)}R_{\mu \nu}
  & =
    \perp \left(
    \p_\lambda {}^{(3)}\Gamma^\lambda{}_{\mu \nu}
    -
    \p_\nu {}^{(3)}\Gamma^\lambda{}_{\mu \lambda}
    +
    {}^{(3)}\Gamma^\lambda{}_{\mu \nu}  {}^{(3)}\Gamma^\sigma{}_{\lambda \sigma}
    -
    {}^{(3)}\Gamma^\lambda{}_{\mu \sigma}  {}^{(3)}\Gamma^\sigma{}_{\nu \lambda}
    \right)
    \,,
\end{align*}
where~$D_\mu$ is the covariant derivative compatible
with~$\gamma_{\mu \nu}$, the symbol~$\perp$ denotes projection
with~$\gamma^\mu{}_\nu$ on every open index and~$S_{\mu \nu}$ denotes
an arbitrary spatial tensor. Imposing~$R_{\mu \nu} = 0$ and focusing
only on the spatial components of~$R^\perp_{\mu \nu}$ one can obtain
the evolution equations for the spatial components of the extrinsic
curvature
\begin{align*}
  \mathcal{K}_{ij}
  & \equiv - \p_t K_{ij}
    -D_i D_j \alpha
    + \alpha
    \left(
    {}^{(3)}R_{ij} + K K_{ij} -2 K_{im} K^{m}{}_j 
    \right)
    \nonumber
  \\
  & \quad \;
    + \beta^m \p_m K_{ij}
    + K_{im} \p_j \beta^m
    + K_{mj} \p_i \beta^m
    = 0
    \,,
\end{align*}
where~$K=g^{\mu \nu} K_{\mu \nu}$. The full projection perpendicular
to~$\Sigma_t$ is
\begin{align*}
  n^\mu n^\nu R_{\mu \nu}
  \equiv R^\parallel
  & =
    \mathcal{L}_n K + \frac{1}{\alpha} D^i D_i \alpha - K_{ij} K^{ij}
    \,.
\end{align*}
Using
\begin{align*}
  \mathcal{L}_n K = \gamma^{ij} \mathcal{L}_n K_{ij} + 2 K_{ij} K^{ij}
  \,,
\end{align*}
Eq.~\eqref{eqn:gamma_gamma_Rdd} and imposing the EFE,~$R^\parallel$
provides the Hamiltonian constraint
\begin{align*}
  H \equiv {}^{(3)}R + K^2 - K_{ij} K^{ij}
  = R^\parallel + \gamma^{ij} R^\perp_{ij} = 0
  \,.
\end{align*}
Finally, the mixed projection is given by the contracted Codazzi
relation
\begin{align*}
  n^\mu \gamma^\lambda{}_\nu R_{\mu \lambda}
  \equiv R^{| \perp}_\nu
  &=
    D_\nu K - D_\mu K^\mu{}_\nu 
    \,,
\end{align*}
with~$n^\mu R^{| \perp}_\mu = 0$. After imposing the EFE it yields the
momentum constraints
\begin{align*}
  M_i \equiv 
  D_j K^j{}_i - D_i K
  = 0
  \,.
\end{align*}
From Eq.~\eqref{eqn:projector_general} and the previous projections we
write
\begin{align}
  \delta^\alpha{}_\mu \delta^\beta{}_\nu R_{\alpha \beta} =
  R_{\mu \nu}^\perp + n_\mu n_\nu R^\parallel
  - n_\mu R_\nu^{| \perp} - n_\nu R_\mu^{| \perp}
  \,.
  \label{eqn:Rmunu_in_ADM_proj}
\end{align}
Using Eq.~\eqref{eqn:Rmunu_in_ADM_proj}, with
Eq.~\eqref{eqn:main_BS_in_ADM_coords} and taking linear combinations
of Eq.~\eqref{eqn:main_BS_sys}, we obtain the ADM system
\begin{align}
  &
    \frac{((f')^2-1)(1+f' \beta^\rho)^2}{(f')^2} \mathcal{K}_{\rho \rho} 
    + \alpha^2 H
    - 2 \alpha f' (1 + f' \beta^\rho) M_\rho
    - 2 \alpha \beta^A M_A
    = 0
    \,,
    \nonumber
  \\
  &
    (1 + f' \beta^\rho) \mathcal{K}_{\rho A} - \alpha f' M_A = 0
    \,,
    \label{eqn:equiv_ADM_to_main_sys}
  \\
  &
    \mathcal{K}_{AB} = 0
    \,, \nonumber
\end{align}
that is equivalent to the main Bondi-like
system~\eqref{eqn:main_BS_sys}, where we have also used
Eq.~\eqref{eqn:gauge_cond_null_gamma_uu},~\eqref{eqn:gauge_cond_null_gamma_dd}
and~\eqref{eqn:gauge_cond_null_gamma_uu_sphere}.

If the slope of~$\Sigma_t$ of the~$3+1$ foliation in the~$u,r$ plane
is~$f' \neq 1$, then the main Bondi-like
system~\eqref{eqn:main_BS_sys} corresponds to evolution equations for
all the components of~$K_{ij}$ with specific addition of the ADM
Hamiltonian and momentum constraints. For~$f'=1$ though, the first
equation of~\eqref{eqn:equiv_ADM_to_main_sys} involves only ADM
constraints. In this foliation the evolution equation
for~$K_{\rho \rho}$ is provided by the trivial equation, which after
imposing~\eqref{eqn:equiv_ADM_to_main_sys} reads
\begin{align*}
  (1+\beta^\rho) \mathcal{K}_{\rho \rho} - \alpha M_\rho
  + \frac{\alpha }{1 + \beta^\rho}  \beta^A M_A
  =0
  \,.
\end{align*}

The lapse and shift are not determined by the Einstein equations, but
in a~$3+1$ formulation are arbitrarily specifiable. In the present
setting, their choice is dictated by the explicit Bondi-like gauge
imposed. Adopting the terminology of~\cite{KhoNov02} we can classify
between algebraic and differential gauge choices:
\begin{description}
\item[\textit{Affine null}]It is a complete algebraic gauge for the lapse and
  shift, which is apparent by
  combining~\eqref{eqn:gauge_cond_null_gamma_dd}
  and
  \begin{align*}
    \beta^\rho =  \alpha^2 - 1/f'
    \,,
  \end{align*}
  which results from~$g^{ur}=-1=1/g_{ur}$. The determinant
  condition~\eqref{eqn:gauge_cond_det} does not act as a constraint
  among the three unknown metric components of the two-sphere, but
  merely relates them to the areal radius~$\hat{R}$ that is an
  unknown. The six equations of the main
  system~\eqref{eqn:main_BS_sys} correspond to the six ADM equations
  for~$K_{ij}$ (if~$f' \neq 1$) with a specific addition of
  Hamiltonian and momentum constraints, as well as the lapse and
  shift.
\item[\textit{Bondi-Sachs proper}] This gauge choice is completed by
  the definition of the determinant~\eqref{eqn:gauge_cond_det}. As we
  show in Sec.~\ref{sec:bondi-hyp:BS_proper} this definition can be
  viewed as providing a differential relation for the shift vector
  component~$\beta^\rho$. In this sense, the Bondi-Sachs gauge proper
  is a mixed algebraic-differential gauge in terms of the lapse and
  shift.
\item[\textit{Double null}] It is also a complete algebraic gauge. The
  complete gauge choice is implied by~$g^{rr}=0$, which combined
  with~$g^{uu}=0$ yields~$\beta^\rho = 0$.
\end{description}

\section{Coordinate light speeds}
\label{sec:bondi_properties:coord_lightspeeds}

Bondi-like gauges are constructed using either incoming or outgoing
null geodesics (or both). It is therefore natural to examine the
coordinate light speeds in these gauges. It is helpful to employ
a~$2+1$ split of the spatial metric~$\gamma_{ij}$ for this purpose. We
briefly review the key elements of this decomposition as necessary for
our discussion. The interested reader can find a complete presentation
in~\cite{Hil15}.

Level sets of constant~$\rho$ are two-spheres. The coordinate~$\rho$
defines an outward pointing normal vector on these spheres
\begin{align}
  s_{(\rho)}^i \equiv \gamma^{ij} L D_j \rho\,,
  \quad
  L^{-2} \equiv \gamma^{ij}(D_i\rho)(D_j\rho)\,.
  \label{eqn:def_s2+1_L}
\end{align}
We call~$L$ the length scalar. The induced metric on two-spheres of
constant~$\rho$ is
\begin{align}
  q_{(\rho)\,ij} \equiv \gamma_{ij} - s_{(\rho)\, i} s_{(\rho)\,j}
  \,,
  \label{eqn:2+1_gamma_split}
\end{align}
where the indices of~$s_{(\rho)}^i$ and~$q_{(\rho)\,ij}$ are lowered
and raised with~$\gamma_{ij}$ and its inverse. Let~$\rho^i$ be the
vector tangent to the lines of constant angular coordinates~$x^A$
i.e.~$\rho^i = (\p_\rho)^i$. Then
\begin{align}
  \rho^i = L s_{(\rho)}^i + b^i
  \,,
  \label{eqn:s_rho_1}
\end{align}
where~$b^i s_{(\rho)i} = 0$ and~$b^i$ is called the slip vector. The
length scalar~$L$ and the slip vector~$b^i$ are analogous to the~$3+1$
lapse and shift. They are not, however, freely specifiable but rather
are pieces of the spatial metric~$\gamma_{ij}$.

\begin{figure}[!t]
  \begin{center}
    \includegraphics[width=0.4\textwidth]{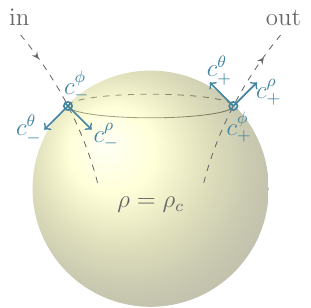}
  \end{center}
  \caption[Coordinate light speeds]
  {The coordinate light speeds for an ingoing and outgoing null ray
    that pass through a surface of constant radius~$\rho_c$ i.e. a
    two-sphere in this example. In an outgoing Bondi-like
    gauge~$c_+^\theta = 0 = c_+^\phi$, i.e. the
    coordinates~$\theta,\phi$ are Lie dragged along the outgoing null
    ray. This ray is orthogonal to the depicted two-sphere, as
    illustrated by the vectors drawn. Notice that the coordinate light
    speeds are scalar quantities.}
  \label{Fig:coord_lightspeeds}
\end{figure}

Let~$\gamma(t)$ be a null curve parameterized by~$t$
and~$L^\mu = \dot{x}^\mu = (1,\dot{x}^i(t))$ a null vector tangent
to~$\gamma(t)$. The coordinate light speeds
are~$C^i \equiv \dot{x}^i(t)$.  Let us further assume that the chosen
null vector obeys the relation
\begin{align}
  L^\mu \propto n^\mu \pm s^\mu_{(\rho)}
  \,.
  \label{eqn:L_mu_1}
\end{align}
From~\eqref{eqn:def_s2+1_L} we get
\begin{align}
  s^\mu_{(\rho)} = (0,L^{-1},L \gamma^{\rho A})
  \,.
  \label{eqn:s_rho_2}
\end{align}
Using~$\rho^\mu = (0,1,0,0)$, solving~\eqref{eqn:s_rho_1}
for~$s^i_{(\rho)}$ and comparing with~\eqref{eqn:s_rho_2} we obtain
\begin{align}
  b^\rho =0 \,, \quad -b^A = L^2 \gamma^{\rho A}\,.
  \label{eqn:slip_vector_relations}
\end{align}
After multiplying~\eqref{eqn:L_mu_1} with~$\alpha$ we have
\begin{align*}
  L^\mu \propto
  (1, -  \beta^\rho \pm \alpha L^{-1}, -  \beta^A \mp \alpha L^{-1} b^A) 
\end{align*}
from which we read off the coordinate light speeds along null curves
orthogonal to level sets of constant~$\rho$.  They are
\begin{align}
  c_{\pm}^\rho = - \beta^\rho \pm \alpha L^{-1}\,,
  \label{eqn:radial_coord_light_speed}
\end{align}
in the radial direction and
\begin{align}
  c_{\pm}^A = - \beta^A \mp b^A \alpha L^{-1}
  \,,
  \label{eqn:transverse_coord_light_speed}
\end{align}
in the angular directions. The subscript~$\pm$ refers to
outgoing/ingoing trajectories.  See Fig.~\ref{Fig:coord_lightspeeds}
for an illustration of the coordinate lightspeeds. For~$f(\rho)=\rho$,
using Eq.~\eqref{eqn:slip_vector_relations}, the gauge
conditions~\eqref{eqn:gauge_cond_null_gamma_uu} yield
\begin{align}
  c_+^\rho = 1 \,, \quad c_+^A = 0
  \,,
  \label{eqn:bondi-like_coord_light_speeds}
\end{align}
which just expresses the fact that transverse coordinates are Lie
dragged along outgoing null geodesics. For an ingoing single-null
Bondi-like characteristic formulation~$c_+^i \rightarrow c_-^i$ and
for double null~$c_\pm^\rho = \pm 1$. Away from spherical symmetry it
is not generally possible to have~$c_\pm^A$ both vanishing.

%% file: sections/gauge_structure.tex
\chapter{Gauge fixing and the principal symbol}
\label{chap:PG}

\minitoc

Following closely~\cite{KhoNov02,HilRic13} we now discuss the
structure of the principal symbol of the systems we
analyze. See~\cite{AbaReu18,Aba21} for interesting related work on
systems with constraints. As shown in~\cite{HilRic13}, working with
the ADM formalism, in this context, one can distinguish between the
gauge, constraint and physical variables of the system. This
distinction is reflected in the structure of the principal symbol and
allows us to understand which gauges can possibly result in SH
systems.

\section{FT$2$S Systems and their principal part}
\label{sec:PG:FT2S_princ_symbol}

According to~\cite{GunGar05,HilRic13a} the general first order in time
and second in space (FT$2$S) linear constant coefficient, system that
admits a standard first order reduction is of the form
\begin{align}
  \p_t \mathbf{v}
  & = \mathbfcal{A}_1^i \p_i \mathbf{v} + \mathbfcal{A}_1 \mathbf{v}
  + \mathbfcal{A}_2 \mathbf{w} + \mathbf{S}_{\mathbf{v}}
    \,,\nonumber
  \\
  \p_t \mathbf{w}
  & =
  \mathbfcal{B}_1^{ij} \p_i \p_j \mathbf{v}
  + \mathbfcal{B}_1^i \p_i \mathbf{v} + \mathbfcal{B}_1 \mathbf{v}
    + \mathbfcal{B}_2^i \p_i \mathbf{w}
     + \mathbfcal{B}_2 \mathbf{w}
     + \mathbf{S}_{\mathbf{w}}
     \,,
     \label{eqn:FT2S}
\end{align}
where~$\mathbf{S}_{\mathbf{v}},\mathbf{S}_{\mathbf{w}}$ are forcing
terms
and~$\mathbfcal{A}_1^i,\, \mathbfcal{A}_2,\,
\mathbfcal{B}_1^{ij},\,\mathbfcal{B}_2^i$ the principal matrices. In
the linear constant coefficient approximation the ADM equations lie in
this category. By \textit{standard first order reduction} we mean one
in which all first order derivatives (temporal and spatial) of
variables that appear with second order derivatives are introduced as
auxiliary variables. We call any first order reduction different than
the aforementioned~\textit{non-standard}. In such a case only a subset
of the first order derivatives of a variable that appears up to second
order is introduced as auxiliary variables. Specific higher
derivatives of certain variables could also be treated as auxiliary
variables in a non-standard reduction, if necessary. Given an
arbitrary unit spatial covector~$s_i$ (not to be confused
with~$s_{(\rho)}^i$ from
Sec.~\ref{sec:bondi_properties:coord_lightspeeds}), the principal
symbol of the system in the~$s_i$ direction is defined as
\begin{align}
  \mathbf{P}^s = \left(
  \begin{array}{cc}
    \mathbfcal{A}_1^s & \mathbfcal{A}_2\\
    \mathbfcal{B}_1^{ss} & \mathbfcal{B}_2^s
  \end{array}\right)
  \,,\label{second_order_principal_symbol}
\end{align}
where~$\mathbfcal{A}_1^s\equiv \mathbfcal{A}_1^is_i$ (and so
forth). Writing~$\mathbf{u}=(\p_s\mathbf{v},\mathbf{w})$, we have
\begin{align}
  \p_t\mathbf{u} \simeq \mathbf{P}^s \p_s\mathbf{u}\,,
  \label{eqn:second_order_principal_symbol_eom_form}
\end{align}
where here we dropped non-principal terms and all derivatives
transverse to~$s^i$. The definitions of weak and strong hyperbolicity
are identical to those discussed for first order systems in
Sec.~\ref{sec:pde_theory:hyp_degree}. Weak hyperbolicity is the
requirement that the eigenvalues of~$\mathbf{P}^s$ are real for
each~$s^i$, while strong hyperbolicity is the requirement that it is
also uniformly diagonalizable in~$s^i$. The second order principal
symbol~\eqref{second_order_principal_symbol} is inherited as a
diagonal block of the principal symbol of any standard first order
reduction, where the latter furthermore takes an upper block
triangular form. Consequently only SH second order systems may admit a
standard first order reduction that is SH. The importance of this is
that~\eqref{eqn:FT2S} has a well-posed initial value problem in the
norm
\begin{align*}
  E_1=\sum_i||\partial_i \mathbf{v}||_{L^2}
  + ||\mathbf{v}||_{L^2} + ||\mathbf{w}||_{L^2}
  \,,
\end{align*}
if and only if it is strongly hyperbolic, where here the norms are
defined over spatial slices of constant~$t$. For our analysis, observe
that the original characteristic form of the equations of motion is
not of the form~\eqref{eqn:FT2S}, even after linearization. We
overcome this issue by working instead with the ADM equivalent
obtained in Sec.~\ref{sec:bondi_properties:BS_to_ADM_eqs}. Working
with the equivalent ADM system
not only provides an invertible time principal matrix, but
has also the advantage that the theory discussed below was developed
in this language, making application straightforward. Due to the
freedom in choosing a time slicing, there is freedom in the
construction of the equivalent ADM formulation. This was parameterized
by~$f'(\rho)$ in the previous section. For brevity we work
assuming~$f'(\rho)=1$, but since the structural properties discussed
above hold true in {\it any} alternative slicing this restriction does
not affect the outcome of the analysis.

\section{Pure gauge and constraint subsystems}
\label{sec:PG:PG_subsys}

The linearized ADM system allows us to identify specific variables
that are associated with degrees of freedom related to pure gauge and
constraint violation. The pure gauge and constraint subsystems are
closed systems on their own, and their principal structure is
inherited in the principal symbol of the linearized ADM system.

\paragraph*{Pure gauge degrees of freedom:} In many cases of physical
interest~FT$2$S systems arise with additional structure in their
principal symbol. In GR for instance, structure arises as a
consequence of gauge freedom. To see this, suppose that we are working
in a coordinate basis with an arbitrary solution to the vacuum field
equations. The field equations are of course invariant under changes
of coordinates~$x^{\mu}\to X^{\underline{\mu}}$, so that both the
metric and curvature transform in the same manner. This invariance has
important consequences on the form of the field equations. Consider an
infinitesimal change to the coordinates
by~$x^{\mu}\to x^{\mu}+\xi^\mu$. Such a change results in a
perturbation to the metric of the form
\begin{align*}
\delta g_{\mu\nu}= -\nabla_\mu \xi_\nu - \nabla_\nu \xi_\mu = -\Lie_\xi g_{\mu\nu}\,.
\end{align*}
This transformation, the linearization of the condition for covariance
in a coordinate basis, simultaneously serves as the gauge freedom of
linearized GR. Working now in the ADM language, and~$3+1$
decomposing~$\xi^a$ by
\begin{align*}
  \Theta \equiv - n_\mu \xi^\mu
  \,, \quad
  \psi^i \equiv - \gamma^i {}_\mu \xi^\mu
  \,, 
\end{align*}
the pure gauge perturbations~$(\Theta,\psi^i)$ satisfy
(see~\cite{KhoNov02})
\begin{align}
\p_t\Theta&=\delta\alpha-\psi^iD_i\alpha+\Lie_\beta\Theta\,,\nonumber\\
\p_t\psi^i&=\delta\beta^i+\alpha D^i\Theta-\Theta D^i\alpha+\Lie_\beta\psi^i\,,
\label{eqn:PG_evolution}
\end{align}
with~$\delta\alpha$ and~$\delta\beta^i$ the perturbation of the lapse
and shift respectively. The resulting perturbations to the metric and
extrinsic curvature can be explicitly computed~\cite{HilRic13}, and
are given by,
\begin{subequations}
  \label{eqn:gauge_transf_2}
  \begin{align}
    \delta \gamma_{ij}
    &= - 2 \Theta K_{ij} + \mathcal{L}_\psi \gamma_{ij}
    \,,
    \label{eqn:delta_gamma_gauge_tranf}
    \\
    \delta K_{ij}
    & =
    - D_i D_j \Theta
    + \Theta \big( R_{ij} - 2 K^k {}_i K_{jk} + K_{ij} K \big)
    + \mathcal{L}_\psi K_{ij}
    \,,
    \label{eqn:delta_K_gauge_tranf}
  \end{align}
\end{subequations}
where~$\gamma_{ij}$ and~$K_{ij}$ are the metric and extrinsic
curvature associated with the background metric. It is a remarkable
fact that these equations are nothing more than the ADM evolution
equations under the replacements~$\alpha\to\Theta$
and~$\beta^i\to\psi^i$, so that the ADM evolution equations can be
interpreted as a local gauge transformation in a coordinate
basis. Given a choice for either the lapse and shift, or an equation
of motion for each, or a combination thereof, we may
combine~\eqref{eqn:PG_evolution} and~\eqref{eqn:gauge_transf_2}, to
obtain a closed system for the pure gauge variables~$(\Theta,\psi^i)$
and~$(\delta\alpha,\delta\psi^i),$ on the background spacetime. We
call this the \textit{pure gauge subsystem}. Suppose for example that
we employed a harmonic time coordinate~($\Box t=0$) with vanishing
shift. In~$3+1$ language this gives
\begin{align*}
\p_t\alpha&=-\alpha^2K.
\end{align*}
The pure gauge subsystem~\eqref{eqn:PG_evolution}
for~$(\Theta,\psi^i),$ is then completed by
\begin{align*}
  \p_t\delta\alpha&\simeq \alpha^2\p^i\p_i\Theta\,,\qquad
  \delta\beta^i=0\,,
\end{align*}
where we have used~\eqref{eqn:gauge_transf_2} and discarded
non-principal terms. The additional structure alluded to above is that
for a given choice of gauge, the principal symbol of the pure gauge
subsystem is inherited as a sub-block of the principal symbol of any
formulation of GR that employs said gauge. This is demonstrated by
using suitable projection operators which are stated explicitly in
Sec.~\ref{sec:PG:projectors}.

\paragraph*{Constraint violating degrees of freedom:} Yet more
structure arises from the constraints. Assuming the ADM evolution
equations hold, the Hamiltonian and Momentum constraints formally
satisfy evolution equations,
\begin{align*}
  \p_tH&=-2\alpha D^iM_i-4M^iD_i\alpha+2\alpha K H+\Lie_\beta H\,,\nonumber\\
  \p_tM_i&= -\tfrac{1}{2}\alpha D_iH+\alpha K M_i-D_i\alpha H+\Lie_\beta M_i\,,
\end{align*}
so given constraint satisfying initial data, the solution in their
future domain of dependence satisfies these constraints as well.
%
%
These equations follow from the contracted Bianchi identities. In
free-evolution formulations of GR however, the ADM evolution equations
need not hold, since combinations of the constraints can be freely
added to the evolution equations. Doing so results in adjusted
evolution equations for the constraints, which nevertheless remain a
closed set of equations. Just as the principal symbol of the full
equations of motion inherits the pure gauge principal symbol, the
principal symbol of the constraint subsystem manifests as a
sub-block. This is again seen using the projection operators stated in
Sec.~\ref{sec:PG:projectors}.

\paragraph*{Linearized ADM:} To apply straightforwardly the theory
described at Sec.~\ref{sec:PG:FT2S_princ_symbol} we linearize about
flat space in global inertial coordinates. The analysis can be carried
out around a general background leading to the same conclusions. In
this setting we obtain for the metric and extrinsic curvature
perturbations the evolution equations,
\begin{subequations}
  \label{eqn:ADM_lin_mink}
  \begin{align}
    \p_t \delta \gamma_{ij}
    & = -2 \delta K_{ij} + \p_{(i} \delta \beta_{j)}
      \,,
      \label{eqn:ADM_lin_mink_gamma}
    \\
    \p_t \delta K_{ij}
    & =
      - \p_i \p_j \delta \alpha
      - \tfrac{1}{2}\p^k \p_k \delta \gamma_{ij}
      -\tfrac{1}{2} \p_i \p_j \delta \gamma
      + \p^k \p_{(i} \delta \gamma_{j)k}
      \,.
      \label{eqn:ADM_lin_mink_K}
  \end{align}
\end{subequations}
The constraints become
\begin{align*}
  \delta H&=\p^i\p^j\delta\gamma_{ij}-\p^i\p_i\delta\gamma \,,\nonumber\\
  \delta M_i&=\p^j\delta K_{ij}-\p_i \delta K  \,,
\end{align*}
and evolve according to
\begin{align}
  \label{eqn:Cons_subsys_lin_mink}
  \begin{aligned}
  \p_t\delta H&= -2 \p^i\delta M_i\,,\\
  \p_t\delta M_i&= -\tfrac{1}{2} \p_i\delta H\,.
  \end{aligned}
\end{align}
About this background the pure gauge
equations~\eqref{eqn:PG_evolution} simplify to
\begin{subequations}
  \label{eqn:PG_subsys_lin_mink}
  \begin{align}
  \p_t \Theta
   & = \delta \alpha
     \,,
     \label{eqn:PG_subsys_lin_mink_theta}
  \\
  \p_t \psi_i
   &=
     \delta \beta_i + \p_i \Theta
     \,.
     \label{eqn:PG_subsys_lin_mink_psi}
  \end{align}
\end{subequations}

\section{Projection operators}
\label{sec:PG:projectors}

To unravel the special structure of the principal symbol of the
linearized ADM system we use projection operators. This structure
distinguishes in the linear, constant coefficient approximation
between the pure gauge, constraint violating and physical ADM
variables, along an arbitrary spatial direction.

\paragraph*{Pure gauge projection operators:} Let~$s^i$ be an arbitrary
constant spatial unit vector. To extract the gauge, constraint and
physical degrees of freedom within the principal symbol in this
direction we must decompose the state vector appropriately. The
induced metric on the surface transverse to~$s^i$ is
\begin{align*}
  q_{ij} \equiv \gamma_{ij} - s_i s_j \,.
\end{align*}
Here we denote by~$\hat{A},\,\hat{B}$ the spatial directions
transverse to~$s^i$, which---since in
general~$s^i \neq s_{(\rho)}^i$---do not necessarily coincide with the
angular directions from
Sec.~\ref{sec:bondi_properties:coord_lightspeeds}. Projections of the
ADM variables that capture pure gauge equations of
motion~\eqref{eqn:PG_subsys_lin_mink} are given by,
\begin{align}
  \label{eqn:gauge_vars}
  \begin{aligned}
    [\p_s^2 \Theta]
    &= - \delta K_{ss}
    \,, \quad
    [\p_s^2 \psi_s] = \tfrac{1}{2} \p_s \delta \gamma_{ss}
    \,, \quad
    [\p_s^2 \psi_{\hat{A}}]
    &= \p_s \delta \gamma_{s \hat{A}}
    \,.
  \end{aligned}
\end{align}
Here the notation~$[\cdots]$ is used to emphasize that the specific
projection of the ADM variables on the right-hand-side shares, within
the principal symbol, the structure of the pure gauge variable named
on the left-hand-side. This is spelled out below. Thus, together
with~$\p_s \delta \alpha,\, \p_s \delta \beta_s,\, \p_s \delta
\beta_{\hat{A}}$ they encode the complete pure gauge variables of the
system, with~$\delta \alpha,\, \delta \beta^i$ the perturbation to the
lapse and shift.

\paragraph*{Constraint projection operators:} Likewise, within the principal
symbol the Hamiltonian and Momentum constraints are encoded by the
projections,
\begin{align}
  \begin{aligned}
    [H]&=-\p_s\delta\gamma_{qq}
    \,,\quad
    [M_s]=-\delta K_{qq}
    \,, \quad
    [M_{\hat{A}}]&=\delta K_{s\hat{A}}\,,
  \end{aligned}
  \label{eqn:constr_vars}
\end{align}
with the naming convention as above. Here and in the following
indices~$qq$ denote that the trace was taken with~$q^{ij}$.

\paragraph*{Physical projection operators:} Finally, the remaining
variables to be taken account of are the trace free
projections. Defining the projection operator familiar from textbook
treatments of linear gravitational waves,
\begin{align}
  {}^P \! \! \! \perp^{kl}_{ij}
  \equiv
  q^k_{\;(i} q^l{}_{j)} -  \tfrac{1}{2} q_{ij} q^{kl}
  \,.
  \label{eqn:phys_projector}
\end{align}
we define 
\begin{align*}
  \begin{aligned}
    \p_s\delta \gamma_{\hat{A} \hat{B}}^{\textrm{TF}}
    &={}^P \! \! \! \perp^{ij}_{\hat{A}\hat{B}} \p_s\delta \gamma_{ij} 
    \,,\quad
    \delta K_{\hat{A} \hat{B}}^{\textrm{TF}}
    ={}^P \! \! \! \perp^{ij}_{\hat{A}\hat{B}} \delta K_{ij} 
    \,.
  \end{aligned}
\end{align*}
The superscript~$\textrm{TF}$ denotes trace free. These variables are
associated with the physical degrees of freedom.

\paragraph*{The principal symbol:} Employing the notation above we can
now write out the principal symbol in the
form~\eqref{eqn:second_order_principal_symbol_eom_form}. Starting with
the pure gauge block, this gives
\begin{align}
  \label{eqn:PG_Ps}
  \begin{aligned}
    \p_t[\p_s^2\Theta] &\simeq \p_s(\p_s\delta\alpha) + \tfrac{1}{2}\p_s[H]\,,\\
    \p_t[\p_s^2\psi_s] &\simeq \p_s(\p_s\delta\beta_s) + \p_s[\p_s^2\Theta]\,,\\
    \p_t[\p_s^2\psi_{\hat{A}}] &\simeq \p_s(\p_s\delta\beta_{\hat{A}})
    -2\p_s[M_{\hat{A}}]\,.
  \end{aligned}
\end{align}
Comparing this with~\eqref{eqn:PG_subsys_lin_mink} it is clear that up
to additions of the ``constraint variables'' there is agreement. Next,
the constraint violating block gives
\begin{align}
  \label{eqn:Cons_Ps}
  \begin{aligned}
    \p_t[H] &\simeq -2\p_s [M_s] \,,\\
    \p_t[M_s] &\simeq - \tfrac{1}{2}\p_s[H]\,,\\
    \p_t[M_{\hat{A}}] &\simeq 0\,.
  \end{aligned}
\end{align}
Comparing this with~\eqref{eqn:Cons_subsys_lin_mink} there is perfect
agreement. Finally the physical block is
\begin{align}
  \label{eqn:physical_subsys}
  \begin{aligned}
    \p_t\p_s\delta \gamma_{\hat{A} \hat{B}}^{\textrm{TF}}
    &\simeq -2 \p_s\delta K_{\hat{A} \hat{B}}^{\textrm{TF}}
    \,,\\
    \p_t \delta K_{\hat{A} \hat{B}}^{\textrm{TF}}
    & \simeq
    - \tfrac{1}{2} \p_s^2\delta \gamma_{\hat{A} \hat{B}}^{\textrm{TF}}
    \,,
  \end{aligned}
\end{align}
which is decoupled from the rest of the equations. These equations are
not yet complete, because we have not yet made a concrete choice of
gauge. Several Bondi-like gauges are treated in detail in
Chap.~\ref{chap:bondi-hyp}.

\section{Discussion}
\label{sec:PG:discussion}

The results of the foregoing discussion follow because GR is a
constrained Hamiltonian system that satisfies the hypotheses
of~\cite{HilRic13}. To make the presentation here somewhat more
standalone however, let us consider a plane wave ansatz
\begin{align}
  \label{eqn:Met_Decomp}
  \begin{aligned}
    \delta\gamma_{ij} &= 2 e^{\kappa^{(\psi_s)}_\mu x^\mu} s_is_j[\p_s\tilde{\psi_s}]
    - \tfrac{1}{2} q_{ij} e^{\kappa^{(H)}_\mu x^\mu} [\tilde{H}]
    \\
    &+2e^{\kappa^{(\psi_A)}_\mu x^\mu}q^{\hat{A}}{}_{(i}s_{j)}[\p_s\tilde{\psi}_{\hat{A}}]
    + e^{\kappa^{(P)}_\mu x^\mu}
    {}^P \! \! \! \perp^{\hat{A}\hat{B}}_{ij}
    \delta\gamma_{\hat{A} \hat{B}}^{\textrm{TF}} \,,\\
    \delta K_{ij} &=  - e^{\kappa^{(\Theta)}_\mu x^\mu} s_is_j[\p^2_s\tilde{\Theta}]
    - \tfrac{1}{2} q_{ij} e^{\kappa^{(M_s)}_\mu x^\mu} [\tilde{M}_s]\\
    &+2e^{\kappa^{(M_A)}_\mu x^\mu}q^{\hat{A}}{}_{(i}s_{j)}[\tilde{M}_{\hat{A}}]
    + e^{\kappa^{(P)}_\mu x^\mu}
    {}^P \! \! \! \perp^{\hat{A}\hat{B}}_{ij} \delta K_{\hat{A} \hat{B}}^{\textrm{TF}}\,,
  \end{aligned}
\end{align}
with each wave vector of the
form~$\kappa_\mu = (\kappa, i\,\omega s_i)$. These solutions travel in
the~$\pm s^i$ directions, although since the lapse and shift are as
yet undetermined, the~$\kappa$'s can not be solved for so
far. Defining the projections exactly as above, the unknowns can be
decomposed {\it explicitly} into their gauge, constraint violating and
gravitational wave pieces as indicated by the naming, and
equations~\eqref{eqn:PG_Ps},~\eqref{eqn:Cons_Ps}
and~\eqref{eqn:physical_subsys} become exact. In the nonlinear setting
it is of course hopeless to try and decompose metric components into
their constituent gauge, constraint violating and physical degrees of
freedom. But even in the linear constant coefficient approximation,
solutions consist in general of a sum over many such plane waves
propagating in different directions, and so the
decomposition~\eqref{eqn:Met_Decomp} is not a sufficient description.
What is important for our purposes however, is that the structure in
the field equations that permits the
decomposition~\eqref{eqn:Met_Decomp} for plane wave solutions is
present regardless of the direction~$s^i$ considered. The principal
symbol sees only this structure and thus, with the
equations~\eqref{eqn:PG_Ps},~\eqref{eqn:Cons_Ps}
and~\eqref{eqn:physical_subsys} above completed with a choice for the
lapse and shift, can be written in the schematic form
\begin{align}
  \mathbf{P}^s =
  \begin{pmatrix}
    \mathbf{P}_G & \mathbf{P}_{GP} & \mathbf{P}_{GC} \\
    0 & \mathbf{P}_P & \mathbf{P}_{PC} \\
    0 & 0 &  \mathbf{P}_C
  \end{pmatrix}
  \,,\label{eqn:princ_symbol_triang}
\end{align}
even upon linearization about an arbitrary
background. Here~$\mathbf{P}_G,\,\mathbf{P}_C,\,\mathbf{P}_P$ denote
the gauge, constraint and physical sub-blocks
and~$\mathbf{P}_{GC},\,\mathbf{P}_{GP},\,\mathbf{P}_{PC}$ parameterize
the coupling between them. As seen in~\cite{HilRic13} there is a very
large class of gauge conditions and natural constraint additions that
result in~$\mathbf{P}_{GP}=\mathbf{P}_{GC}=\mathbf{P}_{PC}=0$.
The affine null gauge analyzed in Sec.~\ref{sec:bondi-hyp:aff_null} is
however an explicit example where both~$\mathbf{P}_{GP}$
and~$\mathbf{P}_{GC}$ are non-vanishing.
Consequently, it follows from~\eqref{eqn:princ_symbol_triang} that a
necessary condition for strong hyperbolicity of the formulation is
that the pure gauge and constraint subsystems are themselves strongly
hyperbolic (see e.g. App. A of~\cite{GunGar05} for
details). Following~\cite{KhoNov02} we may therefore restrict our
attention first to pure gauge systems of interest, which have the
advantage of being smaller, and thus much easier to treat.

\paragraph*{Bondi-like gauges:} The gauges we are concerned with all require
the condition~\eqref{eqn:gauge_cond_null}, which in characteristic
coordinates implies the same for the perturbation to the metric, that
is,
\begin{align*}
  \delta g^{uu} = \delta g^{uA} = 0 \,.
\end{align*}
There remains one gauge condition to be specified, namely the
parameterization along outgoing null surfaces by a radial coordinate.
In Chap.~\ref{chap:bondi-hyp} we study specific instances of this
condition.


%% file: sections/hyperbolicity.tex
\chapter{Hyperbolicity of Bondi-like PDE systems}
\label{chap:bondi-hyp}

\minitoc

We perform hyperbolicity analyses for some popular Bondi-like free
evolution PDE systems. For the affine null gauge we identify the pure
gauge sub-block in the ADM equivalent system, as well as in the
characteristic one for asymptotically flat spacetimes and show that
the system is only WH due to the angular sector of the pure gauge
subsystem. We also demonstrate weak hyperbolicity for a planar
symmetric asymptotically~$AdS_5$ setup based on the affine null
gauge. The Bondi-Sachs pure gauge subsystem is also shown to be WH due
to its angular structure and the inheritance of this is demonstrated
for the ADM equivalent setup. An axisymmetric characteristic setup in
Bondi-Sachs coordinates is shown to be only WH as well due to its
angular structure. After identifying the same type of weak
hyperbolicity for the double-null pure gauge subsystem as well, we
argue in Sec.~\ref{sec:bondi-hyp:double_null} that under certain
conditions all Bondi-like free evolution PDE systems with up to second
order metric derivatives are only WH.

The intrinsic equations of the systems in
Subsecs.~\ref{subsec:bondi-hyp:aff_null:AAdS}
and~\ref{subsec:bondi-hyp:BS_proper:axisym_char} possess a nested
structure that makes them convenient for numerical studies, since they
require less computational resources. It is not clear why this
structure appears and it seems to be affected by the choice of
variables used~\cite{Win13}. This however is not a generic feature of
characteristic PDEs but rather a special structure that can be found
in Bondi-like systems.

The PDE systems under consideration are quasilinear, whereas the
definitions for the degree of hyperbolicity provided in
Sec.~\ref{sec:pde_theory:hyp_degree} refer to first order linear
systems. To apply these definitions we perform a linearization about a
fixed background and a first order reduction. We chose this background
to be Minkowski except for
Subsec.~\ref{subsec:bondi-hyp:aff_null:AAdS} where it is
vacuum~$AdS_5$. In~\cite{GiaHilZil20_public} and in the ancillary
files of~\cite{GiaBisHil21} the same calculations for arbitrary
backgrounds can be found. It turns out that the hyperbolic character
of the analyzed PDEs is unaffected by the choice of background for
this analysis. To determine the degree of hyperbolicity of each system
we work in the frozen coefficient approximation and demand that for a
system to be WH or SH, the definitions of
Sec.~\ref{sec:pde_theory:hyp_degree} are satisfied at each point in
the domain of interest. In~\cite{FriLeh99} and~\cite{GomFri03} the
authors studied existence and uniqueness of the CIBVP for a free
evolution system in Bondi-Sachs coordinates. They considered the
linearized and quasilinear systems, but did not study continuous
dependence on given data. The latter is the main focus here and in
chapters~\ref{chap:bondi_well-p} and \ref{chap:numerics} we provide
more details at the continuum and numerical level, respectively.

\begin{figure}[!t]
  \begin{center}
    \includegraphics[width=0.75\textwidth]{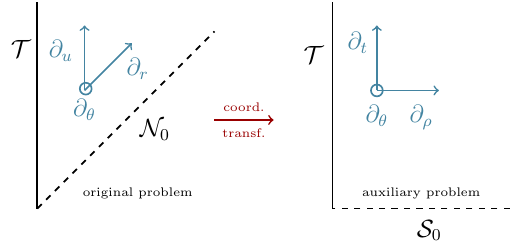}
    \end{center}
    \caption[CIBVP: characteristic to auxiliary Cauchy setup]
    {The original CIBVP is transformed into an auxiliary frame using
      the coordinate transformation~\eqref{flat_coord_transf} as
      shown. This allows us to employ the hyperbolicity definitions of
      Sec.~\ref{sec:pde_theory:hyp_degree} but does not affect the
      solution space.
    \label{flat_CIBVP_to_IBVP}
  }
\end{figure}

The time principal part matrix of characteristic PDE systems is
non-invertible. To construct the principal
symbol~\eqref{eqn:principal_symbol} after linearization and first
order reduction we employ a coordinate transformation to an auxiliary
Cauchy-type setup.  We wish to bring the characteristic system to the
form~\eqref{eqn:gen_hyper_PDE}, with an invertible principal part
matrix~$\mathbfcal{A}^t$. In the following sections we use the
coordinate transformation~\eqref{eqn:coord_transf} with~$f(\rho)=\rho$,
namely
\begin{align}
  u = t-\rho \, , \qquad r=\rho \,,
  \label{flat_coord_transf}
\end{align}
with the angular coordinates unchanged, which yields the following
relation between the old and new basis vectors,
\begin{align*}
  \p_u = \p_t \, , \qquad \p_r  = \p_t
  + \p_\rho  \,,
\end{align*}
with the remaining vectors unaltered. An illustration of the auxiliary
setup is given in Fig.~\ref{flat_CIBVP_to_IBVP}. Applying the
transformation yields
\begin{align*}  
  \mathbfcal{A}^t \, \p_t  \mathbf{u} 
  + \mathbfcal{A}^r \, \p_\rho  \mathbf{u}
  + \mathbfcal{A}^\theta \, \p_\theta \mathbf{u} \, + \mathbfcal{S} = 0
  \, ,
\end{align*}
with~$\mathbfcal{A}^t = \mathbfcal{A}^u + \mathbfcal{A}^r$
invertible. After multiplying from the left
with~$(\mathbfcal{A}^t)^{-1}$ we can bring the system in the form
\begin{align}
  \p_t\mathbf{u} + 
  \mathbf{B}^p\p_p\mathbf{u} + \mathbfcal{S}
  = 0
  \,,
  \label{eqn:prototype_1st_order_lin_const_coef_PDE_2}
\end{align}
where the principal symbol is
simply~$\mathbf{P}^s = \mathbf{B}^p s_p$. In comparison to the
form~\eqref{eqn:prototype_1st_order_lin_const_coef_PDE} notice the
difference in the sign convention,
namely~$\mathbf{B}^p = - \mathbfcal{B}^p$
and~$ \mathbfcal{B} \mathbf{u} = - \mathbfcal{S}$. The solution space
in this frame is equivalent to that of the original one, so in this
sense the character of the PDE is invariant. In
Sec.~\ref{sec:bondi_hyp:frame_indep} we show that the hyperbolic
character of the system is also independent of the auxiliary frame
chosen for the analysis. An auxiliary Cauchy setup was also used
in~\cite{Ren90} to show well-posedness of the CIBVP for a symmetric
hyperbolic characteristic system.

To treat the system in the original higher-order derivative form, we
could follow~\cite{GunGar05,HilRic13a}. But for convenience in
building the principal parts we instead perform an explicit first
order reduction. Since this PDE is built as a reduction, there is the
subtlety of the associated reduction constraints and the specific
choice of reduction, which we discuss in detail making use of
\textit{generalized characteristic variables}. To understand this
notion consider the principal symbol~$\mathbf{P}^s$ and the
generalized eigenvalue problem,
\begin{align*}
  \mathbf{l}_{\lambda_i} \left( \mathbf{P}^{s}
  - \lambda_i \mathbb{1} \right)^m =0\,,
\end{align*}
with~$\lambda_i$ standing for the various eigenvalues,
and~$\mathbf{l}_{\lambda_i}$ representing either a true eigenvector
when~$m=1$ or else a generalized eigenvector when~$m>1$. Defining the
invertible matrix~$\mathbf{T}_{s}^{-1}$ with the
vectors~$\mathbf{l}_{\lambda_i}$, as rows, we obtain the Jordan normal
form of the principal symbol in the~$s$ direction by the similarity
transformation
\begin{align*}
  \mathbf{J}^{s} \equiv \mathbf{T}_{s}^{-1} \,
  \mathbf{P}^{s} \, \mathbf{T}_{s}\,.
\end{align*}
The same matrix can be used to construct the generalized
characteristic variables of the system in the~$s$ direction, namely
the components
of~$\mathbf{v}\equiv \mathbf{T}_{s}^{-1} \, \mathbf{u}$. These are of
course nothing more than the left generalized eigenvectors contracted
with the state vector. Working in the frozen coefficient
approximation, focusing on the~$t,s$ parts
of~\eqref{eqn:prototype_1st_order_lin_const_coef_PDE_2} and
multiplying on the left with~$\mathbf{T}_{s}^{-1}$ one can get
\begin{align}
  \p_t \mathbf{v} + \mathbf{J}_{s}  \, \p_{s} \mathbf{v}
  \simeq 0 \, ,\label{gen_char_system}
\end{align}
with~$\simeq$ denoting here equality up to non-principal terms and
spatial derivatives transverse to~$\p_{s}$. In this form, weak
hyperbolicity can be understood as the failure of a generalized
characteristic variable to satisfy an advection equation along~$\p_s$
with some characteristic speed~$\lambda_s$ which may also be
vanishing.

\section{The affine null gauge}
\label{sec:bondi-hyp:aff_null}

In terms of lapse and shift specification, the complete affine null
gauge fixing is given by
\begin{align}
  \alpha = L^{-1}
  \,, \quad
  \beta^\rho = L^{-2} -1
  \,, \quad
  \beta^A = - b^A L^{-2}
  \,,
  \label{eqn:aff_null_gauge}
\end{align}
where
Eqs.~\eqref{eqn:radial_coord_light_speed}-\eqref{eqn:bondi-like_coord_light_speeds}
and~$g^{ur}=-1$ have been combined and~$f(\rho)=\rho$ in
Eq.~\eqref{eqn:coord_transf} is assumed.

\subsection{Pure gauge subsystem}
\label{subsec:bondi-hyp:aff_null:pg_subsys}

Let us first consider pure gauge metric
perturbations~\eqref{eqn:gauge_transf_2}. To close the
system~\eqref{eqn:PG_subsys_lin_mink} further input for~$\delta
\alpha$ and~$\delta \beta^i$ is needed. For the affine null gauge this
follows from~\eqref{eqn:aff_null_gauge}, which after linearization
about flat space reads
\begin{align}
  \delta \alpha
  &= -\tfrac{1}{2} \delta \gamma_{\rho \rho}
    \,, \quad 
    \delta \beta^\theta
    = - \rho^{-2} \delta \gamma_{\rho \theta}
    \,,
    \label{eqn:aff_null_gauge_lin_mink}
  \\
  \delta \beta^\rho
  &= - \delta \gamma_{\rho \rho}
    \,, \qquad
    \delta \beta^{\phi} = - \rho^{-2} \sin^2\theta \,
    \delta \gamma_{\rho \phi}
    \,.
    \nonumber
\end{align}
Using~$\delta \gamma_{ij} =  \p_i \psi_j + \p_j \psi_i$
and~$\psi^i = \gamma^{ij} \psi_j$ the latter reads
\begin{align}
  \delta \alpha
  &= - \p_\rho \psi^\rho
    \,, \quad \, 
  \delta \beta^\theta
  = - \p_\rho \psi^\theta - \rho^{-2} \p_\theta \psi^\rho
    \,,
    \label{eqn:aff_null_coord_pert}
  \\
  \delta \beta^\rho
  &= - 2 \p_\rho \psi^\rho
    \,, \; \;
    \delta \beta^{\phi} =
    -\p_\rho \psi^\phi  - \rho^{-2} \sin^2\theta \, \p_\phi \psi^\rho
    \,.
    \nonumber
\end{align}
The pure gauge subsystem~\eqref{eqn:PG_subsys_lin_mink} is then
\begin{subequations}
  \label{eqn:aff_null_pg}
  \begin{align}
    &
    (\p_t+\p_\rho) ( \psi^\rho - \Theta) =  0
      \,,
      \label{eqn:aff_null_pg_a}
    \\
    &
    (\p_t + \p_\rho) \psi^\rho + \p_\rho(\psi^\rho - \Theta) =  0
      \,,
      \label{eqn:aff_null_pg_b}
    \\
    &
    (\p_t + \p_\rho) \psi^\theta
      + \rho^{-2} \p_\theta (\psi^\rho - \Theta) =  0
      \,,
      \label{eqn:aff_null_pg_c}
     \\
    &
    (\p_t + \p_\rho) \psi^\phi
    + (\rho \sin \theta)^{-2} \p_\phi (\psi^\rho - \Theta) =  0
      \,,
      \label{eqn:aff_null_pg_d}
  \end{align}
\end{subequations}
where~$\p_t + \p_\rho =\p_r$ is an outgoing null derivative
and~\eqref{eqn:aff_null_pg_a} results from a linear combination
of~\eqref{eqn:PG_subsys_lin_mink_theta}
and~\eqref{eqn:PG_subsys_lin_mink_psi} with~$i=\rho$. The principal
symbol of the pure gauge subsystem~\eqref{eqn:aff_null_pg} is clearly
non-diagonalizable along the~$\rho,\theta, \phi$ directions, and in
fact in any
direction. In~\eqref{eqn:aff_null_pg_b},~\eqref{eqn:aff_null_pg_c}
and~\eqref{eqn:aff_null_pg_d} the
terms~$\p_\rho(\psi^\rho - \Theta)$,~$\p_\theta(\psi^\rho - \Theta)$
and~$\p_\phi(\psi^\rho - \Theta)$ result in~$2 \times 2$ Jordan
blocks, along~$\rho,\theta$ and~$\phi$ respectively. The principal
symbol of the full set of equations of motion for GR has the upper
triangular form~\eqref{eqn:princ_symbol_triang} when a standard first
order reduction is considered. Thus it will possess non-trivial Jordan
blocks along all~$\rho,\theta,\phi$ directions as well. In
Subsecs.~\ref{subsec:bondi-hyp:aff_null:pg_sub-block_rho_dir}
and~\ref{subsec:bondi-hyp:aff_null:pg_sub-block_theta_dir} we show
this explicitly and demonstrate the connection to the PDE system in
characteristic coordinates.

An intriguing observation is that the pure gauge
variable~$(\psi^\rho-\Theta)$ satisfies a transport equation
along~$\p_r$. So, acting from the left on~\eqref{eqn:aff_null_pg}
with~$\p_r$ and commuting the spatial and null derivatives
on~$(\psi^\rho - \Theta)$, one obtains
\begin{subequations}
  \begin{align}
    &
    \p_r^2 (\Theta - \psi^\rho)  = 0
      \,,
      \label{eqn:aff_null_pg_2nd_order_a}
    \\
    &
    \p_r^2 \psi^\rho   = 0
      \,,
      \label{eqn:aff_null_pg_2nd_order_b}
    \\
    &
    \p_r^2 \psi^\theta
      =  0
      \,,
      \label{eqn:aff_null_pg_2nd_order_c}
     \\
    &
    \p_r^2 \psi^\phi
      = 0
      \,.
      \label{eqn:aff_null_pg_2nd_order_d}
  \end{align}
  \label{eqn:aff_null_pg_2nd_order}%
\end{subequations}
This system admits a non-standard reduction to first order which is
strongly hyperbolic. To see this, we introduce only outgoing null
derivatives of the unknowns as auxiliary variables. All of the
variables then satisfy transport equations in the outgoing null
direction. In contrast to this, for a standard first order reduction
both the time and space derivatives of the unknowns would be
introduced as auxiliary variables.

The relevant question is whether or not there exists a formulation of
GR that inherits the structure of the second version of the pure gauge
subsystem~\eqref{eqn:aff_null_pg_2nd_order}, rather than the
first~\eqref{eqn:aff_null_pg}. In view of the results
of~\cite{HilRic13}, if such a formulation exists it would necessarily
admit a non-standard first order reduction. In
Subsec.~\ref{subsec:bondi-hyp:aff_null:pg_sub-block_rho_dir} we show
that there is a convenient combination of ADM variables that allows
one to remove the non-trivial Jordan block along the~$\rho$ direction
that appears in a standard first order reduction. This is true due to
the specific gauge choice and its construction upon outgoing null
geodesics. Crucially however, this special combination is only
possible along the~$\rho$ direction but not~$\theta,\phi$. So, away
from spherical symmetry the EFE in the affine null gauge are only WH.

\subsection{Pure gauge sub-block: radial direction}
\label{subsec:bondi-hyp:aff_null:pg_sub-block_rho_dir}

We now demonstrate how the radial part of the pure gauge
subsystem~\eqref{eqn:aff_null_pg} is inherited by the linearized
EFE. For brevity in this subsection we work in spherical symmetry,
which is sufficient, since the coupled gauge variables in the radial
Jordan block of~\eqref{eqn:aff_null_pg} are present already under this
assumption.

\subsubsection{ADM setup}

In spherical symmetry the principal part of the linearized ADM
equations in outgoing affine null gauge is
\begin{subequations}
  \label{eqn:ADM_lin_mink_aff_null_spher}
  \begin{align}
  \p_t \delta \gamma_{\rho \rho}
  &\simeq -2 \delta K_{\rho \rho}
    - 2 \p_\rho \delta \gamma_{\rho \rho}
    \,,
    \label{eqn:ADM_lin_mink_aff_null_spher_a}
  \\
  \p_t \delta K_{\rho \rho}
  & \simeq
  \tfrac{1}{2} \p_\rho^2 \delta \gamma_{\rho \rho}
  - \rho^{-2} \p_\rho^2
    \delta \gamma_{\theta \theta}
    \,,
    \label{eqn:ADM_lin_mink_aff_null_spher_b}
  \\
  \p_t \delta \gamma_{\theta \theta}
  & \simeq -2 \delta K_{\theta \theta}
    \,,
    \label{eqn:ADM_lin_mink_aff_null_spher_c}
  \\
  \p_t \delta K_{\theta \theta}
  & \simeq
  - \tfrac{1}{2} \p_\rho^2 \delta \gamma_{\theta \theta}
    \,.
    \label{eqn:ADM_lin_mink_aff_null_spher_d}
\end{align}
\end{subequations}
From Eq.~\eqref{eqn:gauge_vars}, the gauge variables along the~$\rho$
direction in spherical symmetry are
\begin{align}
  &
    -\delta K_{\rho \rho} = [\p_\rho^2 \Theta]
  \,,
  \qquad 
    \tfrac{1}{2} \p_\rho \delta \gamma_{\rho \rho}
    = [\p_\rho^2 \psi^\rho]
    \,.
    \label{eqn:gauge_ADM_vars_spher}
\end{align}
To recover the pure gauge structure it suffices to analyze the
coupling between~\eqref{eqn:ADM_lin_mink_aff_null_spher_a}
and~\eqref{eqn:ADM_lin_mink_aff_null_spher_b}
\begin{subequations}
\label{eqn:ADM_lin_mink_aff_null_spher_2}
\begin{align}
 \p_r (\tfrac{1}{2} \delta \gamma_{\rho \rho})
  & \simeq
    - \delta K_{\rho \rho}
    -  \p_\rho (\tfrac{1}{2}\delta \gamma_{\rho \rho})
    \,,
    \label{eqn:ADM_lin_mink_aff_null_spher_2_a}
  \\
    \label{eqn:ADM_lin_mink_aff_null_spher_2_b}
 \p_r ( \delta K_{\rho \rho}
    + \tfrac{1}{2} \p_\rho \delta \gamma_{\rho \rho})
  & \simeq - \rho^{-2} \p_\rho^2 \delta \gamma_{\theta \theta}
    \,,
\end{align}
\end{subequations}
where~$\p_r=\p_t+\p_\rho$ is an outgoing null vector,
and~\eqref{eqn:ADM_lin_mink_aff_null_spher_2_b} results from a linear
combination of~\eqref{eqn:ADM_lin_mink_aff_null_spher_a}
and~\eqref{eqn:ADM_lin_mink_aff_null_spher_b}. The right-hand-side
of~\eqref{eqn:ADM_lin_mink_aff_null_spher_2_b} involves the constraint
variable
\begin{align*}
  [H] = - \p_\rho \delta \gamma_{qq}
  = - 2\rho^{-2} \p_\rho
  \delta \gamma_{\theta \theta}\,.
\end{align*}
In a standard first order reduction, the term~$ (\p_\rho \delta
\gamma_{\rho \rho}) $ would be introduced as an evolved variable
satisfying
\begin{align}
  \p_r  (\tfrac{1}{2}\p_\rho \delta \gamma_{\rho \rho})
  & \simeq
    - \p_\rho \delta K_{\rho \rho}
    - \p_\rho (\tfrac{1}{2}\p_\rho \delta \gamma_{\rho \rho})
    \,.
    \label{eqn:ADM_lin_mink_aff_null_spher_3}  
\end{align}
The above and~\eqref{eqn:ADM_lin_mink_aff_null_spher_2_b} expressed in
terms of gauge and constraint variables read
\begin{align*}
  & \p_r [\p_\rho^2 \psi^\rho]
  + \p_\rho [ \p_\rho^2 \left( \psi^\rho - \Theta \right)]
    \simeq 0
    \,,
    \nonumber
  \\
  & \p_r
    [\p_\rho^2 \left(\Theta - \psi^\rho \right)]
    \simeq \tfrac{1}{2} \p_\rho [H]
    \,.
\end{align*}
As explained in Sec.~\ref{sec:PG:discussion}, this system has a
pure gauge part which consists of the coupling among the gauge
variables~$\Theta$ and~$\psi^\rho$ and a part that captures the
coupling of the gauge to the constraint variables. The pure gauge
part~$\mathbf{P}_G$ is obtained by neglecting the
term~$\p_\rho [H]/2$. This part has the same principal structure as
the pure gauge subsystem~\eqref{eqn:aff_null_pg} in the radial
direction, since it is just an overall~$\p_\rho^2$ derivative of the
latter. This is in accordance with the result of~\cite{HilRic13},
because for a standard first order reduction~$\mathbf{P}_G$ inherits
the structure of the first order system formed
by~$(\Theta, \psi_i,\delta \alpha,\delta \beta_i)$. The
term~$\p_\rho [H]/2$ is encoded in the~$\mathbf{P}_{GC}$ sub-block of
the full principal symbol~$\mathbf{P}^\rho$.

Next, let us consider a reduction in
which~$(\p_r \delta \gamma_{\rho \rho})$ is introduced as an auxiliary
variable rather than~$(\p_\rho \delta \gamma_{\rho
  \rho})$. From~\eqref{eqn:ADM_lin_mink_aff_null_spher_2_a}
and~\eqref{eqn:gauge_ADM_vars_spher} we get
\begin{align}
  \p_r (\tfrac{1}{2}\delta \gamma_{\rho \rho})
  =[ \p_r \p_\rho  \psi^\rho]
  \simeq [\p_\rho^2 (\Theta- \psi^\rho)]\,,
  \label{eqn:gauge_ADM_vars_spher_null_dev_a}
\end{align}
where in the first step we are just using our normal naming convention
with~$[\dots]$, and likewise in the second
Eq.~\eqref{eqn:gauge_ADM_vars_spher}. Similarly, from
Eq.~\eqref{eqn:gauge_ADM_vars_spher} we get
\begin{align}
  \tfrac{1}{2} \p_\rho \delta \gamma_{\rho \rho}
  + \delta K_{\rho \rho}
  = [\p_\rho^2(\psi^\rho - \Theta)] =
  [\p_r \p_\rho  \psi^\rho] =
  - [\p_r \p_\rho  \Theta]
  \,,
  \label{eqn:gauge_ADM_vars_spher_null_dev_b}
\end{align}
where in the second step Eq.~\eqref{eqn:aff_null_pg_b} and in the
third Eq.~\eqref{eqn:aff_null_pg_a} are used. The equation of motion
for the auxiliary variable~$(\p_r\delta \gamma_{\rho\rho})$ results
from~\eqref{eqn:ADM_lin_mink_aff_null_spher_2_a} after acting
with~$\p_r$, namely
\begin{align}
\p_r (\tfrac{1}{2}\p_r\delta \gamma_{\rho \rho})
  \simeq - \p_r ( \delta K_{\rho \rho}
  + \tfrac{1}{2} \p_\rho \delta \gamma_{\rho \rho})
    \simeq
    \rho^{-2} \p_\rho^2 \delta \gamma_{\theta \theta}
    \,,
    \label{eqn:p_r_delta_gamma_rhorho_eom}
\end{align}
where in the second step
Eq.~\eqref{eqn:ADM_lin_mink_aff_null_spher_2_b} is used. The above
together with~Eq.~\eqref{eqn:ADM_lin_mink_aff_null_spher_2_b} in terms
of the gauge and constraint variables read
\begin{subequations}
  \label{eqn:pure_gauge_sub-block_2nd_order_spher}
  \begin{align}
     \p_r [  \p_r \p_\rho \psi^\rho]
    &
      \simeq - \tfrac{1}{2} \p_\rho [H]
  \,,
  \label{eqn:pure_gauge_sub-block_2nd_order_spher_a}
  \\
    \p_r [ \p_r \p_\rho  \Theta]
    &
      \simeq \tfrac{1}{2} \p_\rho [H]
  \,,
  \label{eqn:pure_gauge_sub-block_2nd_order_spher_b}
\end{align}
\end{subequations}
where the
relations~\eqref{eqn:gauge_ADM_vars_spher_null_dev_a},~\eqref{eqn:gauge_ADM_vars_spher_null_dev_b}
have been used. Thus, the
system~\eqref{eqn:ADM_lin_mink_aff_null_spher_2_b},~\eqref{eqn:p_r_delta_gamma_rhorho_eom}
inherits the principal structure
of~\eqref{eqn:aff_null_pg_2nd_order_a}-\eqref{eqn:aff_null_pg_2nd_order_b}
in~$\mathbf{P}_G$. Again the term~$\p_\rho [H]/2$ is in
the~$\mathbf{P}_{GC}$ sub-block. This result does not
contradict~\cite{HilRic13} due to the non-standard first order
reduction considered. In the outgoing affine null gauge the outgoing
null direction possesses a special role as the foundational piece of
the construction. This construction provides the opportunity to group
ADM variables in such a way that we can avoid the non-trivial Jordan
block in the radial direction.

\subsubsection{Characteristic setup}

The ADM analysis above teaches us which variables inherit the
principal structure of the pure gauge degrees of freedom. However, the
original PDE problem is formulated in the characteristic
domain. In~\cite{HilRic13} the pure gauge structure was identified for
a spacelike foliation. Whether or not this is possible in the
characteristic domain is closely related to the existence of the
previous first order reductions in this domain as well. We show here
that both previous first order reductions and their principal
structure can be realized in the characteristic setup directly.

To demonstrate this consider the affine null gauge in an outgoing
characteristic formulation. The complete calculation can be found in
the ancillary files of~\cite{GiaBisHil21}. We first employ the metric
ansatz
\begin{align*}
  ds^2 = g_{uu} du^2 -2dudr + g_{\theta \theta} d\theta^2 +  g_{\phi \phi} d\phi^2
  \,,
\end{align*}
which for flat space reads
\begin{align*}
  g_{uu}=-1\,,
  \quad
  g_{\theta \theta}=r^2\,,
  \quad
  g_{\phi \phi}=r^2 \sin^2 \theta
  \,.
\end{align*}
Analyzing the main equations~$R_{rr} = R_{\theta \theta} = R_{\phi
  \phi}= 0$ linearized about flat space we see the following structure
\begin{subequations}
  \label{eqn:aff_null_spher_jord_block}
  \begin{align}
    &
    \p_r \delta g_{uu} - \frac{1}{2\rho} \p_\rho
    \left(
    \p_r \delta g_{\theta \theta}
    + \sin^{-2} \theta \p_r \delta g_{\phi \phi}
    \right)
    = 0
    \,,
    \label{eqn:aff_null_spher_jord_block_1}
    \\
    &
    \p_r \left(
    \p_r \delta g_{\theta \theta}
    + \sin^{-2} \theta \p_r \delta g_{\phi \phi}
    \right)
    = 0
    \,.
    \label{eqn:aff_null_spher_jord_block_2}
  \end{align}
\end{subequations}
The
variable~$ \left( \p_r \delta g_{\theta \theta} + \sin^{-2} \theta
  \p_r \delta g_{\phi \phi} \right)$
in~\eqref{eqn:aff_null_spher_jord_block_1} prevents~$\delta g_{uu}$
from satisfying just an advection equation along~$\p_r$ and so
provides a non-trivial Jordan block. The combination
of~$\delta g_{\theta \theta}$ and~$\delta g_{\phi \phi}$ in the former
hints that a different choice of variables may be more
appropriate. This combination of variables furthermore appears in the
trivial equation~$R_{ur}=0$ when linearized about flat space, and so
it may be optimal to group them together. We thus next consider the
equations as resulting from the metric ansatz
\begin{align*}
  ds^2 = g_{uu} du^2 -2dudr + \hat{R}(u,r)^2
  \left( d\theta^2 + \sin^2 \theta d\phi^2 \right)
  \,,
\end{align*}
where~$\hat{R}$ is the radius of the two-sphere. This form of the
metric ansatz is used in the spherically symmetric case
of~\cite{Win13}, employed by~\cite{CreOliWin19} in the study of
gravitational collapse of a massless scalar field, as well as
in~\cite{vdWBis12} for cosmological considerations using past null
cones. Upon linearization about flat space the characteristic PDE
system takes the form
\begin{subequations}
  \label{eqn:char_aff_null_spher}
  \begin{align}
  &
    \p_r^2 \delta \hat{R} = 0
  \,,
  \label{eqn:char_aff_null_spher_a}
  \\
  &
    2 r \p_u \p_r \delta \hat{R} + 2 \p_u \delta \hat{R}
    -2 \p_r \delta \hat{R} + r \p_r \delta g_{uu} + \delta g_{uu}
    = 0
    \,,
    \label{eqn:char_aff_null_spher_b}
  \\
  &
    4 \p_u \p_r \delta \hat{R} + r \p_r^2 \delta g_{uu} + 2 \p_r \delta g_{uu}
    =0
    \,.
    \label{eqn:char_aff_null_spher_c}
\end{align}
\end{subequations}
Eq.~\eqref{eqn:char_aff_null_spher_a}
and~\eqref{eqn:char_aff_null_spher_b} correspond to the main
equations~$R_{rr} = 0$ and~$R_{\theta \theta} =0$ respectively, and
Eq.~\eqref{eqn:char_aff_null_spher_c} to the trivial
one~$R_{ur}=0$. The main equation~$R_{\phi \phi}$ is dropped since it
is proportional to~$R_{\theta \theta}$ and the two-sphere is
parameterized only by its radius.

Comparing once more with the ADM form of the problem, including the
trivial equation~\eqref{eqn:char_aff_null_spher_c} in the system
corresponds to including the linearized ADM equation
for~$\delta K_{\rho \rho}$ in the analysis. This is an essential
component in identifying the pure gauge sub-block along the radial
direction. To achieve this we first make the following identification
using Eq.~\eqref{eqn:gen_BS_vars_to_ADM}
\begin{align*}
  g_{uu} = \alpha^{-2} - 2\,,
\end{align*}
which after linearization about flat space yields
\begin{align}
  \delta g_{uu} = -2 \delta \alpha = \delta \gamma_{\rho \rho}
  \,,\label{eqn:dguu_gaugefixing}
\end{align}
where the gauge
condition~$\delta \alpha = - \delta \gamma_{\rho \rho}/2$ is used. We
consider now a first order reduction with
\begin{align*}
  (\p_r \delta \hat{R})
  \,,
  \quad
  (\p_u \delta \hat{R})
  \,,
  \quad
  ( \p_r \delta g_{uu})
\end{align*}
promoted to independent variables where,
by~\eqref{eqn:dguu_gaugefixing}, the latter is equivalent to~$(\p_r
\delta \gamma_{\rho \rho})$ being treated as a reduction
variable. This first order reduction provides a diagonalizable radial
principal part for~\eqref{eqn:char_aff_null_spher} with advection
equations along~$\p_r$ for all variables--original and auxiliary--and
corresponds to the pure gauge
subsystem~\eqref{eqn:aff_null_pg_2nd_order}. More precisely, the
relation between the ADM gauge variables and the characteristic
variables is
\begin{subequations}
  \label{eqn:char_vars_to_gauge_ADM_spher}
  \begin{align}
    \tfrac{1}{2} \p_\rho \delta \gamma_{\rho \rho}
    &
      = \tfrac{1}{2}
      ( \p_r \delta g_{uu} - \p_u \delta g_{uu})
      \,,
      \label{eqn:char_vars_to_gauge_ADM_spher_a}
    \\
    - \delta K_{\rho \rho}
    &
      = \p_r \delta g_{uu}
      - \tfrac{1}{2} \p_u \delta g_{uu}
      \,.
      \label{eqn:char_vars_to_gauge_ADM_spher_b}
  \end{align}
\end{subequations}
Since all characteristic variables satisfy advection equations
along~$\p_r$, combining~\eqref{eqn:char_vars_to_gauge_ADM_spher}
with~\eqref{eqn:gauge_ADM_vars_spher_null_dev_a},~\eqref{eqn:gauge_ADM_vars_spher_null_dev_b}
one recovers~\eqref{eqn:pure_gauge_sub-block_2nd_order_spher}.

If~$(\p_u \delta g_{uu})$ is also taken as an auxiliary variable, then
the first order reduction is of the standard type, since
\begin{align*}
  \p_\rho \delta g_{uu} = \p_r \delta g_{uu} - \p_u \delta g_{uu}
  \,.
\end{align*}
The equation of motion for~$(\p_u \delta g_{uu})$ can be obtained from
\begin{align*}
  \p_r ( \p_u \delta g_{uu})  = \p_u (\p_r \delta g_{uu})
  \,.
\end{align*}
This first order reduction of~\eqref{eqn:char_aff_null_spher}
possesses the following non-trivial Jordan block
\begin{align*}
  (\p_t+\p_\rho) (\p_u \delta g_{uu} ) + \p_\rho (\p_r \delta g_{uu}) = 0
  \,,
  \\
  (\p_t + \p_\rho) ( \p_r \delta g_{uu} ) = 0
  \,,
\end{align*}
and a linear combination yields
\begin{align*}
  (\p_t + \p_\rho) \left[
  ( \p_r \delta g_{uu}) - (\p_u \delta g_{uu})
  \right]
  = \p_\rho ( \p_r \delta g_{uu} )
\,.
\end{align*}
Via the identification~\eqref{eqn:char_vars_to_gauge_ADM_spher} the
latter matches~\eqref{eqn:ADM_lin_mink_aff_null_spher_3}, modulo an
overall factor of~$1/2$. Hence, the Jordan block of the characteristic
PDE with this characteristic standard first order reduction coincides
precisely with the pure gauge principal
part~\eqref{eqn:aff_null_pg_a},~\eqref{eqn:aff_null_pg_b}. This is
merely the characteristic version of the standard first order
reduction in the Cauchy frame. The alternative choice where, instead
of introducing both~$(\p_u \delta g_{uu})$ and~$(\p_r \delta g_{uu})$
as auxiliary variables, only the latter is introduced, renders the
characteristic PDE system in spherical symmetry strongly
hyperbolic. Consequently, the initial value problem of this system is
not well-posed in a norm where both~$(\p_t \delta g_{uu})^2$
and~$(\p_\rho \delta g_{uu})^2$ are included in the integrand, but in
one that involves only~$ (\p_r \delta g_{uu})^2$. Based on this norm,
one can study well-posedness of the CIBVP of the system by seeking
energy estimates. In Chap.~\ref{chap:bondi_well-p} we discuss this
topic further. In~\cite{Bal97} similar type of energy estimates for
the wave and Maxwell equations in a single-null characteristic setup
were provided. Energy estimates for the CIBVP of the wave equation
were also provided in \cite{KreWin11} for asymptotically flat and in
\cite{ZhaWu20} for asymptotically AdS spacetimes.

\subsection{Pure gauge sub-block:
  angular direction~$\theta$}
\label{subsec:bondi-hyp:aff_null:pg_sub-block_theta_dir}

We next expand the previous analysis to a setup without symmetry,
focusing purely on the angular direction~$\theta$. The pure gauge
structure is identified in both the ADM and characteristic setups. In
contrast, however, to the radial direction there is no combination of
variables that allows us to avoid the non-trivial Jordan block of the
pure gauge. We also discuss which choice of variables is most
convenient for the analysis.

\subsubsection{ADM setup}

The partition into gauge, constraint and physical variables along
the~$\theta$ direction is still achieved using
Eq.~\eqref{eqn:gauge_vars},~\eqref{eqn:constr_vars}
and~\eqref{eqn:phys_projector}, respectively. The gauge variables are
\begin{equation}
      \begin{aligned}
  [\p_\theta^2 \Theta]
  &= - \delta K_{\theta \theta}
  \,, \quad \quad \; \;
  [\p_\theta^2 \psi^\rho] = \p_\theta \delta \gamma_{\rho \theta}
  \,, 
  \\
  [\p_\theta^2 \psi^\theta]
  & = \frac{1}{2 \rho^2} \p_\theta \delta \gamma_{\theta \theta}
    \,, \quad
  [\p_\theta^2 \psi^\phi] = \frac{1}{\rho^2 \sin^2 \theta}
    \p_\theta \delta \gamma_{\theta \phi}
      \,.
    \end{aligned}
    \label{eqn:gauge_vars_theta}
  \end{equation}
The constraint variables are
\begin{align}
  &
    [H]
    = -\p_\theta \delta \gamma_{\rho \rho}
    -  \frac{1}{ \rho^2 \sin^2 \theta} \p_\theta \delta \gamma_{\phi \phi}
    \,, \quad
    [M_\rho] = \delta K_{\rho \theta}
    \,,
    \label{eqn:constr_vars_theta}
  \\
  & \! \!
    [M_\theta]
    = - \delta K_{\rho \rho} -  \frac{1}{\rho^2 \sin^2 \theta} \delta K_{\phi \phi}
    \,, \qquad
    [M_\phi] = \delta K_{\theta \phi}
    \,.
    \nonumber
\end{align}
The physical variables are obtained with the action
of~$^P \! \! \!  \perp$ on~$\delta \gamma_{ij}$ and~$\delta
K_{ij}$. As seen from the physical
subsystem~\eqref{eqn:physical_subsys}, the latter is essentially a
time derivative of the former. We work with the physical variables
\begin{subequations}
  \begin{align}
    [h_+]
    &\equiv \frac{1}{2} \delta \gamma_{\rho \rho}
    - \frac{1}{2 \rho^2 \sin^2 \theta} \delta \gamma_{\phi \phi}
    \,, \quad
    [h_\times] \equiv \delta \gamma_{\rho \phi}
    \,,
          \label{eqn:physical_vars_theta_h}
    \\
    [\dot{h}_+] &\equiv
    \frac{1}{ \rho^2 \sin^2 \theta} \delta K_{\phi \phi}
    - \delta K_{\rho \rho}\,, \quad
    [\dot{h}_\times] \equiv -2 \delta K_{\rho \phi}
    \,,
    \label{eqn:physical_vars_theta_hdot}
  \end{align}
  \label{eqn:physical_vars_theta}%
\end{subequations}
which correspond to the two polarizations of the gravitational waves
in GR. In Eq.~\eqref{eqn:physical_vars_theta_hdot} we have multiplied
with an overall factor of~$-2$ for the definitions to be compatible
with the physical subsystem~\eqref{eqn:physical_subsys}
when~$[\dot{h}_+] =\p_t h_+$, and similarly for~$[h_\times]$. As
expected for a gravitational wave that travels along the~$\theta$
direction, the physical variables involve only spatial metric
components that are transverse to this direction. The principal symbol
in the form~\eqref{eqn:second_order_principal_symbol_eom_form} in
the~$\theta$ direction for the linearized ADM formulation is
\begin{subequations}
  \begin{align}
    \p_t \delta \gamma_{\rho \rho}
    &\simeq - 2 \delta K_{\rho \rho}
      \,,
      \label{eqn:delta_gamma_ADM_theta_princ_a}
    \\
    \p_t \delta \gamma_{\rho \theta}
    &\simeq - 2 \delta K_{\rho \theta} -  \p_\theta \delta \gamma_{\rho \rho}
      \,,
      \label{eqn:delta_gamma_ADM_theta_princ_b}
    \\
    \p_t \delta \gamma_{\rho \phi}
    &\simeq - 2 \delta K_{\rho \phi}
      \,,
      \label{eqn:delta_gamma_ADM_theta_princ_c}
    \\
    \p_t \delta \gamma_{\theta \theta}
    &\simeq - 2 \delta K_{\theta \theta} - 2 \p_\theta \delta \gamma_{\rho \theta}
      \,,
      \label{eqn:delta_gamma_ADM_theta_princ_d}
    \\
    \p_t \delta \gamma_{\theta \phi}
    &\simeq - 2 \delta K_{\theta \phi} - \p_\theta \delta \gamma_{\rho \phi}
      \,,
         \label{eqn:delta_gamma_ADM_theta_princ_e}  
    \\
    \p_t \delta \gamma_{\phi \phi}
    &\simeq - 2 \delta K_{\phi \phi}
      \,,
    \label{eqn:delta_gamma_ADM_theta_princ_f}
  \end{align}
  \label{eqn:delta_gamma_ADM_theta_princ}
\end{subequations}
and
\begin{subequations}
  \label{eqn:delta_K_ADM_theta_princ}
  \begin{align}
    \p_t \delta K_{\rho \rho}
    & \simeq - \frac{1}{2 \rho^2} \p_\theta^2 \delta \gamma_{\rho \rho}
      \,,
      \label{eqn:delta_K_ADM_theta_princ_a}
    \\
    \p_t \delta K_{\rho \theta}
    & \simeq 0
      \,,
      \label{eqn:delta_K_ADM_theta_princ_b}
    \\
    \p_t \delta K_{\rho \phi}
    & \simeq - \frac{1}{2 \rho^2} \p_\theta^2 \delta \gamma_{\rho \phi}
      \,,
      \label{eqn:delta_K_ADM_theta_princ_c}
    \\
    \p_t \delta K_{\theta \theta}
    & \simeq - \frac{1}{2 \rho^2 \sin^2 \theta} \p_\theta^2 \delta \gamma_{\phi \phi}
      \,,
      \label{eqn:delta_K_ADM_theta_princ_d}
    \\
    \p_t \delta K_{\theta \phi}
    & \simeq 0
      \,,
      \label{eqn:delta_K_ADM_theta_princ_e}
    \\
    \p_t \delta K_{\phi \phi}
    & \simeq - \frac{1}{2\rho^2} \p_\theta^2 \delta \gamma_{\phi \phi}
      \,.
      \label{eqn:delta_K_ADM_theta_princ_f}
  \end{align}
\end{subequations}
For a standard first order reduction the pure gauge principal
structure along the~$\theta$ direction is inherited by
\begin{subequations}
  \begin{align}
    \p_t (\tfrac{1}{2\rho^2} \p_\theta \delta \gamma_{\theta \theta})
    & \simeq - \rho^{-2} \p_\theta
      \left( \p_\theta \delta \gamma_{\rho \theta}
      + \delta K_{\theta \theta} \right)
      \,,
      \label{eqn:theta_pg_sub-block_jord_a}
    \\
    \p_t \left( \p_\theta \delta \gamma_{\rho \theta} + \delta K_{\theta \theta} \right)
    & \simeq
      - \p_\theta^2 \delta \gamma_{\rho \rho}
      - \tfrac{1}{2 \rho^2 \sin^2 \theta} \p_\theta^2 \delta \gamma_{\phi \phi}
      - 2 \p_\theta \delta K_{\rho \theta}
      \,.
      \label{eqn:theta_pg_sub-block_jord_b}
  \end{align}
  \label{eqn:theta_pg_sub-block_jord}
\end{subequations}
After using
Eq.\eqref{eqn:gauge_vars_theta},~\eqref{eqn:constr_vars_theta},
\eqref{eqn:physical_vars_theta} the
system~\eqref{eqn:theta_pg_sub-block_jord} yields
\begin{equation}
  \begin{aligned}
    & \p_t [\p_\theta^2 \psi^\theta]
    + \rho^{-2}  \p_\theta [\p_\theta^2 \left( \psi^\rho -\Theta \right)]
    \simeq  0
    \,,
    \\
    & \p_t [\p_\theta^2  \left( \psi^\rho - \Theta \right)]
    \simeq \tfrac{3}{4} \p_\theta [H]
    - 2 \p_\theta [M_\theta]
    - \tfrac{1}{2} \p_\theta^2[h_+]
    \,,
  \end{aligned}
  \label{eqn:ADM_pg_subsystem_theta}
\end{equation}
so that, comparing with~\eqref{eqn:aff_null_pg}, the pure gauge
structure of~$\mathbf{P}_G$ is manifest within the full principal
symbol, as too is the coupling between gauge, constraint and physical
variables encoded in~$\mathbf{P}_{GC}$ and~$\mathbf{P}_{GP}$. Here we
have worked with the plain ADM evolution equations. Working with the
ADM equivalent discussed in
Sec.~\ref{sec:bondi_properties:BS_to_ADM_eqs} changes only the
coupling to the constraints. To obtain this result the necessary
conditions were:
\begin{enumerate}
\item Introduction of the quantities
  $(\p_\theta \delta \gamma_{\theta \theta})$
  and~$(\p_\theta \delta \gamma_{\rho \theta})$ as auxiliary
  variables.
\item Inclusion of the equation of motion for~$\delta K_{\theta
  \theta}$ in the analyzed system.
\end{enumerate}
Interestingly, the affine null gauge provides an explicit example
where the sub-block~$\mathbf{P}_{GP}$ of the full principal
symbol~$\mathbf{P}^s$ is non-vanishing, so there is non-trivial
coupling between gauge and physical variables in the principal symbol.

\subsubsection{Characteristic setup}

We repeat now the previous analysis directly in the characteristic
coordinates and variables to demonstrate how the pure gauge structure
is inherited in~$\mathbf{P}^\theta$ for the characteristic setup. The
ADM analysis is again used as guidance in this.  More specifically,
from the equivalent ADM system~\eqref{eqn:equiv_ADM_to_main_sys} we
know that the characteristic system involves the equation of motion
for~$\delta K_{\theta \theta}$, which is one of the two necessary
conditions in order to recover the structure we are looking for. We
parameterize the metric functions simply by
~$g_{uu},\,g_{u\theta},\,g_{u\phi},\,g_{\theta\theta},\,g_{\phi
  \theta},\, g_{\phi \phi}$. For the present calculations this choice,
as opposed to that of~\cite{Win13}, is preferred due to its cleaner
connection to the ADM variables, and allows us to uncover the pure
gauge structure more easily.

With this parameterization the PDE system consisting of the main
equations~\eqref{eqn:main_BS_sys} does not involve terms of the
form~$\p_\theta^2 \delta g_{u \theta}$
and~$\p_\theta^2 \delta g_{\theta \theta}$, which in the ADM language
correspond to~$\p_\theta^2 \delta \gamma_{\rho \theta}$
and~$\p_\theta^2 \delta \gamma_{\theta \theta}$. A minimal first order
reduction of the characteristic system, the details of which can be
found in the ancillary files of~\cite{GiaBisHil21}, exhibits the
following Jordan block in the~$\theta$ direction
\begin{align*}
  &
    \p_t \delta g_{uu}
    + \frac{1}{2 \rho \sin^2 \theta} \p_t (\p_r \delta g_{\theta \theta})
    - \frac{1}{\rho^2} \p_\theta \delta g_{u \theta}
    + \frac{\cot \, \theta}{2 \rho^3} \p_\theta \delta g_{\theta \theta}
    \simeq 0
    \,,
  \\
  & \frac{1}{\rho^2} \p_t \delta g_{u \theta}
    - \frac{\cot \, \theta}{2 \rho^3} \p_t \delta g_{\theta \theta}
    \simeq 0 \,.
\end{align*}
This reduction is minimal in the sense that the minimum number of
auxiliary variables needed to form a complete first order system were
introduced. The above structure motivates the introduction
of~$(\p_\theta \delta g_{u \theta})$
and~$(\p_\theta \delta g_{\theta \theta})$ as auxiliary variables in
addition to the minimum, since they form the non-trivial Jordan
block. But, as we saw earlier, this is the other necessary condition
to recover the pure gauge structure in the full system. Thus in the
new first order reduction the~$2 \times 2$ Jordan block along
the~$\theta$ direction persists, namely
\begin{subequations}
  \begin{align}
    &
      \p_t ( \p_\theta \delta g_{\theta \theta})
    - \rho^2 \p_t (\p_r \delta g_{u \theta})
      - \p_\theta (\p_r \delta g_{\theta \theta})
    - \frac{1}{\sin^2\, \theta} \p_\theta (\p_r \delta g_{\phi \phi})
      \simeq 0
      \,,
      \label{eqn:theta_pg_sub-block_char_aff_null_a}
    \\
    &
      \p_t (\p_r \delta g_{\theta \theta})
      + \frac{1}{\sin^2 \, \theta} \p_t (\p_r \delta g_{\phi \phi})
      \simeq 0
      \,.
      \label{eqn:theta_pg_sub-block_char_aff_null_b}
  \end{align}
  \label{eqn:theta_pg_sub-block_char_aff_null}%
\end{subequations}
The latter is indeed the pure gauge sub-block expected from the ADM
analysis. To realize this explicitly we first express the
characteristic auxiliary variables in terms of the ADM ones
  \begin{align*}
    \p_\theta \delta g_{\theta \theta}
    & =
      \p_\theta \delta \gamma_{\theta \theta}
      \,,
    \\
    \p_r \delta g_{\theta \theta}
    & =
      (\p_t + \p_\rho) \delta \gamma_{\theta \theta}
      \simeq -2 \delta K_{\theta \theta} - 2 \p_\theta \delta \gamma_{\rho \theta}
      \,,
    \\
    \p_r \delta g_{u \theta}
    & = (\p_t + \p_\rho) \delta \gamma_{\rho \theta}
      \simeq -2 \delta K_{\rho \theta} - \p_\theta \delta \gamma_{\rho \rho}
      \,,
    \\
    \p_r \delta g_{\phi \phi}
    & = (\p_t + \p_\rho) \delta \gamma_{\phi \phi}
      \simeq - 2 \delta K_{\phi \phi}
      \,,
  \end{align*}
where we have dropped derivatives transverse
to~$\p_{\theta}$. Then,~\eqref{eqn:theta_pg_sub-block_char_aff_null}
reads
\begin{align*}
  &
    \p_t \p_\theta \delta \gamma_{\theta \theta}
    + 2 \rho^2 \p_t \delta K_{\rho \theta}
    + \rho^2 \p_\theta \p_t \delta \gamma_{\rho \rho}
    + 2 \p_\theta \delta K_{\theta \theta}
    + 2 \p_\theta^2 \delta \gamma_{\rho \theta}
    + \frac{2}{\sin^2 \, \theta} \p_\theta \delta K_{\phi \phi}
    \simeq 0
    \,,
  \\
  &
    \p_t \delta K_{\theta \theta}
    + \p_t \p_\theta \delta \gamma_{\rho \theta}
    + \frac{1}{\sin^2 \, \theta} \p_t \delta K_{\phi \phi}
    \simeq 0
\end{align*}
which after replacing~$\p_t \delta \gamma_{\rho
  \rho}$,~$\p_t \delta K_{\rho \theta}$,~$\p_t \delta K_{\phi \phi}$
with the right-hand-side of \eqref{eqn:delta_gamma_ADM_theta_princ_a},
\eqref{eqn:delta_K_ADM_theta_princ_b},
\eqref{eqn:delta_K_ADM_theta_princ_f} respectively yields
\begin{subequations}
  \begin{align}
    \p_t \left( \frac{1}{2 \rho^2} \p_\theta \delta \gamma_{\theta \theta} \right)
    + \rho^{-2}
    \p_\theta ( \delta K_{\theta \theta} + \p_\theta \delta \gamma_{\rho \theta})
    &\simeq
        \p_\theta \delta K_{\rho \rho}
        - \frac{1}{\rho^2 \sin^2 \, \theta} \p_\theta \delta K_{\phi \phi}
        \,,
        \label{eqn:theta_pg_sub-block_char_aff_null_2_a}
    \\
    \p_t ( \delta K_{\theta \theta} + \p_\theta \delta \gamma_{\rho \theta})
    & \simeq
      \frac{1}{2 \rho^2 \sin^2 \, \theta} \p_\theta^2 \delta \gamma_{\phi \phi}
      \,,
      \label{eqn:theta_pg_sub-block_char_aff_null_2_b}
  \end{align}
  \label{eqn:theta_pg_sub-block_char_aff_null_2}%
\end{subequations}
where in~\eqref{eqn:theta_pg_sub-block_char_aff_null_2_a} we have
multiplied overall with a factor of~$1/2 \rho^2$. The right-hand-side
of~\eqref{eqn:theta_pg_sub-block_char_aff_null_2} involves only
constraint and physical variables along the~$\theta$ direction, while
the left-hand-side shows the coupling only between gauge
variables. Using the
relations~\eqref{eqn:gauge_vars_theta},~\eqref{eqn:constr_vars_theta}
and~\eqref{eqn:physical_vars_theta} the
system~\eqref{eqn:theta_pg_sub-block_char_aff_null_2} reads
\begin{subequations}
  \label{eqn:theta_pg_sub-block_char_aff_null_3}%
  \begin{align}
    &
    \p_t [ \p_\theta^2  \psi^\theta]
    + \rho^{-2}
    \p_\theta [ \p_\theta^2 ( \psi^\rho - \Theta)]
    \simeq
    - \p_\theta [\dot{h}_+]
    \label{eqn:theta_pg_sub-block_char_aff_null_3_a}
    \,,
    \\
    &
    \p_t [ \p_\theta^2  ( \psi^\rho - \Theta)]
    \simeq
    - \frac{1}{4} \p_\theta [H]
    + \frac{1}{2} \p_\theta^2 [h_+]
    \,,
    \label{eqn:theta_pg_sub-block_char_aff_null_3_b}
  \end{align}
\end{subequations}
which again inherits the structure of the pure gauge subsystem, namely
the Jordan block~\eqref{eqn:aff_null_pg_a}, \eqref{eqn:aff_null_pg_c},
and provides non-trivial coupling of gauge to constraint and physical
variables. Hence, the non-trivial Jordan block of~$\mathbf{P}^\theta$
in the characteristic affine null system corresponds precisely to the
non-trivial Jordan block of the pure gauge
subsystem~\eqref{eqn:aff_null_pg} along the same direction. Comparing
the form~\eqref{eqn:theta_pg_sub-block_char_aff_null_3} to the
form~\eqref{eqn:ADM_pg_subsystem_theta} in the ADM setup, the only
difference is in the coupling of gauge variables to constraint and
physical ones.

A different choice of variables that makes use of the
definition~\eqref{eqn:gauge_cond_det} is common in affine null
formulations. Such a choice can however make less clear the
distinction between gauge, constraint and physical variables. In the
ancillary files of~\cite{GiaBisHil21} we include analyses where we
explore such parameterizations. Crucially, the principal symbol of the
characteristic system is still non-diagonalizable along~$\theta,\phi$,
but the choice of variables is inconvenient in identifying the
different sub-blocks.

\subsection{Asymptotically anti-de Sitter spacetimes}
\label{subsec:bondi-hyp:aff_null:AAdS}

The affine-null gauge is particularly popular for evolutions in AAdS
spacetimes within the context of holography. Part II of the thesis is
devoted to this research direction. Here, we treat the specific system
that occurs in the case of five-dimensional AAdS spacetimes with
planar symmetry in the affine null gauge choice, but we expect similar
results in setups with less symmetry.
The metric is written as
\begin{align}
  ds^2 = -A d\varv^2 + \Sigma^2
  \left[ e^B dx_\perp^2 + e^{-2B} dz^2 \right]
      + 2dR \,d\varv + 2 F d\varv dz\, . \label{EF_metric ansatz}
\end{align}
Here~$\varv$ denotes the advanced time, $R$ is called the holographic
coordinate, and increases from the bulk of the spacetime towards the
boundary. All metric components are functions of~$(\varv, R, z)$. We
also denote by~$dx_\perp^2$ the flat metric in the plane spanned
by~$x_\perp$, the two coordinates associated with the symmetry. Using
the convenient definitions
\begin{equation}
\begin{aligned}
  d_z \equiv \p_z -F \p_R
  \, ,
  \quad
  d_+
  \equiv \p_\varv + \tfrac{A}{2} \p_R \,,
\end{aligned}
\label{tilde_dot_def}
\end{equation}
the field equations can be succinctly stated, and are
\begin{equation}
\begin{aligned}
 & \p_R^2\Sigma
  = - \frac{1}{2} \left( \p_R B \right)^2 \Sigma
   \, , \\
 &\Sigma^2 \, \p_R^2 F
 = \Sigma \left( 6 \, d_z \Sigma \, \p_R B + 4 \, \p_R d_z \Sigma
 + 3 \, \p_R F\, \p_R \Sigma \right)
   + \Sigma^2 \left( 3 \, d_z B \, \p_R B + 2\, \p_R d_z B\right)
   - 4\, d_z \Sigma \, \p_R \Sigma
   \, , \\
 & 12 \Sigma^3 \p_R d_+ \Sigma
 = - 8\, \Sigma^2 \left(-3 \Sigma^2  + 3\, d_+\Sigma \, \p_R \Sigma \right)
 + e^{2B}\Big\{ \Sigma^2
   \left[
   4\, d_z \, B \p_RF - 4\, d_z^2 B - 7 \left(d_z B\right)^2 \right.
    \\
    & \qquad \qquad \qquad \;
    \left. + 2\, \p_Rd_z F + \left( \p_RF\right)^2 \right]
   + 4 \left( d_z \Sigma\right)^2
   + 2\, \Sigma \left[ d_z\Sigma \left(\p_R F
     - 8\, d_z B \right) - 4\, d_z^2 \Sigma \right]
     \Big\} \,,  \\
     &6 \Sigma^4 \p_R d_+ B
     =  - 9\, \Sigma^3 \left( \p_R \Sigma \, d_+ B
       + \p_R B \, d_+ \Sigma \right)
   + e^{2B}\Big\{
   \Sigma^2 \big[ \left(d_zB\right)^2 - d_z B\, \p_R F + d_z^2 B
     \\
 &\qquad \qquad \qquad
     - 2\, \p_R d_z F - \left(\p_R F\right)^2 \big]
   -4\, \left( d_z \Sigma \right)^2
  + \Sigma \left[ d_z\Sigma
    \left( d_z B + 4\, \p_R F \right)
   + 2\, d_z^2 \Sigma \right] 
 \Big\}\, , \\
 &6 \Sigma^4 \p_R^2 A = 72 \, \Sigma^2 \, d_+ \Sigma  \, \p_R\Sigma
       - 2 \Sigma^4 \left( 9\, \p_R B \,  d_+ B
         +12 \right)
       +3\, e^{2B}
       \left\{
         \Sigma^2
         \left[ 4\, d_z^2 B +7 \left( d_z B \right)^2
           - \left( \p_R F \right)^2  \right]
       \right.  \\
       & \left.
         \qquad\quad\qquad 
   + \,8 \Sigma \left( 2\, d_z B \, d_z \Sigma
   + d_z^2 \Sigma \right) - 4\left(d_z\Sigma\right)^2
\right\}\,,
\end{aligned}
\label{dr_eqs}
\end{equation}
and finally
\begin{align}
\p_\varv B = d_+ B - \tfrac{A}{2} \p_R B\, . \label{dtB_eq}
\end{align}
We still analyze only the free evolution system in vacuum, but with a
non-vanishing cosmological constant. The vector~$d_+$ points to the
direction of the outgoing null rays and hence equations~\eqref{dr_eqs}
do involve derivatives extrinsic to the hypersurfaces of constant
time. However, if one considers~$d_+ B$ and~$d_+ \Sigma$ as
independent variables of the system, then equations~\eqref{dr_eqs} are
intrinsic to the ingoing null hypersurfaces. Furthermore, this choice
of variables provides intrinsic equations with a nested structure. The
only equation that involves derivatives extrinsic to the hypersurfaces
of constant retarded time is~\eqref{dtB_eq}. To analyze the
hyperbolicity of the resulting PDE system we follow the steps
described in the beginning of the chapter, namely first order
reduction, linearization and coordinate transformation to an auxiliary
Cauchy-type setup.

\subsubsection{First order reduction and Linearization}

The definition~\eqref{tilde_dot_def} was used earlier to write the
field equations in a more compact form, but for the rest of the
analysis we expand out the definition of~$d_z$. Before performing the
first order reduction, we apply the coordinate transformation~$r=1/R$,
drawing the boundary to~$r=0$. The metric components however still
exhibit singular behavior there, so as elsewhere in the
literature,~\cite{CheYaf14}, we apply appropriate field redefinitions
to obtain regular fields on the boundary, namely
\begin{align*}
  A(\varv, r, z)
  &\rightarrow \frac{1}{r^2} + r^2 A(\varv, r, z)
    \,, \quad \,\;
  B(\varv, r, z)
      \rightarrow r^4 B(\varv, r, z)
      \,,
  \\
  \Sigma(\varv, r, z)
  & \rightarrow \frac{1}{r} + r^3 \Sigma (\varv, r, z)
    \,,
    \qquad
  F(\varv, r, z)
      \rightarrow r^2 F(\varv, r, z)
  \,,
\end{align*}
and similarly for derivatives of the above fields. To simplify the
presentation we linearize here about vacuum AdS. Our conclusions are
however unaltered if we work about an arbitrary background. Full
expressions in the general case can be found
in~\cite{GiaHilZil20_public}. We define reduction variables according
to
\begin{align*}
  &A_r = \p_r A
    \,,\quad
    B_r = \p_r B
    \, , \quad
    F_r = \p_r F
    \, ,\quad
    \Sigma_r = \p_r \Sigma \,,\\
  &A_z = \p_z A
    \, ,\quad
    B_z = \p_z B
    \, ,\quad
    F_z = \p_z F
    \, ,\quad
    \Sigma_z = \p_z \Sigma \, ,
  \\
  &\!B_+ = d_+ B
    \, , \;\;
    \Sigma_+ = d_+ \Sigma \, .
\end{align*}
The complete first order system is then
\begin{equation}
\begin{aligned}
    r^4 \p_\varv B &= -S_1 \, , 
    \\
    r^4 \p_\varv B_r&=
    \frac{ r^4}{2} \p_r B_r + r^3 \p_r B_+ - S_2
    \, , 
    \\
    -6 r \p_r B_+
    &= 2 r^2 \p_r F_z + r^2 \p_z B_z + 2 r^2 \p_z \Sigma_z
    - \, S_3 \, ,
    \\
    \p_r B_z
    &= \p_z B_r
    \, , 
    \\
    \p_r \Sigma
    &= -S_5
    \, , 
    \\
    r^7 \p_r \Sigma_r
    &= -S_6 
    \, ,                      
    \\
    12 r  \p_r \Sigma_+
    &=
    2 r^2 \p_r F_z + 4 r^2 \p_z B_z + 8 r^2 \p_z \Sigma_z  - \, S_7
    \, , 
    \\
    \p_r \Sigma_z
    &= \p_z \Sigma_r
    \, ,
    \\
    \p_r F
    &= -S_9
    \, ,
    \\
    r^4 \p_r F_r
    &=
    -4 r^4 \p_r \Sigma_z - 2r^4 \p_r B_z  - \, S_{10} \, , 
    \\
    \p_r F_z
    &= \p_z F_r
    \, ,
    \\
    \p_r A
    &= -S_{12}
    \, ,
    \\
    6 r^2  \p_r A_r
    &=
    12 r^2 \p_z B_z + 24 r^2 \p_z \Sigma_z - \, S_{13} 
    \, ,
    \\
    \p_r A_z
    &= \p_z A_r
    \, ,
  \end{aligned}
  \label{eqn:Lin_Sys_Vac_AdS}
\end{equation}
which can be written as
\begin{align} 
  \mathbfcal{A}^\varv \p_\varv \mathbf{u} + \mathbfcal{A}^r \p_r \mathbf{u}
  + \mathbfcal{A}^z \p_z \mathbf{u}
  + \mathbfcal{S} = 0, \label{AdS_system_matrix_form_1}
\end{align}
with state vector
\begin{align}
  \mathbf{u} = \left( A_r, B_+, \Sigma_+, \Sigma_r, F_r, B_z,
  \Sigma_z, B_r, A_z, F_z, A, F, B, \Sigma \right)^T\,.\nonumber
\end{align}
The principal part matrix associated with the retarded
time~$\mathbfcal{A}^\varv$ is not invertible as expected for a
characteristic setup and hence we proceed with a transformation to an
appropriate auxiliary frame.

\subsubsection{Coordinate transformation}

To obtain a suitable coordinate frame we transform from~$(\varv,r,z)$
to~$(t,\rho,z)$ with
\begin{align*}
  \varv = t-\rho \, , \qquad r=\rho \, ,
\end{align*}
and the remaining coordinates unaltered, which gives
\begin{align*}
  \p_\varv = \p_t \, ,
  \qquad \p_r = \p_t + \p_\rho  \, ,
\end{align*}
with~$\p_z$ unaffected. Applying the transformation yields
\begin{align*}
    \mathbfcal{A}^t \, \p_t  \mathbf{u} 
    + \mathbfcal{A}^r \, \p_\rho  \mathbf{u}
    + \mathbfcal{A}^z \, \p_z  \mathbf{u} \, + \mathbfcal{S} = 0
    \, ,
\end{align*}
where now $\mathbfcal{A}^t = \mathbfcal{A}^\varv + \mathbfcal{A}^r$ is
invertible. After multiplying from the left with the inverse
of~$\mathbfcal{A}^t$ we again bring the system to the form
\begin{align}
   \p_t  \mathbf{u} 
  + \mathbf{B}^\rho \, \p_\rho  \mathbf{u}
  + \mathbf{B}^z \, \p_z  \mathbf{u} \, + \mathbf{S} = 0
  \, , \label{AdS_new_frame_system_2}
\end{align}
where~$\mathbf{B}^\rho = \left( \mathbfcal{A}^t \right)^{-1}
\mathbfcal{A}^r$ and
$\mathbf{B}^z = \left( \mathbfcal{A}^t \right)^{-1}
\mathbfcal{A}^z$. The principal part~$\mathbf{B}^\rho$ is
diagonalizable with real eigenvalues~$0$ and~$\pm 1$. The principal
part~$\mathbf{B}^z$ has the same real eigenvalues but it does not have
a complete set of eigenvectors, so it is not diagonalizable. The
system resulting from this specific first order reduction is thus only
WH. Next, by constructing generalized characteristic variables in
the~$z$ direction we will examine whether or not an appropriate
addition of the reduction constraints can render the reduction
strongly hyperbolic. The reduction constraints are
\begin{equation}
\begin{aligned}
  &\p_z A - A_z = 0 \,, \qquad 
  \p_z B - B_z = 0 \, ,\\
  &\p_z \Sigma - \Sigma_z = 0 \, ,\qquad 
   \p_z F - F_z = 0 \,,\\
  &\p_zB_\rho - \p_\rho B_z
  = \tfrac{1}{2}\p_zB_r - \p_z B_+ - \p_\rho B_z=0 \,,\\
  &\p_z\Sigma_\rho - \p_\rho \Sigma_z
  = \tfrac{1}{2}\p_z\Sigma_r - \p_z\Sigma_+ - \p_\rho \Sigma_z = 0\,.
\end{aligned}
\label{AN_red_constr_all}
\end{equation}

\subsubsection{Generalized characteristic variables}

The eigenvalues of~$\mathbf{B}^z$ are~$\lambda = \pm 1$ with algebraic
multiplicity one and~$\lambda =0$ with algebraic multiplicity
twelve. There is one eigenvector for~$\lambda=1$, one for~$\lambda=-1$
and nine for~$\lambda=0$. Since the algebraic and geometric
multiplicity of~$\lambda=0$ differ by three, the Jordan normal
form
\begin{align*}
  \mathbf{J}^z \equiv \mathbf{T}_z^{-1} \,
  \mathbf{B}^z \, \mathbf{T}_z \,,
\end{align*}
must have some non-trivial block. Let us consider the~$t,z$ part
of~\eqref{AdS_new_frame_system_2} and, use~$\mathbf{T}_z^{-1}$ to
construct the generalized characteristic variables in the~$z$
direction,
\begin{align}
\mathbf{v} = \mathbf{T}_z^{-1} \, \mathbf{u}
\end{align}
satisfying
\begin{align}
  &\p_t \mathbf{v} + \mathbf{J}_z  \, \p_z \mathbf{v}
   \simeq 0 \, , \label{tz_gen_char_system}
\end{align}
with~$\simeq$ here denoting equality up to transverse derivatives and
non-principal terms. The components of~$\mathbf{v}$ begin,
\begin{align*}
 & - B_r - \frac{1}{3} B_z - \frac{2}{3}F_r
   -2 \Sigma_r  -\frac{2}{3} \Sigma_z\, ,\\
 & -B_r + \frac{1}{3} B_z + \frac{2}{3} F_r
   - 2 \Sigma_r + \frac{2}{3}\Sigma_z\,,
\end{align*}
with speeds~$\mp 1$ respectively. Next we have those with vanishing
speeds, which are most naturally presented in three blocks. The first
of these consists of the set of {\it true} characteristic variables,
\begin{align*}
  &  B_+ - \frac{\rho}{2} B_r - \rho \Sigma_r\,,\quad
  \Sigma_+ - \frac{\rho}{8} A_r + \frac{\rho}{4} B_r
  + \frac{\rho}{2} \Sigma_r\,,\\
  &\frac{1}{4}A_r + \frac{3}{2} B_r + F_z + 3 \Sigma_r
   \, ,\quad A \, ,\quad  F \, ,\quad B \, ,\quad \Sigma \, ,
\end{align*}
a coupled pair consisting of one generalized and one characteristic
variable, respectively,
\begin{align}
  & -\frac{4}{3}B_z - \frac{2}{3} F_r - \frac{2}{3}\Sigma_z
        \, ,\qquad - 2 \Sigma_r \, ,\label{AAdS_char_var_pair}
\end{align}
and finally a coupled triplet of two generalized characteristic
variables and one characteristic variable, respectively,
\begin{equation}
\begin{aligned}
& \frac{1}{4}A_z + \frac{1}{6} B_z + \frac{1}{3}F_r
  + \frac{1}{3}\Sigma_z \, ,\quad
  -\frac{1}{4}A_r + \frac{1}{2}B_r + \Sigma_r\, , \\
& \frac{2}{3} B_z + \frac{1}{3} F_r
  + \frac{4}{3} \Sigma_z\, .
\end{aligned}
\label{AAdS_char_var_triple}
\end{equation}
In other words, from the structure of the Jordan blocks
of~$\mathbf{J}^z$, reading off the components
of~\eqref{tz_gen_char_system} the first member of the
pair~\eqref{AAdS_char_var_pair} and the first two members of the
triple~\eqref{AAdS_char_var_triple} we have the schematic form,
\begin{align}
  \p_t v_i + \p_z v_{i+1} \simeq 0 \,, \label{Jordan_pathology}
\end{align}
with~$v_i$ referring to the field and~$v_{i+1}$ the next element of
the pair or triple. The question is whether or not there exists an
appropriate addition of the reduction
constraints~\eqref{AN_red_constr_all} such that equations of the
form~\eqref{Jordan_pathology} are turned into equations of the form
\begin{align}
  \p_t v_i + \lambda_i \, \p_z v_i \simeq 0 \,,
  \label{jordan_pathology_cured}
\end{align}
where we are allowing different first order reductions to adjust also
characteristic speeds. This is a necessary condition for building an
alternative reduction that is SH. This would mean that the generalized
characteristic variable~$v_i$ that is originally coupled
with~$v_{i+1}$ could be decoupled, and the respective generalized
eigenvector replaced by a simple eigenvector. We examine this for the
second two elements of the triplet~\eqref{AAdS_char_var_triple} and
show by contradiction that this necessary condition can not be
fulfilled. With our original, specific reduction we have
\begin{equation}
\begin{aligned}
  & \p_t \left(
    \frac{2}{3} B_z + \frac{1}{3} F_r + \frac{4}{3} \Sigma_z\right)
    \simeq 0 \, , \\
  & \p_t \left( -\frac{1}{4}A_r + \frac{1}{2}B_r + \Sigma_r \right)
  +  \p_z \left(
    \frac{2}{3} B_z + \frac{1}{3} F_r + \frac{4}{3} \Sigma_z\right)
    \simeq 0 \,.
  \end{aligned}
  \label{tz_gen_char_eq12}
\end{equation}
Observe, first of all, that neither of these two equations, nor the
two large terms grouped separately in the second, can be written as a
linear combination (equality taken here in the sense of~$\simeq$) of
the reduction constraints~\eqref{AN_red_constr_all}. The choice of
reduction lies in the freedom to add multiples of the six reduction
constraints~\eqref{AN_red_constr_all} to the evolution
equations. Suppose that some choice of addition of these constraints
did result in a SH first order reduction. Starting with the first
equation of~\eqref{tz_gen_char_eq12}, for our alternative reduction we
have
\begin{align}
 \p_t \left(
 \frac{2}{3} B_z + \frac{1}{3} F_r
 + \frac{4}{3} \Sigma_z\right)
 \simeq \sum_\alpha c_\alpha\,C_\alpha\,,\label{eqn:alternative_reduction_cv}
\end{align}
with the terms on the right-hand-side a linear combination of the
reduction constraints~$C_\alpha$. Since this alternative reduction is
SH we have,
\begin{align*}
  \sum_\alpha c_\alpha\,C_\alpha\simeq \sum_\alpha a^{0}_\alpha \p_zv^0_\alpha
  +\sum_\alpha a^{\pm}_\alpha \p_zv^{\pm}_\alpha\,,
\end{align*}
with~$v^0_\alpha$ denoting the set of~$0$-speed characteristic
variables and~$v^{\pm}_\alpha$ denoting the remaining characteristic
variables.
Using~$\p_tv^{\pm}_\alpha\simeq\lambda_\alpha \p_zv^{\pm}_\alpha$ we
may therefore rewrite~\eqref{eqn:alternative_reduction_cv} as
\begin{align*}
 \p_t \left( \frac{2}{3} B_z + \frac{1}{3} F_r + \frac{4}{3}
 \Sigma_z-\sum_\alpha a^{\pm}_\alpha\lambda_\alpha^{-1}
 v^{\pm}_\alpha\right) \simeq
 \sum_\alpha a^{0}_\alpha \p_zv^0_\alpha.
\end{align*}
Now, by our observation directly after~\eqref{tz_gen_char_eq12}, the
term inside the large bracket can not vanish identically. Therefore we
must have~$a^{0}_\alpha=0$ or we have found, on the left-hand-side, a
non-trivial generalized characteristic variable, in contradiction to
the assumption that our reduction is SH. Moving on to the second
equation of~\eqref{tz_gen_char_eq12}, we can write the equivalent
expression for the alternative first order reduction as,
\begin{align*}
  & \p_t \left( -\frac{1}{4}A_r + \frac{1}{2}B_r + \Sigma_r \right) +
  \p_z \left( \frac{2}{3} B_z + \frac{1}{3} F_r + \frac{4}{3}
    \Sigma_z\right)
         \simeq\sum_\alpha
  c'_\alpha\,C_\alpha\,,
\end{align*}
again with the right-hand-side a linear combination of the reduction
constraints. From here a simple calculation shows that
\begin{align*}
  -\frac{1}{4}A_r + \frac{1}{2}B_r + \Sigma_r
  +\sum_\alpha a'_\alpha \lambda_\alpha^{-1}v^{\pm}_\alpha\,,
\end{align*}
is nevertheless {\it still} a non-trivial generalized characteristic
variable for a suitable choice of~$a'_\alpha$. By contradiction we
have therefore shown that there is no first order reduction that gives
a SH first order PDE system in the~$(t,\rho,z)$ frame used here.

\section{The Bondi-Sachs gauge proper}
\label{sec:bondi-hyp:BS_proper}

In the outgoing Bondi-Sachs proper gauge the coordinate light speed
conditions~$c_+^\rho=1,\,c_+^A=0$ are imposed--as in all outgoing
Bondi-like gauges--and lead to
\begin{align*}
  \alpha L^{-1} - \beta^\rho =1
  \,,
  \qquad
  \beta^A = - b^A \alpha L^{-1}
  \,,
\end{align*}
in terms of lapse and shift. The gauge is closed by setting
\begin{align}
  \rho = \hat{R}  \,.
  \label{eqn:Bondi_Radius}
\end{align} 
In this form the gauge fixing is not so easily expressed in an ADM
setup, since we do not have a complete specification of the lapse and
shift. We can however achieve this by combining the ADM
equations~\eqref{eqn:ADM_lin_mink}, the~$2+1$
split~\eqref{eqn:2+1_gamma_split} of the spatial metric~$\gamma_{ij}$
and the determinant condition~\eqref{eqn:Bondi_Radius}. We basically
want to specify a~$\beta^\rho$ for which the determinant
condition~\eqref{eqn:Bondi_Radius} is satisfied at later
times. Starting from the standard ADM equations on the two-sphere we
get
\begin{align}
  \mathcal{L}_t q_{AB}
    & = - 2 \alpha \, \prescript{(q)\!\!}{}\perp K_{AB}
      + \mathcal{L}_{[\beta^\rho \p_\rho]\,} q_{AB}
      - \mathcal{L}_{[(1+\beta^\rho) b]\,}
      q_{AB}
       \,,
       \label{eqn:angular_ADM_eom}
\end{align} 
where~$\prescript{(q)\!\!}{}\perp$ denotes the projection with respect
to~$q_{AB}$ on every open index and~$b^a$ denotes the slip vector. The
general relation between the derivative of a matrix and the derivative
of its determinant applied to~$q_{AB}$ yields
\begin{align*}
  q^{ab} \mathcal{L}_t q_{ab} = q^{ab} \p_t q_{ab} = \p_t \ln(q)
  \,,
\end{align*}
where~$q \equiv \det(q)$. Imposing the determinant
condition~\eqref{eqn:Bondi_Radius} the latter
yields~$q^{ab} \mathcal{L}_t q_{ab} = 0$. Then,
Eq.~\eqref{eqn:angular_ADM_eom} after tracing with~$q^{AB}$ returns
\begin{align*}
  0
  & =
    -2 \alpha K_{qq}
    + \beta^\rho \left[
    \p_\rho \ln(q) - 2 \slashed{D}_A b^A  
    \right] - 2 \slashed{D}_A b^A
    \,,
\end{align*}
where~$\slashed{D}_A$ is the covariant derivative compatible
with~$q_{AB}$. Using~$c_+^\rho=1=-\beta^\rho+\alpha/L$ we finally
obtain~$\beta^\rho = \rho \, X /(4 - \rho \, X)$ with
\begin{align*}
  X = 2 L K_{qq} + 2 \slashed{D}_a b^a
\end{align*}
and~$\p_\rho \ln(q) = 4/\rho$. In terms of the lapse and shift the
Bondi-Sachs proper gauge can thus be imposed by
\begin{equation}
  \begin{aligned}
    \alpha
    &= L (1 + \beta^\rho)
    \,, \quad
    \beta^\rho
    &= \frac{X \, \rho/4}{1 - X \, \rho/4}
    \,, \quad
    \beta^\theta
    &= - b^\theta \alpha L^{-1}
    \,, \quad
    \beta^\phi
    &= - b^\phi \alpha L^{-1}
    \,,
  \end{aligned}
  \label{eqn:BS_gauge_2}
\end{equation}
which is a mixed algebraic-differential gauge.

\subsection{Pure gauge subsystem}
\label{subsec:bondi-hyp:BS_proper:pg_subsys}

To proceed with our analysis we first need to obtain the pure gauge
subsystem~\eqref{eqn:PG_subsys_lin_mink} for the Bondi-Sachs gauge.
We continue in the linear constant coefficient approximation. Under
this assumption the Bondi-Sachs proper gauge~\eqref{eqn:BS_gauge_2}
reads
\begin{equation}
\label{eqn:BS_gauge_lin_mink}
\begin{aligned}
  \delta \alpha
  &= \delta \beta^\rho+ \frac{1}{2} \delta \gamma_{\rho \rho}
    \,,
  \\
  \delta \beta^\rho
  & = \frac{\delta K_{\theta \theta}}{2\rho}
    + \frac{\delta K_{\phi \phi}}{2 \rho \sin^2 \theta}
    + \frac{\p_\theta \delta \gamma_{\rho \theta}}{2\rho}
    + \frac{\p_\phi \delta \gamma_{\rho \phi}}{2 \rho \sin^2 \theta}
    + \frac{\cot \theta \, \delta \gamma_{\rho \theta}}{2 \rho}
    \,,
  \\
  \delta \beta^\theta
  &= - \rho^{-2} \delta \gamma_{\rho \theta}
    \,,
  \\
  \delta \beta^\phi
  &= - (\rho \sin \theta)^{-2} \delta \gamma_{\rho \phi}
    \,.
  \end{aligned}
\end{equation}
Replacing these in Eq.~\eqref{eqn:PG_subsys_lin_mink} and using the
relations~\eqref{eqn:gauge_vars_theta} to translate to the gauge
variables, the pure gauge subsystem of the Bondi-Sachs proper gauge
reads
\begin{equation}
      \label{eqn:BS_PG_sys_2nd_order}
\begin{aligned}
  & \p_t \Theta
+ \frac{1}{2\rho} \p_\theta^2 \Theta
    + \frac{1}{2 \rho \sin^2 \theta} \p_\phi^2 \Theta
    - \frac{1}{2 \rho} \p_\theta^2 \psi^\rho
    - \frac{1}{2 \rho \sin^2 \theta} \p_\phi^2 \psi^\rho
    \\
  & \quad \; \; \,
    - \frac{\rho}{2} \p_\rho \p_\theta \psi^\theta
    - \frac{\rho}{2} \p_\rho \p_\phi \psi^\phi
    - \p_\rho \psi^\rho
    - \frac{\cot \theta}{2\rho} \p_\theta \psi^\rho
    - \frac{\rho \cot \theta}{2} \p_\rho \psi^\theta
    = 0\,,
  \\
  &
    \p_t \psi^\rho
    + \frac{1}{2\rho} \p_\theta^2 \Theta
    + \frac{1}{2 \rho \sin^2 \theta} \p_\phi^2 \Theta
    - \frac{1}{2 \rho} \p_\theta^2 \psi^\rho
    - \frac{1}{2 \rho \sin^2 \theta} \p_\phi^2 \psi^\rho \\
  & \quad \; \; \; \,
    - \frac{\rho}{2} \p_\rho \p_\theta \psi^\theta
    - \frac{\rho}{2} \p_\rho \p_\phi \psi^\phi
    - \p_\rho \Theta
    - \frac{\cot \theta}{2\rho} \p_\theta \psi^\rho
    - \frac{\rho \cot \theta}{2} \psi_\rho \psi^\theta
    = 0\,, \\
  &
    \p_t \psi^\theta
    + \p_\rho \psi^\theta
    + \rho^{-2} \p_\theta ( \psi^\rho - \Theta) = 0\,, \\
  &\p_t \psi^\phi
    + \p_\rho \psi^\phi
    + (\rho \sin \theta)^{-2} \p_\phi (\psi^\rho -\Theta)
    =0\,.
  \end{aligned}
\end{equation}
To analyze the hyperbolicity of this second order in space system we
consider a first order reduction with variables
\begin{align*}
  &
  \Theta-\psi^\rho
  \,, \quad
  \p_\theta(\Theta-\psi^\rho)
  \,, \quad
  \p_\phi(\Theta-\psi^\rho)
  \,, \quad
     \Theta+\psi^\rho
  \,, \quad
  \psi^\theta
  \,, \quad
  \p_\theta \psi^\theta
  \,, \quad
  \psi^\phi
  \,, \quad
  \p_\phi \psi^\phi
    \,.
\end{align*}
The minimal first order reduction of this system reads
\begingroup \allowdisplaybreaks
\begin{subequations}
  \begin{align}
    &
      \p_t( \Theta - \psi^\rho) + \p_\rho(\Theta - \psi^\rho)
      =0
      \,,
      \label{eqn:BS_PG_sys_1st_order_a}
    \\
    &
      \p_t [\p_\theta( \Theta - \psi^\rho)] + \p_\rho [\p_\theta(\Theta - \psi^\rho)]
      =0
      \,,
      \label{eqn:BS_PG_sys_1st_order_b}
    \\
    &
      \p_t [\p_\phi( \Theta - \psi^\rho)] + \p_\rho [\p_\phi(\Theta - \psi^\rho)]
      =0
      \,,
      \label{eqn:BS_PG_sys_1st_order_c}
    \\
    &
      \p_t( \Theta + \psi^\rho) - \p_\rho( \Theta+\psi^\rho)
      -\frac{\cot \theta}{2 \rho} \p_\theta(\Theta+\psi^\rho)
      \nonumber
    \\
    & \qquad
      +\rho^{-1} \p_\theta [ \p_\theta(\Theta-\psi^\rho)]
      + \rho^{-1} \sin^{-2}\theta \p_\phi [\p_\phi(\Theta - \psi^\rho)]
      +\frac{\cot \theta}{2 \rho} \p_\theta(\Theta-\psi^\rho)
      \nonumber
    \\
    & \qquad
      + \rho \cot \theta \p_\rho \psi^\theta
      - \rho \p_\rho (\p_\theta \psi^\theta)
      - \rho \p_\rho (\p_\phi \psi^\phi)
      = 0
      \,,
      \label{eqn:BS_PG_sys_1st_order_d}
    \\
    &
      \p_t \psi^\theta + \p_\rho \psi^\theta
      -\rho^{-2}[\p_\theta(\Theta-\psi^\rho)]
      = 0
      \,,
      \label{eqn:BS_PG_sys_1st_order_e}
    \\
    &
      \p_t (\p_\theta\psi^\theta) + \p_\rho (\p_\theta \psi^\theta)
      - \rho^{-2}\p_\theta [\p_\theta ( \Theta - \psi^\rho)]
      = 0
      \,,
      \label{eqn:BS_PG_sys_1st_order_f}
    \\
    &
      \p_t \psi^\phi + \p_\rho \psi^\phi
      -\rho^{-2}\sin^{-2}\theta [\p_\phi(\Theta-\psi^\phi)]
      = 0
      \,,
      \label{eqn:BS_PG_sys_1st_order_g}
    \\
    &
      \p_t (\p_\phi \psi^\phi) + \p_\rho (\p_\phi \psi^\phi)
      - (\rho \sin \theta)^{-2} \p_\phi [ \p_\phi ( \Theta - \psi^\rho)]
      = 0
      \,.
      \label{eqn:BS_PG_sys_1st_order_h}
  \end{align}
  \label{eqn:BS_PG_sys_1st_order}%
\end{subequations}
\endgroup
All principal matrices of this system possess real eigenvalues, but
the angular principal matrices are non-diagonalizable. The non-trivial
Jordan block along the~$\theta$ direction is given by (see ancillary
files of~\cite{GiaBisHil21})
\begin{align*}
  &\p_t[\p_\theta(\Theta-\psi^\rho)]
  \simeq 0
  \,,
  \\
  &\p_t (\p_\theta \psi^\theta)
  - \rho^{-2} \p_\theta[\p_\theta(\Theta-\psi^\rho)]
  \simeq 0
  \,,
\end{align*}
and similarly along~$\phi$ by
\begin{align*}
  &\p_t (\p_\phi \psi^\phi)
    - \rho^{-2}\sin^{-2}\theta \p_\phi[\p_\phi(\Theta-\psi^\rho)]
    \simeq 0
    \,,
  \\
  &\p_t[\p_\phi(\Theta-\psi^\rho)]
    \simeq 0
    \,.
\end{align*}
The coupled generalized characteristic variables obtained here
effectively involve second order angular derivatives. Hence, they
cannot be removed with a different first order reduction of the second
order system~\eqref{eqn:BS_PG_sys_2nd_order}. Thus, the analysis based
on the minimal reduction just performed suffices to show that the pure
gauge subsystem of the Bondi-Sachs proper
gauge~\eqref{eqn:BS_PG_sys_2nd_order} is only WH.

\subsection{Pure gauge sub-block: angular direction~$\theta$}
\label{subsec:bondi-hyp:BS_proper:pg_sub-block_theta_dir}

Similarly to
Subsec.~\ref{subsec:bondi-hyp:aff_null:pg_sub-block_theta_dir} we
present here the set of evolution equations that inherit the structure
of the pure gauge subsystem in the ADM setup, for the Bondi-Sachs
gauge proper. The necessary conditions to uncover this structure
remain the same. The system that captures the structure of the pure
gauge subsystem along the~$\theta$ direction is
\begin{subequations}
  \begin{align}
    -\p_t \left(
      \delta K_{\theta \theta} + \p_\theta \delta \gamma_{\rho \theta}
      \right)
      &\simeq
      \frac{1}{2} \p_\theta^2 \delta \gamma_{\rho \rho}
      + 2 \p_\theta \delta K_{\rho \theta}
       + \frac{1}{2} \p_\theta^2 \delta \gamma_{\rho \rho}
       + \frac{1}{2\rho^2 \sin^2 \theta} \p_\theta^2 \delta \gamma_{\phi \phi}
       \,,
       \label{eqn:ADM_theta_PG_sub-block_a}
    \\
    -\p_t \left(
      \delta K_{\theta \theta} - \p_\theta \delta \gamma_{\rho \theta}
      \right)
      &\simeq
      \frac{1}{2} \p_\theta^2 \delta \gamma_{\rho \rho}
      - 2 \p_\theta \delta K_{\rho \theta}
      + \frac{1}{2} \p_\theta^2 \delta \gamma_{\rho \rho}
      + \frac{\p_\theta^2 \delta \gamma_{\phi \phi}}{2\rho^2 \sin^2 \theta} 
      + \p_\theta^2 \delta \beta_\rho
      \,,
     \label{eqn:ADM_theta_PG_sub-block_b}
    \\
    \frac{1}{2\rho^2}  \p_t (\p_\theta \delta \gamma_{\theta \theta})
      &\simeq
      - \frac{1}{\rho^2} \p_\theta \delta K_{\theta \theta}
      + \frac{1}{\rho^2} \p_\theta^2 \delta \beta_\theta
      \,,
      \label{eqn:ADM_theta_PG_sub-block_c}
    \\
    \frac{1}{\rho^2 \sin^2 \theta}  \p_t (\p_\theta \delta \gamma_{\theta \phi})
      &\simeq
      \frac{-2}{\rho^2 \sin^2 \theta} \p_\theta \delta K_{\theta \phi}
      + \frac{1}{\rho^2 \sin^2 \theta} \p_\theta^2  \delta \beta_\phi
       \,,
       \label{eqn:ADM_theta_PG_sub-block_d}
  \end{align}
  \label{eqn:ADM_theta_PG_sub-block}%
\end{subequations}
where spatial derivatives transverse to~$\theta$ are dropped. This
system results from linear combinations of the linearized about flat
space ADM equations and does not include equations outside the main
system~\eqref{eqn:main_BS_sys}. Combining
Eqs.~\eqref{eqn:BS_gauge_lin_mink},~\eqref{eqn:gauge_vars_theta},
\eqref{eqn:constr_vars_theta},~\eqref{eqn:physical_vars_theta}
and~\eqref{eqn:Cons_subsys_lin_mink}, the
system~\eqref{eqn:ADM_theta_PG_sub-block} yields
\begin{subequations}
  \begin{align}
      \p_t [\p_\theta^2 (\Theta - \psi^\rho)]
      &\simeq
      -\frac{3}{4} \p_\theta [H]
      + 2 \p_\theta [M_\rho]
      + \frac{1}{2}\p_\theta^2 [h_+]
      \,,
      \label{eqn:ADM_theta_PG_sub-block_2_a}
    \\
      \p_t [\p_\theta^2  (\Theta + \psi^\rho)]
      &\simeq
      - \rho^{-1} \p_\theta^2 [ \p_\theta^2(\Theta-\psi^\rho)]
      - \frac{\cot \theta}{\rho} \p_\theta [\p_\theta^2 \psi^\rho]
      \nonumber
    \\
    & \quad \,
      - \frac{3}{4} \p_\theta [H]
      - 2 \p_\theta [M_\rho]
       - \frac{3}{2} \p_\theta^2 [M_\theta]
       + \frac{1}{2} \p_\theta^2 [h_+]
       + \frac{1}{2} \p_\theta^2 [\dot{h}_+]
      \,,
      \label{eqn:ADM_theta_PG_sub-block_2_b}
    \\
      \p_t [\p_\theta^2 \psi^\theta]
      &\simeq
      \rho^{-2} \p_\theta [\p_\theta^2(\Theta-\psi^\rho)]
      \,,
      \label{eqn:ADM_theta_PG_sub-block_2_c}
    \\
      \p_t [\p_\theta^2 \psi^\phi]
      &\simeq
      \frac{-2}{\rho^2 \sin^2 \theta} \p_\theta [M_\theta]
      + \frac{1}{\rho^2 \sin^2 \theta}  \p_\theta^2 [h_\times]
      \,.
      \label{eqn:ADM_theta_PG_sub-block_2_d}
  \end{align}
  \label{eqn:ADM_theta_PG_sub-block_2}%
\end{subequations}
To see how this system inherits the structure of the pure gauge
subsystem~\eqref{eqn:BS_PG_sys_1st_order}, let us neglect all
non-gauge variables. Let us furthermore consider adding to the system
the following equations:~$\p_\theta$ of
\eqref{eqn:ADM_theta_PG_sub-block_a},~$\p_\phi$ of
\eqref{eqn:ADM_theta_PG_sub-block_a}, ~$\p_\theta$ of
\eqref{eqn:ADM_theta_PG_sub-block_c} and~$\p_\phi$ of
\eqref{eqn:ADM_theta_PG_sub-block_d}. As seen from the
form~\eqref{eqn:ADM_theta_PG_sub-block_2} these additional equations
provide the identification to
Eq.~\eqref{eqn:BS_PG_sys_1st_order_b},~\eqref{eqn:BS_PG_sys_1st_order_c},
\eqref{eqn:BS_PG_sys_1st_order_f}
and~\eqref{eqn:BS_PG_sys_1st_order_h}, respectively i.e. the equations
of the auxiliary variables introduced by the minimal first order
reduction. The resulting system is an overall~$\p_\theta^2$ derivative
of the first order reduced pure gauge
subsystem~\eqref{eqn:BS_PG_sys_1st_order}. Thus, the hyperbolic
character of the sub-block~$\mathbf{P}_G$ is that of the pure gauge
subsystem, which is WH. Furthermore, from the
form~\eqref{eqn:ADM_theta_PG_sub-block_2} we see another explicit
example of a Bondi-like gauge where~$\mathbf{P}_{GP} \neq
0$. Identification of the pure gauge structure directly in the
characteristic setup is messy with this radial coordinate, so we do
not discuss it in detail. However, in
Subsec.~\ref{subsec:bondi-hyp:BS_proper:axisym_char} we do show weak
hyperbolicity for the original characteristic system in axisymmetry.

\subsection{Axisymmetry in characteristic variables}
\label{subsec:bondi-hyp:BS_proper:axisym_char}

In Bondi-Sachs gauge~\cite{BonBurMet62,Sac62} a generic 4-dimensional
axially symmetric metric can be written as
\begin{align}
  ds^2 &= \left( \frac{V}{r} e^{2 \beta}
  - U^2 r^2 e^{2\gamma} \right) \, du^2 
         + 2 e^{2 \beta} du \, dr
     \label{BS_metric_ansatz}
  \\
  &
     + 2 U r^2 e^{2 \gamma} \, du \, d\theta
  - r^2 \left( e^{2 \gamma} \, d\theta^2
     + e^{-2 \gamma} \sin^2\theta \, d\phi^2 \right)
     \,.
    \nonumber
\end{align}
Here~$u$ denotes the retarded time, $r$ is the areal radius,
and~$\theta,\phi$ give coordinates on the two-sphere in the standard
way. All metric functions are functions of~$(u, r,\theta)$. The
signature convention chosen here is~$(+,-,-,-)$, which is the same as
in~\cite{Win12}. This convention does not affect the degree of
hyperbolicity of the free evolution PDE system. For axially symmetric
spacetimes the PDE system consists of three equations intrinsic to the
hypersurfaces of constant retarded time,
\begin{equation}
\begin{aligned}
  &\beta_{,r} = \frac{1}{2} r \left( \gamma_{,r} \right)^2
  \, ,\\
  &\left[r^4 e^{2 (\gamma-\beta)} U_{,r} \right]_{,r}
  =
   2r^2\left[r^2 \left( \frac{\beta}{r^2} \right)_{,r\theta}
   - \frac{\left(\sin^2 \theta \, \gamma \right)_{,r \theta}}
   {\sin^2 \theta}
   + 2 \gamma_{,r} \, \gamma_{,\theta} \right] \,,\\
  &V_{,r} = - \frac{1}{4} r^4 e^{2(\gamma-\beta)} \left(U_{,r} \right)^2
   + \frac{\left( r^4 \,
     \sin \theta \,  U\right)_{,r \theta}}{2 r^2 \sin \theta}\\
  & \quad \quad \;\,
  + e^{2(\beta-\gamma)} \Big[ 1 - \frac{\left( \sin\theta \, \beta_{,\theta}
    \right)_{, \theta}}{\sin \theta} + \gamma_{, \theta \theta}
  + 3 \cot \theta \, \gamma_{,\theta}
    - \left( \beta_{,\theta} \right)^2 - 2 \gamma_{,\theta}
    \left(\gamma_{,\theta} - \beta_{, \theta} \right)  \Big]\,,
  \end{aligned}
\end{equation}
and one equation that involves extrinsic derivatives,
\begin{align}
  &4r \left( r \gamma \right)_{,u r}
  = \left\{ 2 r \, \gamma_{,r} \, V - r^2 \left[ 2 \gamma_{,\theta} \, U
    + \sin\theta \left( \frac{U}{\sin \theta}
    \right)_{,\theta} \right] \right\}_{,r}
    \label{evol_eq}
  \\
  & \qquad \qquad
  - 2r^2 \frac{\left(\gamma_{,r} \, U \, \sin \theta
    \right)_{,\theta}}{\sin \theta}
  + \frac{1}{2} r^4 e^{2(\gamma-\beta)} \left(U_{,r} \right)^2
  + 2 e^{2(\beta-\gamma)} \left[ \left( \beta_{,\theta} \right)^2
    + \sin \theta \left( \frac{\beta_{,\theta}}{\sin \theta}
     \right)_{,\theta} \right]
     \, .
  \nonumber
\end{align}
The intrinsic equations possess a nested structure. The above free
evolution scheme is formed by the main
equations~\eqref{eqn:main_BS_sys} in axisymmetry and the supplementary
equations are ignored. To determine the degree of hyperbolicity we
again follow a first order reduction, a linearization about Minkowski,
and a coordinate transformation to an auxiliary Cauchy-type
frame. In~\cite{GiaHilZil20_public} the same analysis for an arbitrary
background can be found. The degree of hyperbolicity is the same for
both cases.

\subsubsection{First order reduction and Linearization}

The minimal set of reduction variables are given by
\begin{align*}
  U_r = \p_r U \,,
  \gamma_r = \p_r \gamma \,,
  \gamma_\theta = \p_\theta \gamma \,,
  \beta_\theta = \p_\theta \beta\,.
\end{align*}
The linearized about flat space first order reduced system reads
\begin{equation}
\begin{aligned}
  &\p_r \beta = 0 \, , \\
  &\p_r U_r  - \frac{2}{r^2}
   \p_r\beta_\theta + \frac{2}{r^2}
   \p_r\gamma_\theta + S_2 = 0 \,, \\
  &\p_r V +  \p_\theta \beta_\theta
   - \p_\theta \gamma_\theta
    - 2 r \p_\theta U - \frac{r^2}{2}
    \p_\theta U_r + S_3 = 0 \,,\\
  & 4 r^2 \p_u \gamma_r + 4 r \p_u \gamma -2 r^2 \,
  \p_r \gamma_r
    + 2 r \, \p_\theta U
    + r^2 \p_\theta U_r - 2 \p_\theta
    \beta_\theta + S_4 = 0 \, ,\\
  & \p_rU + S_5 = 0 \,,\\
  & \p_r \gamma + S_6 = 0 \,,\\
  & \p_r \gamma_\theta - \p_\theta \gamma_r = 0 \,,\\
  & \p_r \beta_\theta = 0 \,,
\end{aligned}
  \label{BS_linear_pde}
\end{equation}
where~$S_i$ denotes the various source terms and as earlier we work in
the frozen coefficient approximation, so that~$r$ and so forth must be
treated as constants. The variables can be collected in the state
vector
\begin{align*}
  \mathbf{u} = \left( \beta\, , \gamma \, , U \, , V \, ,\gamma_r \, ,
  U_r \, , \beta_\theta \, ,\gamma_\theta  \right)^T\,,
\end{align*}
and the system can be written in the form~\eqref{eqn:gen_hyper_PDE}
with
\begin{align} 
  \mathbfcal{A}^u \p_u \mathbf{u} + \mathbfcal{A}^r \p_r \mathbf{u}
  + \mathbfcal{A}^\theta \p_\theta \mathbf{u} + \mathbfcal{S} = 0
  \label{flat_system_matrix_form_1}
\end{align}
where the principal part matrix~$\mathbfcal{A}^u$ associated with
retarded time~$u$ is not invertible.

\subsubsection{Coordinate transformation}

After applying the coordinate transformation~\eqref{flat_coord_transf}
and multiplying on the left with the inverse of~$\mathbfcal{A}^t$ we
bring the system to the form,
\begin{align}
  \p_t \mathbf{u} 
  + \mathbf{B}^\rho \, \p_\rho  \mathbf{u}
  + \mathbf{B}^{\hat{\theta}} \, \p_{\hat{\theta}}  \mathbf{u} \, + \mathbf{S} = 0
  \, , \label{flat_new_frame_system_2}
\end{align}
where~$\mathbf{B}^\rho = \left( \mathbfcal{A}^t \right)^{-1} \,
\mathbfcal{A}^r$
and~$\mathbf{B}^{\hat{\theta}} = \rho \left( \mathbfcal{A}^t
\right)^{-1} \, \mathbfcal{A}^\theta$
with~$\p_{\hat{\theta}} \equiv 1/\rho\,\p_\theta \, $,
and~$\mathbf{S}$ was redefined in the obvious manner. For our system,
the principal part matrix~$\mathbf{B}^\rho$ is diagonalizable with
real eigenvalues.  Although~$\mathbf{B}^{\hat{\theta}}$ has real
eigenvalues, it does not have a complete set of eigenvectors, and
hence is not diagonalizable. Therefore the system resulting from the
specific first order reduction we made is only WH. In~\cite{Fri04} a
subsystem of a similar first-order reduction was shown to be symmetric
hyperbolic. Here, however, we are concerned with the best estimates
that can be made for the full system. In
Sec.~\ref{sec:bondi_well-p:algebraic_char} we elaborate further on
this using toy models.

So far we have not ruled out the existence of an alternative first
order reduction that {\it is} SH however. To examine this possibility
we have to understand if any potential addition of reduction
constraints can make the system SH. The reduction constraints are
\begin{align}
  \p_\theta \beta - \beta_\theta
  &  = 0 \, ,\qquad
  \p_\theta \gamma - \gamma_\theta
  = 0 \, ,
    \label{flat_red_constr}
\end{align}
The definitions of the variables~$\gamma_r$ and~$U_r$ are solved
explicitly as time evolution equations within the
system~\eqref{BS_linear_pde} and therefore do not have an associated
constraint. Using the generalized characteristic variables of the
system we examine next this subtlety, along with the form of the
degeneracy.

\subsubsection{Generalized characteristic variables}

The generalized characteristic
variables~$\mathbf{v}\equiv \mathbf{T}_{\hat{\theta}}^{-1} \,
\mathbf{u}$ with speed (eigenvalue) zero are
\begin{align*}
  \rho \, U + \frac{\rho^2}{2} \,U_r -\beta_\theta
  + \gamma_\theta
  \,,\quad
  \beta_\theta \,,
  \quad
  V,\, \quad
  \rho \left(
  -2\rho U - \frac{\rho^2}{2} U_r
  + \beta_\theta - \gamma_\theta
  \right)
  \,,\quad
  \gamma\,,
  \quad
  \beta \,,
\end{align*}
of which the third and fourth are associated with the
non-trivial~$2\times2$ Jordan block
within
\begin{align*}
    \mathbf{J}^{\hat{\theta}} \equiv \mathbf{T}_{\hat{\theta}}^{-1} \,
    \mathbf{B}^{\hat{\theta}} \, \mathbf{T}_{\hat{\theta}}\,.
\end{align*}
Likewise we have
\begin{align*}
  -\frac{\rho}{2}U + \frac{\rho}{2}\gamma_r
  - \frac{\rho^2}{4} U_r + \frac{1}{2} \beta_\theta
  \,,\quad
  -\frac{\rho}{2}U - \frac{\rho}{2}\gamma_r
  - \frac{\rho^2}{4} U_r + \frac{1}{2} \beta_\theta \,,
\end{align*}
with speeds~$\pm 1$ respectively. The structure
of~$\mathbf{J}_{\hat{\theta}}$ and the relation
\begin{align}
    \p_t \mathbf{v} + \mathbf{J}_{\hat{\theta}}  \, \p_{\hat{\theta}} \mathbf{v}
    \simeq 0 \, ,\label{gen_char_system}
\end{align}
obtained in the frozen coefficient approximation and focusing on
the~$t,\theta$ directions yield
\begin{equation}
\begin{aligned}
&- \p_t\left(2\rho U +\frac{\rho^2}{2} U_r
      - \beta_\theta + \gamma_\theta\right) \simeq 0\,,\\
&\p_tV - \rho \, \p_{\hat{\theta}}\left(2\rho U +\frac{\rho^2}{2} U_r
  - \beta_\theta + \gamma_\theta\right) \simeq 0 \,.
\end{aligned}
\label{K3_gen_char_var}
\end{equation}

Strongly hyperbolic systems admit a complete set of characteristic
variables in each direction. In other words, if our system were
strongly hyperbolic then up to non-principal and transverse derivative
terms each component of~$\mathbf{v}$ would satisfy an advection
equation. Presently the best we can achieve for~$V$ however
is~\eqref{K3_gen_char_var}. Physically we may therefore understand
weak hyperbolicity as the failure of~$V$, a {\it generalized}
characteristic variable, to satisfy such an advection equation. As
mentioned earlier, we could try and cure the equations by using a
different first order reduction. Observe that the choice of different
reductions corresponds to the freedom to add (derivatives of) the
reduction constraints to~\eqref{K3_gen_char_var} without introducing
second derivatives. As~$V$ appears at most once differentiated in the
original equations there is no associated constraint, so we must hope
to eradicate the~$\p_\theta$ term from~\eqref{K3_gen_char_var}
using~\eqref{flat_red_constr} without introducing second
derivatives. Even if the variable~$U_\theta=\p_\theta U$ were
introduced in the reduction however, the~$\p_\theta \beta_\theta$
and~$\p_\theta\gamma_\theta$ terms would obviously persist. Thus one
non-trivial {\it generalized} characteristic variable always survives
and prevents the existence of a complete set of characteristic
variables. Hence within the coordinate basis built
from~$(t,\rho,\theta)$, the field equations are at best only weakly
hyperbolic regardless of the specific reduction. In
Sec.~\ref{sec:bondi_hyp:frame_indep} we show that this result carries
over to other auxiliary frames as well. Notice that the structure that
renders the system of this section only WH is essentially the same as
that of the WH pure gauge subsystem in Bondi-Sachs coordinates, namely
the angular sector.

\section{Double-null and more gauges}
\label{sec:bondi-hyp:double_null}

Another common choice is to use double null coordinates. This was used
in~\cite{Ren90, Chr08, Luk11} to construct initial data on
intersecting ingoing and outgoing null hypersurfaces.~\cite{Ren90}
provided a first well-posedness result for the CIVP in the region near
the intersection, using the harmonic gauge though for the evolution
system, which is symmetric hyperbolic.~\cite{Luk11} improved this
result including in the analysis metric derivatives higher than second
order. A similar approach was used in~\cite{Chr08} as well to analyze
the mathematical conditions for black hole formation. Norm-type
estimates are of course central in these studies, but they are
obtained using PDE systems that are not of the free evolution type and
for which the hyperbolic character is not manifest. If instead one is
interested in analyzing a free evolution system--which is the topic of
the current study--then a certain subset of the systems used
in~\cite{Chr08,Luk11} has to be extracted. There are different choices
on how to construct this subsystem, and in~\cite{HilValZha19} a
specific one was shown to provide a symmetric hyperbolic free
evolution scheme in double-null coordinates. To the best of our
knowledge, an evolution scheme with up to second order metric
derivatives using the double null gauge choice has been used
numerically only in spherical symmetry~\cite{Gar95, GunBauHil19}.

Working with~$f(\rho)=\rho$ in the
coordinate transformation~\eqref{eqn:coord_transf}, the
conditions~$g^{uu}=0$ and~$g^{rr}=0$ yield
\begin{align}
  (\beta^\rho+1)^2 = \alpha^2 \gamma^{\rho \rho}
  \,, \quad
  (\beta^\rho-1)^2 = \alpha^2 \gamma^{\rho \rho}
  \,,
  \label{eqn:double_null_condition_ur}
\end{align}
where the first is the former of the
conditions~\eqref{eqn:gauge_cond_null_gamma_uu} with~$f'=1$. The
conditions~$g^{uA}=0$ are still imposed in the double null gauge,
which provide the latter of
conditions~\eqref{eqn:gauge_cond_null_gamma_uu} with~$f'=1$. From the
coordinate light speed
expressions~\eqref{eqn:radial_coord_light_speed} the
conditions~\eqref{eqn:double_null_condition_ur} yield
\begin{align*}
  c_+^\rho = \pm 1 \,, \quad c_-^\rho = \mp 1
  \,.
\end{align*}
We choose to set~$c_+^\rho = 1$ and~$c_-^\rho = -1$.
Then,~$c_+^\rho + c_-^\rho =0 = -2 \beta^\rho$
implies~$\beta^\rho = 0$, which
from~\eqref{eqn:double_null_condition_ur} leads to~$\alpha =
L$. Replacing these in the second of
conditions~\eqref{eqn:double_null_condition_ur} with~$f'=1$ and
using~\eqref{eqn:slip_vector_relations} provides
$\beta^A = - b^A \alpha L^{-1}$. Then, the whole set of the coordinate
light
speeds~\eqref{eqn:radial_coord_light_speed},~\eqref{eqn:transverse_coord_light_speed}
in the double null gauge reads
\begin{align*}
  c_+^\rho = 1 \,, \quad
  c_-^\rho = -1 \,,
  \quad
  c_+^A = 0
  \,.
\end{align*}
After linearization about Minkowski, the lapse and shift perturbations
read
\begin{align*}
  \delta \alpha
  = -\frac{1}{2} \delta \gamma_{\rho \rho}
  \,, \quad
  \delta \beta^\rho
  = 0
  \,, \quad
  \delta \beta^\theta
  = - \rho^{-2} \delta \gamma_{\rho \theta}
  \,, \quad
  \delta \beta^{\phi} = - \rho^{-2} \sin^2\theta
  \, 
  \delta \gamma_{\rho \phi}
  \,.
\end{align*}
In terms of~$\Theta$ and~$\psi^i$ the above is similar
to~\eqref{eqn:aff_null_gauge_lin_mink} with the only difference that
here~$\delta \beta^\rho = 0$. Then, the pure gauge
subsystem~\eqref{eqn:PG_subsys_lin_mink} for the double null gauge
choice reads
  \begin{align*}
    &
    \p_t \Theta - \p_\rho \psi^\rho =  0
      \,,
    \\
    &
    \p_t \psi^\rho - \p_\rho \Theta =  0
      \,,
    \\
    &
    \p_t \psi^\theta + \p_\rho \psi^\theta
      + \rho^{-2} \p_\theta (\psi^\rho - \Theta) =  0
      \,,
     \\
    &
    \p_t \psi^\phi + \p_\rho \psi^\phi
    + (\rho \sin \theta)^{-2} \p_\phi (\psi^\rho - \Theta) =  0
      \,,
  \end{align*}
which again possesses non-trivial Jordan blocks along the~$\theta$
and~$\phi$ directions and so is only WH. This was expected since the
difference among the affine null, Bondi-Sachs proper and double null
cases with respect to the lapse and shift is only in the specification
of the radial coordinate.

This structure in the pure gauge subsystem of the double null gauge
was already discovered in~\cite{Hil15}. We review it here in order to
stress its differences and similarities with other Bondi-like
gauges. We observe that in all three examples that are presented, the
gauge choice~$\beta^A = - b^A \alpha L^{-1}$ renders the pure gauge
subsystem only WH. This choice implies the condition~$c_+^A=0$. Thus
the pure gauge subsystem will also be WH if~$c_-^A=0$ is instead
imposed. In such a case the difference would be a sign change in the
non-trivial Jordan block along the angular directions. Furthermore,
since the specific nature of the angular coordinates (i.e. coordinates
on a two-sphere) is not essential to the WH, we expect that the pure
gauge subsystem would retain this structure if these coordinates
parameterize level sets of a different topology. Our expectation is
the same for higher dimensional spacetimes. The value of the
cosmological constant does not affect the principal part of the EFEs
and so neither their hyperbolic character. An explicit example that
verifies this expectation is the one of
Subsec.~\ref{subsec:bondi-hyp:aff_null:AAdS}.

In summary, we expect that formulations that result from the EFE,
including up to second order metric derivatives will be at best WH if
they are formulated in a Bondi-like gauge. The claim is based on the
following:
\begin{enumerate}
\item The system admits an equivalent ADM setup.
\item The principal symbol~$\mathbf{P}^s$ has the upper triangular
  form~\eqref{eqn:princ_symbol_triang}.
\item The pure gauge sub-block~$\mathbf{P}_G$ inherits the structure
  of the pure gauge subsystem.
\item The pure gauge subsystem is WH.
\end{enumerate}

Notice however the symmetric hyperbolic free evolution systems
of~\cite{Rip21} in affine null and~\cite{HilValZha19} double null
gauge. These systems include equations with higher than second order
metric derivatives and so do not fall into the category analyzed
here. Since these systems are symmetric hyperbolic, their CI(B)VP is
well-posed in the~$L^2$-norm. However, this is not the~$L^2$-norm of
the systems we analyzed here in those gauges. It is interesting to
understand the possible relation between the second and higher order
metric formulations in the same Bondi-like gauge, as well as the
implication on well-posedness.

\section{Frame independence}
\label{sec:bondi_hyp:frame_indep}

Earlier we presented hyperbolicity analyses of widely used Bondi-like
formulations of GR. We worked with a particular auxiliary Cauchy-type
frame with one timelike element and the remainder spacelike. The
auxiliary basis was used to express the original PDEs in a form that
allowed us to utilize the definitions of
Sec.~\ref{sec:pde_theory:hyp_degree} and show weak hyperbolicity. In
this section we argue that this result persists for other auxiliary
frames. Our argument is based on the dual foliation (DF) approach
of~\cite{Hil15} and follows closely Sec.~II.D
of~\cite{SchHilBug17}. Here, Latin letters~$a\dots e$ are used as
abstract indices, Greek letters run from~$0$ to~$d+1$ for
a~$d+1$-dimensional spacetime and a given basis and Latin
indices~$i,j,k$ denote only the spatial components of this basis. We
also use~$p$ as an abstract index for the spatial derivatives
appearing on the right-hand-side of a first order PDE. The
symbol~$\p_\alpha$ stands for the flat covariant derivative naturally
defined by~$x^\mu$.

The idea of the DF approach is to express a region of spacetime in
terms of two different frames, which we call uppercase and lowercase.
Considering a~$d+1$ split of the spacetime, let us denote as~$n^a$
and~$N^a$ the normal vectors on the hypersurfaces of constant time for
the lower and uppercase frames, respectively. We call~$v^a$ and~$V^a$
the boost vectors for each frame, which are spatial with respect to
the corresponding normal vector. The Lorentz factor is~$W=\left(1- v^a
  v_a\right)^{-1/2}=\left(1- V^a V_a\right)^{-1/2}$ and we denote
as~$\gamma_{ab}$ and~$^{(N)}\!\gamma_{ab}$ the lower and uppercase
spatial metrics. The following useful relations hold
\begin{align}
  \delta^a{}_b
  = \gamma^a{}_b-n^a n_b
  = {}^{(N)}\!\gamma^a{}_b - N^aN_b
  \,,\quad
  n^a  = W \left(N^a+V^a\right)
  \,,
  \quad
  N^a = W \left(n^a+v^a\right) \,.
  \label{low_up_projectors}
\end{align}
Let us consider a first order PDE in the compact form
\begin{align*}
  \mathbfcal{A}^b \delta^a{}_b \partial_a \mathbf{u}
  + \mathbfcal{S} = 0 \, ,
\end{align*}
and~$d+1$ split using the lower and uppercase frames,
replacing~$\delta^a{}_b$ by means of~\eqref{low_up_projectors},
giving
\begin{align}
  \mathbfcal{A}^n  \partial_n \mathbf{u}
  &\simeq \mathbfcal{A}^b \gamma^a{}_b \partial_a \mathbf{u}
    \,, \quad
  \mathbfcal{A}^N  \partial_N \mathbf{u}
  \simeq \mathbfcal{A}^b \,^{(N)}\!\gamma^a{}_b \partial_a \mathbf{u}
    \, . \label{lowercase_uppercase_PDE}
\end{align}
We obtain two evolution systems for the variables of~$\mathbf{u}$,
with
\begin{equation}
\begin{aligned}
  \mathbfcal{A}^a n_a
  \equiv \mathbfcal{A}^n \, ,
  \quad
  n^a \partial_a \equiv \partial_n
  \, ,\quad
  \mathbfcal{A}^a N_a
  \equiv \mathbfcal{A}^N \, ,
  \quad
  N^a \partial_a \equiv \partial_N\, .
\end{aligned}
  \label{up_low_frames_matrix_defs}
\end{equation}
Without loss of generality we choose to identify the uppercase frame
with the auxiliary frames used in
Subsec.~\ref{subsec:bondi-hyp:aff_null:AAdS}
and~\ref{subsec:bondi-hyp:BS_proper:axisym_char}. The definitions
\begin{align*}
  \mathbfcal{A}^n \equiv \mathbf{A}^n
  \, ,
  \quad 
  \mathbfcal{A}^a \, \gamma^b{}_a \equiv
  \mathbf{A}^b
  \, , \quad
  \mathbfcal{A}^N  \equiv \mathbf{B}^N
  \, ,
  \quad
  \mathbfcal{A}^a \, ^{(N)}\!\gamma^b{}_a \equiv
  \mathbf{B}^b \, , 
\end{align*}
imply~$\mathbf{B}^b N_b=0$,~$\mathbf{A}^b n_b=0$ and lead to the
following upper and lowercase first order PDE forms
\begin{align}
  \p_N \mathbf{u} &= \mathbf{B}^p \p_p \mathbf{u} - \mathbf{S}
  \,, \quad
  \mathbf{A}^n \p_n \mathbf{u} = \mathbf{A}^p \p_p \mathbf{u} - \mathbf{S}
  \,, \label{uppercase_lowercase_new_frame_PDE}
\end{align}
where~$\mathbf{B}^N = \mathbb{1}$ by assumption. The former is the
same form as in equations~\eqref{flat_new_frame_system_2}
and~\eqref{AdS_new_frame_system_2}. In this form we found the PDE
systems only WH due the~$2\times2$ Jordan blocks of the angular
principal parts. This can be represented in a generalized eigenvalue
problem of the form
\begin{align}
  \mathbf{l}^N_{\lambda_N}
  \left(\mathbf{P}^S - \mathbb{1} \lambda_N\right)^M
  = 0 \, , \label{up_gen_eigenv_prob}
\end{align}
where~$S^a$ is a unit spatial vector,~$\mathbf{P}^S \equiv
\mathbf{B}^a S_a$ the principal symbol and~$M$ is the rank of the
generalized left eigenvector~$\mathbf{l}^N_{\lambda_N}$ with
eigenvalue~$\lambda_N$, with~$M=2$ for the generalized eigenvectors
that correspond to the aforementioned Jordan blocks. We wish to
examine if generalized eigenvalue problems of this form exist also in
the lowercase frame. Hence we need to relate the two equations
of~\eqref{uppercase_lowercase_new_frame_PDE}, obtaining
\begin{equation}
\begin{aligned}
  \mathbf{A}^n=W(\mathbb{1} + \mathbf{B}^V)
  \,,\quad
  \mathbf{A}^p= \mathbf{B}^a(\gamma^p{}_a+WV_av^p)
                -W(\mathbb{1} + \mathbf{B}^V)v^p\,,
\end{aligned}
\label{low_in_up_PDE}
\end{equation}
and
\begin{equation}
\begin{aligned}
  \mathbf{B}^N=\mathbb{1}=W(\mathbf{A}^n + \mathbf{A}^v)
  \,,\quad
  \mathbf{B}^p= \mathbf{A}^a \, ^{(N)}\!\gamma^p{}_a -
                W \mathbf{A}^n V^p \,,
\end{aligned}
\label{up_in_low_PDE}
\end{equation}
where we write~$\mathbf{B}^aV_a\equiv\mathbf{B}^V$. Let us
examine~$\mathbb{1} + \mathbf{B}^V$. In~\cite{SchHilBug17}
invertibility of this matrix was guaranteed by strong
hyperbolicity. Here we want to analyze PDEs that are only WH and so
may not assume that~$\mathbf{B}^V$ is diagonalizable. Hence, let us
denote as
\begin{align*}
\mathbf{J}^{S_V} = \mathbf{T}^{-1}_{S_V} \mathbf{B}^{S_V} \mathbf{T}_{S_V}\,,
\end{align*}
the Jordan normal form of~$\mathbf{B}^{S_V}=\mathbf{B}^a(S_V)_a$,
where~$V^a=|V|S_V^a$ is the uppercase boost vector with norm~$|V|$
pointing in the direction of~$S_V^a$. One can write each
block~$\mathbf{j}$ of the Jordan form~$\mathbf{J}$ with only the
eigenvalue~$\lambda_i$ on the diagonal as
\begin{align*}
\mathbf{j} = \lambda_i \mathbb{1} + \mathbf{N}\,,
\end{align*}
where~$\mathbf{N}$ is a nilpotent matrix of the size of~$\mathbf{j}$
with~$\mathbf{N}^q=0$. Consequently
\begin{align*}
  \mathbf{T}^{-1}_{S_V}\left(\mathbb{1} + \mathbf{B}^V\right)
  \mathbf{T}_{S_V}=\mathbb{1} + \mathbf{J}^{S_V}|V|\,,
\end{align*}
and for each block~$\mathbf{j}^{S_V}$,
\begin{align*}
  \mathbb{1} + \mathbf{j}^{S_V}
  =\tilde{\lambda}^{S_V}_i \left(\mathbb{1}
  + \frac{|V|}{\tilde{\lambda}^{S_V}_i}\mathbf{N}^{S_V}\right)\,,
\end{align*}
assuming that
\begin{align}
  \tilde{\lambda}^{S_V}_i = 1 + |V|\lambda_i^{S_V}
  \neq 0 \,. \label{lambda_tilde_condition}
\end{align}
The inverse of this block is then
\begin{align*}
  \frac{1}{\tilde{\lambda}^{S_V}_i}
  \left[
  \mathbb{1} + \sum_{j=1}^{q-1}
  \left(-\frac{|V|}{\tilde{\lambda}^{S_V}_i}\right)^j
  \left(\mathbf{N}^{S_V}\right)^j
  \right]\,,
\end{align*}
and hence~$\mathbb{1}+\mathbf{B}^V$ is invertible as long as
condition~\eqref{lambda_tilde_condition} is satisfied for
each~$\lambda_i$. Note that in our normalization light-speed
corresponds to~$\lambda=1$. Since~$|V|<1$,
inequality~\eqref{lambda_tilde_condition} is always satisfied for
physical propagation speeds, although could be violated when
superluminal gauge speeds are present. If one considers for instance
the analysis of subsections~\ref{subsec:bondi-hyp:aff_null:AAdS}
and~\ref{subsec:bondi-hyp:BS_proper:axisym_char} on top of vacuum AdS
and Minkowski background respectively, then this condition is
satisfied. We wish to find the equivalent of the uppercase generalized
eigenvalue problem~\eqref{up_gen_eigenv_prob} in the lowercase
frame. Thus, using the second equation of~\eqref{up_in_low_PDE}
and~$S_a=\s_a-W V^Sn_a$~\cite{SchHilBug17,HilSch18} we express the
principal symbol in the lowercase frame, namely
\begin{align*}
\mathbf{P}^S& \equiv \mathbf{B}^a S_a
=\mathbf{A}^a \s_a - \mathbf{A}^nWV^S \,.
\end{align*}
Hence, the equivalent of~\eqref{up_gen_eigenv_prob} in the lowercase
frame is
\begin{align}
  \mathbf{l}^N_{\lambda_N}
  \left[ \mathbf{A}^{(\s-\lambda_N Wv)}
    -W(\lambda_N+V^S)\mathbf{A}^n\right]^M = 0 \,.
  \label{low_gen_eigenv_prob}
\end{align}
Thus if in the uppercase frame the
eigenproblem~\eqref{up_gen_eigenv_prob} with~$M=1$ fails to admit a
complete set of left eigenvectors then so does the lowercase frame,
and so both setups would be at best weakly hyperbolic. To see this we
need only set~$M=1$ in~\eqref{low_gen_eigenv_prob} and note that the
lowercase principal symbol in the~$\s_a-\lambda_N Wv_a$ direction is
proportional to
\begin{align*}
  (\mathbf{A}^n)^{-1}\mathbf{A}^{(\s-\lambda_N Wv)}\,,
\end{align*}
and so deficiency of the lower case principal symbol in this direction
is equivalent to that of the upper case principal symbol stated
before. Unfortunately the relationship between the upper and lowercase
generalized left eigenvectors is more subtle.  Returning to our
specific systems and identifying the uppercase unit spatial
vector~$S^a$ with the unit spatial vectors in the~$\partial_z$
and~$\partial_\theta$ directions of Subsec.
\ref{subsec:bondi-hyp:aff_null:AAdS}
and~\ref{subsec:bondi-hyp:BS_proper:axisym_char} respectively, we
conclude that weak hyperbolicity of those PDEs persists in other
frames.

\section{Conclusions}
\label{sec:bondi_hyp:conclusions}

In this chapter we analyzed the hyperbolic character of some popular
Bondi-like free evolution systems and their pure gauge subsystems. We
found that all of them are weakly hyperbolic due to their structure in
the angular directions. In some examples we were able to explicitly
identify the pure gauge sub-block as the source of the weakly
hyperbolic structure for the principal symbol of the full system. To
show this we had to jump through a number of technical hoops. We
mapped the characteristic free evolution system to an ADM setup so
that the results of~\cite{KhoNov02,HilRic13} could be easily
used. This allowed us to distinguish between the gauge, constraint and
physical degrees in the linear, constant coefficient
approximation. Crucially it is known that weakly hyperbolic pure
gauges give rise to weakly hyperbolic formulations. We were able to
show the former in the affine null, the Bondi-Sachs proper and the
double null gauges. All three have the same degenerate structure
rendering the pure gauge subsystem weakly hyperbolic, which is caused
by the gauge condition~$g^{uA}=0$. We have thus argued that when the
EFE are written in a Bondi-like gauge with at most second derivatives
of the metric and there are non-trivial dynamics in at least two
spatial directions, then, due to the weak hyperbolicity of the pure
gauge subsystem, the resulting PDE system is only WH.

Given the above, the obvious approach to circumvent weak hyperbolicity
is to adopt a different gauge. Yet, symmetric hyperbolic formulations
of GR employing Bondi-like gauges are
known~\cite{Rac13,CabChrTag14,HilValZha19,Rip21}. At first sight this
seems to contradict the claim that any formulation of GR inherits the
pure gauge principal symbol within its own. But these formulations all
promote the curvature to be an evolved variable, so the results
of~\cite{HilRic13} do not apply. As we have seen in
Subsec.~\ref{subsec:bondi-hyp:aff_null:pg_subsys}, taking an outgoing
null derivative of the affine null pure gauge subsystem, we obtain a
strongly hyperbolic PDE. It is thus tempting to revisit the model
of~\cite{HilRic13} to investigate the conjecture that formulations of
GR with evolved curvature can be built that inherit specific
derivatives of the pure gauge subsystem. 
A deeper understanding of this would shed more light into the reason
that these formulations avoid weak hyperbolicity. As mentioned in
Chap.~\ref{chap:pde_theory}, the hyperbolic character of a PDE system
dictates the existence and the form of the norms in which the
respective PDE problems are well-posed. This is the topic of the
following chapter, as well as the implications for applications of
characteristic formulations in precision gravitational wave astronomy.


%% file: sections/well-posedness.tex
\chapter{Well-posedness of Bondi-like PDE problems}
\label{chap:bondi_well-p}

\minitoc

The hyperbolic character of a PDE relates to the well-posedness of its
CIBVP. Here we examine this relation in detail. We first do so by
using toy models that we introduce in
Sec.~\ref{sec:bondi_well-p:toy_PDEs} and which mimic the structure of
Bondi-like characteristic systems. We analyze their algebraic
character for well-posedness as described in
Sec.~\ref{sec:pde_theory:well-posedness}. This determines the
well-posedness of their IVP~\cite{KreLor89,GusKreOli95}, which we
later adjust to the CIBVP. Specifically, we wish to understand what
inequalities, with what norms, can be used to bound solutions in terms
of their given data, and how lower order perturbations affect such
estimates. In Sec.~\ref{sec:bondi_well-p:toy_CCE_CCM} we examine
energy estimates for model CCE and CCM setups based on the toy PDE
systems introduced. In Sec.~\ref{sec:bondi_well-p:axisym_BS} we repeat
the algebraic characterization for the Bondi-Sachs axisymmetric system
of Subsec.~\ref{subsec:bondi-hyp:BS_proper:axisym_char} and comment on
energy estimates. In Sec.~\ref{sec:bondi_well-p:CIBVP_symm_hyp} we
provide an energy estimate for the CIBVP of a model symmetric
hyperbolic PDE system. All the calculations are performed in the
linear constant coefficient approximation.
The material of Secs.~\ref{sec:bondi_well-p:axisym_BS}
and~\ref{sec:bondi_well-p:CIBVP_symm_hyp} is not part of any
publication or preprint and may be included in future ones.

\section{Toy model PDEs}
\label{sec:bondi_well-p:toy_PDEs}

A simple WH PDE model that captures the structure of Bondi-like
systems is the following
\begin{equation}
\begin{aligned}
  & \p_x \phi = -S_\phi
    \, ,\\
  & \p_x \psi_\varv  - \p_z \phi = -S_{\psi_\varv} 
    \, ,\\
  & \p_u \psi - \frac{\left(1-x^2\right)^{3/2}}{2 c_x}  \p_x \psi
    -\p_z\psi = -S_\psi \,,
\end{aligned}
\label{eqn:WH_model}
\end{equation}
where~$x \in [0,1]$,~$z \in [0,2\pi)$ with periodic boundary
conditions, $u \geq u_0$ for some initial time~$u_0$ and~$c_x$ a
constant. This PDE can be written in the form
\begin{align}
  \mathbfcal{A}^u \, \p_u \mathbf{u} + \mathbfcal{A}^x \, \p_x \mathbf{u}
  + \mathbfcal{A}^z \, \p_z \mathbf{u} + \mathbfcal{S} = 0\, ,
\end{align}
where~$\mathbf{u} = (\phi, \psi_\varv, \, \psi )^T$ is the state vector,
and the principal matrices are given by
\begin{align*}
  \mathbfcal{A}^u = \mbox{diag}(0,0,1)
  \,, \quad
  \mathbfcal{A}^x = \mbox{diag}(1,1,\tfrac{-1}{2 c_x}(1-x^2)^{3/2})
  \,,
\end{align*}
and
\begin{align*}
  \mathbfcal{A}^z = \begin{pmatrix}
    0 & 0 & 0\\
   -1 & 0 & 0\\
    0 & 0 & -1 
  \end{pmatrix}\,.
\end{align*}
The source terms are denoted by~$S_\phi, S_{\psi_\varv}$
and~$S_\psi$. The first two Eqs.\ of~\eqref{eqn:WH_model} are
intrinsic to a hypersurface of constant~$u$, whereas the last is the
``evolution equation'' of the system i.e. involves also directional
derivatives pointing outside a null hypersurface. The angular
principal part~$\mathbfcal{A}^z$ is not diagonalizable since it has
a~$2\times2$ Jordan block for the intrinsic equations, mimicking the
core structure of the previously analyzed Bondi-like PDEs. One may
think of this model as a simplified analog of these systems with a
compactified radial coordinate, similar to the way that Bondi-like
formulations are used for characteristic extraction. This role can be
played by the coordinate~$x$ with~$c_x$ a constant involved in the
compactification. More specifically
\begin{align*}
  x = \frac{r-r_\textrm{min}}{\sqrt{c_x^2 + (r-r_\textrm{min})^2}} \,,
\end{align*}
where~$r_\textrm{min}$ is the minimum physical radius that we consider
and the factor~$c_x$ controls the density of points towards~$r
\rightarrow \infty$, if we were to map the compactified grid~$x$ to
the physical radius grid~$r$.

By removing the angular derivative from the second intrinsic
equation~\eqref{eqn:WH_model} we obtain our SH toy model
\begin{equation}
\begin{aligned}
  & \p_x \phi = -S_\phi
        \, ,\\
  & \p_x \psi_\varv  = -S_{\psi_\varv} 
        \, ,\\
  & \p_u \psi - \frac{\left(1-x^2\right)^{3/2}}{2 c_x}  \p_x \psi
    -\p_z\psi = -S_\psi
    \, ,
  \end{aligned}
  \label{eqn:SH_model}
\end{equation}
which has the same principal part matrices~$\mathbfcal{A}^u$
and~$\mathbfcal{A}^x$ as before, but has diagonal~$\mathbfcal{A}^z$. We
employ this model for comparison between numerical results with SH and
WH systems. The PDE problem for both systems~\eqref{eqn:WH_model}
and~\eqref{eqn:SH_model} has as domain
\begin{align*}
  x \in [0,1]
  \,, \quad
  z \in [0,2\pi)
  \,, \quad
  u \in [u_0, u_f]
  \,,
\end{align*}
for some initial and final times~$u_0$ and~$u_f$ respectively. We
apply periodic boundary conditions in the~$z$ direction for
simplicity. The initial and boundary data are
\begin{align}
  \psi_* \equiv \psi(u_0,x,z)
\end{align}
and
\begin{align}
  \hat{\phi} \equiv \phi(u,0,z)
  \,, \qquad
  \hat{\psi_\varv} \equiv \psi_\varv(u,0,z)
  \,,\label{eqn:model_boundary_data}
\end{align}
respectively and are freely specifiable.

\section{Algebraic determination of
  well-posedness}
\label{sec:bondi_well-p:algebraic_char}

We wish to apply the tools of Sec.~\ref{sec:pde_theory:well-posedness}
to the toy models. For this we want to write the system in the
form~\eqref{eqn:prototype_1st_order_lin_const_coef_PDE} where the time
principal part is the identity matrix. We achieve the latter via a
coordinate transformation similar to~\eqref{eqn:coord_transf}, namely
\begin{align*}
  u=t-\rho \, , \qquad  x = \rho \, ,  \qquad  z = z\,.
\end{align*}
Starting from Eqs.~\eqref{eqn:WH_model}, we bring the system to
the form
\begin{align*}
  \p_t \phi & = - \p_\rho \phi  - S_\phi
    \, ,\\
  \p_t \psi_\varv  &= - \p_\rho \psi_\varv + \p_z \phi -  S_{\psi_\varv} 
    \, ,\\
  \p_t \psi &= F \, \p_\rho \psi
    + G  \, \p_z\psi - G \, S_\psi\,,
\end{align*}
where
\begin{align*}
  F = \frac{\left( 1 - \rho^2 \right)^{3/2}}
  {2 c_x - \left( 1 - \rho^2 \right)^{3/2}}\,,
  \quad
  G = \frac{2 c_x}{2 c_x - \left( 1 - \rho^2 \right)^{3/2}}
\end{align*}
are fixed real constants for fixed~$\rho$ and~$c_x$, with non-zero
denominator for our~$\rho$ domain and an appropriately
chosen~$c_x$. In this frame the principal parts
are~$ \mathbfcal{B}^t=\mathbb{1}$ and
\begin{align*}
  \mathbfcal{B}^\rho = \begin{pmatrix}
    -1 & 0 & 0\\
    0 & -1 & 0\\
    0 & 0 & F 
  \end{pmatrix}
  ,\qquad
  \mathbfcal{B}^z = \begin{pmatrix}
    0 & 0 & 0\\
    1 & 0 & 0\\
    0 & 0 & G 
  \end{pmatrix}
  .
\end{align*}
This is the auxiliary Cauchy-type setup for the WH model, similarly to
the earlier Bondi-like PDEs in Chap.~\ref{chap:bondi-hyp}. After
applying a Fourier transformation, the principal symbol for the WH
model is
\begin{align*}
  i\omega_p\mathbfcal{B}^p =
  i \omega_\rho \mathbfcal{B}^\rho
  + i \omega_z \mathbfcal{B}^z \, .
\end{align*}
By algebraic characterization of well-posedness we mean the study of
the inequalities the symbol~$\mathbf{P}(i \omega)$ as defined in
Eq.~\eqref{eqn:symbol_P_Fourier} satisfies. These inequalities inform
us on the existence and the form of the norms that can be used to
control the solution for the IVP of a given PDE system.

\subsection{Homogeneous WH model}
\label{subsec:bondi_well-p:algebraic_char:homo_WH}

Focusing first on the homogeneous WH model
where~$S_\phi = S_{\psi_\varv} = S_\psi =0$, we obtain
\begin{align}
  e^{(i\hat{\omega}_p\mathbfcal{B}^p)|\omega| t}=
  \begin{pmatrix}
    e^{-i |\omega| \hat{\omega}_\rho t} & 0 & 0\\
    i |\omega| \hat{\omega}_z t \, e^{-i |\omega| \hat{\omega}_\rho t}  &
    e^{-i |\omega| \hat{\omega}_\rho t} & 0\\
    0 & 0 & e^{i |\omega| (F \hat{\omega}_\rho + G \hat{\omega}_z) t}
  \end{pmatrix}
            \, , \label{exp_matrix_homogeneous_WH}
\end{align}
where we express the wavevector as
\begin{align*}
  \omega_p = |\omega| \hat{\omega}_p\,,
\end{align*}
with~$|\omega|$ its magnitude so that~$\hat{\omega}_\rho^2 +
\hat{\omega}_z^2 =1$. The norm of~\eqref{exp_matrix_homogeneous_WH} is
(see chapter~2 of~\cite{SarTig12} for useful definitions)
\begin{align}
  \left|e^{(i\hat{\omega}_p\mathbfcal{B}^p)|\omega| t}\right|^2
  = 1
  + \frac{|\omega|^2 \hat{\omega}_z^2 t^2}{2}
  + \left[ \left(1 + \frac{|\omega|^2 \hat{\omega}_z^2 t^2}{2} \right)^2
  - 1  \right]^{1/2}.
  \label{homogeneous_WH_norm}
\end{align}
This norm behaves as~$|\omega| t$ for large~$|\omega|$ and so the
homogeneous WH model obeys an inequality of the
form~\eqref{eqn:weak_wp_inequality}, with~$q=1$. Hence, this PDE is
only weakly well-posed, and so satisfies an estimate in
some~$||\cdot||_q$-norm, which we call~\textit{lopsided}. This norm is
specified for our system in
Sec.~\ref{sec:bondi_well-p:toy_CCE_CCM}. If one would discard from the
previous analysis the equation for~$\psi_\varv$ of the homogeneous WH
model~\eqref{eqn:WH_model} since it is decoupled, the remaining
subsystem would be symmetric hyperbolic and one might expect
well-posedness of the full PDE problem in the~$L^2$-norm. However, it
is important to keep in mind that well-posedness is a property of the
full PDE problem, which means that the PDE system should be treated as
whole when studying its hyperbolic character. As a matter of fact, in
Chap.~\ref{chap:numerics} we present numerical experiments that
demonstrate this point.

\subsection{Inhomogeneous WH model}
\label{subsec:bondi_well-p:algebraic_char:inhomo_WH}

For the homogeneous WH model we computed the norm
of~$e^{(i\hat{\omega}_p\mathbfcal{B}^p)|\omega| t}$ to estimate the
behavior of solutions. However, we could also examine the form of the
eigenvalues of the full symbol~$\mathbf{P}(i \omega)$ for
large~$|\omega|$ to understand if the solutions exhibit exponential
growth in~$\omega_p$ (see lemma 2.3.1 of~\cite{KreLor89}). If there is
any eigenvalue~$\lambda$ of~$\mathbf{P}(i \omega)$ such that
\begin{align*}
  \Re[\lambda] \sim |\omega|^s > 0 \; \text{with} \; s>0 \, ,
\end{align*}
for large~$|\omega|$, then solutions of the PDE may exhibit frequency
dependent exponential growth, and the PDE problem is ill-posed in any
sense. For the inhomogeneous WH model we consider the following
possible lower order source terms
\begin{align*}
  \mathbfcal{B}_1 =
  \begin{pmatrix}
    0 & 0 & 1\\
    1 & 0 & 1\\
    1 & 0 & 0 
  \end{pmatrix}
  ,\, \quad
  \mathbfcal{B}_2 =
  \begin{pmatrix}
    1 & 0 & 1\\
    1 & 1 & 1\\
    1 & 1 & 1 
  \end{pmatrix}
  ,\, \quad
  \mathbfcal{B}_3 =
  \begin{pmatrix}
    0 & 1 & 0\\
    0 & 0 & 0\\
    0 & 0 & 0 
  \end{pmatrix}
  \, ,
\end{align*}
where~$-\mathbfcal{S} = \mathbfcal{B}\mathbf{u}$. The
choice~$\mathbfcal{B}_1$ is motivated by analogy with the linearized
Bondi-Sachs system with~$\phi \sim \beta$,~$\psi_\varv \sim V$
and~$\psi \sim \gamma_r$. In~$\mathbfcal{B}_2$ we include all possible
source terms that do not break the nested structure of the intrinsic
equations and finally in the choice~$\mathbfcal{B}_3$ we introduce
source terms that violate the nested structure, thus rendering the
intrinsic subsystem a coupled PDE. For both~$\mathbfcal{B}_1$
and~$\mathbfcal{B}_2$ the eigenvalues of~$\mathbf{P}(i \omega)$ are
\begin{align*}
  \lambda_1 = \lambda_2 = - i |\omega| \, \hat{\omega}_\rho\,,
  \quad
  \lambda_3 = i |\omega|
  \left( F\, \hat{\omega}_\rho + G \,\hat{\omega}_z \right) \, ,
\end{align*}
as~$|\omega| \rightarrow \infty$, with the next terms appearing at
order~$|\omega|^0$. For these choices of lower order source terms the
inhomogeneous WH model remains well-posed in the lopsided norm. On the
other hand if~$\mathbfcal{B} = \mathbfcal{B}_3$ the eigenvalues of the
symbol are
\begin{align*}
  \lambda_1
  & =  - i |\omega| \hat{\omega}_\rho
    - (-1)^{1/4} \sqrt{|\omega| \hat{\omega}_z} + O(|\omega|^0)
    \,,\\
  \lambda_2
  & = - i |\omega| \hat{\omega}_\rho
    + (-1)^{1/4} \sqrt{|\omega| \hat{\omega}_z} + O(|\omega|^0)
  \,,\\
  \lambda_3
  & = i |\omega| \left( F\, \hat{\omega}_\rho
    + G\,\hat{\omega}_z \right)
    + O(|\omega|^0)\, ,
\end{align*}
for large~$|\omega|$. Since~$\Re[\lambda] \sim |\omega|^{1/2}$, we
conclude that when the nested structure of the intrinsic equations is
broken, the solution of the inhomogeneous WH exhibits frequency
dependent exponential growth. Consequently, the IVP with this system
is no longer weakly well-posed but ill-posed. Note, in contrast, that
for the homogeneous SH model we have
\begin{align*}
  | e^{\mathbf{P}(i \omega) t} | = 1.
\end{align*}
Hence for this model, the IVP is well-posed already in
the~$L^2$-norm. Unlike the WH model, well-posedness for this model is
not affected by source terms. The detailed calculations for both
models can be found in~\cite{GiaHilZil20_public}.

\section{Toy CCE and CCM energy estimates}
\label{sec:bondi_well-p:toy_CCE_CCM}

\begin{figure}[t]
  \begin{center}
    \includegraphics[width=0.55\textwidth]{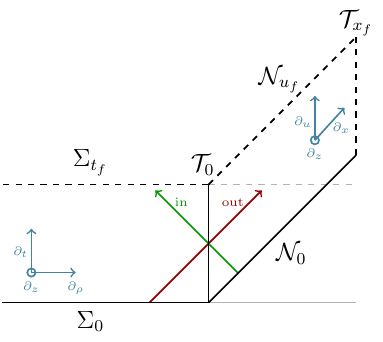}
  \end{center}
  \caption[Model Cauchy-Characteristic extraction (CCE) and matching
  (CCM) setups]{The IBVP (left) and the CIBVP (right) setups.  For CCE
    outgoing data from the IBVP serve as boundary data
    on~$\mathcal{T}_0$ for the CIBVP, which can be viewed as an
    independent PDE problem. In this case the IBVP's spatial domain is
    more extended such that data on $\mathcal{T}_0$ are unaffected by
    the boundary conditions chosen for the problem.  For CCM the IBVP
    and CIBVP are solved simultaneously and out/ingoing data are
    communicated from one to the other
    via~$\mathcal{T}_0$. Effectively, the two problems are viewed as
    one.  \label{FIG:CCM} }
\end{figure}

The previous analysis was performed in Fourier space and yielded that
an IVP based on the homogeneous WH model may be well-posed in an
appropriate lopsided norm, whereas one on the SH model is (strongly)
well-posed in the~$L^2$-norm. We now present our energy estimates for
solutions to the IBVP and CIBVP by working in position space. For
concreteness and simplicity the PDE system for the IBVP is a
homogeneous SH model (which is furthermore symmetric hyperbolic)
\begin{align}
  \begin{aligned}
    & \p_t \bar{\phi} + \p_\rho \bar{\phi} + \p_z\bar{\psi_\varv} = 0 \,,
    \\
    & \p_t \bar{\psi_\varv} + \p_\rho \bar{\psi_\varv}+\p_z\bar{\phi}  = 0 \,,
    \\
    & \p_t \bar{\psi} - \tfrac{1}{2} \p_\rho \bar{\psi} - \p_z \bar{\psi}  = 0 \,,
  \end{aligned}\label{eqn:CCM_IBVP_sys}
\end{align}
with initial data~$\bar{\phi}_*,\, \bar{\psi_\varv}_*,\, \bar{\psi}_*$
on~$\Sigma_0$, boundary data~$\hat{\bar{\psi}}$ on~$\mathcal{T}_0$ and
domain~$t \in[0,t_f],\, \rho \in (-\infty,0]$ and the
compact~$z \in [0, 2\pi)$, and for the CIBVP the homogeneous WH model
\begin{subequations}
  \begin{align}
    &  \p_x \phi  = 0 \,,   \label{eqn:CCM_CIBVP_sys_a}
    \\
    & \p_x \psi_\varv - \p_z\phi  = 0 \,,   \label{eqn:CCM_CIBVP_sys_b}
    \\
    & \p_u \psi - \tfrac{1}{2} \p_x \psi - \p_z \psi  = 0 \,,
    \label{eqn:CCM_CIBVP_sys_c}
  \end{align}
  \label{eqn:CCM_CIBVP_sys}
\end{subequations}
with initial data~$\psi_*$ on~$\mathcal{N}_0$, boundary
data~$\hat{\phi}$ and~$\hat{\psi_\varv}$ on~$\mathcal{T}_0$ and
domain~$u \in[0,u_f],\, x \in [0,x_f]$ and the aforementioned~$z$. The
domains of the two problems are illustrated in Fig.~\ref{FIG:CCM}. We
view the IBVP as a simplified analog of GR in strongly (here even
symmetric) hyperbolic formulations widely used in Cauchy-type
problems, with the CIBVP formulated in the Bondi-like gauges used in
characteristic evolutions. We wish to understand whether or not
problems with these features can be successfully used for CCE or CCM
in principle.

For the IBVP estimate our starting point is
\begin{align*}
  \p_t ||\mathbf{\bar{u}}||^2_{L^2(\Sigma_t)} =
  \p_t \int_{\Sigma_t} \mathbf{\bar{u}}^T \mathbf{\bar{u}} =
  \p_t \int_{\Sigma_t} \left(\bar{\phi}^2 + \bar{\psi_\varv}^2
  + \bar{\psi}^2 \right)
  \,,
\end{align*}
which after using~\eqref{eqn:CCM_IBVP_sys} reads
\begin{align*}
  \p_t ||\mathbf{\bar{u}}||^2_{L^2(\Sigma_t)} =
  2 \int_{\Sigma_t}
  \left(
  -\bar{\phi} \p_\rho \bar{\phi}
  -\bar{\phi} \p_z \bar{\psi_\varv}
  - \bar{\psi_\varv} \p_\rho \bar{\psi_\varv}
  - \bar{\psi_\varv} \p_z \bar{\phi}
  + \frac{1}{2} \bar{\psi} \p_\rho \bar{\psi}
  + \bar{\psi} \p_z \bar{\psi}
  \right)
  \,.
\end{align*}
Applying here the divergence theorem (see
App.~\ref{sec:appA:divergence_thm} for more details)
assuming~$\mathbf{\bar{u}} \rightarrow \mathbf{0}$
as~$\rho \rightarrow - \infty$ yields
\begin{align*}
  \p_t ||\mathbf{\bar{u}}||^2_{L^2(\Sigma_t)} =
  \int
  \left(
  -\bar{\phi}^2|_{\rho=0} 
  -{\psi_\varv}^2|_{\rho=0}
  +\frac{1}{2}{\psi}^2|_{\rho=0}
  \right)
  \; dz
  \,,
\end{align*}
where the
terms~$\bar{\phi} \bar{\psi_\varv}|_{z=0}-\bar{\phi}
\bar{\psi_\varv}|_{z=2\pi}$
and~$\bar{\psi}^2|_{z=2\pi}-\bar{\psi}^2|_{z=0}$ vanish due to
periodicity in~$z$. Finally, and integrating in the~$t$ domain returns
\begin{align}
  ||\mathbf{\bar{u}}||^2_{L^2(\Sigma_{t_f})}
  + ||\mathbf{\bar{u}}||^2_{L^2_\textrm{out} (\mathcal{T}_0)}
  =
  ||\mathbf{\bar{u}}||^2_{L^2(\Sigma_0)}
  + \frac{1}{2} ||\mathbf{\bar{u}}||^2_{L^2_\textrm{in} (\mathcal{T}_0)}
  \,,
  \label{eqn:IBVP_estimate}
\end{align}
where~$||\mathbf{\bar{u}}||^2_{L^2_\textrm{out}(\mathcal{T}_0)}$
denotes integral over~$\mathcal{T}_0$ that contains only the outgoing
fields~$\bar{\phi},\,\bar{\psi_\varv}$, and similarly for the
ingoing~$\bar{\psi}$. The estimate~\eqref{eqn:IBVP_estimate} states
that the energy of the solution equals the energy of its given data,
so that the solution is controlled by the given data.

In a Cauchy-type setup we specify all fields on the initial spacelike
hypersurface and, by solving the system we obtain all of them on
spacelike hypersurfaces to the future. On the contrary, in a
single-null characteristic setup, fields with ``evolution'' equations
are chosen on the initial null hypersurface and those that satisfy
equations intrinsic to the null hypersurfaces are specified as
boundary data. As we will see in the following, this has a natural
consequence on the type of estimates that we can hope to demonstrate,
both in terms of the domain on which we integrate and the particular
fields that appear. This is due to the geometry of the setup.

Motivated from the IVP estimates in Fourier space of
Subsec.~\ref{subsec:bondi_well-p:algebraic_char:homo_WH}
and~\ref{subsec:bondi_well-p:algebraic_char:inhomo_WH} we might
naively first consider for the CIBVP the lopsided norm
\begin{align*}
  ||\mathbf{u}||^2_{q (\mathcal{D})} = \int_{\mathcal{D}}
  \left(
  \phi^2 + \psi_\varv^2 + \psi^2 + \left( \p_z \phi \right)^2
  \right)
  \,,
\end{align*}
in some domain~$\mathcal{D}$, where only~$\p_z \phi$ is added to the
integrand of the~$L^2$-norm, because precisely this term causes the
pathological structure in the angular principal part of the WH
model. Following our previous discussion however, it is more
appropriate to split the integrand into separate pieces for the
ingoing and outgoing variables. The domain~$\mathcal{D}$
becomes~$\mathcal{N}_{u}$ and $\mathcal{T}_{x}$ respectively for
each. For the ingoing variables we start from
\begin{align*}
  \p_u ||\mathbf{u}||^2 _{q_\textrm{in}(\mathcal{N}_u)} =
  \p_u \int_{\mathcal{N}_u} \psi^2
  \,,
\end{align*}
since there are no~$\p_u$ equations for the outgoing ones. We assume
that~$\psi \rightarrow 0$ as~$x \rightarrow x_f$ in the given data,
which is the analog in our model to requiring no incoming
gravitational waves from future null infinity, working on a
compactified radial domain. After using~\eqref{eqn:CCM_CIBVP_sys_c},
the divergence theorem and integrating in the~$u$ domain we obtain
\begin{align}
  2 ||\mathbf{u}||^2 _{q_\textrm{in}(\mathcal{N}_{u_f})} +
  ||\mathbf{u}||^2 _{q_\textrm{in}(\mathcal{T}_0)}
  =
  2 ||\mathbf{u}||^2 _{q_\textrm{in}(\mathcal{N}_0)}
  \,.
  \label{eqn:CIBVP_in_estimate}
\end{align}
For the outgoing variables the starting point is
\begin{align*}
  \p_x ||\mathbf{u}||^2_{q_\textrm{out}(\mathcal{T}_x)} =
  \p_x \int_{\mathcal{T}_x}
  \left(
  \phi^2 + \psi_\varv^2 +
  \left( \p_z \phi \right)^2
  \right)
  \,,
\end{align*}
and by using~\eqref{eqn:CCM_CIBVP_sys_a} and \eqref{eqn:CCM_CIBVP_sys_b}, the
divergence theorem and integrating in the~$x$ domain up to some
arbitrary~$x^\prime$ we obtain
\begin{align}
  ||\mathbf{u}||^2_{q_\textrm{out}(\mathcal{T}_{x^\prime})}
  =
  ||\mathbf{u}||^2_{q_\textrm{out}(\mathcal{T}_0)}
  +
  \int_0^{x^\prime}
  \left(
  \int_{\mathcal{T}_x} 2 \psi_\varv \p_z \phi
  \right)
  \, dx
  \,,
  \label{eqn:CIBVP_out_estimate_1}
\end{align}
where the last term is due to the hyperbolicity of the system and
would not appear for our SH example. Using
$ 2 \psi_\varv \p_z \phi \leq \phi^2 + \psi_\varv^2 + \left( \p_z \phi
\right)^2$ the latter reads
\begin{align*}
  ||\mathbf{u}||^2_{q_\textrm{out}(\mathcal{T}_{x^\prime})}
  \leq
  ||\mathbf{u}||^2_{q_\textrm{out}(\mathcal{T}_0)}
  +
  \int_0^{x^\prime} ||\mathbf{u}||^2_{q_\textrm{out}(\mathcal{T}_x)} \, dx
  \,,
\end{align*}
and by applying Gr\"onwall's inequality (see
App.~\ref{sec:appA:Gronwall} for more details) we obtain
\begin{align}
  ||\mathbf{u}||^2_{q_\textrm{out}(\mathcal{T}_{x^\prime})}
  \leq
  e^{x^\prime}
  ||\mathbf{u}||^2_{q_\textrm{out}(\mathcal{T}_0)}
  \,.
  \label{eqn:CIBVP_out_estimate_2}
\end{align}
Hence, the energy of the outgoing fields at each arbitrary timelike
hypersurface~$\mathcal{T}_{x^\prime}$ in the characteristic domain is
bounded. The sum of~\ref{eqn:CIBVP_in_estimate}
and~~\ref{eqn:CIBVP_out_estimate_2} is the complete energy estimate
for the CIBVP and yields
\begin{align}
  2||\mathbf{u}||^2 _{q_\textrm{in}(\mathcal{N}_{u_f})} +
  ||\mathbf{u}||^2 _{q_\textrm{in}(\mathcal{T}_0)} +
  \textrm{sup}_{x'} ||\mathbf{u}||^2_{q_\textrm{out}(\mathcal{T}_{x^\prime})}
  \leq
  2 ||\mathbf{u}||^2 _{q_\textrm{in}(\mathcal{N}_0)}
  +
  e^{x_f}
  ||\mathbf{u}||^2_{q_\textrm{out}(\mathcal{T}_0)}
  \,,
  \label{eqn:CIBVP_estimate_full}
\end{align}
where we used that~$e^{x^\prime} \leq e^{x_f}$
for~$x^\prime \in [0,x_f]$ and chose the supremum
of~$||\mathbf{u}||^2_{q_\textrm{out}(\mathcal{T}_{x^\prime})}$ to
obtain the largest possible bounded left-hand-side, since the outgoing
lopsided norm is not necessarily monotonically increasing with~$x$.
Thus, the energy of the solution to the CIBVP is controlled by the
given data on~$\mathcal{N}_0$ and~$\mathcal{T}_0$.

We first interpret these estimates in the framework of CCE. Choosing
suitable data, our estimate for the IBVP shows that one obtains a
smooth solution in the domain of the Cauchy-type setup. One can then
use this solution to provide boundary data on~$\mathcal{T}_0$ for the
CIBVP that are finite also in the lopsided norm, and the solution to
this characteristic problem has a good energy estimate as shown
earlier too. Hence the CCE process is perfectly valid for our model,
and provided analogous estimates for GR in the Bondi-like gauges used,
would be in that context too. One question that arises for GR is
whether or not this procedure excludes any data of interest. Effort in
this direction is currently ongoing, but there is no clear answer at
the time of writing of this thesis. In
Sec.~\ref{sec:bondi_well-p:axisym_BS},
\ref{sec:bondi_well-p:CIBVP_symm_hyp} we collect the current status of
this work.

For CCM the discussion is rather different, since IBVP and CIBVP are
solved simultaneously and data are communicated between
domains. Effectively, one joins the PDE problems and they may be
viewed as one. Hence, let us try to obtain an energy estimate for the
joint PDE problem, by adding~\eqref{eqn:IBVP_estimate}\footnote{A way
  to get the exact coefficients that appear in~\eqref{CCM_estimate} is
  to
  add~$\frac{1}{2}||\mathbf{u}||^2_{L^2_\textrm{in} (\mathcal{T}_0)}$
  on the right-hand-side of~\eqref{eqn:IBVP_estimate} and change~$=$
  to~$\leq$.}{} and~\eqref{eqn:CIBVP_estimate_full}:
\begin{align}
  &
    ||\mathbf{u}||^2_{L^2(\Sigma_{t_f})} +
    ||\mathbf{u}||^2_{L^2_\textrm{out} (\mathcal{T}_0)} +
    2 ||\mathbf{u}||^2 _{q_\textrm{in}(\mathcal{N}_{u_f})} +
    ||\mathbf{u}||^2 _{q_\textrm{in}(\mathcal{T}_0)} +
    \textrm{sup}_{x'} ||\mathbf{u}||^2_{q_\textrm{out}(\mathcal{T}_{x^\prime})}
  \nonumber \\
  & \qquad \qquad \; \;
    \leq \,
    ||\mathbf{u}||^2_{L^2(\Sigma_0)} +
    ||\mathbf{u}||^2_{L^2_\textrm{in} (\mathcal{T}_0)} +
    2 ||\mathbf{u}||^2 _{q_\textrm{in}(\mathcal{N}_0)}
    +
    e^{x_f}
    ||\mathbf{u}||^2_{q_\textrm{out}(\mathcal{T}_0)}
    \,,
    \label{CCM_estimate}
\end{align}
where now~$\mathbf{\bar{u}} = \mathbf{u}$. For the joint problem there
is ``effectively'' no boundary~$\mathcal{T}_0$ at which we are free to
choose data, and hence any estimate should not involve integrals over
this domain. The relevant terms can however cancel each other only if
the two norms that appear coincide. This requires either that the
CIBVP relies on a symmetric hyperbolic PDE system and hence is
well-posed in the~$L^2$-norm (see for instance~\cite{BisGomHol96,
  BisGomHol97, Cal06}), or that the IBVP relies on a system that is
well-posed in the same lopsided norm as the CIBVP. But this requires
special structure, above and beyond symmetric hyperbolicity, on the
equations used in the IBVP. Regarding GR, the first option would
translate into developing a SH (hopefully also symmetric hyperbolic)
single-null formulation and the second to building a formulation that
is well-posed in the same lopsided norm that Bondi-like gauges
(perhaps) are. Given the long search for formulations that {\it work}
for practical evolution however, such an artisanal construction seems
poorly motivated. In summary; unless special structure is present in
the field equations solved for the IBVP, the solution to the weakly
hyperbolic CIBVP cannot be combined with that of an IBVP of a
symmetric hyperbolic system in such a way as to provide a solution to
the whole problem which has an energy bounded by that of the given
data.

\section{Energy estimates for the axisymmetric Bondi-Sachs system}
\label{sec:bondi_well-p:axisym_BS}

Working again with the linearized about flat space equations and
within the constant coefficient approximation, we explore energy
estimates for the Bondi-Sachs setup in axisymmetry. All the
calculations can be found in~\cite{Gia22_public}. We work with the
minimal first order reduction of
Subsec.~\ref{subsec:bondi-hyp:BS_proper:axisym_char} and the PDE
system reads
\begin{equation}
\begin{aligned}
  &\p_r \beta = 0
  \, , \\
  &\p_r U_r - \frac{2}{r^2} \p_r\beta_\theta + \frac{2}{r^2}
  \p_r\gamma_\theta + S_2 = 0
  \,, \\
  &\p_r V + \p_\theta \beta_\theta - \p_\theta \gamma_\theta - 2 r
  \p_\theta U - \frac{r^2}{2}
  \p_\theta U_r + S_3 = 0
  \,,\\
  & 4 r^2 \p_u \gamma_r + 4 r \p_u \gamma -2 r^2 \, \p_r \gamma_r + 2
  r \, \p_\theta U + r^2 \p_\theta U_r - 2 \p_\theta
  \beta_\theta + S_4 = 0
  \, ,\\
  & \p_rU + S_5 = 0
  \,,\\
  & \p_r \gamma + S_6 = 0
  \,,\\
  & \p_r \gamma_\theta - \p_\theta \gamma_r = 0
  \,,\\
  & \p_r \beta_\theta = 0
  \,,
\end{aligned}
  \label{BS_linear_pde}
\end{equation}
We focus on the IVP in the auxiliary Cauchy-type frame, in order to
understand whether the WH system~\eqref{BS_linear_pde} has any chance
to be weakly well-posed in some lopsided norm. This approach may be
viewed as a preliminary exercise prior to that of an energy estimate
attempt in the characteristic setup. We expect it to be informative of
possible shortcomings due to the weak hyperbolicity of the specific
system. A similar approach was effectively used in
Sec.~\ref{sec:bondi_well-p:toy_CCE_CCM} where the lopsided norm was
inspired by the algebraic characterization of the WH toy model
performed in Sec.~\ref{sec:bondi_well-p:algebraic_char}. The algebraic
characterization of the homogeneous version of
Eq.~\eqref{BS_linear_pde} suggests that there exists a lopsided norm
for a weakly well-posed IVP. Unfortunately, we do not have such a
result for the inhomogenous version of Eq.~\eqref{BS_linear_pde}.

\subsubsection{The auxiliary Cauchy
  frame} \label{Subsubsection:energy_estimates_Cauchy}

We consider the Cauchy-type version of the
system~\eqref{BS_linear_pde} written in the form
\begin{align}
  \p_t \mathbf{u} + \mathbf{B}^p \p_p \mathbf{u} \simeq 0
  \,,
\end{align}
with the index~$p$ denoting an arbitrary spatial coordinate
and~$\simeq$ equality up to principal terms. Driven by the structure
of the Bondi-Sachs system and its pathology along the angular
direction~$\theta$, we employ its generalized characteristic
variables~$\mathbf{v} \equiv \mathbf{T}_{\hat{\theta}}^{-1}
\mathbf{u}$ to define an energy density. For convenience let us first
repeat the angular generalized characteristic variables as given in
Subsec.~\ref{subsec:bondi-hyp:BS_proper:axisym_char}. The ones with
speed zero are
\begin{align*}
  \rho \, U + \frac{\rho^2}{2} \,U_r -\beta_\theta
  + \gamma_\theta
  \,,\quad
  \beta_\theta
  \,,\quad
  V
  \,,\quad
  \rho \left(-2\rho U - \frac{\rho^2}{2} U_r
    + \beta_\theta - \gamma_\theta \right)
  \,,\quad
  \gamma
  \,,\quad
  \beta \,,
\end{align*}
of which the third and fourth are associated with the
non-trivial~$2\times2$ Jordan block
within~$\mathbf{J}^{\hat{\theta}}$. Likewise we have
\begin{align*}
  -\frac{\rho}{2}U + \frac{\rho}{2}\gamma_r
  - \frac{\rho^2}{4} U_r + \frac{1}{2} \beta_\theta
  \,,\quad
  -\frac{\rho}{2}U - \frac{\rho}{2}\gamma_r
  - \frac{\rho^2}{4} U_r + \frac{1}{2} \beta_\theta \,,
\end{align*}
with speeds~$\pm 1$ respectively. The structure
of~$\mathbf{J}_{\hat{\theta}}$ yields
\begin{equation}
\begin{aligned}
&- \p_t\left(2\rho U +\frac{\rho^2}{2} U_r
      - \beta_\theta + \gamma_\theta\right) \simeq 0\,,\\
&\p_tV - \rho \, \p_{\hat{\theta}}\left(2\rho U +\frac{\rho^2}{2} U_r
  - \beta_\theta + \gamma_\theta\right) \simeq 0 \,,
\end{aligned}
\label{chap6:K3_gen_char_var}
\end{equation}
where~$\p_{\hat{\theta}} = 1/\rho \, \p_\theta$,
with~$\p_{\hat{\theta}}$ unit spatial vector. We define the
energy~$E_{\Sigma_t}$ contained within a spacelike
hypersurface~$\Sigma_t$ as

\begin{align}
  E^2_{\Sigma_t}
  &= \int_{\Sigma_t} \epsilon^2
  \nonumber
  \\
  &= \int_{\Sigma_t} \left( \rho \, U +
    \frac{\rho^2}{2} \,U_r -\beta_\theta + \gamma_\theta \right)^2 +
    \beta_\theta^2 + V^2 + \rho^2 \left( -2\rho U - \frac{\rho^2}{2}
    U_r + \beta_\theta - \gamma_\theta \right)^2 + \gamma^2 +
    \beta^2
  \nonumber \\
  & \qquad + \left( -\frac{\rho}{2}U + \frac{\rho}{2}\gamma_r -
    \frac{\rho^2}{4} U_r + \frac{1}{2} \beta_\theta \right)^2 +
    \left( -\frac{\rho}{2}U - \frac{\rho}{2}\gamma_r -
    \frac{\rho^2}{4} U_r + \frac{1}{2} \beta_\theta \right)^2
  \nonumber \\
  & \qquad + \rho^2 \left( -2\rho \p_{\hat{\theta}} U -
    \frac{\rho^2}{2} \p_{\hat{\theta}} U_r + \p_{\hat{\theta}}
    \beta_\theta - \p_{\hat{\theta}} \gamma_\theta \right)^2
  \nonumber \\
  & \qquad + \left( \p_\rho \gamma \right) ^2
    \,,
    \label{eq:BS_lin_homo_Cauchy_energy}
\end{align}
where the second to last term is motivated by the non-trivial Jordan
block structure~\eqref{chap6:K3_gen_char_var} and the last term by the
homogeneous analysis. To simplify the analysis we focus on the IVP and
neglect any boundaries i.e. integrals over worldtubes of constant
radius~$\rho$ are assumed to vanish. The main goal here is to
understand if the specific form of weak hyperbolicity leads to any
bulk integrals in the energy estimate calculation that prevents us
from bounding the solution at future times from the initial data.

\subsubsection{The homogeneous setup}

We take a~$\p_t$ derivative of Eq~\eqref{eq:BS_lin_homo_Cauchy_energy}
and replace with the right-hand-side of the
Eq.~\eqref{BS_linear_pde}. The total~$\p_\rho$ terms are neglected and
the~$\p_\theta$ ones, namely
\begin{equation}
  \begin{aligned}
    \rho_c \p_{\hat{\theta}} \left( \beta_\theta \gamma_r \right)
    \,,
    \quad
    - \rho_c^2 \p_{\hat{\theta}} \gamma_r U
    \,
    \quad
    - \frac{\rho_c^3}{2} \p_{\hat{\theta}} \left( \gamma_r U_r \right)
    \,,
  \end{aligned}
  \label{Cauchy_angular_total_devs}
\end{equation}
vanish due to the periodicity in the angular direction. After
integrating in~$t \in [t_0,t_f]$ we obtain
\begin{align}
  E^2_{\Sigma_{t_f}} 
  = E^2_{\Sigma_{t_0}}
  + \int_{t_0}^{t_f} \int_{\Sigma_t} \Xi_0 \,,
  \label{BS_Cauchy_homo_estimate_1}
\end{align}
where the bulk integrands is
\begin{align}
  \Xi_0 = 2 \rho_c V \left(\p_{\hat{\theta}} \gamma_\theta -
  \p_{\hat{\theta}} \beta_\theta\right) + \boxed{2 \rho_c \gamma_r
  \p_\rho \gamma} + 4 \rho_c^2 V \p_{\hat{\theta}} U + \rho_c^3
  \left( V - \gamma_r \right) \p_{\hat{\theta}}U_r \,.
  \label{Cauchy_bulk_homog}
\end{align}
The boxed term in the above expression is the reason we introduced the
last term in the energy
definition~\eqref{eq:BS_lin_homo_Cauchy_energy}. More specifically, if
the energy density did not include~$\left( \p_\rho \gamma \right)^2$,
then the quantity~$\Xi_0$ could not be bounded by the energy density
exactly because of the boxed term. The rest of the bulk terms already
appear in the energy density due to the generalized characteristic
variables. The next step is to prove that energy of the solution can
be bounded by the energy of the initial data. By using
that~$ \Xi_0 \lesssim \epsilon^2$ we arrive at
\begin{align*}
  &
    E^2_{\Sigma_{t_f}} 
    \lesssim
    E^2_{\Sigma_{t_0}}
    + \int_{t_0}^{t_f} E^2_{\Sigma_{t}}
    \,,
\end{align*}
where~$\lesssim$ denotes smaller or equal up to an overall
constant. Notice
that~$E^2_{\Sigma_{t_0}}$,~$E^2_{\mathcal{T}_{\rho_0}}$ is initial
data. We apply here the integral version of Gr\"onwall's inequality to
arrive at
\begin{align*}
  E^2_{\Sigma_{t_f}}
  \lesssim
  e^{t_f} 
  E^2_{\Sigma_{t_0}} 
  \,,
\end{align*}
which states that the energy of the solution is bounded by the energy
of the initial data. So the homogeneous linearized axisymmetric
Bondi-Sachs system~\eqref{BS_linear_pde} has a weakly well-posed IVP
in the lopsided norm~\eqref{eq:BS_lin_homo_Cauchy_energy}. This result
is compatible with the algebraic characterization of this system,
which is not shown here but follows the same method as in
Sec.~\eqref{sec:bondi_well-p:algebraic_char} and can be found
in~\cite{Gia22_public}.

\subsubsection{The inhomogeneous setup}

However, this result is only valid for the homogeneous case. The same
energy density definition fails if one considers the inhomogeneous
system of the auxiliary Cauchy frame. The equivalent energy estimate
in this case reads
\begin{align}
  E^2_{\Sigma_{t_f}}
  =  E^2_{\Sigma_{t_0}}
  +
  \int_{t_0}^{t_f} \int_{\Sigma_t}
  \Xi_0 + \Xi_1 + \Xi_2
  \,,
  \label{Cauchy_inhomo_estimate_1}
\end{align}
where
\begin{align}
  \Xi_1
  & =
    4 V \beta - 6 U \beta_\theta - 4 V \gamma + 
    2 \gamma \gamma_r + \frac{6 \beta_\theta^2}{\rho_c}
    - \frac{4 \beta_\theta \gamma_\theta}{\rho_c} + 
    4 \rho_c \beta_\theta^2  - 2  \rho_c U_r \gamma_\theta - 
    4  \rho_c \beta_\theta \gamma_\theta - 3 \rho_c^2 U U_r  
    \nonumber
  \\
  & \quad
    - 8 \rho_c^2 U \beta_\theta  - \frac{3 \rho_c^3 U_r^2}{2} - 
    2 \rho_c^3 U_r \beta_\theta - 
    2 \cot(\theta_c) V \beta_\theta  + 
    5  \cot(\theta_c) \beta_\theta \gamma_r + 
    6 \cot(\theta_c) V \gamma_\theta
    \nonumber \\
  & \quad 
    - 4 \cot(\theta_c) \gamma_r \gamma_\theta  + 
    4  \rho_c \cot(\theta_c) U V - 
    5  \rho_c \cot(\theta_c) U \gamma_r + 
    \rho_c^2 \cot(\theta_c)U_r V  - 
    \frac{5 \rho_c^2 \cot(\theta_c) U_r \gamma_r }{2} 
    \nonumber
  \\
  & \quad
    + 4 \rho_c^2 \cot(\theta_c) \beta_\theta \gamma_r  - 
    4  \rho_c^2 \cot(\theta_c) \gamma_r \gamma_\theta - 
    8 \rho_c^3 \cot(\theta_c) U \gamma_r  - 
    2 \rho_c^4 \cot(\theta_c) U_r \gamma_r \,,
  \\
  \Xi_2
  & =
    2 \, (\p_\rho \gamma) \, (\p_\rho \gamma_r) + 
    4 \rho_c \, (\p_{\hat{\theta}}\beta_\theta)^2  - 
    4 \rho_c \, (\p_{\hat{\theta}} \beta_\theta) \, (\p_{\hat{\theta}} \gamma_\theta)  - 
    8 \rho_c^2  \, (\p_{\hat{\theta}} U) \, (\p_{\hat{\theta}} \beta_\theta) 
    \nonumber \\
  & \quad
    - 2 \rho_c^3 \, (\p_{\hat{\theta}} U_r) \, (\p_{\hat{\theta}} \beta_\theta)
    + 4 \rho_c^2 \cot(\theta_c) \, (\p_{\hat{\theta}} \beta_\theta) \, (\p_{\hat{\theta}} \gamma_r)
    - 4  \rho_c^2 \cot(\theta_c) \, (\p_{\hat{\theta}} \gamma_r) \, (\p_{\hat{\theta}} \gamma_\theta)
    \nonumber \\
  & \quad
    - 8  \rho_c^3 \cot(\theta_c) \, (\p_{\hat{\theta}} U) \, (\p_{\hat{\theta}} \gamma_r) - 
    2  \rho_c^4 \cot(\theta_c) \, (\p_{\hat{\theta}} U_r) \, (\p_{\hat{\theta}} \gamma_r)
    \,.
\end{align}
The terms~$\Xi_1$ and~$\Xi_2$ appear due to the source terms of the
linearized system~\eqref{BS_linear_pde}. More specifically, the
terms~$\Xi_2$ are a result of acting with~$\p_t$ on
\begin{align*}
  \rho^2 \left( -2\rho \p_{\hat{\theta}} U -
  \frac{\rho^2}{2} \p_{\hat{\theta}} U_r + \p_{\hat{\theta}}
  \beta_\theta - \p_{\hat{\theta}} \gamma_\theta \right)^2
\end{align*}
and replacing the right-hand-side of the appropriate inhomogeneous
equations of motion. Note that the above combination of variables
together with~$\p_\rho \gamma$ in
Eq.~\eqref{eq:BS_lin_homo_Cauchy_energy} control the terms
of~$\Xi_0$. In~$\Xi_2$ however there are terms of the form
\begin{align*}
  ~ \p_\rho \gamma_r \,, \quad \p_\theta \gamma_r
  \,,
\end{align*}
that cannot be controlled by the
energy~\eqref{eq:BS_lin_homo_Cauchy_energy}. This is an explicit
example where weak well-posedness in a specific lopsided norm is
broken by lower order (source) terms. One first attempt to modify the
previous energy definition to accommodate for the additional terms is
to add in it terms of the form
\begin{align*}
  ~ (\p_\rho \gamma_r)^2 \,, \quad (\p_\theta \gamma_r)^2
  \,.  
\end{align*}
It turns out though that such a change is not sufficient. Including
these terms in the integrand results in terms that do not form total
derivatives and are not controlled by the initial data in this
norm. It becomes clear that finding an appropriate norm (if it exists)
that provides an energy estimate for the IVP of a weakly hyperbolic
PDE system is far from trivial. The fact that we have not found such a
norm yet for the specific system analyzed here does not necessarily
mean that it does not exist. Completing the algebraic characterization
of this inhomogeneous system can answer whether such a norm
exists. However, this analysis has its own challenges. Even if such a
norm is found, this weak well-posedness may break by lower order
perturbations due to non-linearities, when considering the original
non-linear Bondi-Sachs system.

\section{CIBVP energy estimates for symmetric hyperbolic PDEs}
\label{sec:bondi_well-p:CIBVP_symm_hyp}

A symmetric hyperbolic PDE system has a well-posed IVP in
the~$L^2$-norm (see \cite{Hil13} for a brief discussion and references
therein for more details). We assume that there exists a unique
solution to the CIBVP of a symmetric hyperbolic PDE and examine
continuous dependence on the given data by means of energy
estimates. More specifically, we study how the geometric setup of the
CIBVP affects the specific form of the norm that is appropriate for
the problem. We use standard methods of PDE analysis. More
sophisticated mathematical tools that take full advantage of the
geometric setup may be more appropriate, but are beyond the scope of
this thesis. See \cite{Rin09b} for an overview and references therein
for more details.

The PDE system under consideration can be written as
\begin{align*}
  \mathbfcal{A}^\mu \p_\mu \mathbf{u} + \mathbfcal{A} \mathbf{u} = 0
  \,,
\end{align*}
with~$\mathbfcal{A}^\mu$,~$\mathbfcal{A}$,~$\mathbf{u}$ complex
valued,~$\mathbfcal{A}^\mu \xi_\mu = (\mathbfcal{A}^\mu \xi_\mu)^\dag$
for all spacetime vectors~$\xi^\mu$ and the hypersurface of constant
time being an outgoing null hypersurface. Furthermore, there exists a
timelike vector~$t^\mu$ such
that~$\mathbfcal{A}^t \equiv \mathbfcal{A}^\mu t_\mu$ is positive
definite. The system of~\cite{Rip21} after linearization is such an
example, which is based on the Newman-Penrose formalism and uses the
affine-null gauge. In App.~\ref{sec:appA:Rip21} we review its basic
features. This system avoids the pure gauge structure we identified
earlier due to the promotion of the curvature into an independent
variable. This system effectively includes third order metric
derivatives and does not fall into the class of systems investigated
earlier. The following analysis is quite generic and the specific
system is used mainly as a motivation. We highlight the difference
between the homogeneous and inhomogeneous setup.

\subsection{Domain~$\mathcal{D}_1$}
\label{subsec:bondi_well-p:CIBVP_symm_hyp:D1}

\begin{figure}[!t]
  \begin{center}
    \includegraphics[width=0.45\textwidth]{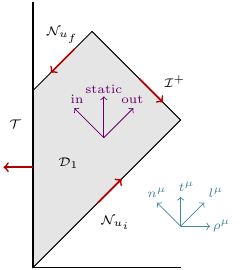}
  \end{center}
  \caption[CIBVP energy estimates for symmetric hyperbolic PDEs:
  domain~$\mathcal{D}_1$]{The geometry of the domain~$\mathcal{D}_1$
    of the CIBVP is illustrated, as well as its
    boundaries~$\mathcal{T}$,~$\mathcal{N}_{u_i}$,~$\mathcal{N}_{u_f}$
    and~$\mathcal{I}^+$. The following vectors are shown: outgoing
    null~$l^\mu$, ingoing null~$n^\mu$, future pointing
    timelike~$t^\mu$, spacelike~$\rho^\mu$ pointing towards increasing
    radius. The vectors normal to the boundary segments
    of~$\mathcal{D}_1$ are also shown, see
    App.~\ref{sec:appA:divergence_thm} for details.}
  \label{Fig:domain_1}
\end{figure}

Our main goal is to provide an energy estimate for a geometric setup
related to Cauchy-Characteristic extraction. The
domain~$\mathcal{D}_1$ for which the CIBVP typically provides a
solution is shown in Fig.~\ref{Fig:domain_1}. We assume that the PDE
system has variables that are ingoing, outgoing and static (with zero
speed) as illustrated in Fig.~\ref{Fig:domain_1}. The
boundary~$\p \mathcal{D}_1$ of this domain can be split into the
following segments:
\begin{description}
\item[\textit{Lower boundary}] $\mathcal{N}_{u_i}$ is the outgoing
  null hypersurface of initial time~$u_i$, with~$l^\mu$ its normal
  vector. The ingoing and static variables are provided here as
  initial data.
\item[\textit{Inner boundary}] $\mathcal{T}$ is a worldtube of
  constant affine parameter. The vector normal to~$\mathcal{T}$
  is~$\rho^\mu$. The outgoing variables are given here as boundary
  data, whereas the ingoing ones are obtained as part of the
  solution. The static variables are known on~$\mathcal{T}$ provided
  that they are given on~$\mathcal{N}_{u_i} \cap \mathcal{T}$ i.e. are
  provided by the initial data.
\item[\textit{Outer boundary}] $\mathcal{I}^+$ is a part of future
  null infinity and is an ingoing null hypersurface with~$n^\mu$ its
  normal vector. The outgoing and static variables are obtained as
  part of the solution on~$\mathcal{I}^+$, whereas the ingoing are
  known on~$\mathcal{I}^+$ provided that they are given
  on~$\mathcal{N}_{u_i} \cap \mathcal{I}^+$ i.e. are provided by the
  initial data.
\item[\textit{Upper boundary}] $\mathcal{N}_{u_f}$ is the outgoing
  null hypersurface at final time~$u_f$, with normal
  vector~$l^\mu$. The ingoing and static variables are provided here
  as part of the solution.
 
\end{description}

Since the system under consideration is symmetric hyperbolic, it is
convenient to define the flux \footnote{The flux defined in
  Eq.~\eqref{eqn:flux_def} trivially provides a total derivative for a
  symmetric hyperbolic system, so that we can conveniently apply the
  divergence theorem. For the SH and WH systems of
  Sec.~\ref{sec:bondi_well-p:toy_CCE_CCM},
  \ref{sec:bondi_well-p:axisym_BS} additional care was needed to form
  total derivatives.}{}
\begin{align}
  f \equiv \mathbf{u}^\dag \mathbfcal{A}^\mu \p_\mu \mathbf{u}
  \,.
  \label{eqn:flux_def}
\end{align}
Considering the integral of~$f$ in~$\mathcal{D}_1$ and applying the
divergence theorem yields
\begin{align}
  \int_{\mathcal{D}_1} f
  & = \int_{\p \mathcal{D}_1} (\mathbf{u}^\dag \mathbfcal{A}^\mu \mathbf{u})
    (g_{\mu \nu} \mathbf{N}^\nu)
    \nonumber
  \\
  & =
    \int_{\mathcal{N}_{u_i}} \mathbf{u}^\dag \mathbfcal{A}^l \mathbf{u}
    -\int_{\mathcal{T}} \mathbf{u}^\dag \mathbfcal{A}^\rho \mathbf{u}
    -\int_{\mathcal{N}_{u_f}} \mathbf{u}^\dag \mathbfcal{A}^l \mathbf{u}
    -\int_{\mathcal{I^+}} \mathbf{u}^\dag \mathbfcal{A}^n \mathbf{u}
    \,,
    \label{eqn:general_div_thm_on_D}
\end{align}
where~$\mathbf{N}^\mu$ is the vector normal to~$\p \mathcal{D}_1$
shown for the different segments in Fig.~\ref{Fig:domain_1} and
\begin{align*}
  \mathbfcal{A}^l \equiv \mathbfcal{A}^\mu l_\mu
  \,, \quad
  \mathbfcal{A}^\rho \equiv \mathbfcal{A}^\mu \rho_\mu
  \,, \quad
  \mathbfcal{A}^n \equiv \mathbfcal{A}^\mu n_\mu
  \,.
\end{align*}

\subsubsection{The homogeneous setup}

We first consider the homogeneous setup
\begin{align*}
  \mathbfcal{A}^\mu \p_\mu \mathbf{u} = 0
  \,,
\end{align*}
to obtain some insight of the problem. In this case the
relation~\eqref{eqn:general_div_thm_on_D} yields
\begin{align}
  \int_{\mathcal{N}_{u_i}} \mathbf{u}^\dag \mathbfcal{A}^l \mathbf{u}
  -\int_{\mathcal{T}} \mathbf{u}^\dag \mathbfcal{A}^\rho \mathbf{u}
  =
  \int_{\mathcal{N}_{u_f}} \mathbf{u}^\dag \mathbfcal{A}^l \mathbf{u}
  +\int_{\mathcal{I^+}} \mathbf{u}^\dag \mathbfcal{A}^n \mathbf{u}
  \,,
  \label{eqn:homo_NP_est_1}
\end{align}
which already provides the desired energy estimate. To make this more
apparent, let us split~$\mathbf{u}^\dag \mathbfcal{A}^\rho \mathbf{u}$
into the ingoing and outgoing
parts~$\mathbfcal{A}^\rho = \mathbfcal{A}^\rho_{\textrm{in}} -
\mathbfcal{A}^\rho_{\textrm{out}}$, which for the symmetric affine
null system of~\cite{Rip21} reads
\begin{align*}
  d_n \equiv \mathbb{1}_{n \times n}
  \,, \quad
  \mathbfcal{A}^\rho_{\textrm{in}}
  \equiv \textrm{diag}(1,0 \times
  d_{19})
  \,, \quad
  \mathbfcal{A}^\rho_{\textrm{out}}
  \equiv \textrm{diag}(0 \times d_4,
  d_{16})
  \,,
  \quad
\end{align*}
Considering this partition Eq.~\eqref{eqn:homo_NP_est_1} reads
\begin{align}
    \int_{\mathcal{N}_{u_i}} \mathbf{u}^\dag \mathbfcal{A}^l \mathbf{u}
    +\int_{\mathcal{T}} \mathbf{u}^\dag \mathbfcal{A}^\rho_{\textrm{out}} \mathbf{u}
    =
    \int_{\mathcal{N}_{u_f}} \mathbf{u}^\dag \mathbfcal{A}^l \mathbf{u}
    +\int_{\mathcal{T}} \mathbf{u}^\dag \mathbfcal{A}^\rho_{\textrm{in}} \mathbf{u}
    +\int_{\mathcal{I^+}} \mathbf{u}^\dag \mathbfcal{A}^n \mathbf{u}
    \,,
    \label{eqn:homo_NP_est_2}
\end{align}
where the left-hand-side includes only given data and the
right-hand-side only the solution. This manifests that the solution is
completely controlled by the given data in this setup.

\subsubsection{The inhomogeneous setup}

For the inhomogenous case
\begin{align*}
  \mathbfcal{A}^\mu \p_\mu \mathbf{u}
  + \mathbfcal{A} \mathbf{u} = 0
  \,,
\end{align*}
the relation~\eqref{eqn:general_div_thm_on_D} yields
\begin{align*}
  \int_{\mathcal{N}_{u_i}} \mathbf{u}^\dag \mathbfcal{A}^l \mathbf{u}
    +\int_{\mathcal{T}} \mathbf{u}^\dag \mathbfcal{A}^\rho_{\textrm{out}} \mathbf{u}
    + \int_{\mathcal{D}_1} \mathbf{u}^\dag \mathbfcal{A} \mathbf{u}
    =
    \int_{\mathcal{N}_{u_f}} \mathbf{u}^\dag \mathbfcal{A}^l \mathbf{u}
    +\int_{\mathcal{T}} \mathbf{u}^\dag \mathbfcal{A}^\rho_{\textrm{in}} \mathbf{u}
    +\int_{\mathcal{I^+}} \mathbf{u}^\dag \mathbfcal{A}^n \mathbf{u}
    \,.
\end{align*}
The difference in comparison to the homogenous case is the bulk
term~$\int_{\mathcal{D}_1} \mathbf{u}^\dag \mathbfcal{A}
\mathbf{u}$. The goal is to show that the solution
on~$\mathcal{N}_{u_f}, \mathcal{T},\mathcal{I}^+$ is controlled by the
given data on~$\mathcal{N}_{u_i},\mathcal{T}$. The idea is to use the
Gr\"onwall inequality. To make use of it we want to express the bulk
term as a double integral over a hypersurface and an appropriate
parameter such that the whole domain~$\mathcal{D}_1$ is covered. The
specific structure of the matrices~$\mathbfcal{A}^l$,
$\mathbfcal{A}^n$, $\mathbfcal{A}^\rho_{\textrm{in}}$,
$\mathbfcal{A}^\rho_{\textrm{out}}$ and $\mathbfcal{A}$ is crucial
here. In characteristic setups it is common that the
matrices~$\mathbfcal{A}^l$, $\mathbfcal{A}^n$,
$\mathbfcal{A}^\rho_{\textrm{in}}$,
$\mathbfcal{A}^\rho_{\textrm{out}}$ are degenerate, meaning that some
variables do not appear at all in the respective integrands. On the
contrary, a generic assumption is that the
integrand~$\mathbf{u}^\dag \mathbfcal{A} \mathbf{u}$ provides coupling
between all of the variables, or at least it does not provide coupling
solely between variables of one of the ingoing, outgoing and static
classes. With this assumption, we can write
\begin{align}
  &
    \int_{\mathcal{N}_{u_f}} \mathbf{u}^\dag \mathbfcal{A}^l \mathbf{u}
    +\int_{\mathcal{T}} \mathbf{u}^\dag \mathbfcal{A}^\rho_{\textrm{in}} \mathbf{u}
    +\int_{\mathcal{I^+}} \mathbf{u}^\dag \mathbfcal{A}^n \mathbf{u}
    \; \leq
    \nonumber
  \\
  &
    \int_{\mathcal{N}_{u_i}} \mathbf{u}^\dag \mathbfcal{A}^l \mathbf{u}
    +\int_{\mathcal{T}} \mathbf{u}^\dag \mathbfcal{A}^\rho_{\textrm{out}} \mathbf{u}
    + C \int_{\mathcal{D}_1} \mathbf{u}^\dag d_{\textrm{length}(\mathbf{u})} \mathbf{u}
    \,,
    \label{eqn:inhomo_NP_ineq_1}
\end{align}
for some constant~$C>0$. To see this, let us focus on the bulk
term. The integrand of the bulk
term~$\int_{\mathcal{D}_1} \mathbf{u}^\dag \mathbfcal{A} \mathbf{u}$
is real valued and involves a sum of terms of the form
\begin{align*}
  \pm (c_{ij} v_i \bar{v_j} + \bar{c}_{ij} \bar{v}_i v_j)
  = \pm \Re(c_{ij} v_i \bar{v}_j)
  \,,
\end{align*}
where~$v_i,\bar{v}_i \in \mathbf{u}$ and~$c_{ij}$ complex valued. For
each such term the following holds:
\begin{align*}
    |c_{ij}|^2|v_i|^2 + |v_j|^2
    \geq
    \mp \Re (c_{ij} v_i \bar{v}_j) 
  \,.
\end{align*}
Choosing a big enough real constant $C \geq \max |c_{ij}|^2$ leads
to~\eqref{eqn:inhomo_NP_ineq_1}.

In~\cite{Rip21} the integrand
$\mathbf{u}^\dag \mathbfcal{A} \mathbf{u}$ provides coupling between
ingoing and outgoing variables, which does not appear in any of the
other integrands. This is why we cannot use the Gr\"onwall inequality
like we described earlier for this setup approach and we fail to
obtain an energy estimate for the inhomogeneous case for the
domain~$\mathcal{D}_1$. However, from the
relation~\eqref{eqn:inhomo_NP_ineq_1} we see that if there was a
hypersurface integral with integrand of the
form~$\sim \mathbf{u}^\dag d_{\textrm{length}(\mathbf{u})} \mathbf{u}$
i.e. without a degeneracy, then we could indeed apply the method
previously explained.

\subsection{Domain~$\mathcal{D}_2$}
\label{subec:bondi_well-p:CIBVP_symm_hyp:D2}

\begin{figure}[t]
  \begin{center}
    \includegraphics[width=0.4\textwidth]{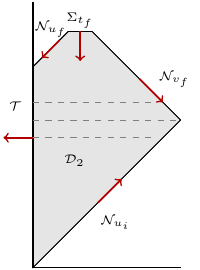}
  \end{center}
  \caption[CIBVP energy estimates for symmetric hyperbolic PDEs:
  domain~$\mathcal{D}_2$]{The geometry of the
    domain~$\mathcal{D}_2$. The upper boundary~$\mathcal{N}_{u_f}$ of
    domain~$\mathcal{D}_1$, which is an outgoing null hypersurface, is
    interrupted by the spacelike
    hypersurface~$\Sigma_{t_f}$. Furthermore, for simplicity this
    domain does not extend all the way to infinity, but is truncated
    to some finite radius~$r_{out}$.}
  \label{Fig:domain_2}
\end{figure}

We repeat the analysis for an inhomogeneous symmetric hyperbolic
characteristic PDE system, but in a different domain, which we
call~$\mathcal{D}_2$. The domain extends between the radii~$r_{in}$
and~$r_{out} < r_\infty$. The upper boundary is the previous outgoing
null hypersurface~$\mathcal{N}_{u_f}$ together with the spacelike
hypersurface~$\Sigma_{t_f}$. The outer boundary is a null hypersurface
of constant retarded time~$\mathcal{N}_{v_f}$. If we wish to extend
the domain to infinity, then~$\mathcal{N}_{v_f}$
becomes~$\mathcal{I}^+$ and~$\Sigma_{t_f}$ a hyperboloid i.e. a
hypersurface that is everywhere spacelike but becomes null at
infinity.

The divergence theorem in~$\mathcal{D}_2$ returns
\begin{align}
  &
    \int_{\mathcal{N}_{u_i}} \mathbf{u}^\dag \mathbfcal{A}^l \mathbf{u}
    +\int_{\mathcal{T}} \mathbf{u}^\dag \mathbfcal{A}^\rho_{\textrm{out}} \mathbf{u}
    + \int_{\mathcal{D}_2} \mathbf{u}^\dag \mathbfcal{A} \mathbf{u}
    =
    \nonumber
  \\
  &
    \int_{\Sigma_{t_f}} \mathbf{u}^\dag \mathbfcal{A}^t \mathbf{u}
    +\int_{\mathcal{N}_{u_f}} \mathbf{u}^\dag \mathbfcal{A}^l \mathbf{u}
    +\int_{\mathcal{T}} \mathbf{u}^\dag \mathbfcal{A}^\rho_{\textrm{in}} \mathbf{u}
    +\int_{\mathcal{N}_{v_f}} \mathbf{u}^\dag \mathbfcal{A}^n \mathbf{u}
    \,,
    \label{eqn:inhomo_NP_est_2}
\end{align}
where~$\mathbfcal{A}^t = d_{20} + \textrm{diag}(0,1,1,1,0 \times
d_{16})$ in~\cite{Rip21}. Like earlier the above yields
\begin{align}
  &
    \int_{\Sigma_{t_f}} \mathbf{u}^\dag \mathbfcal{A}^t \mathbf{u}
    +\int_{\mathcal{N}_{u_f}} \mathbf{u}^\dag \mathbfcal{A}^l \mathbf{u}
    +\int_{\mathcal{T}} \mathbf{u}^\dag \mathbfcal{A}^\rho_{\textrm{in}} \mathbf{u}
    +\int_{\mathcal{N}_{v_f}} \mathbf{u}^\dag \mathbfcal{A}^n \mathbf{u}
    \; \leq
    \nonumber
  \\
  & 
    \int_{\mathcal{N}_{u_i}} \mathbf{u}^\dag \mathbfcal{A}^l \mathbf{u}
    +\int_{\mathcal{T}} \mathbf{u}^\dag \mathbfcal{A}^\rho_{\textrm{out}} \mathbf{u}
    + C \int_{\mathcal{D}_2} \mathbf{u}^\dag d_{20} \mathbf{u}
    \; \leq
    \nonumber
  \\
  & 
    \int_{\mathcal{N}_{u_i}} \mathbf{u}^\dag \mathbfcal{A}^l \mathbf{u}
    +\int_{\mathcal{T}} \mathbf{u}^\dag \mathbfcal{A}^\rho_{\textrm{out}} \mathbf{u}
    + C \int_{\mathcal{D}_2} \mathbf{u}^\dag \mathbfcal{A}^t \mathbf{u}
       \,,
    \label{eqn:inhomo_NP_ineq_2}
\end{align}
Next, consider foliating~$\mathcal{D}_2$ with spacelike
hypersurfaces~$\Sigma_t$ (gray dashed lines in
Fig.~\ref{Fig:domain_2}) i.e.
\begin{align*}
  \int_{\mathcal{D}_2} \mathbf{u}^\dag \mathbfcal{A}^t \mathbf{u}
  =
  \int_{t_i}^{t_f}
  \left(
  \int_{\Sigma_t} \mathbf{u}^\dag \mathbfcal{A}^t \mathbf{u}
  \right) dt
  \,.
\end{align*}
Then,~\eqref{eqn:inhomo_NP_ineq_2} can be rearranged into
\begin{align}
  &
    \int_{\Sigma_{t_f}} \mathbf{u}^\dag \mathbfcal{A}^t \mathbf{u}
    \leq
    F
    + C \int_{t_i}^{t_f}
    \left(
    \int_{\Sigma_t} \mathbf{u}^\dag \mathbfcal{A}^t \mathbf{u}
    \right) dt
    \,,
    \label{eqn:inhomo_NP_ineq_3}
\end{align}
with
\begin{align*}
  F
  \equiv
  \int_{\mathcal{N}_{u_i}} \mathbf{u}^\dag \mathbfcal{A}^l \mathbf{u}
  -\int_{\mathcal{N}_{u_f}} \mathbf{u}^\dag \mathbfcal{A}^l \mathbf{u}
  +\int_{\mathcal{T}}
  \mathbf{u}^\dag
  \left(
  \mathbfcal{A}^\rho_{\textrm{out}} - \mathbfcal{A}^\rho_{\textrm{in}}
  \right)
  \mathbf{u}
  -\int_{\mathcal{N}_{v_f}} \mathbf{u}^\dag \mathbfcal{A}^n \mathbf{u}
  \,.
\end{align*}
Without any assumption on the monotonicity of~$F=F(t)$, we can apply
the first version of the Gr\"onwall inequality as given in
App.~\ref{sec:appA:Gronwall} on~\eqref{eqn:inhomo_NP_ineq_3} to obtain
\begin{align*}
  \int_{\Sigma_{t_f}} \mathbf{u}^\dag \mathbfcal{A}^t \mathbf{u}
    \leq
    F
    +
    \int_{t_i}^{t_f} F(t) C \exp
    \left( \int_t^{t_f} C ds \right) dt
    =
    F + C \int_{t_i}^{t_f} F(t) e^{C(t_f -t)} dt 
    \,,
\end{align*}
which after expanding out~$F$ yields
\begin{align}
  &
    \int_{\Sigma_{t_f}} \mathbf{u}^\dag \mathbfcal{A}^t \mathbf{u}
    +
    \nonumber
   \\
   & 
    \int_{\mathcal{N}_{u_f}}
    \mathbf{u}^\dag \mathbfcal{A}^l \mathbf{u}
    +
    C \int_{t_i}^{t_f} e^{C(t_f-t)}
    \left(
    \int_{\mathcal{N}_{u_f}} \mathbf{u}^\dag \mathbfcal{A}^l \mathbf{u}
    \right) dt
    +
    \nonumber
  \\
  & 
    \int_{\mathcal{T}}
    \mathbf{u}^\dag \mathbfcal{A}^\rho_{\textrm{in}} \mathbf{u}
    +
    C \int_{t_i}^{t_f} e^{C(t_f-t)}
    \left(
    \int_{\mathcal{T}} \mathbf{u}^\dag \mathbfcal{A}^\rho_{\textrm{in}} \mathbf{u}
    \right) dt
    +
    \nonumber
  \\
  &
    \int_{\mathcal{N}_{v_f}} \mathbf{u}^\dag \mathbfcal{A}^n \mathbf{u}
    +
    C \int_{t_i}^{t_f} e^{C(t_f-t)}
    \left(
    \int_{\mathcal{N}_{v_f}} \mathbf{u}^\dag \mathbfcal{A}^n \mathbf{u}
    \right) dt
    \;
    \leq
     \nonumber
  \\
    &
    \int_{\mathcal{N}_{u_i}} \mathbf{u}^\dag \mathbfcal{A}^l \mathbf{u}
    +
    C \int_{t_i}^{t_f} e^{C(t_f-t)}
    \left(
    \int_{\mathcal{N}_{u_i}} \mathbf{u}^\dag \mathbfcal{A}^l \mathbf{u}
    \right) dt
    +
    \nonumber
  \\
  &
    \int_{\mathcal{T}}
    \mathbf{u}^\dag \mathbfcal{A}^\rho_{\textrm{out}} \mathbf{u}
    +
    C \int_{t_i}^{t_f} e^{C(t_f-t)}
    \left(
    \int_{\mathcal{T}} \mathbf{u}^\dag \mathbfcal{A}^\rho_{\textrm{out}} \mathbf{u}
    \right) dt
    \,,
    \label{eqn:inhomo_NP_ineq_4}
\end{align}
where terms have been rearranged such that all the solution parts
appear on the left-hand-side of the inequality. Since~$C>0$
and~$t_f \geq t$ the following holds for the terms of the
left-hand-side of~\eqref{eqn:inhomo_NP_ineq_4}
\begin{align*}
  &
    \int_{\Sigma_{t_f}} \mathbf{u}^\dag \mathbfcal{A}^t \mathbf{u}
    +  \int_{\mathcal{N}_{u_f}}
    \mathbf{u}^\dag \mathbfcal{A}^l \mathbf{u}
    +  \int_{\mathcal{T}}
    \mathbf{u}^\dag \mathbfcal{A}^\rho_{\textrm{in}} \mathbf{u}
    + \int_{\mathcal{N}_{v_f}} \mathbf{u}^\dag \mathbfcal{A}^n \mathbf{u}
    \;
    \leq
  \\
  & 
    \int_{\Sigma_{t_f}} \mathbf{u}^\dag \mathbfcal{A}^t \mathbf{u}
    +
  \\
  & 
    \int_{\mathcal{N}_{u_f}}
    \mathbf{u}^\dag \mathbfcal{A}^l \mathbf{u}
    +
    C \int_{t_i}^{t_f} e^{C(t_f-t)}
    \left(
    \int_{\mathcal{N}_{u_f}} \mathbf{u}^\dag \mathbfcal{A}^l \mathbf{u}
    \right) dt
    +
  \\
  & 
    \int_{\mathcal{T}}
    \mathbf{u}^\dag \mathbfcal{A}^\rho_{\textrm{in}} \mathbf{u}
     +
    C \int_{t_i}^{t_f} e^{C(t_f-t)}
    \left(
    \int_{\mathcal{T}} \mathbf{u}^\dag \mathbfcal{A}^\rho_{\textrm{in}} \mathbf{u}
    \right) dt
    +
  \\
  &
    \int_{\mathcal{N}_{v_f}} \mathbf{u}^\dag \mathbfcal{A}^n \mathbf{u}
    +
    C \int_{t_i}^{t_f} e^{C(t_f-t)}
    \left(
    \int_{\mathcal{N}_{v_f}} \mathbf{u}^\dag \mathbfcal{A}^n \mathbf{u}
    \right) dt
    \,,
\end{align*}
which when combined with~\eqref{eqn:inhomo_NP_ineq_4} yields
\begin{align}
  &
    \int_{\Sigma_{t_f}} \mathbf{u}^\dag \mathbfcal{A}^t \mathbf{u}
    +
    \int_{\mathcal{N}_{u_f}}
    \mathbf{u}^\dag \mathbfcal{A}^l \mathbf{u}
    +
    \int_{\mathcal{T}}
    \mathbf{u}^\dag \mathbfcal{A}^\rho_{\textrm{in}} \mathbf{u}
    +
    \int_{\mathcal{N}_{v_f}} \mathbf{u}^\dag \mathbfcal{A}^n \mathbf{u}
    \;
    \leq
    \nonumber
  \\
  &
    \int_{\mathcal{N}_{u_i}} \mathbf{u}^\dag \mathbfcal{A}^l \mathbf{u}
    +
    C \int_{t_i}^{t_f} e^{C(t_f-t)}
    \left(
    \int_{\mathcal{N}_{u_i}} \mathbf{u}^\dag \mathbfcal{A}^l \mathbf{u}
    \right) dt
    +
    \nonumber
  \\
  &
    \int_{\mathcal{T}}
    \mathbf{u}^\dag \mathbfcal{A}^\rho_{\textrm{out}} \mathbf{u}
    +
    C \int_{t_i}^{t_f} e^{C(t_f-t)}
    \left(
    \int_{\mathcal{T}} \mathbf{u}^\dag \mathbfcal{A}^\rho_{\textrm{out}} \mathbf{u}
    \right) dt
    \,.
    \label{eqn:inhomo_NP_ineq_5}
\end{align}
Hence, the solution to the CIBVP (the left-hand-side
of~\eqref{eqn:inhomo_NP_ineq_5}) is controlled by the given data (the
right-hand-side of~\eqref{eqn:inhomo_NP_ineq_5}).

\section{Conclusions}
\label{sec:bondi_well-p:conclusions}

Popular Bondi-like systems are weakly hyperbolic and textbook results
on these systems then show that they are ill-posed in the~$L^2$-norm
or its obvious derivatives. Considering model problems of a similar
structure we saw that the same result naturally carries over to the
CIBVP. In the latter case care is needed not to confuse the usual
degeneracy of the norms that appear naturally in characteristic
problems with high-frequency blow-up of solutions. This degeneracy is
apparent in the calculations of
Sec.~\ref{sec:bondi_well-p:CIBVP_symm_hyp} and leads to the change of
domain for the energy estimate as shown in Figs.~\ref{Fig:domain_1}
and~\ref{Fig:domain_2}.

Although our weakly hyperbolic toy model is ill-posed in~$L^2$, it may
be well-posed in a lopsided norm in which the angular derivative of
some specific components of the state vector are included. Thus in
such a case one must be able to control the size of not only the
elements of the state vector in the given data, but also some of their
derivatives. This weaker notion of well-posedness is sensitive to the
presence of lower order source terms. For example, our weakly
hyperbolic model is well-posed in a (specific) lopsided norm if it is
homogeneous, or inhomogeneous with sources that respect the nested
structure of the equations intrinsic to the characteristic
hypersurfaces. If this nested structure is broken by the source terms,
it becomes ill-posed in any sense.

Bringing our attention back to the characteristic initial boundary
value problem for GR, which covers both CCE and applications in
numerical holography, it is clear that the Bondi-like formulations we
considered are ill-posed in~$L^2$-norm. It is not clear however, in
general, if they will admit estimates in suitable lopsided norms. But
since the field equations {\it do} have a nested structure, and our
weakly hyperbolic model problem turned out to admit estimates in
lopsided norms whenever this structure was present, there is reason to
be hopeful. More importantly, Bondi-like formulations where the
curvature is an evolved variable provide symmetric hyperbolic
setups~\cite{CabChrTag14,HilValZha19,HilValZha20, Rip21}. Promoting
the curvature to an evolved variable effectively translates into
including specific combinations of second order metric derivatives as
independent variables. In addition, in these setups, the Bianchi
identities provide equations of motion for some variables.
In Sec.~\ref{sec:bondi_well-p:CIBVP_symm_hyp} we present work towards
obtaining energy estimates with model symmetric hyperbolic systems for
the CIBVP. If this system is one of the latter, then it could assist
us in finding a lopsided norm for the weakly hyperbolic Bondi-like
systems analyzed earlier. Comparing with the results of
Sec.~\ref{sec:bondi_well-p:axisym_BS} it seems that constructing an
appropriate lopsided norm directly within the weakly hyperbolic
formulations could be cumbersome. The hope is that by performing an
explicit mapping between the variables of the different formalisms, we
can construct an appropriate lopsided norm for the weakly hyperbolic
one guided by the~$L^2$-norm of the symmetric hyperbolic system. Work
in this direction is undergoing.

A true {\it principle} solution to wave-extraction would be one where
the PDE problem solved is manifestly well-posed and the GW signal is
computed at null infinity. A robust scheme for CCM could be such a
solution, meaning a CCM setup where the composite PDE problem is
well-posed. An alternative to CCM is the use of compactified
hyperboloidal slices, which is also an active research topic for full
GR~\cite{BarSarBuc11,Zen10,VanHusHil14,Van15,DouFra16,HilHarBug16,
  VanHus17,GasHil18,GasGauHil19,BeyFraHen20}. Clearly, a well-posed
PDE problem is essential for the hyperboloidal approach as well. To
understand the consequences of our findings for CCM, in
Sec.~\ref{sec:bondi_well-p:toy_CCE_CCM} we considered a model in which
the IBVP is solved for a symmetric hyperbolic system, and the
solutions are then glued through boundary conditions to those of a
weakly hyperbolic system accepting estimates in lopsided norms. The
former of these two sets of equations is viewed as a model for the
formulation used in the strong-field region, the latter for a WH
Bondi-like formulation used on the outer characteristic domain. With
this setup, we found that the fundamental incompatibility of the norms
naturally associated with the two domains prohibits their combined use
in building estimates. But if the weakly hyperbolic system were made
symmetric hyperbolic progress could be made. A less appealing
possibility would be to demonstrate that the formulation in the Cauchy
domain, or some suitable replacement, admits estimates in a lopsided
norm compatible with that of the characteristic region. Since this
relies on very special structure in the field equations, the outlook
for a complete proof of well-posedness of CCM using existing
Bondi-like gauges is, unfortunately, not rosy. Furthermore, the fact
that numerical approximation to weakly hyperbolic systems (using
lopsided norms) is poorly understood, it is desirable to obtain and
adopt strongly or ideally symmetric hyperbolic alternatives.


%% file: sections/numerical_experiments.tex
\chapter{Numerical Experiments}
\label{chap:numerics}

\minitoc

When a PDE problem is well-posed, then its numerical approximation
converges to the true solution of the problem with increasing
resolution. In this convergence process, it is necessary to specify
the norm in which the problem is well-posed. Following the analysis of
the previous chapter, we put to the test some of the norms presented
there. The tests are performed both for the toy models, and in full
GR.

\section{Toy models}
\label{sec:numerics:toys}

First, we use the toy models introduced in
Sec. \ref{sec:bondi_well-p:toy_PDEs} to diagnose the effects of weak
hyperbolicity at the numerical level. We perform convergence tests in
the single-null setup for both the WH and SH models in a discrete
approximation to the~$L^2$-norm, for smooth and noisy given data. We
also perform convergence tests with noisy given data in the lopsided
norm, for the different versions of the WH model analyzed in the
previous section.

\subsection{Implementation}
\label{subsec:numerics:toys:implementation}

As in other schemes to solve the CIBVP, several different ingredients
are needed in the algorithm. These can be summarized for our
models~\eqref{eqn:WH_model} and~\eqref{eqn:SH_model} as follows:
\begin{enumerate}

\item The domain of the PDE problem is~$x \in [0,1]$,~$z \in[0,2\pi)$
  with periodic boundary conditions and~$u \in[u_0, u_f]$, with~$u_0$
  and~$u_f$ the initial and final times respectively. We always
  include the point~$x=1$ in the computational domain so that we do
  not need to impose boundary conditions at the outer boundary, since
  there are no incoming characteristic variables there.

\item For the initial time~$u_0$ provide initial data~$\psi(u_0,x,z)$
  on the surface~$u=u_0$ and boundary data~$\phi(u_0,0,z)$
  and~$\psi_\varv(u_0,0,z)$.

\item Integrate the intrinsic equations of each model to
  obtain~$\phi(u_0,x,z)$ and~$\psi_\varv(u_0,x,z)$. We perform this
  integration using the two-stage, second order strong stability
  preserving method of Shu and Osher (SSPRK22)~\cite{ShuOsh88}.
  
\item Integrate the evolution equation of each model to
  obtain~$\psi(u_1,x,z)$ at the surface~$u=u_1=u_0 + \Delta u$. We
  choose~$\Delta u = 0.25 \Delta x$ to satisfy the
  Courant-Friedrichs-Lewy (CFL) condition and the numerical
  integration is performed using the fourth order Runge-Kutta (RK4)
  method.

\item Any derivative appearing in the right-hand-sides of these
  integrations is approximated using second order accurate centered
  finite difference operators, except at the boundaries, where second
  order accurate forward and backward difference operators are used
  respectively. 
  
\item Providing boundary data~$\phi(u,0,z)$ and~$\psi_\varv(u,0,z)$ as
  in the PDE specification~\eqref{eqn:model_boundary_data}, we repeat
  steps~$2$ and~$3$ to obtain~$\phi(u,x,z)$,~$\psi_\varv(u,x,z)$
  and~$\psi(u,x,z)$ until the final time~$u_f$. This is the solution
  of the PDE.
  
\end{enumerate}
No artificial dissipation is introduced. The implementation was made
using the Julia language~\cite{BezEdeKar17} with the
DifferentialEquations.jl package~\cite{RacNie17} to integrate the
equations. Our code is freely available~\cite{GiaHilZil20_public}. We
apply convergence tests to our numerical scheme for both toy
models. The tests are performed for smooth, as well as for noisy given
data. The latter are often called robust stability tests. They form
part of the Mexico-city testbed for numerical
relativity~\cite{AlcAllBau03}. These tests have been performed widely
in the literature~\cite{CalHinHus05,Hin05,BoyLinPfe06,BabHusHin07,
  WitHilSpe10, CaoHil11}, often, as in our case, with adaptations for
the setup under consideration.

\subsection{Convergence tests}
\label{subsec:numerics:toys:convergence}

By~\textit{convergence} we mean the requirement that the difference
between the numerical approximation provided by a finite difference
scheme and the exact solution of the continuum PDE system tends to
zero as the grid spacing is increased. The finite difference scheme is
called~\textit{consistent} when it approximates the correct PDE system
and the degree to which this is achieved is its~\textit{accuracy}. The
scheme is called~\textit{stable} if it satisfies a discretized version
of~\eqref{eqn:wp_inequality} or~\eqref{eqn:weak_wp_inequality}. In
this context versions of each continuum norm are replaced by a
suitable discrete analog. Here we replace the~$L^2$-norm for the
characteristic setup with
\begin{align}
  ||\mathbf{u}||_{h_u,h_x,h_z}^2 =
  \sum_{x,z} \, \psi^2 \, h_x \, h_z
  +
  \textrm{max}_x \sum_{u,z} \, \left(\phi^2 + \psi_\varv^2  \right)
  h_u \, h_z
  \,,
  \label{discrete_L2}
\end{align}
with the first sum taken over all points on the grid, with~$h_x$
and~$h_z$ the grid-spacing in the~$x$ and~$z$ directions respectively,
and the second sum over all points in the~$z$ and~$u$ directions
($h_u = 0.25 h_x$ for our setup), for all~$x$ grid points and keeping
the maximum in the $x$ direction. The first sum involves only ingoing
and the second only outgoing variables. When, as will be the case in
what follows, we have~$h_x=h_z=h$ we label the norm simply
with~$h$. Our discrete approximation to the lopsided norm is,
\begin{align}
  ||\mathbf{u}||^2_{\textrm{q}(h_u, h_x , h_z)}
  =
  \sum_{x,z} \, \psi^2 \, h_x \, h_z
  +
  \textrm{max}_x \sum_{u,z} \, \left(\phi^2 + \psi_\varv^2 +
  \left(D_z \phi\right)^2
  \right)
  h_u \, h_z
  \,,
  \label{discrete_lopsided}
\end{align}
where~$D_z$ is the second order accurate, centered, finite difference
operator that replaces the continuum operator~$\p_z$, by
\begin{align}
  D_z f_{h}(x_i) = \frac{f_{h}(x_{i+1}) - f_{h}(x_{i-1})}{2h_z}
  \, ,\label{FT_stencil}
\end{align}
for a grid function~$f_h$ on a grid with spacing~$h_z$. When the two
grid spacings are equal we again label the norm simply with~$h$. This
approximation to the continuum lopsided norm is not unique. If we were
attempting to prove that a particular discretization converged, it
might be necessary to take another. Denoting by~$f$ the solution to
the continuum system and as~$f_h$ the numerical approximation at
resolution~$h$ provided by a convergent finite difference scheme of
accuracy~$n$, then
\begin{align}
  f = f_h + O\left( h^n\right) \,,
  \label{numerical_approx}
\end{align}
and hence
\begin{align}
  ||f - f_h|| = O(h^n)\, ,
  \label{exact_convergence_relation}
\end{align}
in some appropriate norm~$|| \cdot ||$ on the grid, with the
understanding that the exact solution should be evaluated on said
grid. Full definitions of the notions of consistency, stability and
convergence for the IVP can be found, for example,
in~\cite{GusKreOli95,Tho98c,Hin05}.

We use a second order accurate numerical approximation, so
that~$n=2$. Considering numerical evolutions with coarse, medium and
fine grid spacings~$h_c$,~$h_m$ and~$h_f$ respectively, we can
construct a useful quantity for these tests
\begin{align}
  Q \equiv \frac{h_c^n -h_m^n}{h_m^n - h_f^n}
  \,, \label{convergence_factor}
\end{align}
which we call~\textit{convergence factor}. In our convergence tests we
solve the same discretized PDE problem for different resolutions and
every time we want to increase resolution we halve the grid-spacing in
all directions i.e.
\begin{align*}
  h_m = h_c/2 \,, \quad h_f = h_c/4 \,.
\end{align*}
Following this approach the convergence factor
is~$Q=4$. Combining~\eqref{numerical_approx}
and~\eqref{convergence_factor} one can obtain the relation
\begin{align}
  f_{h_c} - f_{h_c/2} = Q \left(f_{h_c/2} - f_{h_c/4} \right)
  \,, \label{pointwise_conv_relation}
\end{align}
understood on shared grid-points in the obvious way, which is used to
investigate pointwise convergence. In what follows the different
resolutions are denoted as
\begin{align*}
  h_q = h_0/2^q \,.
\end{align*}
The lowest resolution~$h_0$ has $N_x=17$ points in the $x$-grid
and~$N_z=16$ in the $z$-grid. We work in units of the code in the
entire section.

\subsubsection{Smooth data}

For the simulations with smooth given data the initial and final times
are~$u_0=0$ and~$u_f=1$ respectively. For both toy models we provide as
initial data
\begin{align*}
  \psi(0,x,z)=e^{-100 \left( x-1/2 \right)^2} \, \sin(z)\,,
\end{align*}
and as boundary data
\begin{align*}
\phi(u,0,z)= 3 \, e^{-100 \left( u-1/2 \right)^2} \, \sin(z)\,,
\end{align*}
and
\begin{align*}
\psi_\varv(u,0,z)=e^{-100 \left( u-1/2 \right)^2} \,\sin(z)\,.
\end{align*}
For the SH model we choose the following source terms
\begin{align}
  -S_\phi = \psi \,,
  \quad -S_{\psi_\varv} = \phi + \psi \,,
  \quad -S_\psi = \phi\,,
\end{align}
and for the WH model we choose the homogeneous case. As discussed in
Sec.~\ref{sec:pde_theory:well-posedness}, well-posedness of the SH
model is unaffected by lower order source terms, so the specific
choice of source terms here is not vital. However, we choose to work
with the homogeneous WH model, because weakly well-posed problems are
sensitive to lower order perturbations.

\begin{figure}[t]
  \begin{center}
  {\includegraphics[width=0.32\textwidth]{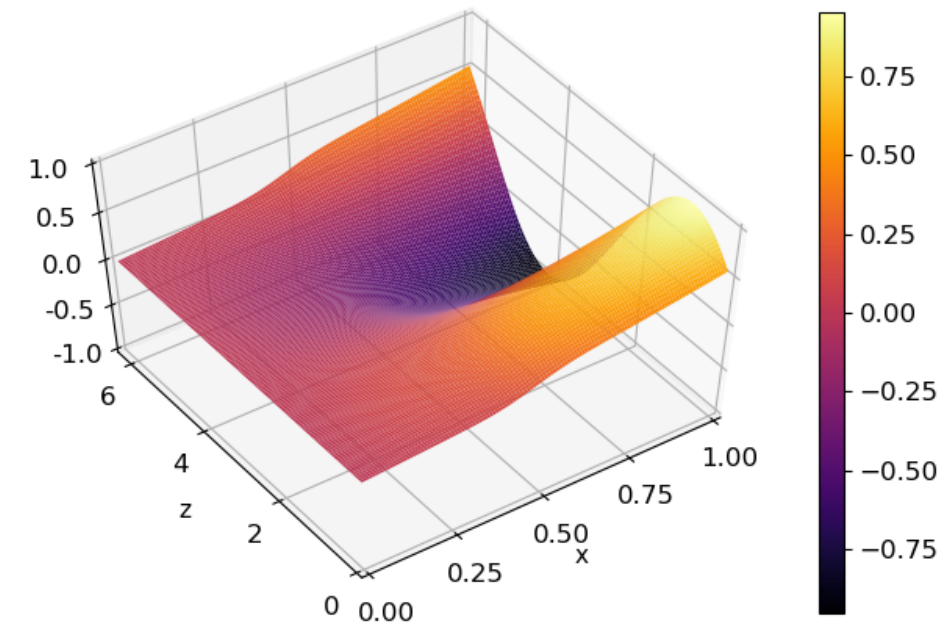}}
  {\includegraphics[width=0.32\textwidth]{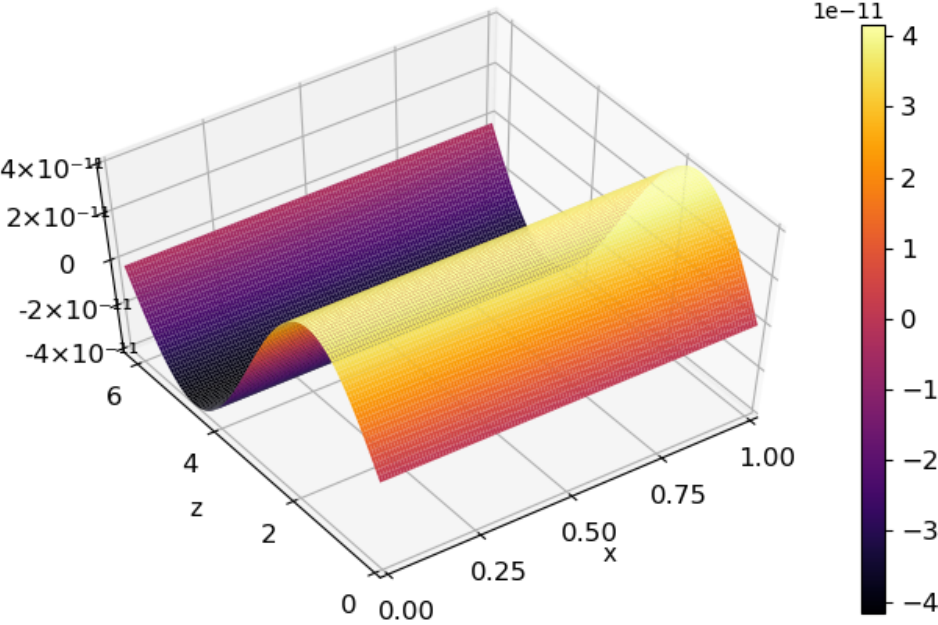}}
  {\includegraphics[width=0.32\textwidth]{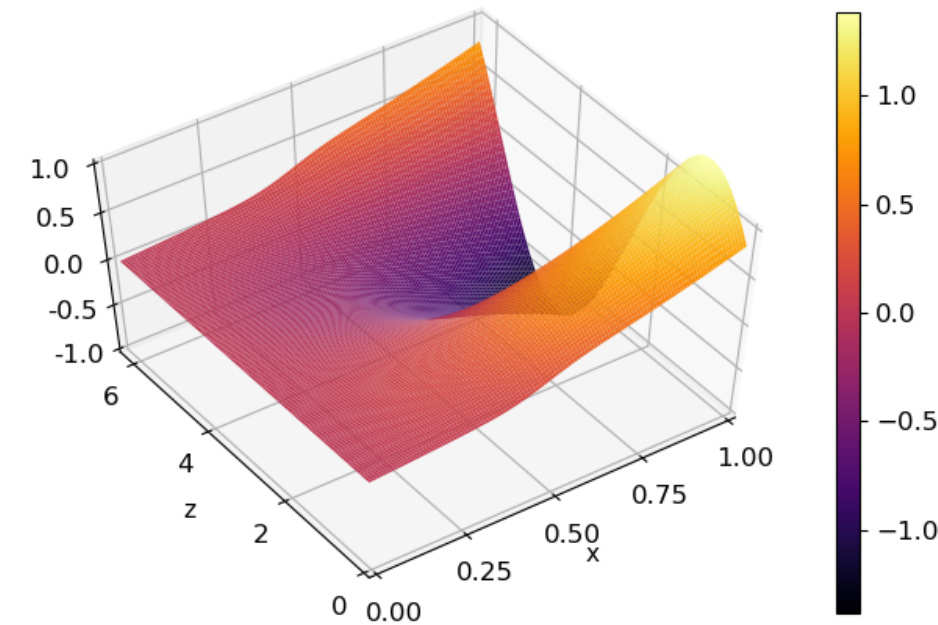}}
  \\
  \vspace{0.2cm}
  {\includegraphics[width=0.32\textwidth]{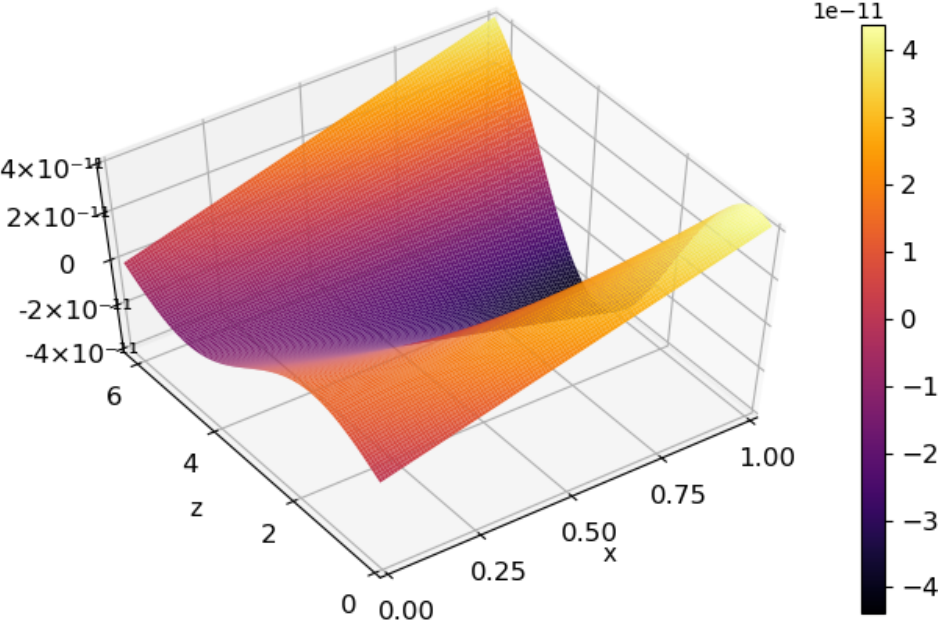}}
  {\includegraphics[width=0.32\textwidth]{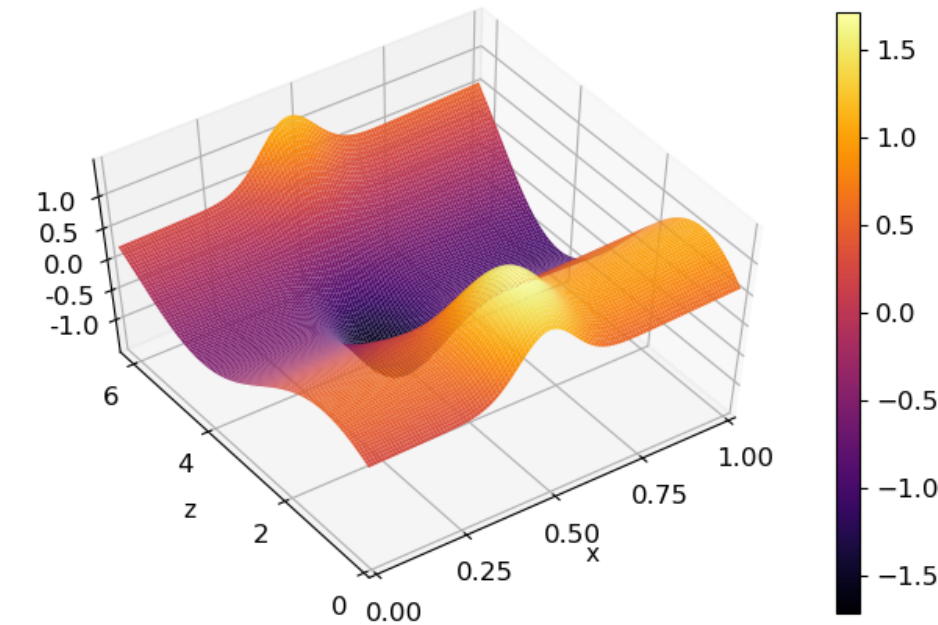}}
  {\includegraphics[width=0.32\textwidth]{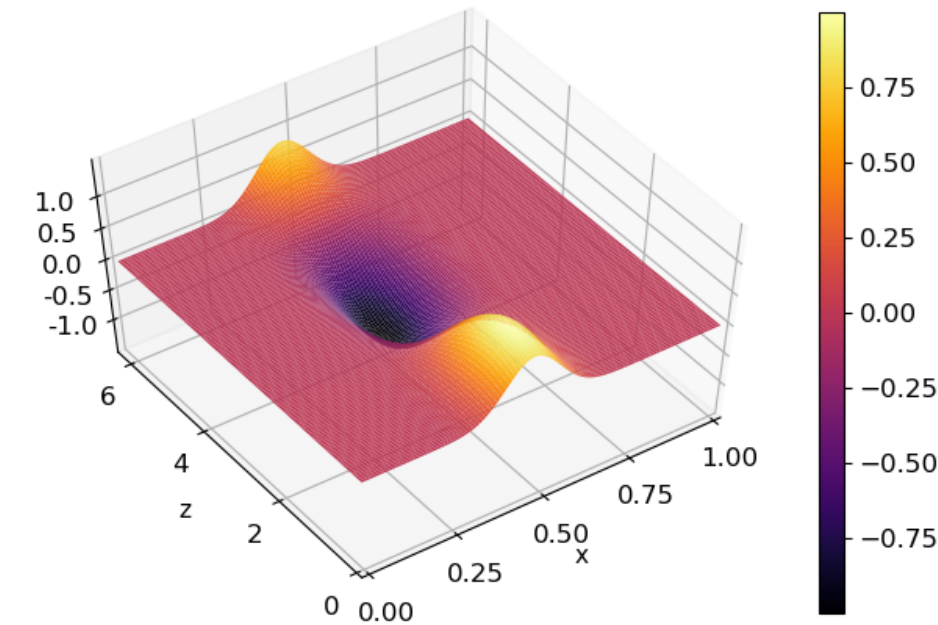}}
\end{center}
\caption[The final profiles of fields~$\phi$, $\psi_\varv$ and~$\psi$
for the toy model numerical experiments]{The fields~$\phi$,
  $\psi_\varv$ and~$\psi$ at final evolution time~$u = 1$, for the SH
  model (top) and the homogeneous WH model (bottom), with the same
  smooth given data.  Observe that the fields~$\phi$ and~$\psi_\varv$
  in the WH case are still of the same magnitude~$\sim 10^{-11}$ as
  the boundary data at the retarded time~$u=1$. This is not true once
  generic source terms are taken. \label{Fig:smooth_sols}}
\end{figure}

\begin{figure}[t]
  \begin{center}
    \includegraphics[width=0.85\textwidth]{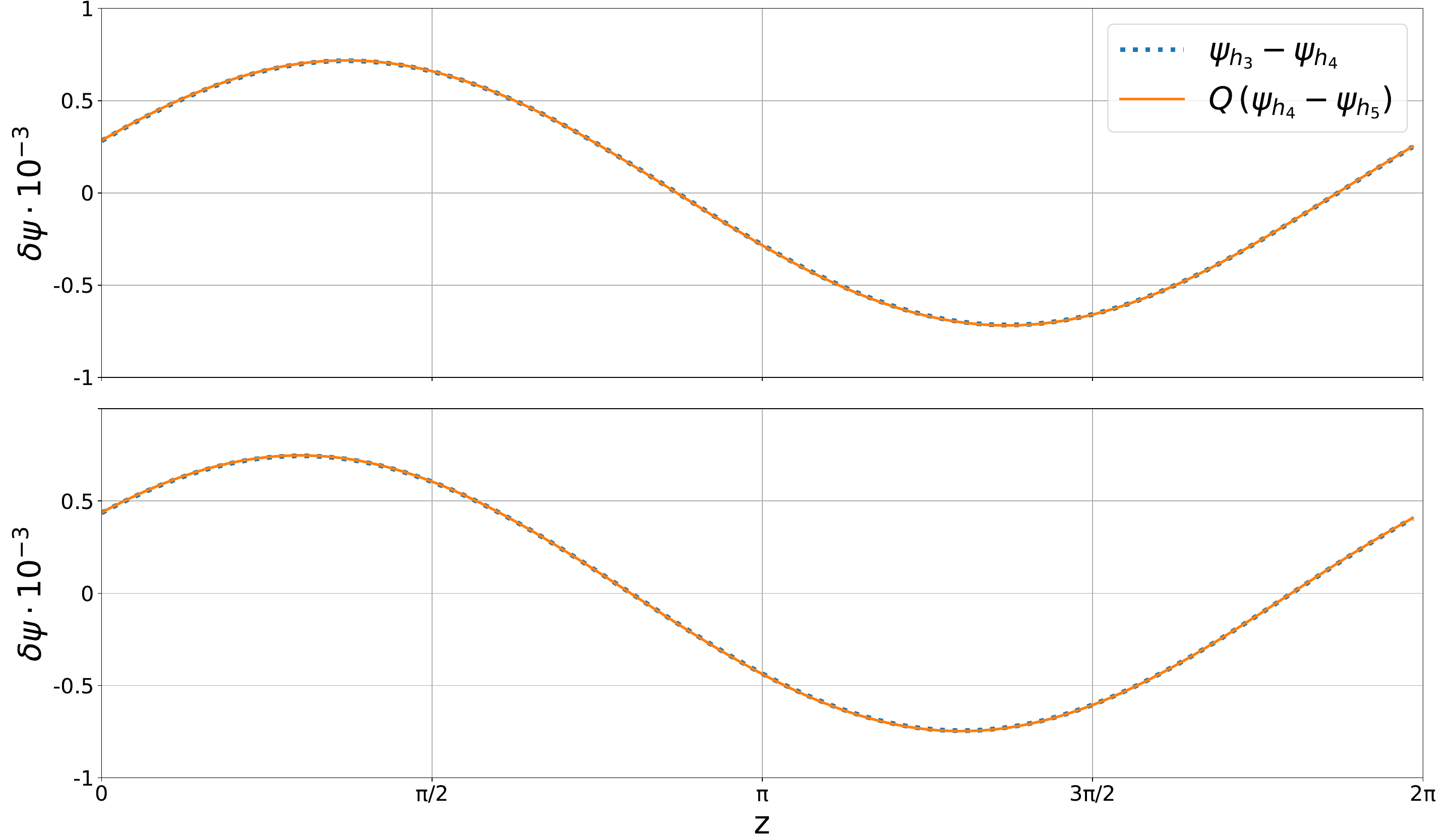}
  \end{center}
  \centering{
    \caption[Pointwise convergence for toy model numerical
    experiments]{Here we plot simultaneously~$\psi_{h_3} - \psi_{h_4}$
      and~$Q \left( \psi_{h_4} - \psi_{h_5}\right)$, for the SH (top)
      and the WH (bottom) toy models. We fix~$x=0.5$. Since our scheme
      is second order and we are doubling resolution we effectively
      fix~$Q=4$. The results for fixed~$z$ are similar. The plot is
      compatible with perfect second order pointwise convergence.}
  \label{Fig:pointwise_convergence}}
\end{figure}

Runs with resolutions~$h_0,\, h_1,\, h_2 ,\, h_3,\, h_4$ and $h_5$
were performed. In Fig.~\ref{Fig:smooth_sols} the basic dynamics are
plotted with each model. To first verify that the numerical scheme is
implemented successfully we performed pointwise convergence tests for
both models. We focus specifically here on the highest three
resolutions. The algorithm is the following:
\begin{enumerate}

\item Consider~$h_3$, $h_4$ and~$h_5$ as coarse, medium and fine
  resolutions, respectively.
  
\item Calculate $\psi_{h_3} - \psi_{h_4}$ and
  $\psi_{h_4} - \psi_{h_5}$ for the gridpoints of $h_3$, for the final
  timestep of the evolution.

\item Plot simultaneously $\psi_{h_3} - \psi_{h_4}$ and
  $Q \left( \psi_{h_4} - \psi_{h_5}\right)$. As indicated
  from~\eqref{pointwise_conv_relation}, for a convergent numerical
  scheme the two quantities should overlap, when multiplying the
  latter with the appropriate convergence factor.
  
\end{enumerate}
In Fig.~\ref{Fig:pointwise_convergence} we illustrate the results of
this test for the aforementioned smooth given data for both models. At
this resolution one clearly observes perfect pointwise convergence in
both cases.

\begin{figure*}[t]
  \begin{center}
    \hspace*{-1.2cm} \includegraphics[width=1.1\textwidth]{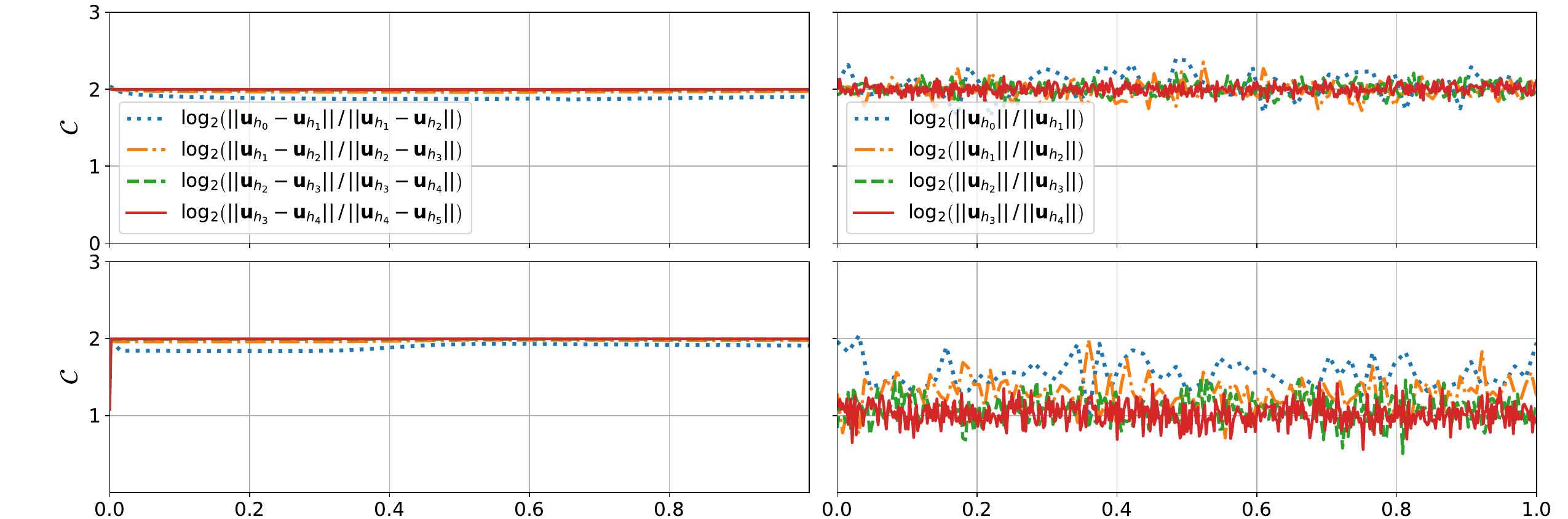}
  \end{center}
  \centering{
    \caption[The self convergence ratio in the~$L^2$-norm for toy
    model numerical experiments]{The convergence ratio in
      the~$L^2$-norm, for the strongly (above) and the weakly (below)
      toy models, for smooth (left) and noisy (right) given data, as a
      function of the simulation time. All plots have the same scale
      on the~$y$-axis. For smooth given data we consider the self
      convergence ratio~\eqref{self_conv_ratio_log2} while for noisy
      given data the exact convergence
      ration~\eqref{exact_conv_ratio}. If we consider the self
      convergence ratio also for the noisy case the results are
      qualitatively the same.\label{fig:L2_norms_all}}}
\end{figure*}

We also wish to examine convergence of our numerical solutions in
discrete approximations of the aforementioned norms. Given that the
exact solution to the PDE problem is unknown and that each time we
increase resolution we decrease the grid spacing in all directions by
a factor of~$d$, we can build the following useful quantity
\begin{align}
  \mathcal{C}_\textrm{self} =
  \log_d
  \frac{|| \mathbf{u}_{h_c} - \perp^{h_c/d}_{h_c}\mathbf{u}_{h_c/d} ||_{h_c} }
       { || \perp^{h_c/d}_{h_c} \mathbf{u}_{h_c/d}
         - \perp^{h_c/d^2}_{h_c}\mathbf{u}_{h_c/d^2} ||_{h_c} }\, ,
       \label{self_conv_ratio}
\end{align}
which we call~\textit{self-convergence ratio}, with
~$\mathbf{u}=\left(\phi, \psi_\varv, \psi \right)^T$ the state vector
of the PDE system and~$\phi$,~$\psi_\varv$,~$\psi$ grid
functions. Here~$\perp^{h_c/d}_{h_c}$ denotes the projection (in our
setup injection) operator from the~$h_c/d$ grid onto the~$h_c$ grid.
We calculate~$\mathcal{C}_\textrm{self}$ for a discrete analog of
the~$L^2$-norm. However, if one wishes to examine convergence in a
different norm, $L^2$ can be replaced with that. The theoretical value
of~$\mathcal{C}_\textrm{self}$ equals the accuracy~$n$ of the
numerical scheme, and in our specific setup
\begin{align}
  \mathcal{C}_\textrm{self} = \textrm{log}_2 \frac{|| \mathbf{u}_{h_c}
    - \perp^{h_c/2}_{h_c}\mathbf{u}_{h_c/2} ||_{h_c} } { ||
    \perp^{h_c/2}_{h_c}\mathbf{u}_{h_c/2} -
    \perp^{h_c/4}_{h_c}\mathbf{u}_{h_c/4} ||_{h_c} } = 2 \,.
  \label{self_conv_ratio_log2}
\end{align}
We obtain numerical solutions for the same smooth given data for both
models at the various resolutions mentioned before. For triple
resolution, double resolution and quadruple resolution, we project all
gridfunctions onto the coarse grid, and compute~$C_\textrm{self}$ at
its timesteps. In the left panel of Fig.~\ref{fig:L2_norms_all} we
collect the results of these norm convergence tests. Both models show
similar behavior. At low resolutions curve drifts from the desired
rate at early times, but the situation improves as we increase
resolution, with~$C_\textrm{self}$ approaching the expected value. The
trend with increasing resolution is the essential behavior we are
looking at in these tests. By limiting ourselves to convergence tests
with smooth given data we could be misled that the WH toy model
provides a well-posed CIBVP in the~$L^2$-norm, since the numerical
solutions appear to converge in this norm during our simulations. In
other words, were we ignorant of the hyperbolicity of the system, it
would be impossible to distinguish strongly and weakly hyperbolic PDEs
with this test.
 
\subsubsection{Noisy data}

One can also perform norm convergence tests with random noise as given
data, which is a strategy to simulate numerical error in an
exaggerated form. Since it is expected that numerical error decreases
as resolution increases, when performing simulations for these tests
one must scale appropriately the amplitude of the noise as resolution
improves. This scaling is important to construct a sequence of initial
data that converges in a suitable norm to initial data appropriate for
the continuum system. The choice of norm here is essential, and should
be one which, if possible, provides a bound for the solution of a
(weakly) well-posed PDE problem, in the sense
of~\eqref{eqn:wp_inequality} and~\eqref{eqn:weak_wp_inequality}.

For these tests we perform simulations where the smooth part of the
given data is trivial (zero), and hence the exact solution for every
PDE problem based on our models vanishes identically. Knowing the
exact solution, in addition to the self convergence
rate~\eqref{self_conv_ratio}, we can also construct the exact
convergence ratio
\begin{align}
  \mathcal{C}_\textrm{exact} = \log_d
  \frac{|| \mathbf{u}_{h_c} - \mathbf{u}_\textrm{exact} ||_{h_c} }
       {|| \perp^{h_c/d}_{h_c}\mathbf{u}_{h_c/d}
         - \mathbf{u}_\textrm{exact} ||_{h_c} }\, ,
  \label{exact_conv_ratio}
\end{align}
where we decrease grid spacing by a factor of~$d$ when increasing
resolution. $\mathcal{C}_\textrm{exact}$ is cheaper numerically
than~$\mathcal{C}_\textrm{self}$ since only two different resolutions
are required to build it, and again the exact solution is understood
to be evaluated on the grid itself. It is possible for a scheme to be
self-convergent but fail to be convergent, for example if one were to
implement the wrong field equations. Therefore one would like to
compare the numerical solution to an exact solution wherever (rarely)
possible. To calculate~$\mathcal{C}_{\textrm{exact}}$ we compute the
discretized approximation to a suitable continuum norm at two
resolutions, one twice the other. Each are computed on the naturally
associated grid. We then take the ratio of the two at shared
timesteps, corresponding to those of the coarse grid~$h_c$. In our
setup~$\mathbf{u}_\textrm{exact} = \mathbf{0}$ and~$d=2$, hence
\begin{align}
  \mathcal{C}_\textrm{exact} = \log_2
  \frac{|| \mathbf{u}_{h_c} ||_{h_c} }
  {|| \perp^{h_c/2}_{h_c}\mathbf{u}_{h_c/2} ||_{h_c} } 
  \,,\label{exact_conv_ratio_log2}
\end{align}
which again equals two for perfect convergence. As previously
mentioned appropriate scaling of the random noise amplitude is crucial
and is determined by the norm in which we wish to test convergence. To
realize the proper scaling in our setup, let us consider the exact
convergence ratio~\eqref{exact_conv_ratio_log2} and denote
as~$A_{h_c}$ and~$A_{h_c/2}$ the amplitude of the random noise for
simulations with resolution~$h_c$ and~$h_c/2$ respectively
\begin{align*}
  \mathcal{C}_\textrm{exact} = \log_2
  \frac{||\mathbf{u}_{h_c}||_{h_c}}{||\perp^{h_c/2}_{h_c}
    \mathbf{u}_{h_c/2}||_{h_c}} \sim \log_2
  \frac{O(A_{h_c})}{O(A_{h_c/2})} \, .
\end{align*}
The above suggests that to construct noisy data that converge in the
discretized version of the~$L^2$-norm~\eqref{discrete_L2} for our
second order accurate numerical scheme, we need to drop the amplitude
of the random noise by a quarter every time we double resolution. For
convergence tests in the lopsided norm the scaling factor is
different, due to the~$D_z \phi$ term that appears in the discretized
version of the lopsided norm~\eqref{discrete_lopsided}. By replacing
the~$L^2$ with the lopsided norm in~\eqref{exact_conv_ratio_log2} we
get
\begin{align*}
  \mathcal{C}_{\textrm{exact}} =
  \log_2
  \frac{||\mathbf{u}_{h_c}||_{q (h_c)}}{||
  \perp^{h_c/2}_{h_c}\mathbf{u}_{h_c/2}||_{q (h_c)}}
  \sim \log_2 \frac{O(A_{h_c})}{2 \, O(A_{h_c/2})}\, ,
\end{align*}
where now the norm estimate is dominated by the~$D_z \phi$
term. Hence, to construct noisy data that converge in the lopsided
norm for our second order accurate numerical scheme, we need to
multiply the amplitude of the random noise with a factor of one eighth
every time we double resolution. This discussion would be more
complicated if we were using either pseudospectral approximation or
some hybrid scheme, which is why we focus exclusively on a
straightforward finite differencing setup.

The results for norm convergence tests with appropriately scaled noisy
data for the~$L^2$-norm, for both SH and WH models, are collected in
the right column of Fig.~\ref{fig:L2_norms_all}. As illustrated there,
the inhomogeneous SH model still exhibits convergence since with
increasing resolution the exact convergence ratio tends closer to the
desired value of two at all times of the evolution. On the contrary,
the homogeneous WH model does not converge, and it becomes clear that
with increasing resolution the exact convergence ratio of this model
moves further away from two at all times.

\begin{figure*}[t]
  \hspace*{-1.25cm}
  \includegraphics[width=1.1\textwidth]{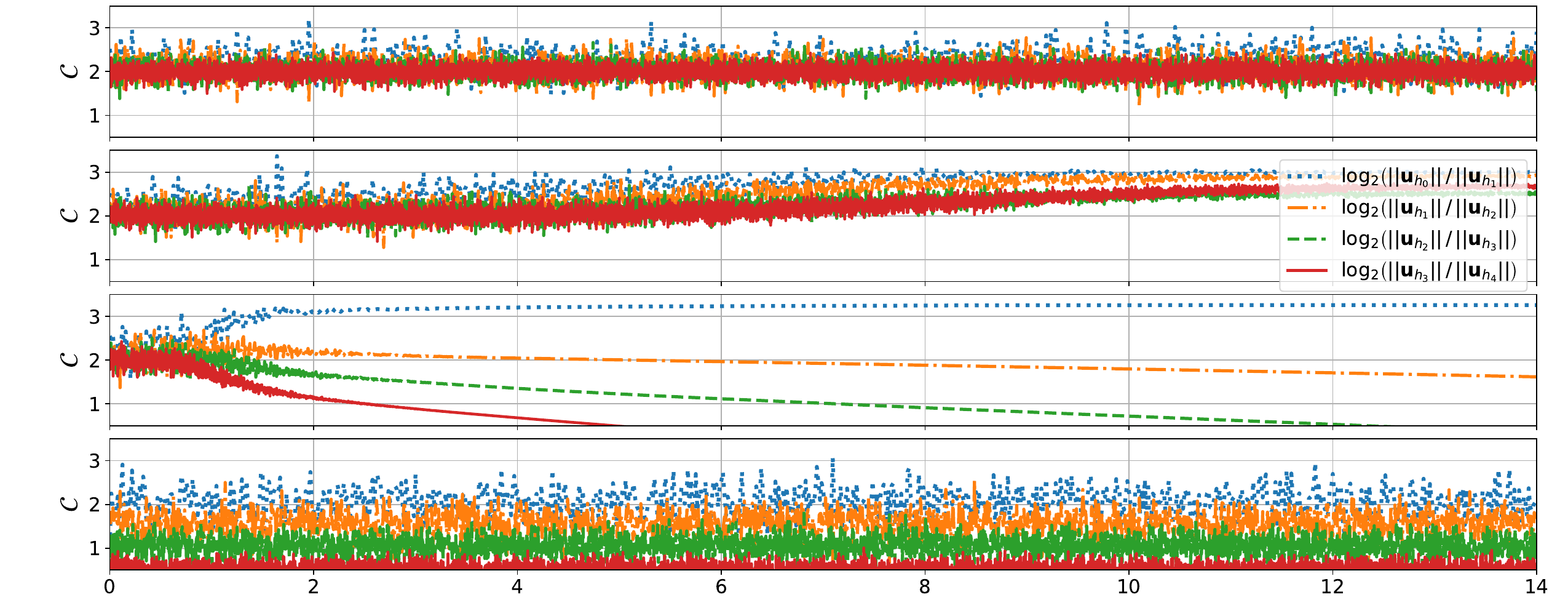}
  \centering{
    \caption[The exact convergence ratio in the lopsided norm for the
    weakly hyperbolic toy model numerical experiments]{The exact
      convergence ratio in the lopsided norm~\eqref{discrete_lopsided}
      for the different WH models. From top to bottom we plot the
      homogeneous WH model, then the inhomogeneous adjustments, in
      order~$B_1,B_2$ and~$B_3$. Overall we conclude that the
      homogeneous model and~$B_1$ models are converging in the limit
      of infinite resolution, with the others failing to do so. Of
      these, all but the third panel, with source~$B_2$, agree with
      our expectation from continuum considerations. In this one case
      our method appears to have an honest numerical instability,
      which could be understood properly by careful consideration of
      the scheme.\label{fig:lopsided_norms_all}}}
\end{figure*}

To appreciate {\it intuitively} why noisy data allow us to diagnose a
lack of strong hyperbolicity, consider the systems in frequency space
as in Sec.~\ref{sec:pde_theory:well-posedness}, which we may think of
as momentum space. In practical terms, Eq.~\eqref{homogeneous_WH_norm}
states that the homogeneous WH model does not satisfy
condition~\eqref{eqn:wp_inequality}, and so high frequency modes can
grow arbitrarily fast. Considering smooth data however, predominantly
low frequency modes are excited, and so using our discretized
approximation the violation of inequality~\eqref{eqn:wp_inequality} is
not visible at the limited resolutions we employ. Noisy data on the
contrary excite substantially both high and low frequency modes, with
the former crucial to illustrate the violation.

We also perform convergence tests in the lopsided
norm~\eqref{discrete_lopsided} to examine the behavior of the
different WH models. As in the previous setup, in these tests we
monitor the exact convergence ratio as a function of the simulation
time. As illustrated in Fig.~\ref{fig:lopsided_norms_all}, our
expectations from Sec.~\ref{sec:pde_theory:well-posedness} for the
homogeneous model are verified. The homogeneous WH model converges at
all times in the lopsided norm, provided of course that the given data
are restricted to converge at second order to the trivial solution in
the same norm. As also expected, the inhomogeneous case with~$B_3$
fails to converge whatsoever during the evolution, exhibiting behavior
similar to the homogeneous WH model in the~$L^2$-norm tests. In fact,
in this test the exact convergence ratio diverges further from two
with increasing resolution and at earlier times. The discussion for
the inhomogeneous WH models with sources~$B_1$ and~$B_2$ is more
subtle. Both cases initially exhibit convergence, with the~$B_1$ case
maintaining this behavior for longer. The difference lies in their
late time behavior and their trend with increasing resolution.  In
particular, the~$B_1$ case converges for longer with increasing
resolution whereas~$B_2$ does the opposite. At late times in the~$B_1$
case~$\mathcal{C}_\textrm{exact}$ reaches a plateau that converges to
two with increasing resolution, which is not true with
sources~$B_2$. Thus our numerical evidence seems to indicate that
the~$B_1$ inhomogeneous WH model converges in the lopsided norm, but
to disagree with the theoretical expectation at the continuum that
the~$B_2$ case does so too. This is not in contradiction with our
earlier calculations however, because, as a careful examination of the
approximation could reveal, purely algorithmic shortcomings may render
a scheme nonconvergent.

\section{GR in the Bondi-Sachs proper gauge}
\label{sec:numerics:GR}

Similarly to the previous section, here we present convergence tests
of the publicly available characteristic code
\texttt{PITTNULL}~\cite{HanSzi14} which employs the Bondi-Sachs
formalism and is part of the~\texttt{Einstein
  Toolkit}~\cite{einsteintoolkitzenodo}. Although similar tests have
been successfully performed in the past~\cite{ZloGomHus03,
  BabBisSzi09, BabSziWin11, HanSzi14}, the novelty here is that we
examine convergence of solutions to the full discretized PDE problem
and not just the individual grid functions. The motivation for this
comes from the fact that well-posedness is a property of the full PDE
problem. We examine the practical consequence of the foregoing results
by performing convergence tests in a discretized version of
the~$L^2$-norm. The specific form of that norm plays a key role,
depends on the geometric setup and is inspired by a hyperbolicity
analysis of the PDE system solved. The details of this analysis can be
found in the ancillary files of~\cite{GiaBisHil21}. The data
illustrated in Figs. \ref{Fig:rescaled_norms_all} and
\ref{Fig:null_noisy_rescaled_norms} can be found
in~\cite{GiaBisHil21_public}.

\subsection{The setup}
\label{subsec:numerics:GR:setup}

Here we collect the fundamental elements on which
the~\texttt{PITTNULL} code is based. The interested reader can find
more details e.g.\ in~\cite{BisGomLeh97a, HanSzi14}. The Bondi-Sachs
metric ansatz~\cite{BonBurMet62,Sac62} used has the form
\begin{align}
  ds^2
  & =
    - \left( e^{2\beta} \frac{V}{r} - r^2 h_{AB} U^A U^B \right)du^2
    - 2 e^{2\beta} du dr
  \nonumber \\
  & \quad
    - 2 r^2 h_{AB} U^B du dx^A
    + r^2 h_{AB} dx^A dx^B
    \,,
    \label{eqn:BS_stereo_metric}
\end{align}
where~$h^{AB} h_{BC} = \delta^A_C$,~$\textrm{det}(h_{AB}) =
\textrm{det}(q_{AB}) = q$, with~$q_{AB}$ the metric on the unit
sphere. The sphere is parameterized using the stereographic
coordinates~$x^A = (q,p)$ following~\cite{BisGomLeh97a}, though
see~\cite{GomLehPap97, Szi00} for a different but equivalent
choice. The metric of the unit sphere reads
\begin{align*}
q_{AB}dx^A dx^B = \frac{4}{P^2} \left(dq^2 + dp^2 \right)\,,
\end{align*}
where~$P=1+q^2+p^2$. One can introduce a complex basis vector~$q^A$
(dyad)
\begin{align*}
  q^A = \frac{P}{2}\left(1,i\right)
  \,,
\end{align*}
and then the metric of the unit sphere can be written as
\begin{align*}
  q_{AB} = \frac{1}{2} \left(q_A \bar{q}_B + \bar{q}_A q_B \right)
  \,.
\end{align*}

Using the complex dyad, a tensor
field~$F_{A_1 \dots A_n}$ on the sphere can be represented as
\begin{align*}
  F = q^{A_1} \dots q^{A_p} \bar{q}^{A_{p+1}} \dots \bar{q}^{A_n} F_{A_1 \dots A_n}
  \,,
\end{align*}
which obeys the relation~$F \rightarrow e^{i s \psi}F$, with spin
weight~$s=2p-n$. The eth operators for this quantity are defined as
\begin{align*}
  \eth F \equiv q^A \nabla_A F
  &= q^A \p_A F + \Gamma s F
    \,,
  \\
  \bar{\eth} F \equiv \bar{q}^A \nabla_A F
  &= \bar{q}^A \p_A F - \bar{\Gamma} s F
    \,,
\end{align*}
with spin~$s \pm 1$ respectively and~$\nabla_A$ the covariant
derivative associated with~$q_{AB}$
i.e.~$\Gamma = -\frac{1}{2}q^a \bar{q}^b \nabla_a q_b$. In the chosen
stereographic coordinates the above reads
\begin{align*}
  \eth F
  &=
    \frac{P}{2}
    \p_q F
    +
    i \frac{P}{2}
    \p_{p} F
    + \left(q + i p \right) s F
    \,,
  \\
  \bar{\eth} F
  &=
    \frac{P}{2}
    \p_q F
    -
    i \frac{P}{2}
    \p_p F
    - \left(q - i p \right) s F
    \,.
\end{align*}

It is convenient to introduce the following complex spin-weighted
quantities
\begin{align*}
  J \equiv \frac{h_{AB} q^A q^B}{2} \,,
  \quad
  K \equiv \frac{h_{AB} q^A \bar{q}^B}{2}
  \, ,
  \quad
  U \equiv U^A q_A
  \,,
\end{align*}
as well as the real variable
\begin{align*}
  W \equiv \frac{V-r}{r^2}
  \,.
\end{align*}
Due to the determinant
condition~$\textrm{det}(h_{AB}) = \textrm{det}(q_{AB})$ the
quantities~$K$ and $J$ are related via~$1 = K^2 - J \bar{J}$.~$J$ has
spin-weight two,~$U$ one and~$K$,~$W$,~$\beta$ zero. The spin-weight
of the complex conjugate is equal in magnitude and opposite in
sign. To eliminate second radial derivatives of~$U$ the following
intermediate quantity is introduced
\begin{align*}
  Q_A \equiv r^2 e^{-2 \beta} h_{AB} U^B_{,r}
  \,.
\end{align*}
Using these variables, the implemented vacuum EFE consist of the
hypersurface equations
\begin{subequations}
\begin{align}
  \beta_{,r}
  & = N_\beta
  \,,
  \label{eqn:hypersurf_beta}
  \\
  \left( r^2 Q \right)_{,r}
  & =
    -r^2 \left( \bar{\eth} J + \eth K \right)_{,r}
    +2 r^4 \eth \left( r^{-2} \beta \right)_{,r}
    + N_Q \,,
    \label{eqn:hypersurf_Q}
  \\
  U_{,r}
  & =
  r^{-2} e^{2 \beta} Q 
  + N_U \,,
    \label{eqn:hypersurf_U}
  \\
  W_{,r}
  & =
    \frac{1}{2} e^{2\beta} \mathcal{R}
    -1 - e^\beta \eth \bar{\eth} e^\beta
    + \frac{1}{4} r^{-2} \left[
    r^4 \left(\eth \bar{U} + \bar{\eth} U \right)
    \right]_{,r}
  + N_W \,,
    \label{eqn:hypersurf_W}
\end{align}
\label{eqn:hypersurf_sys}%
\end{subequations}
where~$Q\equiv Q_A q^A$ and
\begin{align*}
  \mathcal{R}
  &= 2 K - \eth \bar{\eth} K
    + \frac{1}{2} \left(\bar{\eth}^2 J + \eth^2 \bar{J} \right)
    + \frac{1}{4K} \left(
    \bar{\eth}\bar{J} \eth J - \bar{\eth}J \eth \bar{J}
    \right)
    \,,
\end{align*}
the curvature scalar for surfaces of constant~$u$ and~$r$. The
evolution equation of the system is
\begin{align}
  &
    2 \left(r J\right)_{,ur} -
  \left[ \frac{r+W}{r} \left( rJ \right)_{,r} \right]_{,r}
    =
    -r^{-1} \left( r^2 \eth U \right)_{,r}
    + 2r^{-1} e^\beta \eth^2 e^\beta
    - J \left( r^{-1} W \right)_{,r}
    + N_J \,.
    \label{eqn:evol_J}
\end{align}
The complete form of~$N_\beta\,, N_Q\,, N_U\,, N_J$ in terms of the
eth formalism can be found in~\cite{ReiBisPol12}. The
system~\eqref{eqn:hypersurf_sys},~\eqref{eqn:evol_J}. corresponds to
the main equations~\eqref{eqn:main_BS_sys} in the Bondi-Sachs proper
gauge~\eqref{eqn:BS_stereo_metric}. A pure gauge analysis of this
system was presented in Sec.~\ref{sec:bondi-hyp:BS_proper}. For
comparison purposes we employ also the following {\it artificial}
symmetric hyperbolic system
\begin{subequations}
\begin{align}
  \beta_{,r}
  & = N_\beta
  \,,
  \\
  \left( r^2 Q \right)_{,r}
  & =
    0 \,,
  \\
  U_{,r}
  & =
  r^{-2} e^{2 \beta} Q 
  + N_U \,,
  \\
  W_{,r}
  & =
    0 \,,
  \\
    2 \left(r J\right)_{,ur}
    &=
      \left[ \frac{r+W}{r} \left( rJ \right)_{,r} \right]_{,r}
        \,.
\end{align}
\label{eqn:SH_artificial_sys}%
\end{subequations}

Equations~\eqref{eqn:hypersurf_W} and~\eqref{eqn:evol_J} involve the
conjugate variables~$\bar{U}$ and~$\bar{J}$, for which the
system~\eqref{eqn:hypersurf_sys},~\eqref{eqn:evol_J} does not
explicitly possess evolution equations. For the hyperbolicity analysis
provided in the ancillary files we need to complete the system in the
sense of having one equation for each variable. We obtain the
equations for~$\bar{U}$,~$\bar{Q}$ and~$\bar{J}$ by taking the complex
conjugate of~\eqref{eqn:hypersurf_Q},~\eqref{eqn:hypersurf_U}
and~\eqref{eqn:evol_J}, respectively. The state vector of the
linearized about Minkowski and first order reduced system is
\begin{align*}
  \mathbf{u}
  &= \left(
  \beta\,, \beta_q \,, \beta_p \,,
  Q\,, \bar{Q}\,,
  U\,, U_q\,, U_p\,,
  \bar{U}\,, \bar{U}_q\,, \bar{U}_p\,,
    W\,,
  J\,, J_r\,, J_q\,, J_p\,,
  \bar{J}\,, \bar{J}_r\,, \bar{J}_q\,, \bar{J}_p\,,
  \right)^T\,,
\end{align*}
where
\begin{align*}
    \beta_q \equiv \p_q \beta\,,
    \quad
    \beta_p \equiv \p_p \beta\,,
    \quad
    U_q \equiv \p_q U \,,
    \quad
    U_p \equiv \p_p U \,,
     J_q \equiv \p_q J \,,
    \quad
    J_p \equiv \p_p J \,,
    \quad
    J_r \equiv \p_r J
    \,,
\end{align*}
and the complex conjugates are defined in the obvious way. In the ADM
coordinates~$(t,\rho,p,q)$ with
\begin{align*}
  u = t-\rho \, , &\qquad r=\rho \,,
\end{align*}
the system can be written in the form
\begin{align*}
  \p_t \mathbf{u} 
  + \mathbf{B}^\rho \, \p_\rho  \mathbf{u}
  + \mathbf{B}^q \, \p_q  \mathbf{u} \,
  + \mathbf{B}^p \, \p_p  \mathbf{u}
  + \mathbfcal{S} = 0
  \, ,
\end{align*}
and it is only WH due to the non-diagonalizability of the principal
symbol along the angular directions~$q$ and~$p$. This result is
expected from the analysis of Sec.~\ref{sec:bondi-hyp:BS_proper},
since the only difference here is the parameterization of the
two-sphere. The characteristic variables along the radial direction
with speed~$-1$ are ingoing and consist of
\begin{align*}
  \frac{J}{r} + J_r
  \,,
\end{align*}
and its complex conjugate. The outgoing variables are those with
speed~$1$, namely
\begin{align*}
  &- \frac{J}{r}
    \,,
    \quad
    J_q
    \,,
    \quad
    J_p
    \,,
    \quad
    U
    \,,
    \quad
    U_q
    \,,
    \quad
    U_p
    \,,
    \quad
    Q
    \,,
    \quad
    W
    \,,
    \quad
    \beta
    \,,
    \quad
    \beta_q
    \,,
    \quad
    \beta_p
    \,,
\end{align*}
and their appropriate complex conjugates.

As for the toy models earlier, we perform norm convergence tests where
the ingoing variables are integrated over a null hypersurface and the
outgoing ones over a worldtube of constant radius. The code works with
the compactified radial coordinate
\begin{align*}
  z = \frac{r}{R_E + r}
  \,,
\end{align*}
where~$R_E$ is a constant that denotes the extraction radius and for
our tests we set it equal to one. If the grid spacing is denoted
as~$h_z,h_q,h_p$ for the coordinates~$z,q,p$ respectively and the
timestep as~$h_u$, then the discretized version of the~$L^2$-norm that
we use is
\begin{align}
  &
    ||\mathbf{u}_h|| =
    \left\{
    \sum_{z,q,p} \,
    \left[
    \left(\frac{J}{r} + J_r\right) \left(\frac{\bar{J}}{r} + \bar{J}_r\right) 
    \right]
    \, h_z \, h_q \, h_p 
    \right\}^{1/2}
    +
    \label{eqn:discrete_L2_car_vars}
  \\
  &  
    \textrm{max}_z
    \left\{
    \sum_{u,q,p}
    \left(
    \beta^2  + \beta_q^2 + \beta_p^2 + W^2+
    Q \bar{Q} +
    U \bar{U}
    + U_q \bar{U_q} + U_p \bar{U_p}
    + \frac{J\bar{J}}{r^2} + J_q \bar{J}_q + J_p \bar{J}_p
    \right)
    h_u \, h_q \, h_p
    \right\}^{1/2}
    ,
    \nonumber
\end{align}  
where the functions in the sums are to be understood as grid
functions. All the outgoing variables of the artificial SH
system~\eqref{eqn:SH_artificial_sys} satisfy advection equations
towards future null infinity. We further introduce
\begin{align*}
  U_q\, \quad U_p \,, \quad
  \beta_q\,, \quad \beta_p\,,
\end{align*}
as well as the appropriate complex conjugates as independent
variables, even though it is not necessary, in order to include in the
norm terms with angular derivatives. These variables are also outgoing
and their equations of motion are obtained by acting with the
appropriate derivatives to those of~$U$,~$\bar{U}$
and~$\beta$. Consequently, the appropriate~$L^2$-norm for this system
is~\eqref{eqn:discrete_L2_car_vars} without the terms~$J_q \bar{J}_q$
and~$J_p \bar{J}_p$.

\subsection{Convergence tests}
\label{subsec:numerics:GR:convergence}

\begin{figure*}[!t]
  \hspace*{-0.5cm}\includegraphics[width=1.05\textwidth]{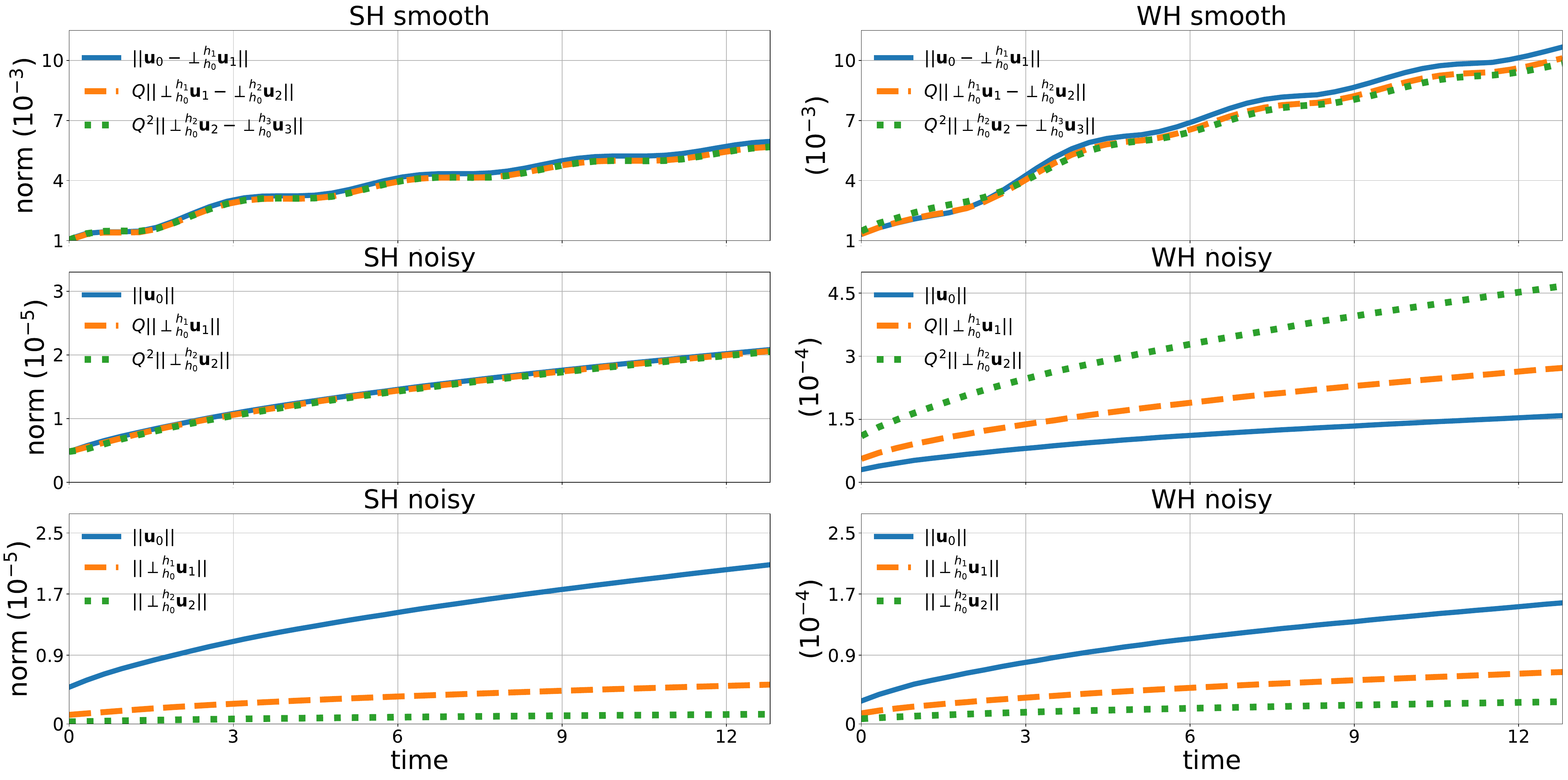}
  \caption[Self and exact convergence tests with the \texttt{PITTNULL}
  code]{Self (above) and exact (below) convergence tests for the
    artificial SH system and the full Bondi-Sachs system that is
    WH. In the top and middle rows the rescaled norms are shown, with
    rescaling factor~$Q=4$. The overlap of the rescaled norms is
    understood as convergence and the lack of overlap as
    non-convergence. The tests are performed in the
    norm~\eqref{eqn:discrete_L2_car_vars} for the WH and the
    norm~\eqref{eqn:discrete_L2_car_vars} without
    the~$J_q\bar{J}_p + J_p\bar{J}_p$ term for the SH system. The self
    convergence tests with smooth data are passed by both systems. The
    exact convergence tests with noisy data are passed only by the SH
    system. In the middle right subfigure we see the failure of
    convergence of the full Bondi-Sachs system, as expected by
    theory. In the bottom row the original norms without rescaling are
    shown. This illustrates that even though the numerical error
    converges to zero with increasing resolution also for the WH case,
    the rate at which this happens is not the expected one and this is
    understood as loss of convergence.}
  \label{Fig:rescaled_norms_all}
\end{figure*}

In the convergence tests we solve the same PDE problem with increasing
resolution and we monitor the behavior of the numerical error. The
numerical domain is
\begin{align*}
  u \in [0,12.8]
  \,, \quad
  z \in [0.45,1]
  \,, \quad
  p,q \in [-2,2]
  \,,
\end{align*}
where~$u$ denotes time,~$z$ is the compactified radial coordinate,
and~$p,q$ the angular coordinates. The two-sphere is covered by
overlapping north and south patches. In the parameter files included
in the supplementary material of~\cite{GiaBisHil21} the
variables~$y,x$ correspond to the~$p,q$ angular coordinates. These
variables refer to the \texttt{Einstein Toolkit} thorn
\texttt{CartGrid3D} and their domain size is different. The grid they
provide corresponds to the grid for~$p,q$. As described
in~\cite{BabBisSzi09}, the~$p,q$ grid points are
\begin{align*}
  p_i
  = -1 + \Delta (i - O - 1)
  \,,
  \quad
  q_j
  = -1 + \Delta (j - O - 1)
  \,,
\end{align*}
where~$O$ denotes the number of overlapping points beyond the
equator. The range of the indices is
\begin{align*}
  1 \leq i,j \leq M + 1 + 2O
  \,,
\end{align*}
where~$M^2$ is the total number of $p,q$ grid points inside the
equator and~$\Delta = 2/M$ is the grid spacing. The physical part of
the stereographic domain consists of the grid points for which
\begin{align*}
  p^2 + q^2 \leq 1
  \,,
\end{align*}
and these are the only points considered in our tests. We
label the different resolutions as~$h_0,\, h_1,\, h_2,\, h_3$ with
\begin{align*}
  h_0: \quad
  N_z,N_p,N_q
  &= 33
    \,,  \; \; \, \,
    h_u = 0.04
    \,,
  &
    h_1: \quad
    N_z,N_p,N_q
  &= 65
    \,, \; \; \,
    h_u = 0.02
    \,,\\
  h_2: \quad
  N_z,N_p,N_q
  &= 129
    \,, \; \,
    h_u = 0.01
    \,,
  &
    h_3: \quad
    N_z,N_p,N_q
  &= 257
    \,, \;
    h_u = 0.005
    \,,
\end{align*}
and~$N_z,N_p,N_q$ the number of points in the~$z,p,q$ numerical grids.
$N_p,N_q$ refer to the total number of grid points (overlapping and
non-overlapping regions together). By construction the grid
points and timesteps of~$h_0$ are common for all resolutions.

We perform convergence tests using both smooth and noisy given
data. The former are based upon the linearized gravitational wave
solutions derived in~\cite{Bis05} and adapted to the notation used
here in~\cite{ReiBisLai06, BabBisSzi09}, namely
\begin{align*}
  J
  &= \sqrt{(l-1) l (l+1) (l+2)} {}_2 R_{lm} \Re (J_l(r) e^{i \nu u})
  \,,
    \quad \,
  U
    = \sqrt{l (l+1)} {}_1 R_{lm} \Re (U_l(r) e^{i \nu u})
    \,,
  \\
  \beta
  &= R_{lm} \Re ( \beta_l e^{i \nu u})
    \,,
  \quad \qquad \qquad \qquad \qquad \qquad \quad
  W_c
    = R_{lm} \Re ( W_{cl}(r) e^{i \nu u})
    \,,
\end{align*}
where~$W_c$ gives the perturbation to~$V$ and for~$l=2$
\begin{align*}
  \beta_2
  & = \beta_0
    \,
  \\
  J_2(r)
  & = \frac{24 \beta_0 + 3 i \nu C_1 - i \nu^3 C_2}{36}
    + \frac{C_1}{4r}
    - \frac{C_2}{12 r^3}
    \,,
  \\
  U_2(r)
  &= \frac{-24 i \nu \beta_0 + 3 \nu^2 C_1 - \nu^4 C_2}{36}
    + \frac{2 \beta_0}{r}
    + \frac{C_1}{2 r^2}
    + \frac{i \nu C_2}{3 r^3}
    + \frac{C_2}{4 r^4}
    \,,
  \\
  W_{c2}(r)
  & =
    \frac{24 i \nu \beta_0 - 3 \nu^2 C_1 + \nu^4 C_2}{6}
    - \frac{\nu^2 C_2}{r^2}
    + \frac{3 i \nu C_1 - 6 \beta_0 - i \nu^3 C_2}{3 r}
    + \frac{i \nu C_2}{r^3}
    + \frac{C_2}{2 r^4}
    \,.
\end{align*}
We fix the parameters of these solutions to
\begin{align*}
  \nu = 1
  \,, \quad
  l = 2
  \,, \quad
  m = 0
  \,,   \quad
  C_1 = 3 \cdot 10^{-3}
  \,, \quad
  C_2 = 10^{-3}
  \,, \quad
  \beta_0 = i \cdot 10^{-3}
  \,.
\end{align*}
The constant~$\nu$ controls the frequency of the solution,~$l,m$ refer
to the spin-weighted spherical harmonics and~$C_1, C_2, \beta_0$ are
integration constants.

For the noisy tests we set all the initial and boundary data to their
Minkowski values, perturbed with random noise of amplitude~$A$ with
\begin{align*}
  A(h_0)
  = 4096 \cdot 10^{-10}
  \,, \quad
  A(h_1)
  = 512 \cdot 10^{-10}
  \,, \quad
  A(h_2)
  = 64 \cdot 10^{-10}
  \,,
\end{align*}
on all the given data. The scaling of the amplitude by a factor of
eight every time we double resolution is due to the first order
derivatives in the norm~\eqref{eqn:discrete_L2_car_vars}, as explained
in Subsec.~\ref{subsec:numerics:toys:convergence}. The amplitude of
the noise is low enough for the non-linear terms to be negligible with
the precision at which we work. The complete parameter files used in
the simulations can be found in the ancillary files
of~\cite{GiaBisHil21}. We call self convergence the tests in which we
obtain an error estimate by taking the difference between two
numerical solutions. This is useful when an exact solution is not
known, as for instance for the artificial SH
system~\eqref{eqn:SH_artificial_sys} when smooth data are
given. Hence, we perform self convergence tests in the smooth setup
for both WH and SH systems. On the contrary, the noisy tests consist
of random noise on top of vanishing given data for both systems and
zero is a solution for both cases. So, for this case we perform exact
convergence tests, i.e. the error estimate is provided by a comparison
between the numerical and the exact solution. We use the
operator~$\perp_{h_0}^{h_i}$ to denote that we consider only the
common grid points of the resolution~$h_i$ with the coarse
resolution~$h_0$, as well as the common time steps. For the self
convergence tests we monitor
\begin{align*}
    ||\mathbf{u}_{h_0} - \perp_{h_0}^{h_1} \mathbf{u}_{h_1}||
  \,, \qquad
  || \perp_{h_0}^{h_1}\mathbf{u}_{h_1} - \perp_{h_0}^{h_2} \mathbf{u}_{h_2}||
  \,,
  \quad
  || \perp_{h_0}^{h_2}\mathbf{u}_{h_2} - \perp_{h_0}^{h_3} \mathbf{u}_{h_3}||
  \,,
\end{align*}
and for the exact convergence
\begin{align*}
  ||\mathbf{u}_{h_0}||
  \,, \quad
  || \perp_{h_0}^{h_1} \mathbf{u}_{h_1}||
  \,, \quad
  || \perp_{h_0}^{h_2} \mathbf{u}_{h_2}||
  \,.
\end{align*}
The code uses finite difference operators that are second order
accurate. This, combined with the doubling of grid points every time
we increase resolution provides again a convergence factor~$Q=4$.

\begin{figure*}[!t]
  \includegraphics[width=0.985\textwidth]{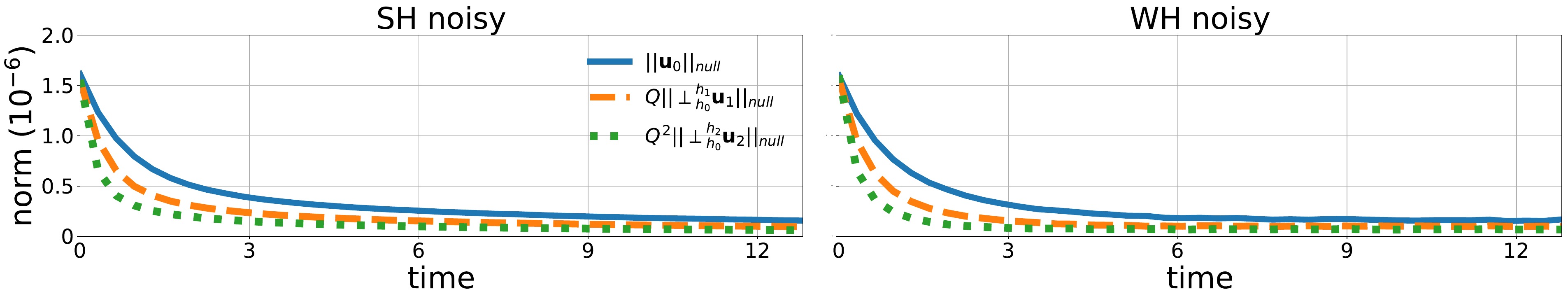}
  \caption[Exact convergence tests with noisy data with the
  \texttt{PITTNULL} code, using only the null part of the norm]{Exact
    convergence test with noisy data for both PDE systems, using only
    the null part of the norm~\eqref{eqn:discrete_L2_car_vars}. The WH
    system does not manifest a clear loss of convergence. Similarly
    to~\cite{HanSzi14} there is no evidence of exponential growth.}
  \label{Fig:null_noisy_rescaled_norms}
\end{figure*}

In Fig.~\ref{Fig:rescaled_norms_all} the rescaled norms for both
smooth and noisy tests, for the artificial
SH~\eqref{eqn:SH_artificial_sys} and the full Bondi-Sachs
system~\eqref{eqn:hypersurf_sys},~\eqref{eqn:evol_J} that is WH are
illustrated. The overlap of the rescaled norms indicates good second
order convergence, whereas the lack of overlap suggests
non-convergence.  For smooth given data both the SH and WH systems
exhibit good second order convergence. However, for noisy given data
only the SH has the appropriate convergence. This feature is expected,
as noisy given data are important to demonstrate WH in numerical
experiments~\cite{CalHinHus05, GiaHilZil20}. These results are
compatible with earlier tests with random noise that demonstrated the
lack of exponential growth in the solution~\cite{HanSzi14}. In
Fig.~\ref{Fig:null_noisy_rescaled_norms} the sum only over the null
hypersurface from~\eqref{eqn:discrete_L2_car_vars} is shown, that is
similar to earlier tests. The loss of convergence in the WH system is
less severe than for the full norm~\eqref{eqn:discrete_L2_car_vars}
and there is no sign of exponential growth in the solution. This fact
alone may be evidence for numerical stability in the colloquial sense
that the code does not crash but, as we demonstrate in
Fig.~\ref{Fig:rescaled_norms_all}, is not enough evidence for
convergence. It becomes apparent then that the choice of norm in which
the convergence tests are performed is crucial. A norm that is
compatible with the PDE system under consideration should be used.

\section{Conclusions}
\label{sec:numerics:conclusions}

A numerical approximation cannot converge to the exact solution of
these PDE problems in any discrete approximation to~$L^2$, if the PDE
systems is only WH. We demonstrated this shortcoming numerically using
our toy models and adapting the well-known robust-stability
test. Spotting this shortcoming in practice is subtle because smooth
data may, and often does, give misleading results.

For our WH toy model, if the nested structure is broken by the source
terms, it becomes ill-posed in any sense, as shown in
Sec.~\ref{sec:bondi_well-p:algebraic_char}. Using random noise for
initial data, our numerical experiments are consistent with this
analytic result. There is one case in which convergence is not
apparent in our approximation, despite the well-posedness of the
continuum equations in the lopsided norm. This is our only example of
a {\it pure} numerical instability, and is important as it highlights
the fact that for weakly hyperbolic systems numerical methods are not
well-developed, and are not guaranteed to converge, even when using
appropriate lopsided norms.

When the numerical experiments are performed in full GR, the same
conclusions carry over; ill-posedness of the continuum PDE (in the
natural equivalent of~$L^2$) for the characteristic problem serves as
an obstruction to convergence of the numerics (in a discrete
approximation to the same norm).
The implication of weak hyperbolicity is that the CIVP and CIBVP of GR
are ill-posed in the natural equivalent of~$L^2$ on these geometric
setups. Therefore we carried out convergence tests in a discretized
version of such a norm. The tests are performed on the Bondi-Sachs
gauge system~\eqref{eqn:hypersurf_sys},~\eqref{eqn:evol_J} implemented
in the~\texttt{PITTNull} thorn of the~\texttt{Einstein Toolkit}, as
well as on the artificial strongly hyperbolic
system~\eqref{eqn:SH_artificial_sys}. The norm used is compatible with
the strongly hyperbolic model in the characteristic domain. The tests
are performed with smooth and with noisy given data. For smooth data
both the strongly and weakly hyperbolic systems exhibit good
convergence. But with noisy data only the strongly hyperbolic model
retains this behavior. This highlights again that noisy given data
are essential to reveal weak hyperbolicity in numerical
experiments. We have furthermore seen that even with noisy data one
might overlook this behavior if tests are performed in a norm that is
not suited to the particular problem.


%% file: sections/jecco.tex
\chapter{Numerical holography with \texttt{Jecco}}
\label{chap:jecco}

\minitoc

The first part of the thesis focused on the hyperbolicity and
well-posedness of characteristic formulations of GR and implications
to accurate gravitational waveform modeling. In this part we discuss
applications of these formulations in the study of out-of-equilibrium
strongly coupled systems via holography. Note that the main standpoint
of the thesis is gravity and more specifically the interplay of PDE
analysis and numerical simulations. Hence, we cannot do full justice
to the vast and growing topic of holography and we rather discuss it
as a tool to model strongly coupled systems by solving the equations
of motion of appropriate gravitational setups.

By the term~\textit{holography} here we mean the duality between a
strongly coupled, non-Abelian, four-dimensional gauge theory on fixed
Minkowski background and a gravitational theory coupled to a scalar
field that resides in a five-dimensional asymptotically AdS
spacetime~\cite{Mal97, Wit98}. We call~\textit{numerical holography}
the process of 1) using standard numerical relativity techniques to
obtain approximate solutions to the complicated and dynamical
gravitational dual setups of interest and 2) mapping them to
quantities of the strongly coupled gauge theory via the holographic
dictionary~\cite{CheYaf14}. The specific dictionary we use here is
given in Eq.~\eqref{ec:boundary_VEVs}.

We consider gravitational setups in the Poincar\'e patch of AdS
spacetimes and always include a non-compact planar horizon. The latter
is effectively acting as an infrared cut-off, which removes caustic
formation from the computational domain. Gravitational constructions
of this type have facilitated through holography the study of
far-from-equilibrium dynamics of strongly coupled gauge theories,
allowing for studies of isotropization \cite{CheYaf09, HelMatSch13,
  GurJanSyb16}, collisions of gravitational shockwaves (used as models
for heavy-ion collisions) \cite{CheYaf11, CasHelMat14, CheYaf15},
momentum relaxation \cite{BalHer14}, turbulence \cite{GreCarLeh13,
  AdaCheLiu14}, collisions in non-conformal theories
\cite{AttCasMat17, AttCasMat17b}, phase transitions and dynamics of
phase separation \cite{AttBeaCas17, JanJanSol17, AttBeaCas19,
  BelJanJan19, BeaDiaGia21, JanJarSon21, BeaCasGia21a}, collisions in
theories with phase transitions \cite{AttBeaCas18}, dynamical
instabilities \cite{GurJanSch16}, and even applications to
gravitational-wave physics \cite{AhmBit17, AhmBit18, BigCadCot21,
  AreHinHoy20, AreHenHin21b} and bubble dynamics~\cite{BigCadCot20,
  BeaCasGia21, BigCadCan21, AreHenHin21}.
Characteristic formulations have also been employed to study the
superradiant instability in asymptotically AdS
spacetimes~\cite{BosGreLeh16, Che21}. See~\cite{CheYaf14} for more
references and a comprehensive overview of the techniques
involved. Cauchy-type evolutions in asymptotically AdS spacetimes can
provide an alternative approach for numerical holography; see for
example~\cite{BanPreGub12, BanFigMat20, BanFigRos21}. This approach
comes with its own complications, which are not discussed in this
thesis.

This chapter presents a new 3+1 code called \texttt{Jecco} (Julia
Einstein Characteristic Code) that solves Einstein's equations in the
characteristic formulation in asymptotically AdS spaces.
\texttt{Jecco} is written in the Julia programming language and comes
with several tools (such as arbitrary-order finite-difference
operators as well as Chebyshev and Fourier differentiation matrices)
useful for generic numerical evolutions. The code is publicly
available and can be obtained from
github~\url{https://github.com/mzilhao/Jecco.jl} and
Zenodo~\cite{jecco-2022}.
To the best of our knowledge, this is the first such freely available
code apart from the \texttt{PittNull} code used in
Sec.~\ref{sec:numerics:GR} for convergence tests in asymptotically
flat spacetimes.

In Subsec.~\ref{subsec:jecco_eqs} we introduce the class of models to
which our code can currently be applied, as well as the corresponding
equations of motion. In Subsec.~\ref{subsec:jecco_implementation} we
discuss the implementation of these equations in the code and the
numerical methods that we use.
In Sec.~\ref{sec:jecco_tests} we present validation tests of the
code. In particular, Subsec.~\ref{subsec:jecco_test_1} discusses
numerical error estimates of the code when reproducing a static
configuration, Subsec.~\ref{subsec:jecco_test_2}
compares~\texttt{Jecco} against \texttt{SWEC}, its precursor
introduced in~\cite{AttCasMat17b} and used also in~\cite{BeaDiaGia21,
  BeaCasGia21, BeaCasGia21a}. In Subsec.~\ref{subsec:jecco_test_3}
convergence tests solely within \texttt{Jecco} are presented. More
validation tests of the code can be found
in~\cite{BeaCasGia22}. Finally, in Sec.~\ref{sec:jecco_results} we
present simulations of the dynamics of phase transitions in some of
our models. The results may be relevant for primordial GW production
scenarios.
In addition to the geometric units $G=c=1$, we also set~$\hbar=1=L$,
where the latter is the AdS radius.

\section{\texttt{Jecco}: a new characteristic code for
  numerical holography}

\subsection{Equations}
\label{subsec:jecco_eqs}

In this section we outline the theoretical background and equations
that are implemented in \texttt{Jecco}. Our approach is similar to
that of~\cite{CheYaf14} and generalizes the code presented
in~\cite{AttCasMat17b} to the 3+1 dimensional case. See
also~\cite{Win12} for an overview of the approaches and codes used in
the asymptotically flat setting.

\subsubsection{Equations of motion and characteristic formulation}

We consider a five-dimensional action consisting of gravity coupled to
a scalar field $\phi$ with a non-trivial potential $V(\phi)$. The
action for this Einstein-scalar model is
\begin{equation}
  S = \frac{2}{\kappa}
  \int d^5x \sqrt{-g}
  \left[
    \frac{1}{4} R - \frac{1}{2} \left( \partial \phi \right)^2 -V(\phi)
  \right],
  \label{eqn:Einstein-scalar_model}
\end{equation}
where $\kappa = 8 \pi$ in our units. The resulting dynamical
equations of motion read
\begin{equation}
\begin{aligned}
E_{\mu\nu}& \equiv R_{\mu\nu}-\frac{R}{2}g_{\mu\nu}-8\pi T_{\mu\nu}=0,\\
\Phi & \equiv \square\phi -\partial_{\phi}V(\phi)=0,
\label{eq:eom}
\end{aligned}
\end{equation}
where
\begin{align*}
  8\pi \, T_{\mu\nu} =
  2 \, \p_{\mu} \phi \, \p_{\nu} \phi -
  g_{\mu\nu}
  \left( g^{\alpha \beta} \, \p_{\alpha} \phi \, \p_{\beta} \phi
  + 2V(\phi) \right)
  \,.
\end{align*}
Our potential $V(\phi)$ comes from a superpotential $W(\phi)$ with the
form
\begin{align}
  W(\phi) =
  -\frac{3}{2} - \frac{\phi ^2}{2} + \lambda_4
  \, \phi^4
  + \lambda_6 \, \phi^6 \,,
  \label{WW}
\end{align}
and its explicit expression can be derived via
\begin{align*}
  V = -\frac{4}{3} W^2 + \frac{1}{2} W'{}^2
  \,,
\end{align*}
resulting in
\begin{align}
  L^2 V(\phi)
  & = -3-\frac{3}{2} \phi^2 - \frac{1}{3} \phi^4
    + \left( \frac{4 \lambda_4}{3} + 8 \lambda_4^2 - 2\lambda_6 \right) \phi^6
    +\left( -\frac{4 \lambda_4^2}{3}
    + \frac{4}{3} \lambda_6 + 24 \lambda_4 \lambda_6 \right) \phi^8
    \nonumber
  \\
  & \quad
    {} + \left(
    18\lambda_6^2
    - \frac{8}{3} \lambda_4 \lambda_6
    \right)
    \phi^{10} - \frac{4}{3} \lambda_6^2
     \, \phi^{12}
    \,.
    \label{eq:potential}
\end{align}
In these equations $\lambda_4$ and $\lambda_6$ are freely specifiable
dimensionless parameters related to the parameters $\phi_M$ and
$\phi_Q$ used in e.g.~\cite{BeaMat18, BeaDiaGia21} through
\begin{align}
\label{param}
\lambda_4 = - \frac{1}{4\phi_M^2}  \, \quad \lambda_6 = \frac{1}{\phi_Q}
 \,.
\end{align}
This potential has a maximum at $\phi=0$, where it admits an exact AdS
solution of radius $L$, here set equal to 1. The holographic dual
field theory corresponds to a 3+1 dimensional conformal field theory
which is deformed by a source $\Lambda$ for the dimension-three scalar
operator $\mathcal{O}_{\phi}$ dual to the scalar field $\phi$. The
thermodynamical and near-equilibrium properties of this model were
presented in~\cite{AttCasMat16, AttCasMat17, AttBeaCas17} for
$\lambda_6 = 0$ and in~\cite{BeaMat18, BeaDiaGia21} for
$\lambda_6 \neq 0$.

Let us point out that even if here we will always make use of the
particular potential~(\ref{eq:potential}), the code implementation is
such that more generic potentials can be used provided that, for low
values of the scalar field, they behave as
\begin{equation}
L^2 V(\phi) = -3-\frac{3}{2}\phi^2-\frac{\phi^4}{3}+\mathcal{O}\left(\phi^6\right).
\label{ec:pot_near_bdry}
\end{equation}
The constant term is fixed by the 4+1 dimensional AdS asymptotics and
the quadratic one is in correspondence with the scaling dimension of
the dual scalar operator $\mathcal{O}_{\phi}$. The quartic term,
determined by the other two in our case, ensures the absence of a
conformal anomaly, which would give rise to logarithms in the
asymptotic expansions. A change in this near boundary behavior of the
potential would alter the hard-coded asymptotic expansions and
variable redefinitions that are introduced later.

We consider the following 5-dimensional ansatz for the metric in
ingoing Eddington-Finkelstein (EF) coordinates, which falls into the
affine null class as described in
Sec.~\ref{sec:bondi_properties:main_features}, with the line element
\begin{equation}
\begin{aligned}
  ds^2= g_{\mu\nu} dx^{\mu} dx^{\nu} & =
  -Adt^2+2dt\left(dr+F_xdx+F_ydy\right)
  +S^2\Big[e^{-B_1-B_2}\cosh(G)dx^2 \\[2mm]
  & \quad { }+e^{B_1-B_2}\cosh(G)dy^2+2e^{-B_2}\sinh(G)dxdy+e^{2
    B_2}dz^2\Big],
  \label{eq:metric}
\end{aligned}
\end{equation}
where all functions depend on the radial coordinate $r$, time $t$ and
transverse directions $x$ and $y$. Nothing depends on the coordinate
$z$, so this is effectively a 3+1 system. Physically, this means that
in the gauge theory we impose translation invariance along the
$z$-direction but allow for completely general dynamics in the
$(t,x,y)$-directions. Note that hypersurfaces of constant $t$ are
ingoing null. This coordinate is often labeled by $v$ in EF
coordinates. This particular gauge choice is the affine null
since~$g_{tr}=1$ in these coordinates, but now in five spacetime
dimensions. At the boundary, $t$ becomes the usual Minkowski time
coordinate. The spatial part of the metric is written such that $S$
encodes the area of constant $t$ and $r$ slices,
\[
  \sqrt{g|_{dt,dr = 0}}=S^3.
\]
We can recover the 2+1 system of~\cite{AttCasMat17b} by setting
\begin{equation}
\begin{aligned}
  &F_y =G=0, & B_1 & =\frac{3}{2}B, & B_2 & = \frac{1}{2}B, \qquad
  \mathrm{or}
  \\[2mm]
  &F_x=G=0, & B_1 & =-\frac{3}{2}B, & B_2 & = \frac{1}{2}B,
\end{aligned}
\label{ec:to_2+1}
\end{equation}
for non-trivial dependence only along the $x$ or $y$ direction
respectively.

Notice that the latter returns the weakly hyperbolic system analyzed
in Subsec.~\ref{subsec:bondi-hyp:aff_null:AAdS} and so we expect the
resulting system here to be only weakly hyperbolic as well. Even
though our main goal with numerical holography is to obtain
qualitative results and universal behaviors for the strongly coupled
models we analyze, it is desirable to work with strongly or even
symmetric hyperbolic characteristic setups, such that we can have
robust error estimates. Since there are no such characteristic
constructions at the moment--to the best of our knowledge--we work
with the standard Bondi-like setups and aim to provide better
alternatives in the future. In fact, part of the results presented
in~\cite{BeaDiaGia21} have been obtained by fully non-linear numerical
evolutions of the setup of
Sec.~\ref{subsec:bondi-hyp:aff_null:AAdS}. The end states of these
dynamical scenarios match the near-equilibrium configurations computed
with completely different methods and presented also
in~\cite{BeaDiaGia21}. This compatibility suggests that possible
errors due to the weak hyperbolicity of the system would affect the
validity of the error estimates of the numerical solutions, but not
their qualitative behavior. Of course further investigation is needed
in order to address this expectation and alternative characteristic
formulations would be necessary. The setup of
Sec.~\ref{subsec:bondi-hyp:aff_null:AAdS} has also been used for the
non-linear numerical evolutions presented in~\cite{BeaCasGia21,
  BeaCasGia21a}.

The metric~\eqref{eq:metric} is invariant under
\begin{equation}
\begin{aligned}
r & \rightarrow \bar{r}=r+\xi(t,x,y) \,, \\
S & \rightarrow\bar{S}=S \,, \\
B_1 & \rightarrow\bar{B}_1=B_1 \,, \\
B_2 & \rightarrow\bar{B}_2=B_2 \,, \\
A & \rightarrow\bar{A}=A+2\partial_t\xi(t,x,y) \,, \\
F_x & \rightarrow\bar{F}_x=F_x-\partial_x\xi(t,x,y) \,,\\
F_y & \rightarrow\bar{F}_y=F_y-\partial_y\xi(t,x,y) \,.
\end{aligned}
\label{ec:gauge_trafo}
\end{equation}
Plugging the ansatz~(\ref{eq:metric}) into~\eqref{eq:eom} results in a
nested system of equations, where some of them can be effectively
viewed as ODEs in the radial (holographic) direction $r$ at each
constant $t$ that can be solved sequentially. As discussed in
Sec.~\ref{sec:bondi_properties:main_features} however, to determine
the degree of hyperbolicity of the full system we still need to treat
it as an actual PDE in its entirety.

We illustrate the system solved here in Table~\ref{tab:system}.  Each
row in the table represents an equation, obtained from the particular
combination of the equations of motion (\ref{eq:eom}) as indicated,
that takes the form
\begin{equation}
  \left[
    A_f(t,u,x,y)
    \, \partial^2_u + B_f(t,u,x,y)
    \, \partial_u + C_f(t,u,x,y)
  \right] f(t,u,x,y)
  = -S_f(t,u,x,y),
  \label{eq:ODEs}
\end{equation}
where $u \equiv 1/r$, $f$ is the corresponding function to be solved
for and the coefficients $A_f$, $B_f$, $C_f$ and $S_f$ are fully
determined once the preceding equations have been solved.
\textit{Dotted} functions denote an operation defined as
\begin{align}
  \dot{f} \equiv
  \left(\p_t + \frac{A}{2} \p_r \right) f
  \,,
  \label{eqn:dot_vars_def}
\end{align}
which are necessary to obtain this nested structure.

\begin{table}[t]
  \begin{center}
    \caption{Nested structure of the equations of motion.}
    \label{tab:system}
\begin{tabular}{c | c}
  Function & Combination\\
  \hline\hline
  \small
  $S$ & $E_{rr}$\\
  \hline
  $F_x$ & $E_{rx}-g_{tx}E_{rr}$\\
  $F_y$ & $E_{ry}-g_{ty}E_{rr}$\\
  \hline
$\dot{S}$ & 	$E_{tr}-\frac{1}{2}g_{	tt}E_{rr}$\\
\hline
$\dot{\phi}$ & $\Phi$\\
\hline
  $A$ & $\frac{E_{zz}}{g_{zz}} +
        \left(g^{ry}g_{ty}+g^{rx}g_{tx}\right)
        E_{rr} + 2g^{rx}
        \left(E_{rx}-g_{tx}E_{rr}\right)
        + 2g^{ry} \left(E_{ry}-g_{ty}E_{rr}\right)$ \\
              &$-4\left(E_{tr}-\frac{1}{2}g_{tt}E_{rr}\right)
+2\frac{E_{xy}}{g_{xy}}+g_{xx}g^{xx}
\left(\frac{E_{yy}}{g_{yy}}+\frac{E_{xx}}{g_{xx}}-2\frac{E_{xy}}{g_{xy}}\right)$\\
\hline
$\dot{B_2}$ & $E_{zz}$\\
\hline
 $\dot{G}$ & $E_{xy}$\\
 $\dot{B_1}$ & $E_{yy}$\\
 \hline
  $\ddot{S}$ & $E_{tt}-\frac{1}{2}g_{tt}E_{tr}
-\frac{1}{2}g_{tt}\left(E_{tr}-\frac{1}{2}g_{tt}E_{rr}\right)$\\
  \hline
 $\dot{F_x}$&$E_{tx}-\frac{1}{2}g_{tt}E_{rx}-g_{tx}\left(E_{tr}-\frac{1}{2}g_{tt}E_{rr}\right)$\\
 $\dot{F_y}$&$E_{ty}-\frac{1}{2}g_{tt}E_{ry}-g_{ty}\left(E_{tr}-\frac{1}{2}g_{tt}E_{rr}\right)$\\
 \hline
\end{tabular}
\end{center}
\end{table}

There are three sets of (two) coupled equations, indicated in the
table by the absence of a separating line.  These still take the form
of (\ref{eq:ODEs}), but now $f$ should be thought of as a vector of
the two functions involved, as is the source term $S_f$, while $A_f$,
$B_f$ and $C_f$ become $2 \times 2$ matrices.  The equations
themselves are lengthy and given in
Eqs.~\eqref{ec:nested_first}-\eqref{ec:nested_last}.
These equations need to be supplemented with boundary conditions
specified at the AdS boundary $u \equiv 1/r = 0$, which are made
explicit later.  In addition, the functions $B_1(t_0,u,x,y)$,
$B_2(t_0,u,x,y)$, $G(t_0,u,x,y)$ and $\phi(t_0,u,x,y)$ should be
thought of as initial data which can be freely specified provided they
are consistent with AdS asymptotics.

\subsubsection{Asymptotic expansions}

The study of the near-boundary behavior ($u\rightarrow 0$) of the
functions is relevant for two reasons. The first is that, as usually
for AAdS spacetimes, some metric components diverge as one approaches
the boundary, and their expansion in powers of $u$ is useful to
redefine the variables in terms of new, finite ones. The second is
that it allows us to understand which boundary conditions to impose on
the Eqs.~\eqref{eq:ODEs}.


For this purpose, we start with an ansatz that is compatible with the
AAdS condition
\begin{equation}
  \begin{aligned}
    A(t,u,x,y)
    & = \frac{1}{u^2}+\sum_{n=-1}^{\infty}a_{(2n)}(t,x,y)
    u^n
    \,,
    &
    B_1(t,u,x,y)
    & = \sum_{n=1}^{\infty}b_{1n}(t,x,y) u^n
    \,, \\
    B_2(t,u,x,y)
    & = \sum_{n=1}^{\infty}b_{2n}(t,x,y) u^n
    \,,
    &
    G(t,u,x,y)
    & = \sum_{n=1}^{\infty}g_n(t,x,y) u^n
    \,,
    \\
    S(t,u,x,y)
    & =\frac{1}{u} +\sum_{n=0}^{\infty}s_{n}(t,x,y) u^n
    \,,
    &
    F_x(t,u,x,y)
    & = \sum_{n=0}^{\infty}f_{xn}(t,x,y) u^n
    \,,
    \\
    F_y(t,u,x,y)
    & = \sum_{n=0}^{\infty}f_{yn}(t,x,y) u^n
    \,, &
    \phi(t,u,x,y)
    & = \sum_{n=1}^{\infty}\phi_{n-1}(t,x,y)u^n
    \,.
\end{aligned}
\label{ec:AAdS_general}
\end{equation}
Substituting into Eqs.~\eqref{ec:nested_first}-\eqref{ec:nested_last}
and solving order by order, we obtain
%
\begin{subequations}
  \begin{align}
    A(t,u,x,y)
    & = \frac{1}{u^2} + \frac{2}{u} \xi
      + \xi^2 - 2 \p_t \xi - \frac{2\phi_0^2}{3}
      + u^2a_4 -
      \nonumber
    \\
    &\quad
      \frac{2}{3}u^3
      \left( 3 \xi a_4 +
      \p_x f_{x2} + \p_y f_{y2} + \phi_0 \p_t \phi_2 \right)
      + \mathcal{O} \left(u^4\right)
      \,,
    \\
    B_1(t,u,x,y)
    & = u^4 b_{14} + \mathcal{O} \left(u^5\right)
      \,,
      \label{eq:asympt-B1}
    \\
    B_2(t,u,x,y)
    & = u^4 b_{24} + \mathcal{O} \left(u^5\right)
      \,,
      \label{eq:asympt-B2}
    \\
    G(t,u,x,y)
    & = u^4 g_4 + \mathcal{O} \left(u^5\right)
      \,,
      \label{eq:asympt-G}
    \\
    S(t,u,x,y)
    & = \frac{1}{u} + \xi - \frac{\phi_0^2}{3} u
      + \frac{1}{3} \xi \phi_0^2 u^2
      + \frac{1}{54} u^3
      \left( -18 \xi^2 \phi_0^2 + \phi_0^4 - 18 \phi_0 \phi_2 \right) +
      \nonumber
    \\
    & \quad
      \frac{\phi_0}{90} u^4
      \left( 30 \xi^3 \phi_0 - 5 \xi \phi_0^3 + 90 \xi \phi_2
      - 24 \p_t \phi_2 \right) + \mathcal{O} \left(u^5\right)
      \,,
    \\
    F_x(t,u,x,y)
    & = \p_x \xi +u^2 f_{x2} -
      \nonumber
    \\
    & \quad
      \frac{2}{15} u^3
      \left(15 \xi f_{x2} + 6 \p_x b_{14} + 6 \p_x b_{24} - \p_y g_4
      -2 \phi_0 \p_x \phi_2 \right)
      + \mathcal{O} \left(u^4\right)
      \,,
    \\
    F_y(t,u,x,y)
    & = \p_y \xi + u^2 f_{y2} -
      \nonumber
    \\
    & \quad
      \frac{2}{15} u^3
      \left(15 \xi f_{y2} - 6 \p_y b_{14}
       + 6 \p_y b_{24} - \p_x g_4
      -2 \phi_0 \p_y \phi_2 \right)
      + \mathcal{O} \left(u^4\right)
      \,,
    \\
    \phi(t,u,x,y)
    & = \phi_0 u - \xi \phi_0 u^2 + u^3
      \left( \xi^2 \phi_0 + \phi_2 \right) +
      \nonumber
    \\
    & \quad
      u^4 \left( \p_t \phi_2 -3 \xi \phi_2 - \xi^3 \phi_0 \right)
      + \mathcal{O} \left(u^5\right)
      \,,
      \label{eq:asympt-phi}
  \end{align}
  \label{eq:asympt}
\end{subequations}
%
where $\phi_2$ is \emph{not} the one in (\ref{ec:AAdS_general}), but
redefined as
\begin{equation}
\phi_2(t,x,y)\rightarrow\phi_2(t,x,y)+\xi^2(t,x,y)\phi_0.
\end{equation}
Note that $\phi_0$ is a constant, while the remaining variables in
this expansion are functions of~$(t,x,y)$. Furthermore, note that the
redefinitions must be modified if \eqref{ec:pot_near_bdry} does not
hold.

We also need the expansions of dotted variables, defined
in Eq.~\eqref{eqn:dot_vars_def}, which take the form
\begin{subequations}
  \begin{align}
    \dot{B_1}(t,u,x,y)
    & = -2b_{14}
      u^3 + \mathcal{O}
      \left(u^4\right)
      \,,
    \\
    \dot{B_2}(t,u,x,y)
    & =-2b_{24}
      u^3 + \mathcal{O} \left(u^4\right)
      \,,
    \\
    \dot{G}(t,u,x,y)
    & =- 2 g_{4}
      u^3 + \mathcal{O} \left(u^4\right)
      \,,
    \\
    \dot{S}(t,u,x,y)
    & =\frac{1}{2u^2} + \frac{\xi }{u}
      + \frac{\xi^2}{2}
      - \frac{\phi_0^2}{6}
      + \frac{1}{36} u^2
      \left(10a_4 -5 \phi_0^4
      + 18 \phi_0 \phi_2 \right)
      + \mathcal{O} \left(u^3\right)
      \,,
    \\
    \dot{F_x}(t,u,x,y)
    & = \p_t \p_x \xi - u f_{x2}
      + \mathcal{O} \left(u^2\right)
      \,,
    \\
    \dot{F_y}(t,u,x,y)
    & = \p_t \p_y \xi - u f_{y2}
      + \mathcal{O} \left(u^2\right)
      \,,
    \\
    \dot{\phi}(t,u,x,y)
    & = -\frac{\phi_0}{2}
      + u^2 \left( \frac{\phi_0^3}{3}
      -\frac{3}{2}\phi_2
      \right)
      +\mathcal{O}\left(u^3\right)
      \, .
  \end{align}
\end{subequations}
The function $\xi(t,x,y)$ encodes our residual gauge freedom, and the
functions $a_4(t,x,y)$, $f_{x2}(t,x,y)$, $f_{y2}(t,x,y)$ are further
constrained to obey
\begin{subequations}
  \label{ec:boundary_evol}
  \begin{align}
    \p_t a_4
    & = -\frac{4}{3}
      \left(\p_x f_{x2} + \p_y f_{y2} + \phi_0 \p_t \phi_2 \right)
      \,,
    \\
    \p_t f_{x2}
    & = -\frac{1}{4} \p_x a_4 - \p_x b_{14}
      - \p_x b_{24} + \p_y g_4
      + \frac{1}{3} \phi_0 \p_x \phi_2
      \,,
    \\
    \p_t f_{y2}
    & = -\frac{1}{4} \p_y a_4
      + \p_y b_{14}
      - \p_y b_{24}
      + \p_x g_4
      + \frac{1}{3}
      \phi_0 \p_y \phi_2
      \,,
  \end{align}
\end{subequations}
where $b_{14}(t,x,y)$, $b_{24}(t,x,y)$, $g_{4}(t,x,y)$,
$\phi_2(t,x,y)$, and $\p_t \phi_2(t,x,y)$ are understood to be read
off from the asymptotic behavior of $B_{1}(t,r,x,y)$,
$B_{2}(t,r,x,y)$, $G(t,r,x,y)$, and $\phi(t,r,x,y)$ in
Eqs.~\eqref{eq:asympt-B1}, \eqref{eq:asympt-B2}, \eqref{eq:asympt-G}
and \eqref{eq:asympt-phi}. The functions $a_4(t_0,x,y)$,
$f_{x2}(t_0,x,y)$, $f_{y2}(t_0,x,y)$, and $\xi(t_0,x,y)$ should also
be thought of as initial data, which can be freely specified. $\phi_0$
is a parameter that must also be specified and corresponds to the
energy scale $\Lambda$ of the dual boundary theory.

\subsubsection{Field redefinitions and boundary conditions}

For the numerical implementation we split the numerical grid in two
parts: the outer grid region (deep bulk) and the inner grid region.
The latter includes the AdS boundary, where boundary conditions are
imposed and the gauge-theory variables are read off. Since some of the
metric functions diverge at the AdS boundary while others vanish, we
employ field redefinitions inspired by the asymptotic field behavior
so that the variables employed in the inner grid remain of order
unity. For the outer grid we choose to make simpler redefinitions,
which is helpful for the equation used to fix the gauge variable
$\xi$. Denoting with the $g1$ ($g2$) subscript the variables defined
in the inner (outer) grid, the redefinitions that we choose to make
are
\begin{align}
    A(t,u,x,y)
    & = \frac{1}{u^2}
      + \frac{2}{u} \xi(t,x,y) + \xi^2(t,x,y)
      -2 \p_t \xi(t,x,y) -\frac{2\phi_0^2}{3}
      + u^2A_{g1}(t,u,x,y)
      \nonumber
    \\
    &
      = -2 \p_t \xi(t,x,y)
      + A_{g2}(t,u,x,y)
      \,,
      \nonumber
    \\
    B_1(t,u,x,y)
    & = u^4B_{1g1}(t,u,x,y)
        = B_{1g2}(t,u,x,y)
        \,,
        \nonumber
    \\
    B_2(t,u,x,y)
    & = u^4 B_{2g1}(t,u,x,y)
        = B_{2g2}(t,u,x,y)
        \,,
        \nonumber
    \\
    G(t,u,x,y)
    & = u^4G_{g1}(t,u,x,y)
        = G_{g2}(t,u,x,y)
        \,,
        \nonumber
    \\
    S(t,u,x,y)
    & = \frac{1}{u} + \xi(t,x,y)
      -\frac{\phi_0^2}{3} u
      +\frac{1}{3} \xi \phi_0^2 u^2
      + u^3 S_{g1}(t,u,x,y)
          = S_{g2}(t,u,x,y)
          \,,
          \nonumber
    \\
    F_x(t,u,x,y)
    & = \p_x \xi(t,x,y)
      + u^2F_{xg1}(t,u,x,y)
          = \p_x \xi(t,x,y)
          +F_{xg2}(t,u,x,y)
          \,,
          \nonumber
    \\
    F_y(t,u,x,y)
    & = \p_y \xi(t,x,y)
      + u^2 F_{yg1}(t,u,x,y)
          = \p_y \xi(t,x,y)
          + F_{yg2}(t,u,x,y)
          \,,
          \nonumber
    \\
    \phi(t,u,x,y)
    & = \phi_0 u -\xi(t,x,y) \phi_0 u^2
      + u^3 \phi_0^3 \phi_{g1}(t,u,x,y)
          = \phi_{g2}(t,u,x,y)
          \,,
          \nonumber
    \\
    \dot{B_1}(t,u,x,y)
    & = u^3 \dot{B}_{1g1}(t,u,x,y)
        = \dot{B}_{1g2}(t,u,x,y)
        \,,
        \nonumber
    \\
    \dot{B_2}(t,u,x,y)
    & = u^3 \dot{B}_{2g1}(t,u,x,y)
        = \dot{B}_{2g2}(t,u,x,y)
        \,,
        \label{eqn:jecco_field_redefs}
    \\
    \dot{G}(t,u,x,y)
    & = u^3 \dot{G}_{g1}(t,u,x,y)
        = \dot{G}_{g2}(t,u,x,y)
        \,,
        \nonumber
    \\
    \dot{S}(t,u,x,y)
    & = \frac{1}{2u^2}
      + \frac{\xi(t,x,y)}{u}
      + \frac{\xi^2(t,x,y)}{2}
      -\frac{\phi_0^2}{6}
      +u^2\dot{S}_{g1}(t,u,x,y)
          = \dot{S}_{g2}(t,u,x,y)
          \,,
          \nonumber
    \\
    \dot{F_x}(t,u,x,y)
    & = \p_t \p_x \xi(t,x,y)
      + u\dot{F}_{xg1}(t,u,x,y)
          = \p_t \p_x \xi(t,x,y)
          + \dot{F}_{xg2}(t,u,x,y)
          \,,
          \nonumber
    \\
    \dot{F_y}(t,u,x,y)
    & = \p_t \p_y \xi(t,x,y)
      + u \dot{F}_{yg1}(t,u,x,y)
          = \p_t \p_y \xi(t,x,y)
          + \dot{F}_{yg2}(t,u,x,y)
          \,,
          \nonumber
    \\
    \dot{\phi}(t,u,x,y)
    & = -\frac{\phi_0}{2}
      + u^2\phi_0^3\dot{\phi}_{g1}(t,u,x,y)
          =\dot{\phi}_{g2}(t,u,x,y).
          \nonumber
\end{align}
After substituting these redefined variables into
Eqs.~\eqref{ec:nested_first}-\eqref{ec:nested_last}, we obtain two
versions of the nested system, one for the near boundary region (inner
grid), and one for the bulk region (outer grid). The boundary
conditions are provided on the timelike boundary of the asymptotically
AdS spacetime which is part of the inner grid ($g1$) and are given by
\begingroup \allowdisplaybreaks
\begin{subequations}
\begin{align}
S_{g1}|_{u=0}&=\frac{1}{54}\left(-18\xi^2\phi_0^2+\phi_0^4-18\phi_0\phi_2\right),\\[2mm]
\partial_uS_{g1}|_{u=0}&=\frac{\phi_0}{90}\left(30\xi^3\phi_0-5\xi\phi_0^3+90\xi\phi_2-24\partial_t\phi_2\right),\\[2mm]
F_{xg1}|_{u=0}&=f_{x2},\\[2mm]
\partial_uF_{xg1}|_{u=0}&=-\frac{2}{15}\left(15\xi
f_{x2}+6\partial_xb_{14}+6\partial_xb_{24}-\partial_yg_4-2\phi_0\partial_x\phi_2\right),\\[2mm]
F_{yg1}|_{u=0}&=f_{y2},\\[2mm]
\partial_uF_{yg1}|_{u=0}&=-\frac{2}{15}\left(15\xi
f_{y2}+6\partial_yb_{14}+6\partial_yb_{24}-\partial_xg_4-2\phi_0\partial_y\phi_2\right),\\[2mm]
\dot{S}_{g1}|_{u=0}&=\frac{1}{36}\left(10a_4-5\phi_0^4+18\phi_0\phi_2\right),\\[2mm]
\dot{B}_{1g1}|_{u=0}&=-2b_{14},\\[2mm]
\dot{B}_{2g1}|_{u=0}&=-2b_{24},\\[2mm]
\dot{G}_{g1}|_{u=0}&=-2g_{4},\\[2mm]
\dot{\phi}_{g1}|_{u=0}&=\frac{1}{3}-\frac{3\phi_2}{2\phi_0^3},\\[2mm]
A_{g1}|_{u=0}&=a_4,\\ \partial_uA_{g1}|_{u=0}&=-\frac{2}{3}\left(3\xi
a_4+\partial_xf_{x2}+\partial_yf_{y2}+\phi_0\partial_t\phi_2\right).
\end{align}
\label{ec:bc_inner}%
\end{subequations}
\endgroup
The functions $B_1$, $B_2$, $G$, $\phi$, $a_4$, $f_{x2}$,
$f_{y2}$ and $\xi$ encode the freely-specifiable initial and boundary
data. Once the inner grid system is integrated, we evaluate each
function at the interface of the inner and outer grids to obtain the
boundary conditions for the $g2$ variables and subsequently integrate
the corresponding equations.

\subsubsection{Gauge fixing}

To fully close our system we still need to fix the residual gauge
freedom as described in Eq.~\eqref{ec:gauge_trafo}. It is advantageous
for the numerical implementation to have the Apparent Horizon (AH) lie
at constant radial slice $r=r_H$ for the whole numerical evolution,
and this guides our choice of gauge fixing. More specifically, we
impose that $\Theta|_{r=r_{H}} = 0$ at all times, where $\Theta$ is
the expansion of outgoing null rays. The explicit expression for our
metric ansatz~\eqref{eq:metric} is given in App.~\ref{sec:AH}.

A simple way to enforce $\Theta|_{r=r_{H}} = 0$ at all times during
the numerical evolution is to impose a diffusion-like equation of the
form
\begin{equation}
\left(\partial_t\Theta+\kappa\Theta\right)|_{u=u_H}=0 
\label{ec:gauge_fixing}
\end{equation}
with $\kappa>0$, ensuring that the expansion $\Theta$ is driven
towards the fix point $\Theta|_{u=u_H}=0$ as the time evolution runs,
pushing the AH surface to $u=u_H=\mathrm{constant}$. To implement
this, we expand Eq.~\eqref{ec:gauge_fixing} using~\eqref{ec:expansion}
as well as the equations of motion for $\ddot{S}$ and $\dot{F}_{x,y}$,
to get rid of these variables. Then, the variables $\dot{F}'$ and
$\dot{F}$ vanish and for every time step we need to solve the nested
system until the equation for $A$ to be able to solve
Eq.~\eqref{ec:gauge_fixing}.
Then, we substitute all the variables by the outer grid redefinitions,
and evaluate them at $u=u_H$. We obtain a linear PDE for~$\p_t\xi$ of
the type
\begin{align}
  \left(
  A^{(\xi)}_{xx} \p^2_x
  + A^{(\xi)}_{xy} \p_x \p_y
  + A^{(\xi)}_{yy} \p^2_y
  + B^{(\xi)}_x \p_x
  + B^{(\xi)}_y \p_y
  + C^{(\xi)}
  \right)
  \p_t \xi(t,x,y)
  = - S^{(\xi)}
  \,, 
  \label{ec:xi_evol}
\end{align}
which can be readily integrated with periodic boundary conditions in
$x$ and $y$.

\subsubsection{Evolution algorithm}

\begin{figure}[tbp]
\centerline{\includegraphics[width=0.55\textwidth]{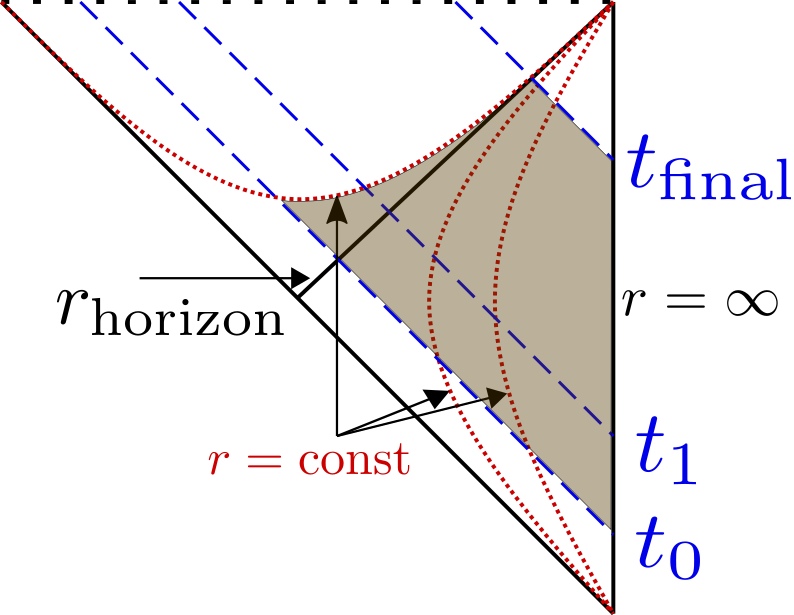}}
\caption[Illustration of \texttt{Jecco}'s evolution algorithm]
{The conformal diagram of the evolution procedure, at constant $x,y$
  slices.  The shaded region represents the region covered by the
  computational domain.
  \label{fig:jecco_penrose_diag}
}
\end{figure}

After integrating the intrinsic nested system
\eqref{ec:nested_first}-\eqref{ec:nested_last}, we use the definition
of the ``dot'' operator~\eqref{ec:new_der} to write
\begin{equation}
  \begin{aligned}
    \partial_t B_1(t,u,x,y) = \dot B_1(t,u,x,y)
    + \frac{u^2}{2} A(t,u,x,y) \partial_u B_1(t,u,x,y) \,,
\end{aligned}
  \label{eq:bulk_t}
\end{equation}
and analogously for $B_2$, $G$ and $\phi$, which are all quantities
that need to be specified by the initial data. This tells us how to
march them forward in time. In practice we write explicitly the
evolution equations in terms of the redefined $g1$ and $g2$ functions.
Schematically, the evolution algorithm is the following:
\begin{enumerate}
\item Initial conditions $B_1(t_0,u,x,y)$, $B_2(t_0,u,x,y)$,
  $G(t_0,u,x,y)$, $\phi(t_0,u,x,y)$, $a_4(t_0,x,y)$,
  $f_{x2}(t_0,x,y)$, $f_{y2}(t_0,x,y)$ and $\xi(t_0,x,y)$ are provided
  for some initial time $t_0$ on an ingoing null
  hypersurface. \label{enum:1}
\item The intrinsic nested system
  \eqref{ec:nested_first}-\eqref{ec:nested_last} is solved for the
  redefined variables in the inner grid $g1$, imposing the boundary
  conditions \eqref{ec:bc_inner}.
\item The value of the~$g1$ variables at the outer end of the inner
  grid is used as boundary condition to solve the nested system in the
  outer grid. The nested system is integrated again in the outer grid
  for the~$g2$ variables.
\item Eq.~\eqref{ec:xi_evol} is solved to find $\p_t \xi(t_0,x,y)$.
  Eq.~\eqref{eq:bulk_t} is then used to evaluate
  $\p_t B_1(t_0,u,x,y)$, $\p_t B_2(t_0,u,x,y)$, $\p_t G(t_0,u,x,y)$,
  $\p_t \phi(t_0,u,x,y)$.
\item Obtain $\partial_ta_4(t_0,x,y)$, $\partial_tf_{x2}(t_0,x,y)$ and
  $\partial_tf_{y2}(t_0,x,y)$ through Eq.~\eqref{ec:boundary_evol}.
\item Advance $B_1$, $B_2$, $G$, $\phi$, $a_4$, $f_{x2}$, $f_{y2}$ and $\xi$ to
  time $t_1$.
\end{enumerate}
In Fig.~\ref{fig:jecco_penrose_diag} an illustration of the geometric setup of
the evolution algorithm can be found.

\subsubsection{Gauge theory expectation values}

The gauge theory expectation values can be obtained from the
asymptotic behavior of the bulk variables in a way similar to
\cite{AttCasMat17b}. These relations are what we call the holographic
dictionary here and their derivation is beyond the scope of this
thesis. The result is:
\begin{equation}
  \begin{alignedat}{2}
    \mathcal{E}
    & =\tfrac{\kappa}{2L^3}
    \, \langle T^{tt}\rangle
    &&
    =
    -\frac{3}{4}a_4-\phi_0\phi_2
    +\left(\frac{7}{36}-\lambda_4\right)\phi_0^4
    \,,
    \\
    \mathcal{P}_x
    &= \tfrac{\kappa}{2L^3} \, \langle T^{xx}\rangle
    && = 
    -\frac{a_4}{4}-b_{14}-b_{24}
    +\frac{\phi_0\phi_2}{3}
    +\left(\frac{-5}{108}+\lambda_4\right)\phi_0^4
    \,,
    \\
    \mathcal{P}_{xy}
    &= \tfrac{\kappa}{2L^3}
    \,\langle T^{xy}\rangle
    && = -g_4,
    \\
    \mathcal{P}_y
    &= \tfrac{\kappa}{2L^3}
    \,\langle T^{yy}\rangle
    && =
    -\frac{a_4}{4}+b_{14}-b_{24}
    +\frac{\phi_0\phi_2}{3}
    +\left(\frac{-5}{108}
      +\lambda_4\right)\phi_0^4
    \,,
    \\
    \mathcal{P}_z
    &=\tfrac{\kappa}{2L^3}
    \,\langle T^{zz}\rangle
    && =
    -\frac{a_4}{4}+2b_{24}
    +\frac{\phi_0\phi_2}{3}
    +\left(\frac{-5}{108}+\lambda_4\right)\phi_0^4
    \,,
    \\
    \mathcal{J}_x
    &=-\tfrac{\kappa}{2L^3}
    \,\langle T^{tx}\rangle
    && = f_{x2}
    \,,
    \\
    \mathcal{J}_y
    &=-\tfrac{\kappa}{2L^3}
    \,\langle T^{ty}\rangle
    && = f_{y2}
    \,,
    \\
    \mathcal{V}
    &= \tfrac{\kappa}{2L^3}
    \,\langle \mathcal{O}_{\phi}\rangle
    && =
    -2\phi_2+\left(\frac{1}{3}
      -4\lambda_4\right)\phi_0^3
    \,.
  \end{alignedat}
  \label{ec:boundary_VEVs}
\end{equation}
For an $SU(N)$ gauge theory the prefactor $\kappa/2L^3$ in these
equations typically scales as $N^{-2}$, whereas the stress tensor
scales as $N^2$. The rescaled quantities are therefore finite in the
large-$N$ limit. The stress tensor and the expectation of the scalar
operator are related through the Ward identity
\begin{align}
\label{ward}
  \langle T^\mu_\mu \rangle = -\Lambda \langle \mathcal{O_\phi} \rangle \,.
\end{align}

\subsection{Implementation}
\label{subsec:jecco_implementation}

The evolution algorithm presented earlier is implemented in a new
numerical code called \texttt{Jecco}~\cite{jecco-2022}, written in
Julia~\cite{Jul17}. Julia is a dynamically-typed language with good
support for interactive use and with runtime performance approaching
that of statically-typed languages such as C or Fortran. Even though a
relative newcomer to the field of scientific computing, its popularity
has been steadily growing in the last few years. It boasts a friendly
community of users and developers and a rapidly growing package
ecosystem.

\texttt{Jecco} was developed as a Julia module and is freely available
at \url{https://github.com/mzilhao/Jecco.jl}. This code is a
generalization of the 2+1 C code introduced in~\cite{AttCasMat17b},
and completely written from scratch. The codebase is neatly divided
into generic infrastructure, such as general derivative operators,
filters, and input/output routines (which are defined in the main
\texttt{Jecco} module) and physics, such as initial data, evolution
equations, and diagnostic routines (which are defined in submodules).

In \texttt{Jecco} we have implemented finite-difference operators of
arbitrary order through the Fornberg algorithm~\cite{For98b} as well
as Chebyshev and Fourier differentiation matrices. These methods are
completely general and can be used with any Julia multidimensional
array. We have also implemented output methods that roughly follow the
openPMD standard~\cite{openpmd18} for writing data.

\subsubsection{Discretization}

For the numerical implementation we have discretized the $x$ and $y$
directions on uniform grids where periodic boundary conditions are
imposed, while along the $u$~direction we break the computational
domain into several (touching) subdomains with $N_u$ points. In each
subdomain a \textit{Lobatto-Chebyshev} grid is used where the
collocation points, given by
\begin{equation}
  X_{i+1} =  -\cos\left(\frac{\pi\,i}{N_u}\right)
  \qquad
  (i=0,1,\ldots,N_u-1)
  \,,
\end{equation}
are defined in the range $[-1:+1]$, and can be mapped to the physical grid by
\begin{equation}
  u_i = \frac{u_R + u_L}{2} + \frac{u_R - u_L}{2} X^{}_i
  \qquad
  (i=1,\ldots,N_u)
  \,,
\end{equation}
where $u_L$ and $u_R$ are the limits of each subdomain. For the
subdomain that includes the AdS boundary ($u=0$), the inner grid
variables of Eq.~\eqref{eqn:jecco_field_redefs} are used; all
remaining subdomains use the outer grid variables.

Derivatives along the $x$ and $y$ directions are approximated by
(central) finite differences. Although in \texttt{Jecco} operators of
arbitrary order are available, we have mostly made use of fourth-order
accurate ones for our applications.  In the radial direction $u$, the
use of the Chebyshev-Lobatto grid allow us to use pseudo-spectral
collocation methods~\cite{Boy01}.  These methods are based on
approximating solutions in a basis of Chebyshev polynomials $T_n(X)$
but, in addition to the spectral basis, we have an additional
\textit{physical} representation--the values that functions take on
each grid point--and therefore we can perform operations in one basis
or the other depending on our needs. Discretization using the
pseudo-spectral method consists in the exact imposition of our
equations at the collocation points of the Chebyshev-Lobatto grid.

The radial equations that determine our grid functions have the
schematic form of equation~\eqref{eq:ODEs}, where $f$ represents the
metric coefficients and scalar field $\phi$. Once our coordinate $u$
is discretized, the differential operator becomes an algebraic one
acting over the values of the functions in the collocation points
taking the form (at every point in the transverse directions $x,y$)
\begin{equation}
  \sum_{j=1}^{N_u}
  \left[
    A_f^i(t,x,y) \mathcal D_{uu}^{ij} + B_f^i(t,x,y) \mathcal D_u^{ij}
    + C_f^i(t,x,y) \mathbb{1}^{ij}
  \right]
  f^j(t,x,y) = -S_f^i(t,x,y)
  \label{eq:linsystem-discrete}
\end{equation}
(no sum in $i$), where $\mathcal D_{uu}$, $\mathcal D_{u}$ represent
the derivative operators for a Chebyshev-Lobatto grid in the physical
representation (see for instance~\cite{Tre00} for the explicit
expression) and $i$, $j$ indices in the $u$ coordinate.  Boundary
conditions are imposed by replacing full rows in this operator by the
values we need to fix: at the inner grid $g1$, we impose the boundary
conditions in (\ref{ec:bc_inner}); at the outer grids these are read
off from the obtained values in the previous subdomain.

The resulting operators are then factorized through an LU
decomposition and the linear systems~\eqref{eq:linsystem-discrete} are
subsequently solved using Julia's left division (\texttt{ldiv!})
operation. Recall that we need to solve one such radial equation per
grid point in the $x,y$ transverse directions. Since these equations
are independent of each other, we can trivially parallelize the
procedure using Julia's \texttt{Threads.@threads} macro.

Equation~(\ref{ec:xi_evol}) for $\p_t\xi$ is a linear PDE in $x,y$. To
solve it, after discretizing in a $N_x \times N_y$ grid, we flatten
the solution vector using lexicographic ordering
\[
  \bm{g} \equiv
  \begin{pmatrix}
    \p_t\xi(t,x_1, y_1 )    \\
    \p_t\xi(t,x_2, y_1 )    \\
    \vdots                        \\
    \p_t\xi(t,x_{N_x}, y_1 ) \\
    \p_t\xi(t,x_{1}, y_2 )   \\
    \vdots                         \\
    \p_t\xi(t,x_{N_x}, y_{N_y} )
  \end{pmatrix}
\]
and introduce enlarged differentiation matrices, which can be
conveniently built as Kronecker products
\begin{equation}
  \begin{aligned}
  \hat{\mathcal D}_x & = \mathbb{1}_{N_y\times N_y} \otimes \mathcal D_x, &
  \qquad
  \hat{\mathcal D}_y & = \mathcal D_y \otimes \mathbb{1}_{N_x\times N_x}, \\
  \hat{\mathcal D}_{xx} & = \mathbb{1}_{N_y\times N_y} \otimes \mathcal D_{xx}, &
  \qquad
  \hat{\mathcal D}_{yy} & = \mathcal D_{yy} \otimes \mathbb{1}_{N_x\times N_x}, 
\end{aligned}
  \label{eq:2DD}
\end{equation}
where $\mathcal D_x$, $\mathcal D_y$, $\mathcal D_{xx}$,
$\mathcal D_{yy}$ are the first and second derivative
finite-difference operators. The cross derivative operator is built as
a matrix product,
$\hat{\mathcal D}_{xy} = \hat{\mathcal D}_x \hat{\mathcal D}_y$. The
PDE~(\ref{ec:xi_evol}) then takes the algebraic form
\begin{equation}
  \sum_{J=1}^{N_x \times N_y}
\left[
  A_{xx}^I \hat{\mathcal D}_{xx}^{IJ}
  + A_{xy}^I \hat{\mathcal D}_{xy}^{IJ}
  + A_{yy}^I \hat{\mathcal D}_{yy}^{IJ}
  + B_{x}^I  \hat{\mathcal D}_x^{IJ}
  + B_{y}^I \hat{\mathcal D}_y^{IJ}
  + C^I \mathbb{1}^{IJ}
\right] \bm{g}^J = -S_{\bm{g}}^I
\label{eq:PDE-discrete}
\end{equation}
(no sum in $I$), where $I,J = 1, \ldots,N_x \times N_y$. The
$x$ and $y$ directions are periodic, so no boundary conditions need to be
imposed. See for example \cite{Kri18} for a pedagogical overview of these
techniques.

As before, the operator defined inside the square brackets is factorized
through an LU decomposition and the linear system~\eqref{eq:PDE-discrete} is
then solved with the left division operation. Since all the matrices are
sparse, we store them in the Compressed Sparse Column format using the type
\texttt{SparseMatrixCSC}.

\subsubsection{Time evolution}

For the time evolution we use a method of lines procedure, where we
find it convenient to pack all evolved variables (across all
subdomains) into one single state vector. This state vector is then
marched forwarded in time with the previously described evolution
algorithm using the \texttt{ODEProblem} interface from the
DifferentialEquations.jl Julia package~\cite{RacNie17}. This package
provides a very long and complete list of integration methods. For our
applications, since evaluating the time derivative of our state vector
is an expensive operation, we find it convenient for reasons of speed
and accuracy to use the Adams-Bashforth and Adams-Moulton family of
multistep methods. Depending on the application, we find that the
(third order) fixed step method \texttt{AB3} and the adaptive step
size ones \texttt{VCAB3} and \texttt{VCABM3} seem to work particularly
well.  The integration package automatically takes care of the
starting values by using a lower-order method initially.

We use Kreiss-Oliger dissipation~\cite{KreOli73a} to remove spurious
high-frequency noise common to finite-difference schemes. In
particular, when using finite-difference operators of order $p-1$, we
add Kreiss-Oliger dissipation of order $p$ to all evolved quantities
$f$ as
\begin{equation}
  \label{eq:KO-op}
  f \leftarrow f +  \sigma \frac{(-1)^{(p+3)/2}}{2^{p+1}} \left(
    h_x^{p+1}\frac{\partial^{(p+1)}}{\partial x^{(p+1)}}
    + h_y^{p+1} \frac{\partial^{(p+1)}}{\partial y^{(p+1)}}
  \right) f
  \,,
\end{equation}
after each time step, where $h_x$ and $h_y$ are the grid spacings and
$\sigma$ is a tuneable dissipation parameter which we typically set to
0.2 unless explicitly stated otherwise. This procedure effectively
works as a low-pass filter.

Along the $u$-direction we can damp high order modes directly in the
spectral representation. After each time step, we apply an exponential
filter to the spectral coefficients of our $u$-dependent evolved
quantities $f$ (see for instance~\cite{KanCarHes06}). The complete
scheme is
\begin{equation}
  \left \{  f_i\   \right\}
  \stackrel{\rm FFT}{\longrightarrow}
  \left \{  \hat f_k \right\}
  \stackrel{}{\longrightarrow}
  \left \{ \hat f_k\; e^{ -\alpha (k/M)^{\gamma M} } \right\}
  \stackrel{\rm FFT}{\longrightarrow}
  \left \{f_i \right\}
  \,,
  \label{eq:u-filter}
\end{equation}
where $M\equiv N_u - 1$, $k=0,\ldots,M$, $\alpha = \log \epsilon$
where $\epsilon$ is the machine epsilon (for the standard choice of
$\epsilon = 2^{-52}$, $\alpha = 36.0437$) and $\gamma$ is a tuneable
parameter which we typically fix to $\gamma = 8$. This effectively
dampens the coefficients of the higher-order Chebyshev polynomials.

\section{Testing the code}
\label{sec:jecco_tests}

To gauge the performance, accuracy and reliability of \texttt{Jecco}
we conduct a number of tests. These tests include comparing the data
from numerical simulations against known analytical results, as well
as those from the 2+1 \texttt{SWEC} code introduced
in~\cite{AttCasMat17b} and convergence tests solely
within~\texttt{Jecco}. We note that the PDE system we solve is
expected to be only weakly hyperbolic. We thus restrict our tests to
smooth data, where the effect of weak hyperbolicity is not expected to
be manifested. In~\cite{BeaCasGia22} we perform more tests where we
contrast obtained results against expected physical quantities and
properties of our model systems, such as the black brane entropy
density and the frequencies of its quasi-normal modes.

\subsection{Analytical black brane}
\label{subsec:jecco_test_1}

In these tests the code is initiated in a homogeneous black brane
configuration, which is a static exact solution of the equations of
motion with $\phi_0=0$ (conformal case). The functions specified in
the initial data vanish and the only non-vanishing boundary data
are~$a_4 = -4/3$. For most of these tests, we do not perform a time
evolution but instead we just solve the whole nested system at $t=0$
and compare the last bulk function to be computed, that is~$A$,
against its analytic form:
\begin{align}
  A = \frac{1}{u^2} + \frac{2 \xi}{u} + \xi^2
  + \frac{a_4 \, u^2}{1 + 2 \, \xi \, u + \xi^2 \, u^2 }
  \,,
  \label{eqn:A_exact_homo_BB}
\end{align}
using the field redefinitions of Eq.~\eqref{eqn:jecco_field_redefs}
appropriately. From~\eqref{eqn:A_exact_homo_BB} we see that the gauge
fixing can be performed via
\begin{align}
  \xi = \left(-a_4\right)^{-1/4} - 1/u_{H}
  \,,
  \label{eqn:horizon_fix_homo_BB}
\end{align}
with~$u_{H}=1$ the gauge fixed position of the apparent horizon for
the tested configuration. Since~\texttt{Jecco} provides us with the
possibility of multiple outer spectral domains, we wish to understand
to what extent faster configurations compromise the accuracy of the
numerical solution. We vary the number of nodes in the $u$-domains, as
well as the number of outer $u$-domains, to examine the accuracy of the
code for different configurations of the spectral grid. The 
inner~$u$-domain discretizes the region~$[0,0.1]$ and the outer one the
region~$[0.1, 1.0]$. The domain of both the transverse directions~$x$
and~$y$ is~$[-5,5)$ and is discretized uniformly with 128 nodes in
each case.

\begin{figure}[t]
  \includegraphics[width=1.\textwidth, height = 3.25cm]
  {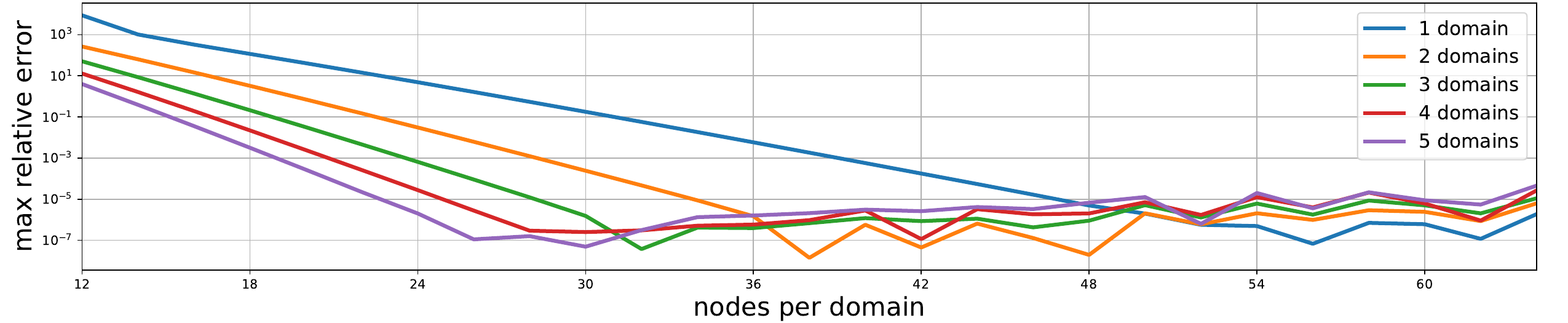}
  \caption[\texttt{Jecco} relative errors for outer spectral domains]
  {The maximum relative errors for the bulk function~$A$, in the outer
    radial domains, for different configurations of the test against
    the analytical homogeneous black brane static solution. The same
    accuracy for this test is achieved e.g.~by three outer radial
    domains with 32 nodes per domain, and a single domain with 56
    nodes. The former configuration is faster.
    \label{Fig:max_rel_errors_Aout}
  }
\end{figure}

The maximum relative error of~$A$ for the inner spectral domain
remains below~$O(10^{-10})$ for a range of nodes between 12 and
36. The respective error for different configurations of outer
spectral domains is shown in Fig.~\ref{Fig:max_rel_errors_Aout}. A
maximum relative error below~$O(10^{-5})$ in the outer region can be
achieved with one or multiple domains, where the latter typically
provides faster configurations. The orders of magnitude difference
between the maximum relative error of the inner and outer domains is
due to the near boundary field redefinition. This redefinition factors
out the near boundary radial dependence of the field and allows for a
more accurate numerical solution. For completeness, we perform a time
evolution for one of the aforementioned configurations, even if the
evolution is expected to be trivial since we are investigating a
static setup. For a configuration with 12 nodes in the inner domain
and 28 nodes on each of the three outer domains we have verified that
the maximum error maintains its expected value even after 550
timesteps, which corresponds to~$t_f=2$ in code units. For the time
integration the third order Adams-Moulton method with adaptive step is
used.

For a generic physical setup we find that some experimentation may be
required to find the optimal numerical parameters, like the number of
outer domains and nodes per domain, the choice of time integrator,
etc. For instance, if accuracy of temporal derivatives of the solution
is important one might consider chosing a fixed timestep integrator
with a small timestep instead of an adaptive one. If the main focus is
the late-time behavior of the solution, perhaps an adaptive step
integrator is preferable.

\subsection{Comparison with \texttt{SWEC}}
\label{subsec:jecco_test_2}

For this test the code is initialized with an $x$-dependent
perturbation on top of a homogeneous black brane configuration. The
initial data are
\begin{equation}
  \begin{aligned}
  B_1(0,u,x,y) & = 0.01 u^4 \,,\\
  a_4(0,x,y) & = -\frac{3}{4} \left[1
  +  \delta a_{4}
  \cos
  \left(
  2 \pi k_x \frac{x-x_{\textrm{mid}}}{x_{\textrm{max}}-x_{\textrm{min}}}
  \right)
\right] \,,\\
  \xi(0,x,y) & = \left(\frac{4}{3}\right)^{1/4} - 1\,,
\end{aligned}
\end{equation}
where~$\delta a_4 = 5 \cdot 10^{-4}$, and the remaining free data
functions ($B_2$, $G$, $\phi$, $f_{x2}$, $f_{y2}$) are set to zero. We
compare the error of the numerical solution provided by~\texttt{Jecco}
against that of the \texttt{SWEC} code used in~\cite{AttCasMat17b},
for the same setup.

We use one inner radial domain spanning the region~$ u \in [0,0.1]$
discretized with 12 grid points, and another (outer) domain spanning
the region~$u \in [0.1, 1.01]$ with 48 grid points. The transverse
direction $x$ spans $x \in [-10, 10)$, which is discretized with 128
grid points, while the $y$ has trivial dynamics for this setup (and 6
grid points are used so that the finite difference operator fits in
the domain). The time evolution is performed using the fourth-order
accurate Adams-Bashforth method. The evolution is performed for a
total of 2000 time steps.  The choice of a single outer radial domain
in~\texttt{Jecco} is made for a more explicit comparison
against~\texttt{SWEC}, since the latter does not offer the possibility
of multiple outer radial domains. It is worth noticing, however, that
there are still differences between the setups in the two codes. For
instance, the inner and outer domains of~\texttt{Jecco} share only one
common radial point, whereas in~\texttt{SWEC} there is an
overlapping~$u$-region between them.

\begin{figure}[t]
  {\includegraphics[width=0.47\textwidth
    ]{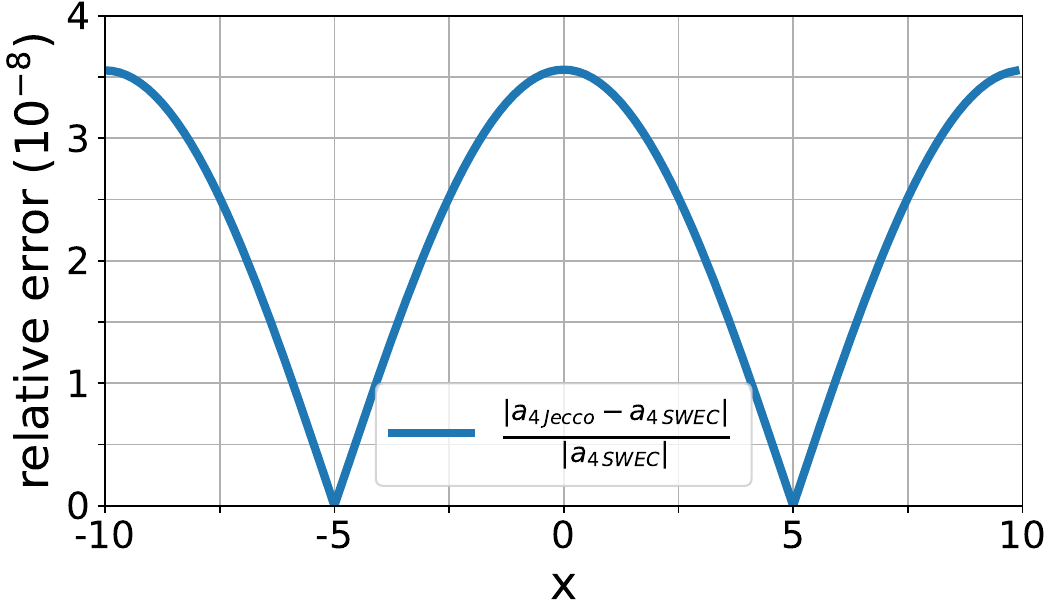}}
  \hfill
  {\includegraphics[width=0.47\textwidth
    ]{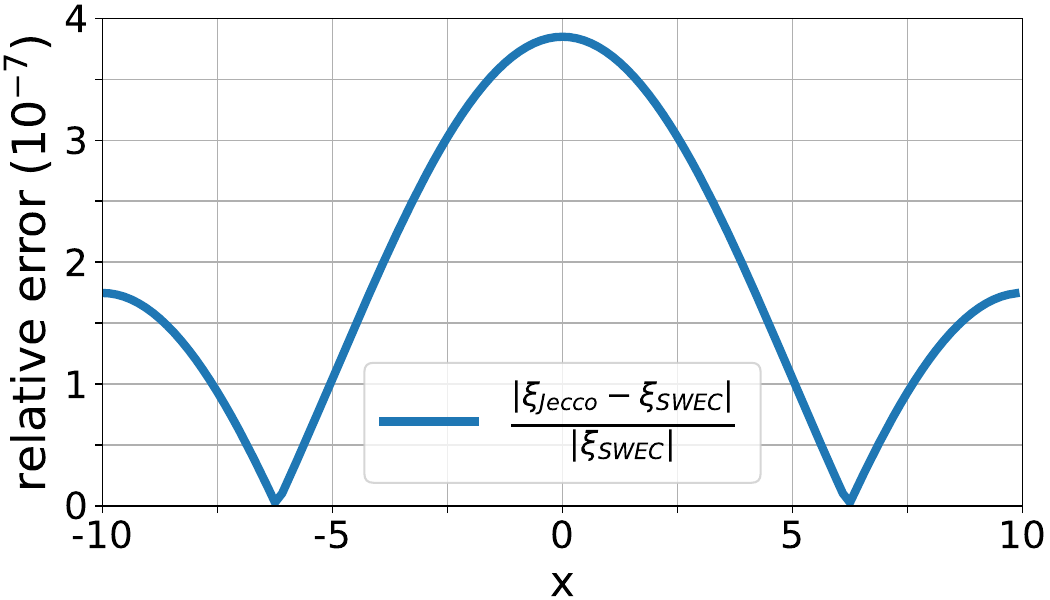}}
  \caption[\texttt{Jecco}'s benchmark against \texttt{SWEC}]
  {Relative errors for the $a_4$ and $\xi$ functions at the end of the
    evolution. Results obtained with the \texttt{SWEC} code are used
    as benchmark.
    \label{Fig:benchmark_jecco_swec}
  }
\end{figure}

We show relative differences between the $a_4$ and $\xi$ functions
obtained in the two codes in Fig.~\ref{Fig:benchmark_jecco_swec}.  The
pattern observed was similar for the metric function $B_1$.  To
compare the output of the two codes exactly on the same grid points we
perform cubic spline interpolation on the data and use the values of
the interpolated functions for the comparison. It is reassuring that
the results from the two codes agree so well.

\subsection{Convergence tests}
\label{subsec:jecco_test_3}

We now show \textit{convergence} tests using numerical solutions
obtained only from~\texttt{Jecco}. For this, we solve the same
physical setup with increasing resolution and inspect the rate at
which the numerical solution tends to the exact one. The rate at which
numerical error tends to zero with increasing resolution is determined
by the approximation~\textit{accuracy}. The latter is the degree to
which a discretized version of a PDE system approximates the correct
continuum PDE system, and such a discretized version is
called~\textit{consistent}. If its numerical solution is bounded at
some arbitrary finite time by the given data of the problem in a
discretized version of a suitable norm, it is furthermore
called~\textit{stable}. The Lax equivalence theorem states that
consistency of the finite difference scheme and stability with respect
to a specific norm guarantee convergence for linear problems (and the
converse)~\cite{LaxRic56}.

For our present case, since the spatial discretization is performed
with a mixture of finite-difference and pseudo-spectral techniques, we
fix the number of grid points along the spectral direction and vary
only the number of grid points in the uniform grid along the
transverse directions $x,y$. The finite-difference operators dominate
the numerical error, so the expected convergence rate is controlled by
the rate at which we increase the resolution in the uniform grid, as
well as the approximation order of the operators.

As in Chap.~\ref{chap:numerics} we denote by~$f$ the solution to the
continuum PDE problem and by~$f_h$ its numerical approximation. We
have
\begin{align}
  f = f_h + O(h^n)
  \,,
  \label{eqn:numerical_error}
\end{align}
where $h$ is the grid spacing and $n$ the accuracy of the
finite-difference operators. Performing numerical evolutions with
coarse and medium resolutions~$h_c$ and~$ h_m$ respectively, we
construct again the exact convergence factor
\begin{align*}
  Q 
  = \frac{f_{h_c}-f}{f_{h_m}-f}
  \,,
\end{align*}
which informs us about the rate at which the numerical error induced
by the finite-difference scheme converges to zero. Comparison of grid
functions corresponding to different resolutions is to be understood
by the use of the common grid points among the different resolutions.

Using a physical setup with known exact solution provides a clear
benchmark to compare with, and we can prepare such a setup by evolving
a homogeneous black brane where the apparent horizon is \emph{not}
fixed at a constant position~$u_{H}$ but is allowed to move, with only
gauge dynamics. This can be achieved by using a different choice for
the evolution of the gauge function~$\xi$ than the one specified
earlier. In particular, we impose the advection equation
\begin{align}
  \partial_t \xi(t,x,y) = - v_x \, \partial_x \xi(t,x,y)
  \,, \label{eq:advect-xi}
\end{align}
which introduces non-trivial dynamics to the numerical evolution. We
choose this function to be a sine with small enough amplitude, such
that the apparent horizon is guaranteed to remain within the
computational domain. Furthermore,~$\xi$ satisfies an advection
equation along the transverse direction~$x$, which makes the numerical
solution time dependent and the comparison between exact and numerical
values non-trivial for later simulation times.

The only non-vanishing initial data for this setup is the boundary
function~$a_4$, which we set to $a_4(t,x,y) = -1$, and the gauge
function~$\xi$, which we initialize to
\begin{align}
  \xi(0,x,y) = \xi_0 + A_x \sin\left( \frac{2 \pi \, n_x}{L_x}
  \left( x_{\textrm{max}} - x \right) \right)
  \,,
  \label{eqn:advect_xi_initial_xi}
\end{align}
where $L_x \equiv x_{\textrm{max}} - x_{\textrm{min}} $. For such a
configuration, the solution to equation~(\ref{eq:advect-xi}) is
\begin{align}
  \xi(t,x,y) = \xi_0 + A_x \sin\left( \frac{2 \pi \, n_x}{L_x}
  \left( x_{\textrm{max}} - x + v_x t \right) \right)
  \,,
  \label{eqn:advect_xi_sol}
\end{align}
and the exact solution of the metric function~$A$ is given
by~\eqref{eqn:A_exact_homo_BB}, where $\xi$ is now provided
by~\eqref{eqn:advect_xi_sol}.

For the tests presented herein we have fixed
\begin{align*}
  \xi_0 = 0
  \,, \quad A_x = 0.1
  \,, \quad n_x = 1
  \,, \quad
  x_{\textrm{max}} = 5
  \, \quad x_{\textrm{min}} = -5
  \,.
\end{align*}
For the numerical discretization we have employed one inner radial
domain with 12 grid points (spanning the region~$u \in [0,0.1]$) and
three equal-sized outer domains for the region~$u \in [0.1, 1.2]$ with
28 grid points each. For the transverse directions we use 16, 32, and
64 grid points for coarse, medium and fine resolution respectively.
The time integration is performed with the third-order accurate
Adams-Moulton method, with adaptive timestep. The (periodic) finite
difference operators are second order accurate and Kreiss-Oliger
dissipation is used with the prescription of equation~(\ref{eq:KO-op})
with~$\sigma = 0.01$. We run the tests on a laptop with 16GB RAM
memory and Intel Core i7-10510U at 1.80GHz CPU. For the forth order
accurate finite difference case, the coarse resolution is performed
with a single thread and is completed within 36 minutes. For the same
finite difference accuracy, the medium and high resolution tests are
performed with two threads running in parallel and are completed
within 66 and 271 minutes, respectively.

\begin{figure}[t]
  {
    \includegraphics[width=0.235\textwidth
    ]
    {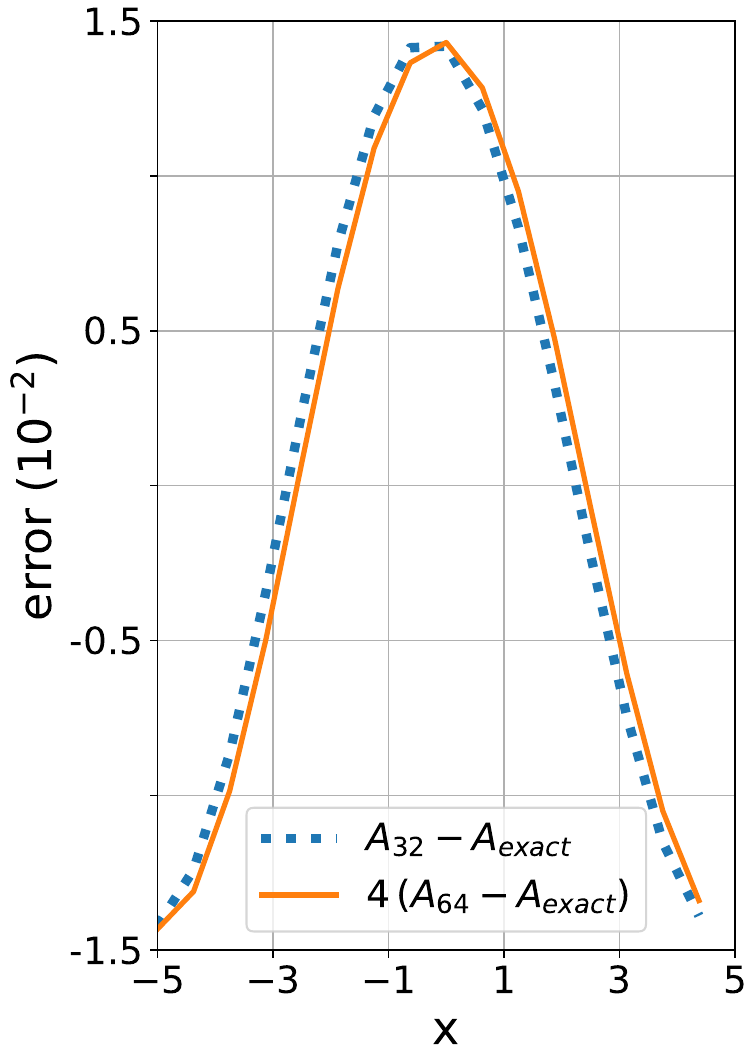}
  }
  {
    \includegraphics[width=0.76\textwidth
    ]
    {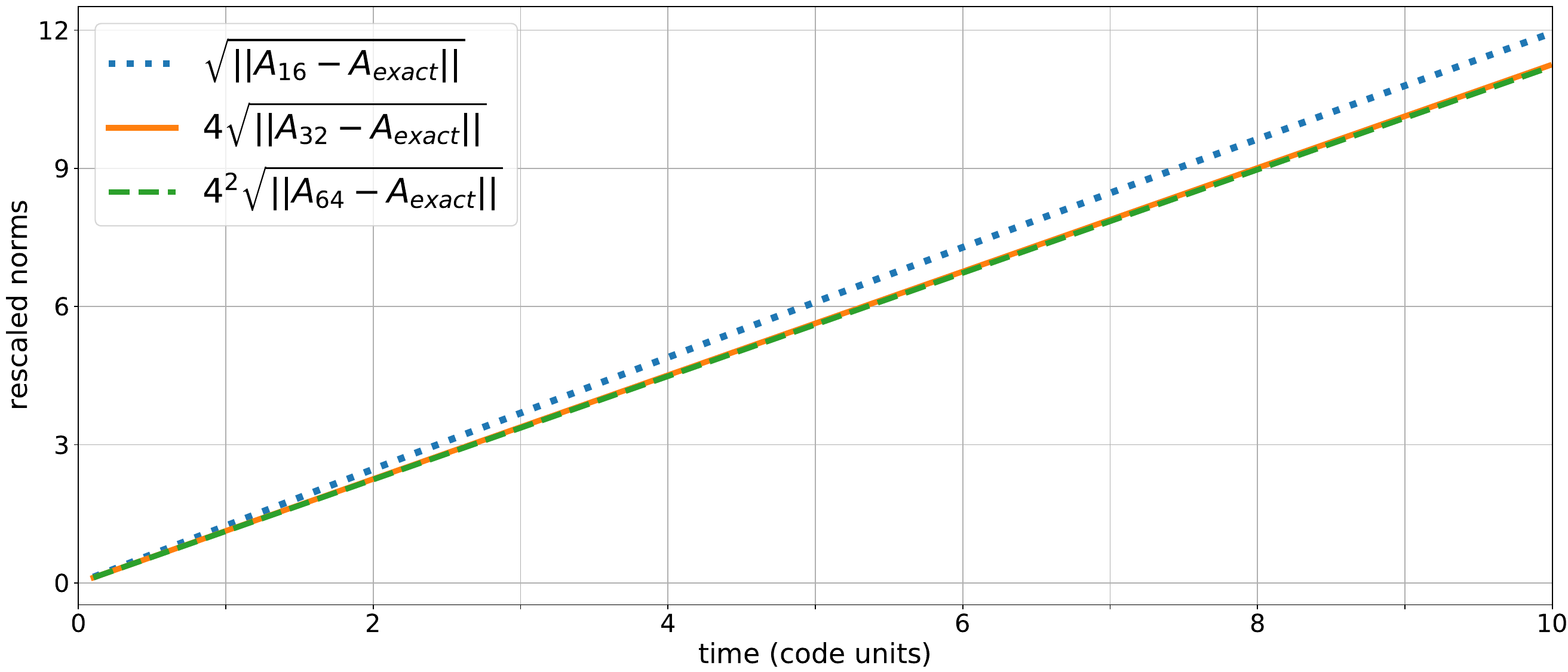}}
  \\
  {
    \includegraphics[width=0.235\textwidth
    ]
    {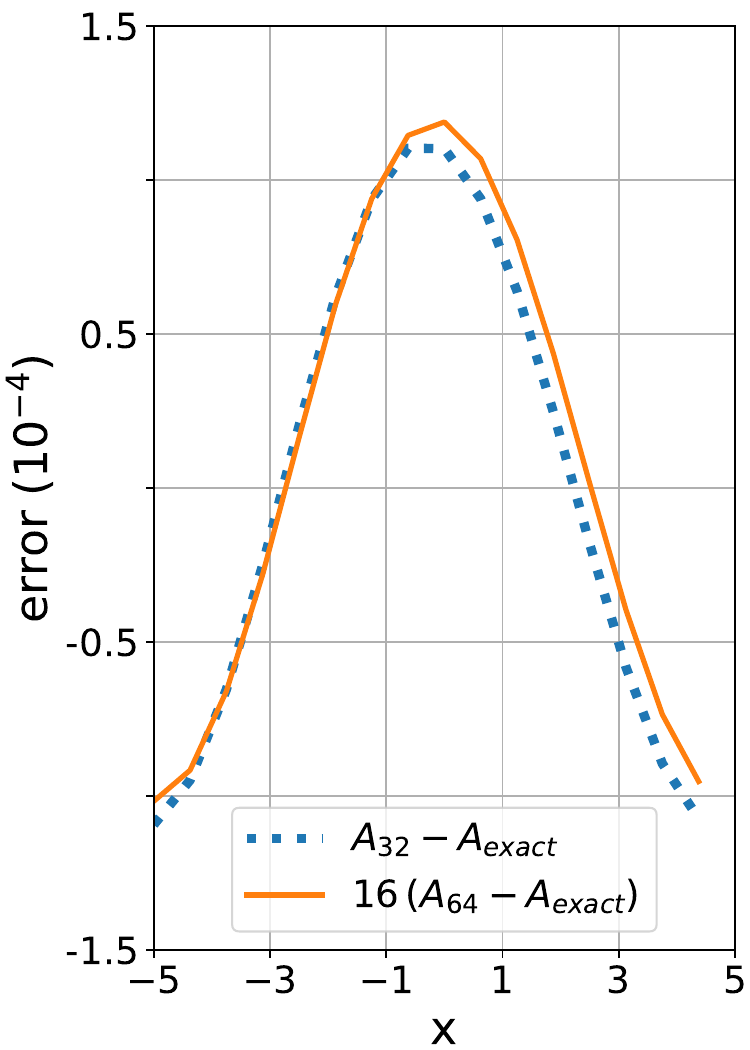}
  }
  {
    \includegraphics[width=0.76\textwidth
    ]
    {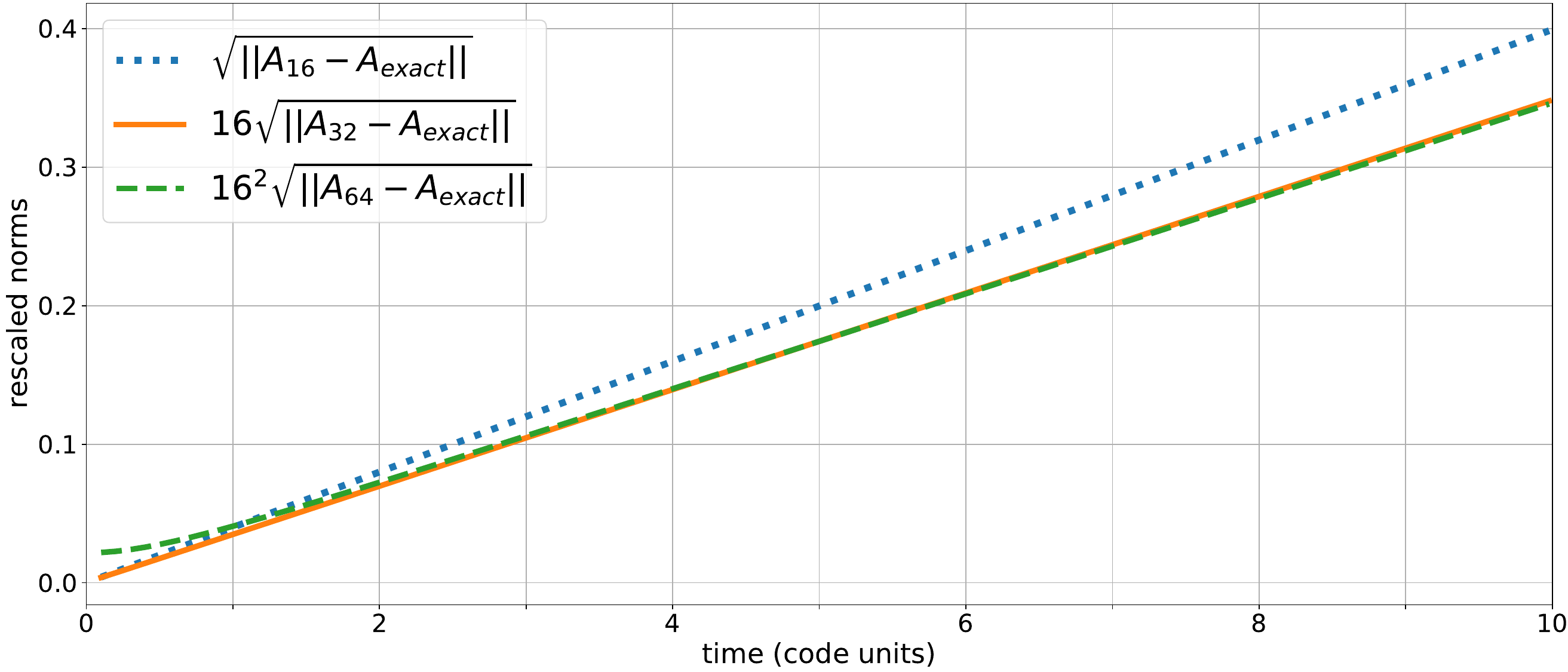}} 
  \caption[Pointwise and norm convergence tests solely with
  \texttt{Jecco}]{(Left) Pointwise convergence of the metric
    function~$A$ along the~$x$ direction, at~$t=9.98$ (code
    units),~$u=0.83$ and~$y=0.625$, for the medium and fine
    resolutions.  
    (Right) Convergence rate 
    for the metric function~$A$ in terms of rescaled norms. Perfect
    overlap of curves should be understood as perfect
    convergence. Second order finite difference approximation
    corresponds to the top and forth order to the bottom row. The
    ideal convergence factor for the former is~$Q=4$ and the
    latter~$Q=16$ for the specific
    tests.
    \label{Fig:conv_plots_BB_advectxi}
  }
\end{figure}

Convergence tests for the $A$ metric function can be seen in
Fig.~\ref{Fig:conv_plots_BB_advectxi}.  As mentioned above, the
comparison of the grid functions against the exact solution is
performed only on grid points that are common to all three
resolutions. The expected convergence factor for this setup is~$Q=4$
for second order finite difference operators and~$Q=16$ for forth
order, which is indeed what we observe in the left column. The same
convergence rate is expected when we perform a norm comparison. The
discretized version of the~$L^2$-norm that we employ here is simply
the square root of the sum of the squared grid function under
consideration (over all domains). In the right column of the figure we
again see very good agreement for the norm convergence rate.

\begin{figure}[h]
  {
    \includegraphics[width=1\textwidth
    ,height = 5.85 cm
    ]
    {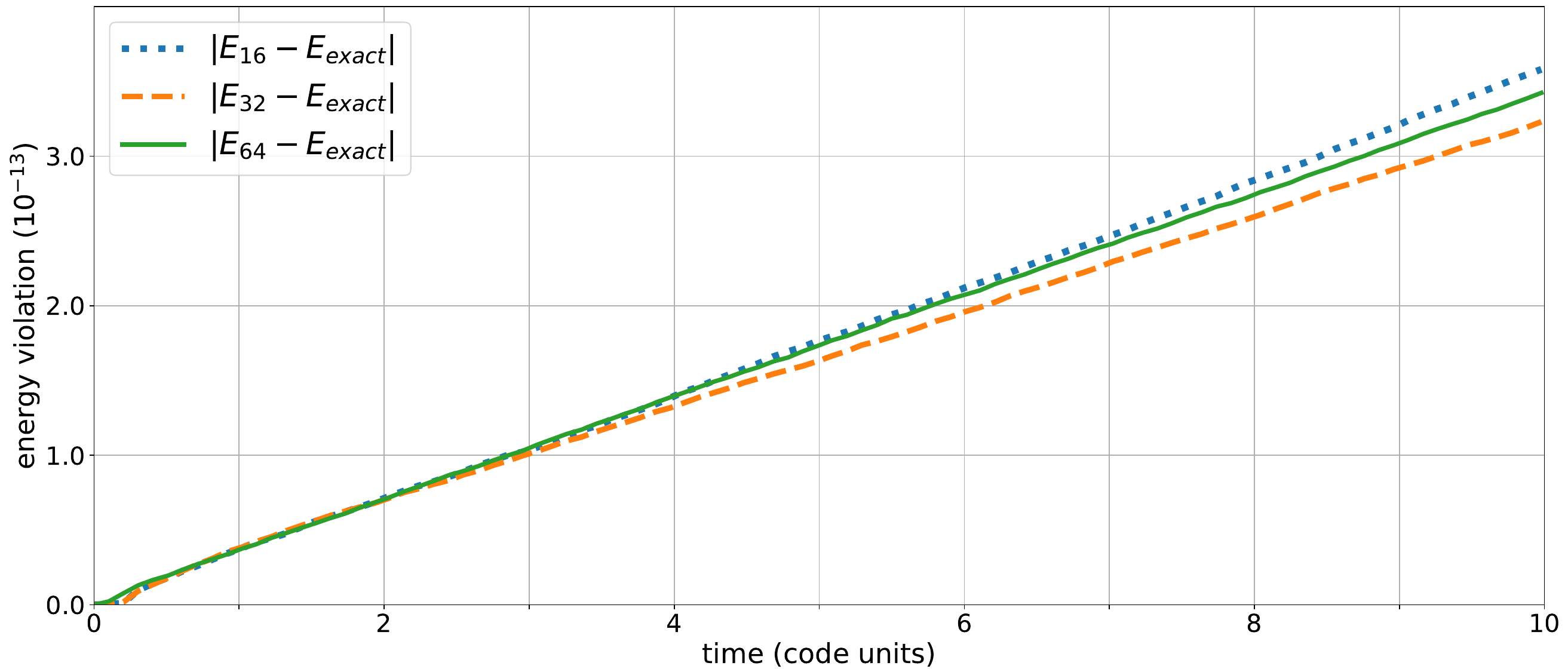}
    \\
    \includegraphics[width=1\textwidth
    ,height = 5.85 cm
    ]
    {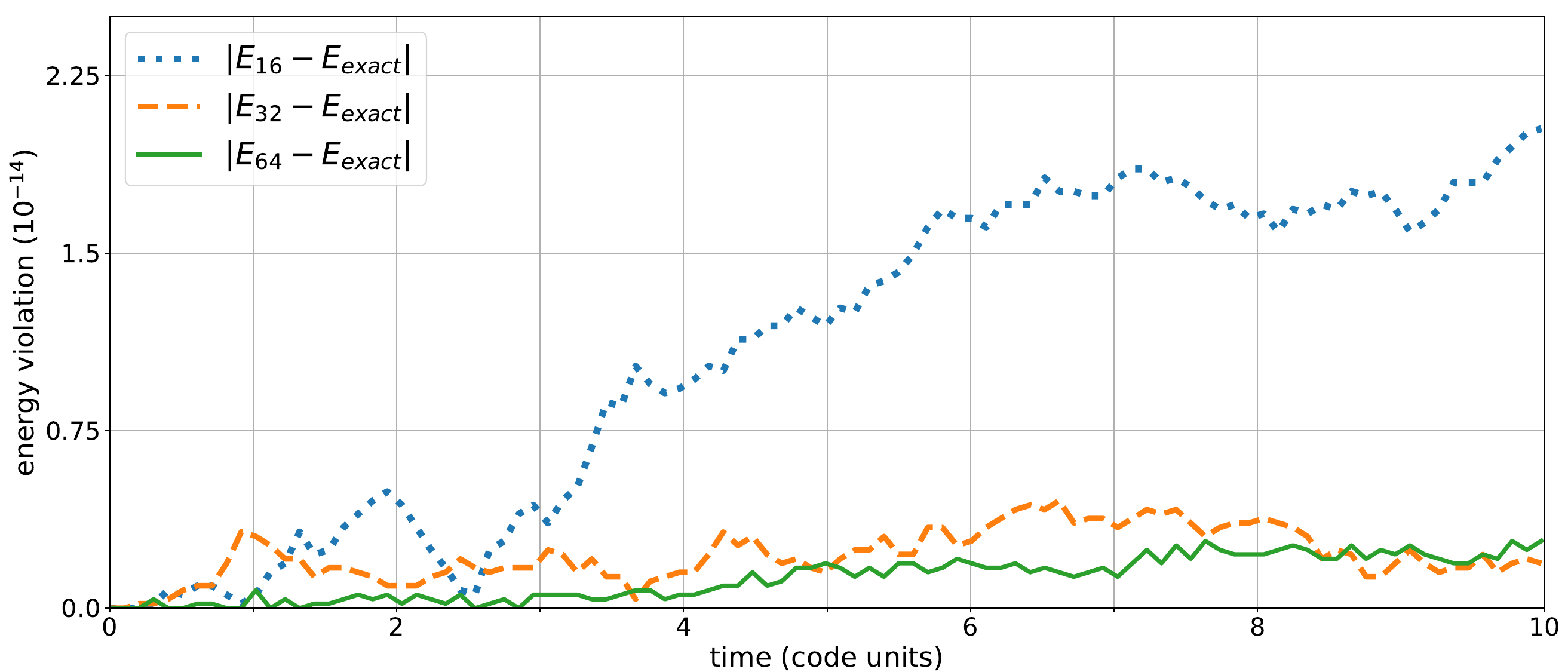}
  } 
  \caption[Numerical violation of the total energy of gauge theory box
  in \texttt{Jecco} simulation]
  {The relative error of the total energy of the box for the numerical
    simulations of the gauge dynamics described by
    Eq.~\eqref{eqn:advect_xi_sol}. The total energy of the gauge
    theory for this setup is~$75$ in code units. The numerical
    violation is within accepted numerical error as illustrated
    here. The top figure corresponds to second order finite difference
    operators, whereas the bottom to forth order. As expected the
    numerical violation is smaller when higher order operators are
    used.
\label{Fig:jecco_energy_error}
}
\end{figure}

Finally, the total energy of the boundary theory is expected to be
constant throughout the numerical evolution, which is indeed the case
up to numerical errors, as illustrated in
Fig.~\ref{Fig:jecco_energy_error}. The case illustrated corresponds to
the setup used for the convergence tests of
Fig.~\ref{Fig:conv_plots_BB_advectxi}, namely the gauge dynamics
of~$\xi$ as described by the exact solution of
Eq.~\eqref{eqn:advect_xi_sol}. By box we mean the region of the
boundary theory is understood to reside, given the periodic boundary
conditions in the~$x,y$ directions.

\section{Simulating strongly coupled systems}
\label{sec:jecco_results}

By strongly coupled systems here we mean matter under extremely high
pressure and temperature. Such scenarios can occur for instance during
the early universe or inside neutron stars, as well as in terrestrial
experiments that mimic these conditions e.g. heavy ion collisions. For
these conditions the fundamental force of the strong interactions is
dominant and the theory that describes it is Quantum Chromodynamics
(QCD). The gauge theory we study here is not QCD however, but it is a
strongly coupled, non-Abelian, gauge theory that exhibits phase
transitions. The gravitational dual setups that we construct and
evolve allow us to follow the dynamics of these phase transitions. We
hope that these models can provide us with valuable insights for the
qualitative behavior of strongly coupled matter in similar physical
scenarios.

A very promising arena where this type of studies can be fruitful is
in GWs. GW detectors like LISA~\cite{Caprini:2019egz} may be able to
observe signals that carry distinct imprints of phase transitions,
either originating from the early universe~\cite{Hindmarsh:2020hop} or
from events that involve Neutron stars~\cite{Kalogera:2021bya}. Such
imprints would inform us about the behavior of matter under extreme
conditions. Regarding early universe scenarios, within the Standard
Model of particle physics the universe cools down from its original
hot dense state via a smooth crossover~\cite{Aoki:2006we,
  Kajantie:1996mn, Laine:1998vn, Rummukainen:1998as}. This process is
not expected to produce any GWs. However, several beyond the Standard
Model scenarios predict that this cooling down can happen via
different channels that involve more abrupt phase transitions, which
could produce GWs~\cite{Carena:1996wj, Delepine:1996vn, Laine:1998qk,
  Huber:2000mg, Grojean:2004xa, Huber:2006ma, Profumo:2007wc,
  Barger:2007im, Laine:2012jy, Dorsch:2013wja,
  Damgaard:2015con}. Detection of such patterns in primordial GWs would
strongly suggest paths to expand our picture of the fundamental
interactions.

\texttt{SWEC} is the code progenitor of~\texttt{Jecco} that was
introduced in~\cite{AttCasMat17b} and used among others
in~\cite{BeaDiaGia21, BeaCasGia21, BeaCasGia21a}. Studying GW
production scenarios was not possible with~\texttt{SWEC} due to the
high symmetry along the spatial directions of the boundary
theory. In~\texttt{Jecco} translational invariance is imposed only
along one of the three spatial directions and thus we can simulate
processes that can produce GWs. The equations of motion for
fully~$3+1$ dynamical setups of the boundary theory are not yet
implemented, but is a desired feature for the future.

In addition to the lower degree of symmetry, another aspect that
improves our ability to simulate the dynamics of phase transitions is
the scalar potential implemented. In~\cite{AttCasMat16, AttCasMat17,
  AttBeaCas17} the scalar potential~\eqref{eq:potential}
with~$\lambda_6=0$ was chosen. Even though the model still exhibited
phase transitions, the separation between the high and low energy
density states was very large, which resulted in very slow
dynamics. Consequently, more computational time was necessary in order
to capture the evolution of the phase transition. Including a
non-vanishing~$\lambda_6$ parameter in the scalar potential allows for
a smaller separation of scales between the different phases, and thus
in faster dynamics. With this setup it is more convenient to explore
the parameter space of the model and search for interesting
phenomena. To understand whether the model exhibits phase transitions,
a phase diagram has to be constructed. For a scalar potential with
non-vanishing~$\lambda_6$ parameter this is done in detail
in~\cite{BeaMat18, BeaDiaGia21} and includes the construction of
various static black brane configurations. A typical shape of a phase
diagram with a phase transition is shown in Fig.~\ref{fig:phase_diag}.

The dynamical scenarios that have so far been explored
with~\texttt{Jecco} are the evolution of the spinodal instability and
bubbles of low energy density phase within a bath of high energy
density phase. In Subsec.~\ref{subsec:spinodal} we provide further
details on the former and present a brief overview of its study
in~\cite{BeaCasGia21b}. In Subsec.~\ref{subsec:bubbles} we give a
short description of the holographic bubble dynamics as presented
in~\cite{BeaCasGia22}. Work is already undergoing into implementing
different types of initial data generating routines such as those
relevant for gravitational shockwaves, which are used to model heavy
ion collisions.

In the following figures, all quantities are shown in units of
~$\Lambda$, which is a characteristic energy scale of the dual theory
and is tuned by the choice of the scalar quantity~$\phi_0$.

\subsection{Spinodal instability}
\label{subsec:spinodal}

The thermodynamics of the gauge theory are extracted by building
various homogeneous black brane configurations on the gravitational
side of the duality (see e.g.~\cite{Gubser:2008ny}). In thermal
equilibrium all the pressures~$\mathcal{P}$ are equal and the free
energy density is~$\mathcal{F}=-\mathcal{P}$. The free parameters of
the scalar potential~\eqref{eq:potential} are fixed to
\begin{align*}
  \phi_M = 1 \,, \quad
  \phi_Q = 10
  \,.
\end{align*}

\begin{figure}[thp]
\begin{center}
\includegraphics[width=.8\textwidth]{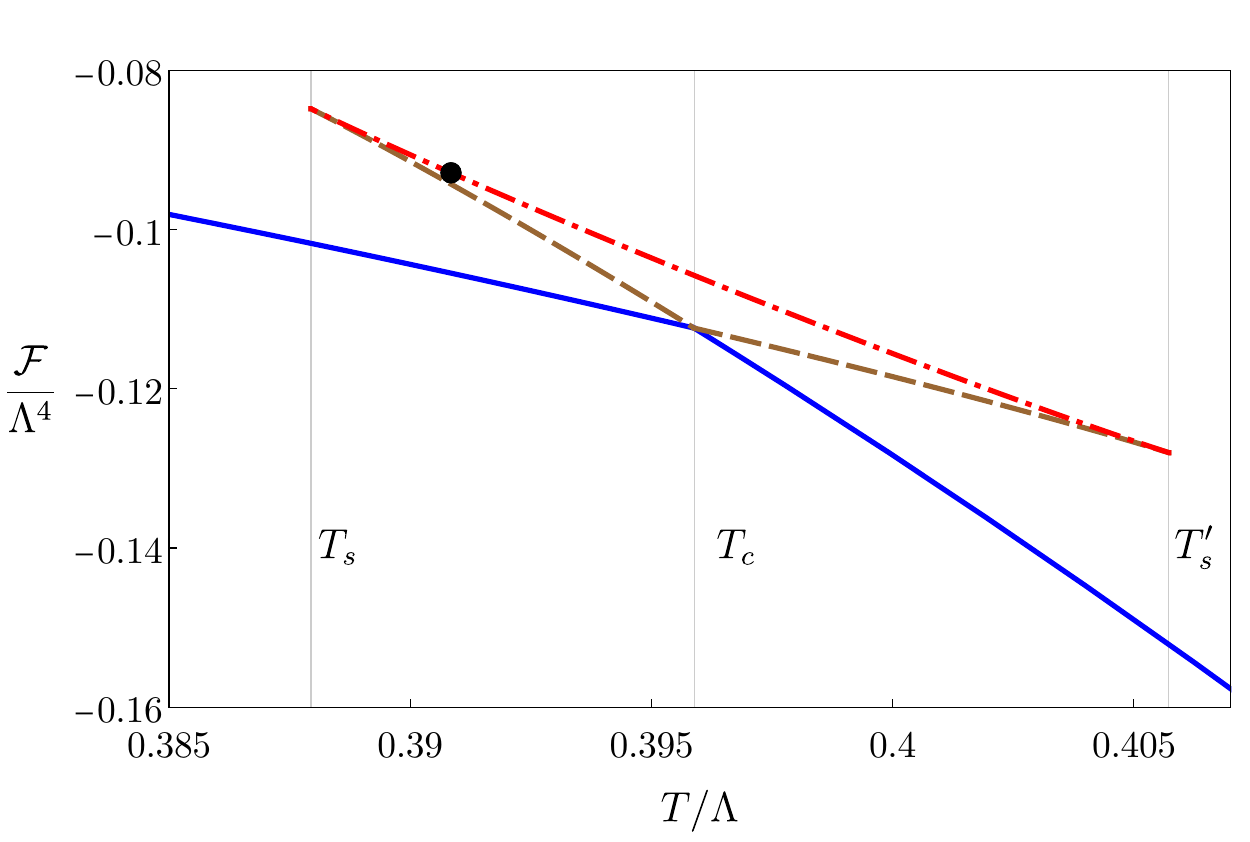}
\includegraphics[width=.8\textwidth]{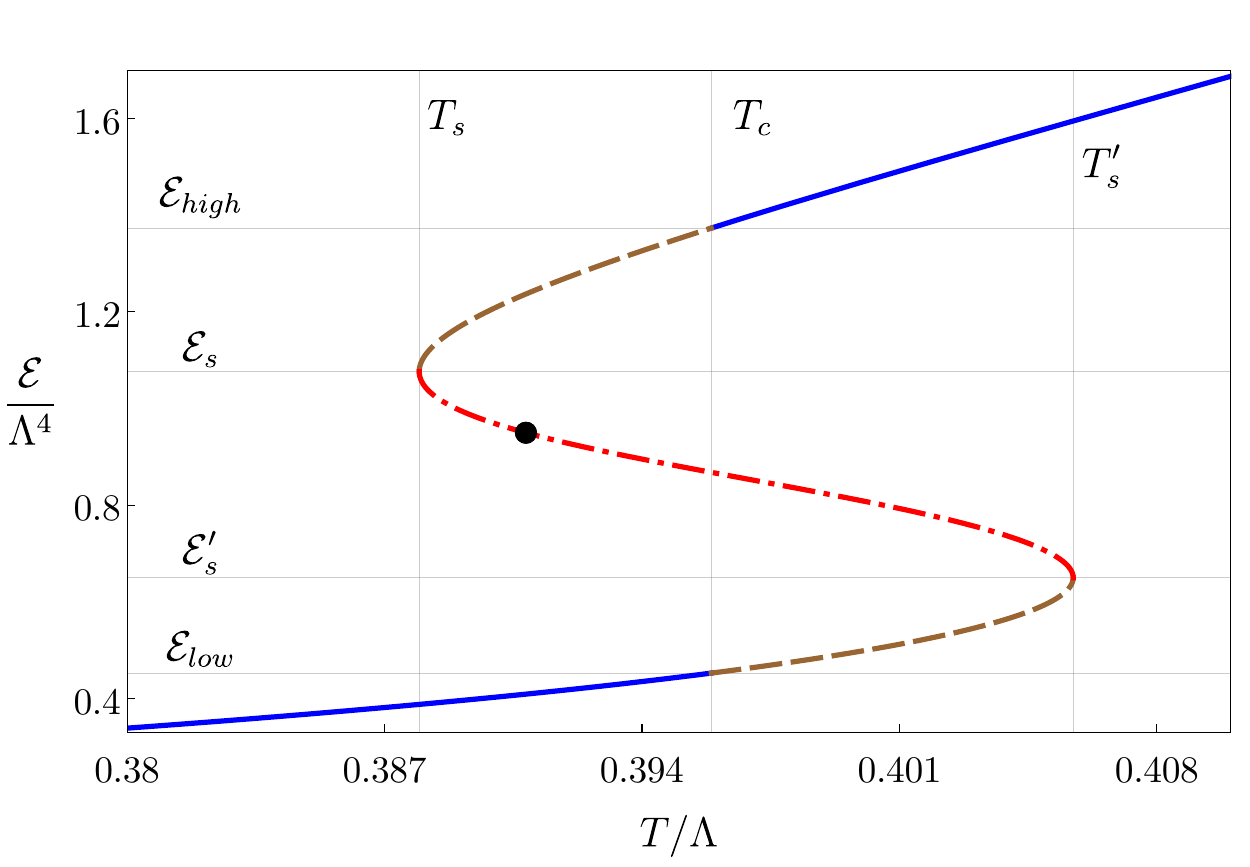}
\end{center}
\caption[The spinodal region on phase diagrams]{Free energy density
  (top) and energy density (bottom) of the four-dimensional gauge
  theory dual. States on the solid, blue curves are thermodynamically
  stable. States on the dashed, brown curves are metastable. States on
  the dashed-dotted, red curve are unstable. The black dot with
  $T=0.3908\Lambda$ indicates the initial state on which we will focus
  here and is within the spinodal unstable region. }
\label{fig:phase_diag}
\end{figure}

The discontinuity of the free energy as a function of temperature as
seen in the top of Fig.~\ref{fig:phase_diag} is indicative of a
first-order phase transition. This behavior leads to multivaluedness
for the energy density as a function of the temperature, as see in the
bottom of Fig.~\ref{fig:phase_diag}. The critical
temperature~$T_c=0.396 \, \Lambda$ is defined as the point in the top
of Fig.~\ref{fig:phase_diag} where the two curves cross. There, the
state that minimizes the free energy changes branch. The solid blue
curves indicate the thermodynamically stable branches and the
difference in their energy density is called the latent heat. These
are the high and low energy density phases and both correspond to
deconfined plasma phases and are dual to homogeneous black brane
geometries. The dashed brown curves are metastable, which means that
are locally thermodynamically stable, but not globally. Finally, the
dashed-dotted red curve is locally unstable and defines the spinodal
region. Initial states within this region are affected by the spinodal
instability, where small amplitude and long wavelength perturbations
grown exponentially with time.

\begin{figure}[thp]
  \begin{center}
    \begin{tabular}{cc}
      \includegraphics[width=.485\textwidth]{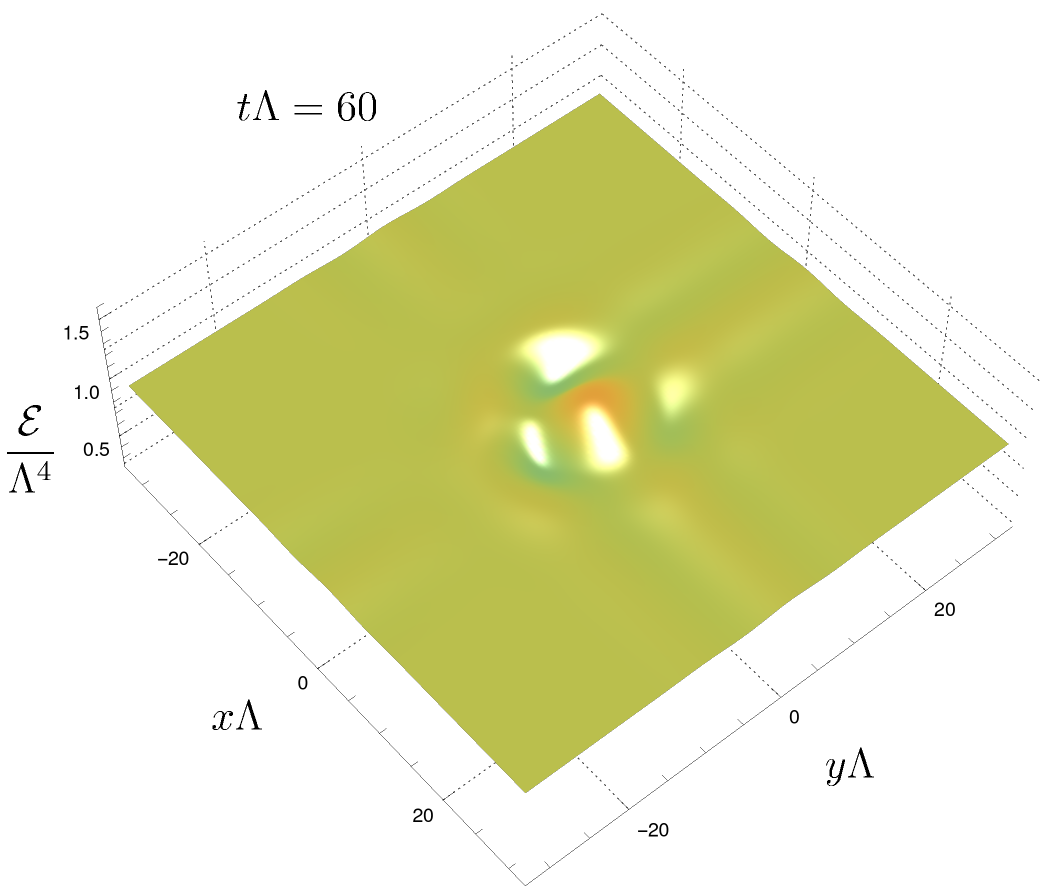}
      &
        \includegraphics[width=.485\textwidth]{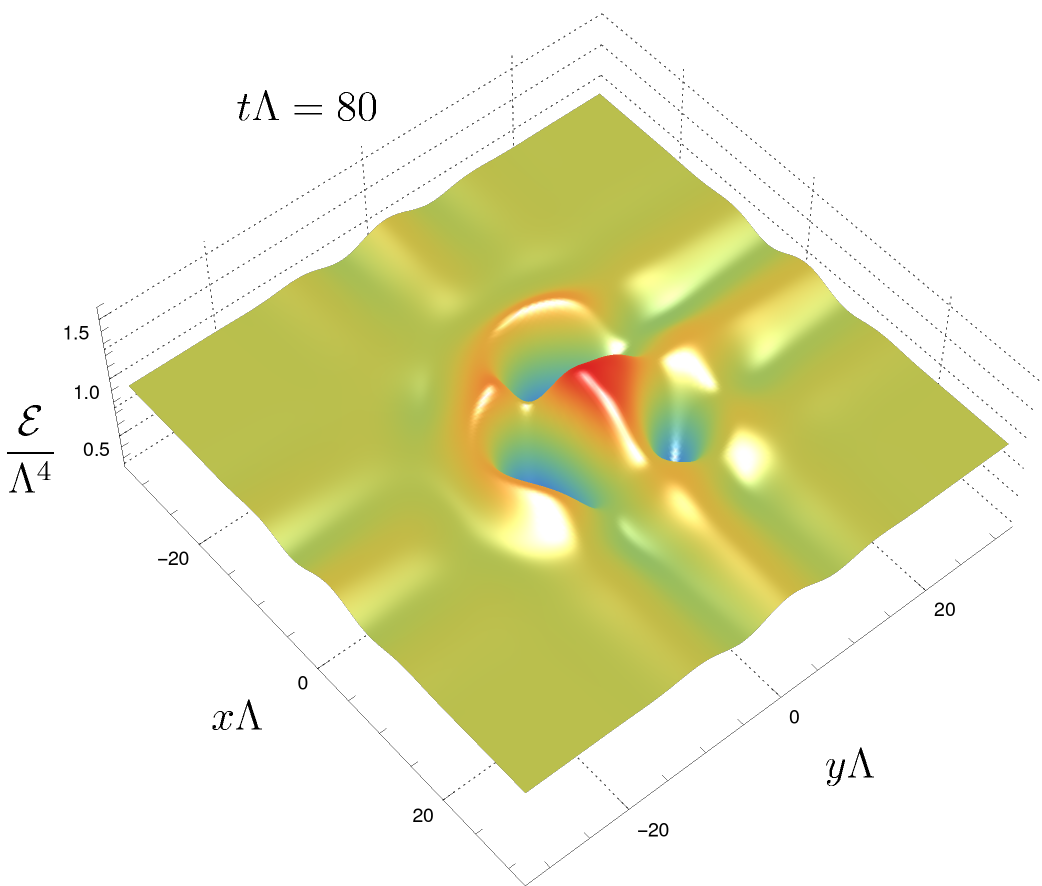}
      \\[0.65cm]
      \includegraphics[width=.485\textwidth]{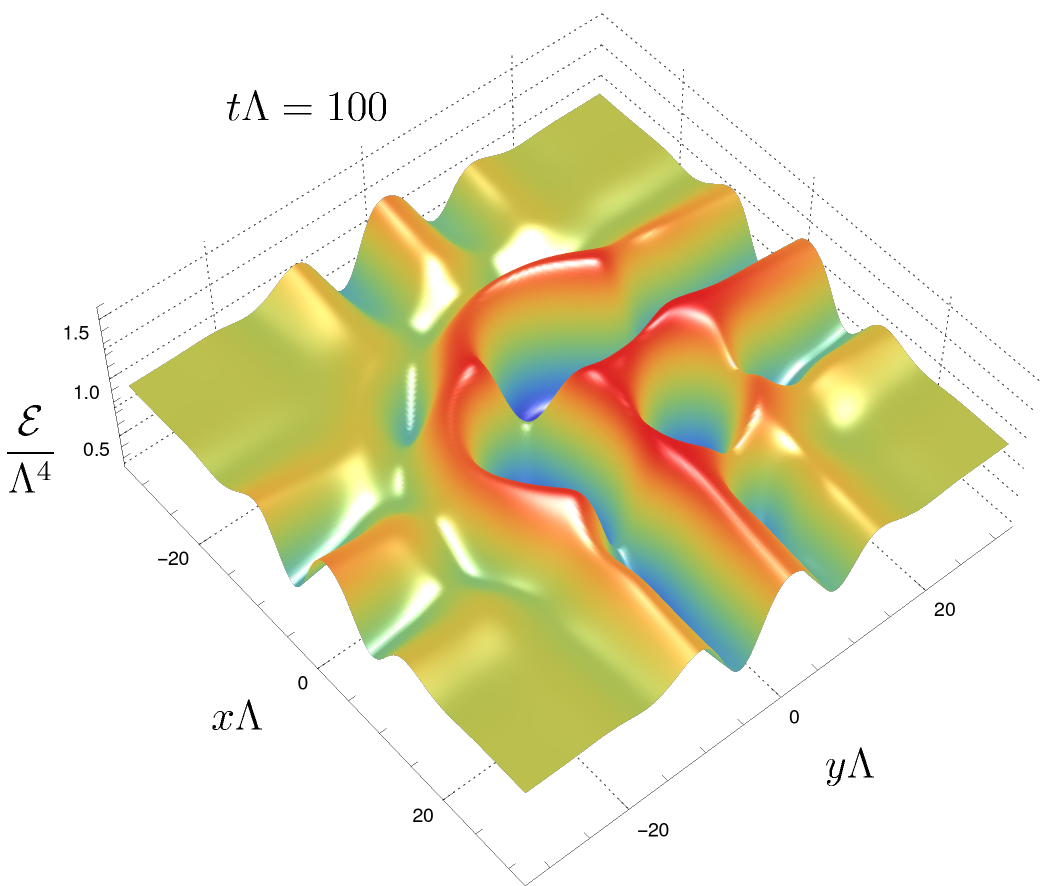}
      &
        \includegraphics[width=.485\textwidth]{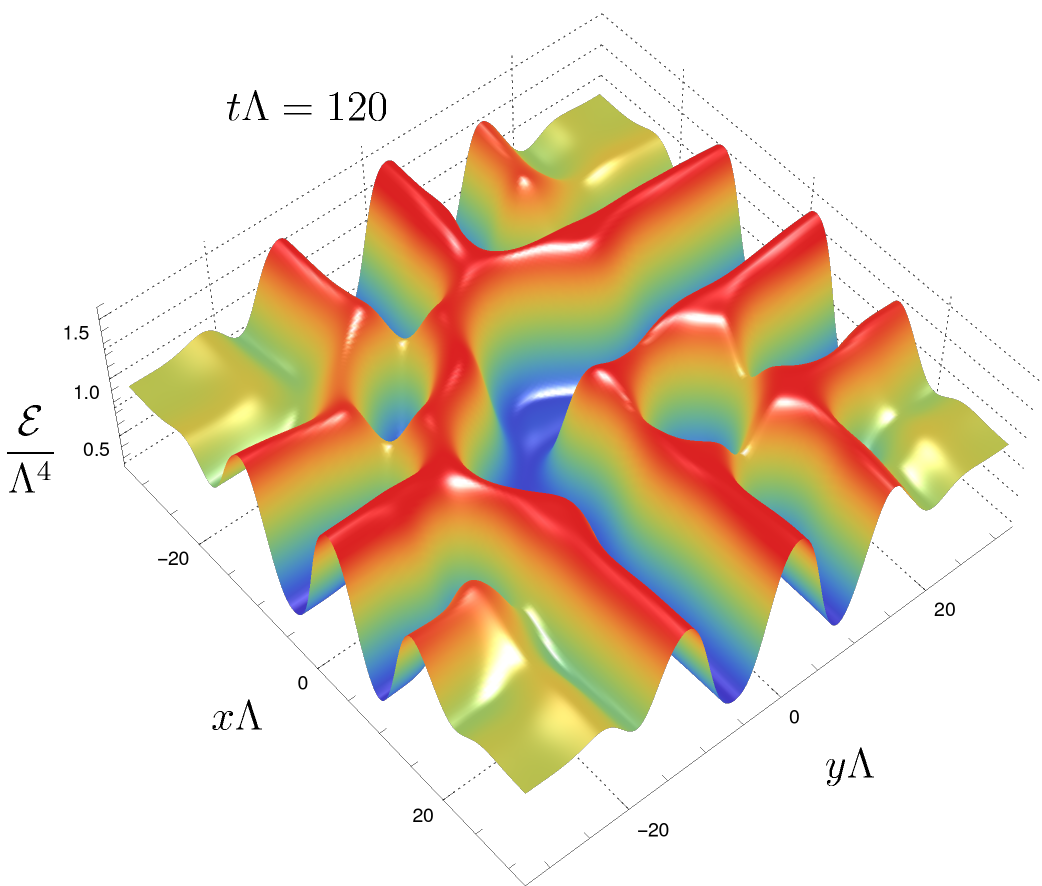}
      \\[0.65cm]
      \includegraphics[width=.485\textwidth]{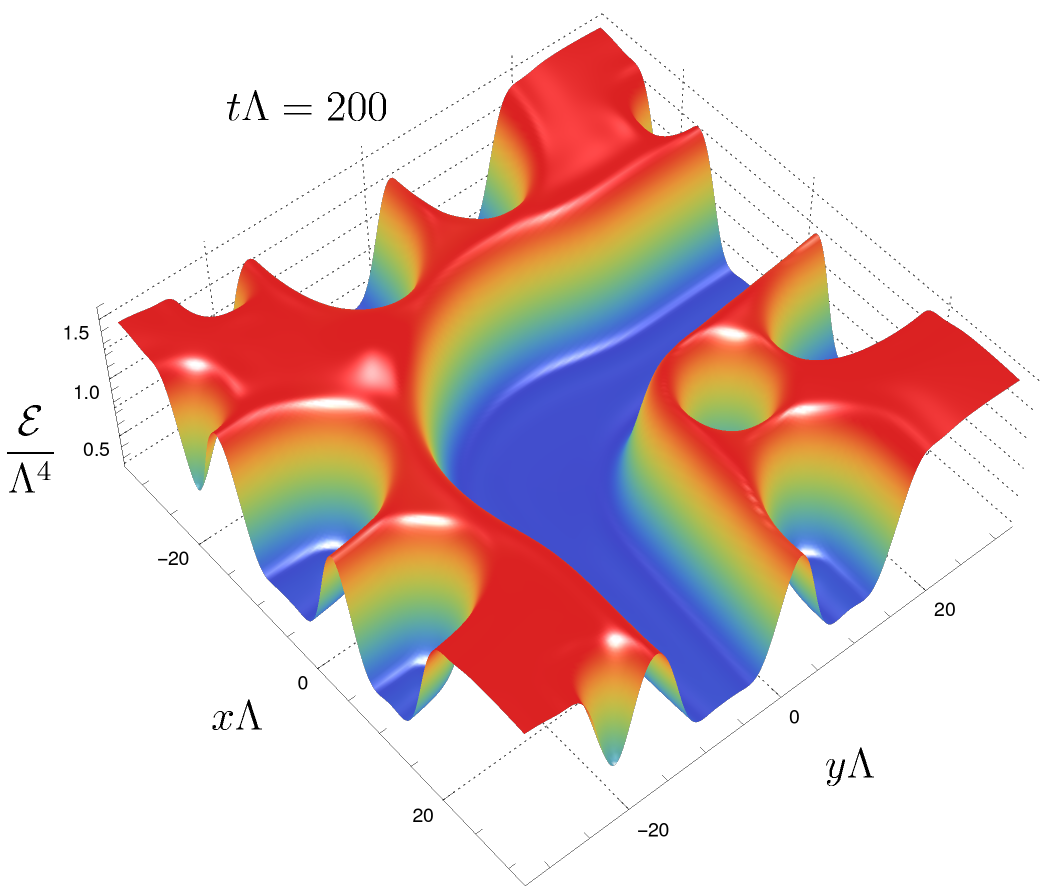}
      &
        \includegraphics[width=.485\textwidth]{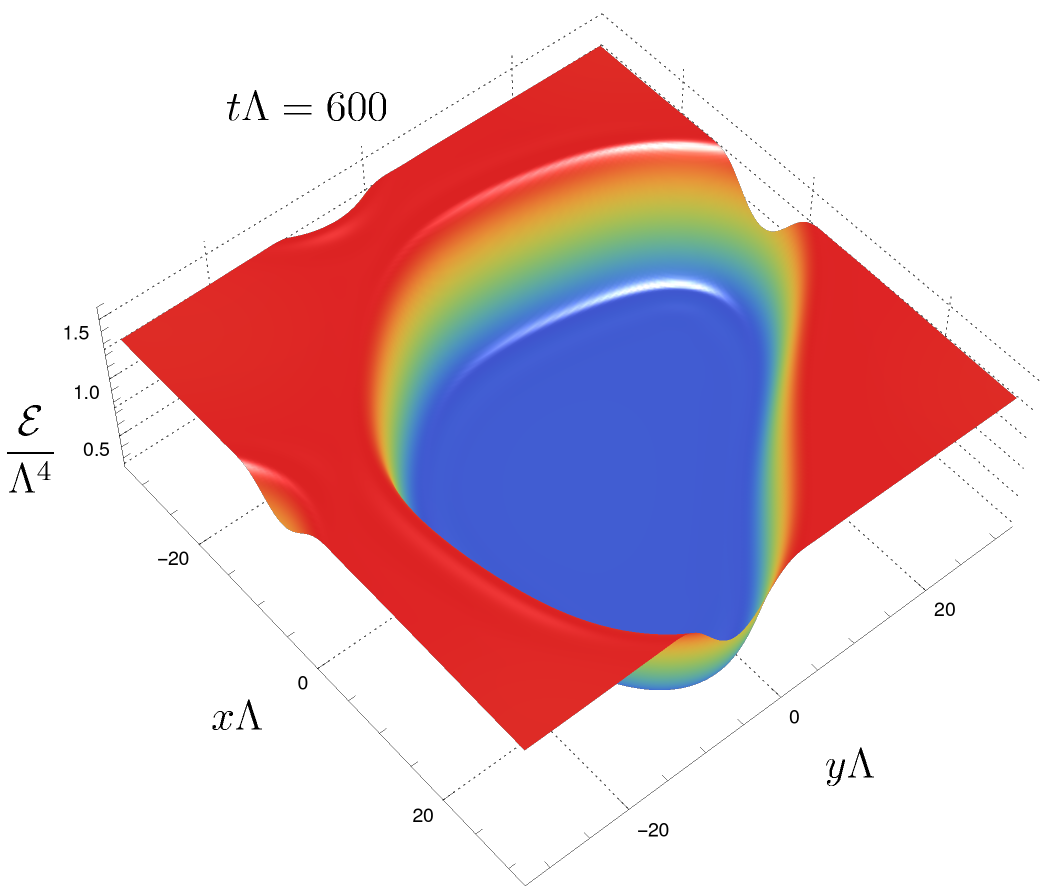}
    \end{tabular}
    \caption[Evolution of the spinodal instability]{ Spacetime
      evolution of the energy density for the initial homogeneous
      state in the spinodal region, perturbed with small
      fluctuations. A video of the evolution can be found at
      \href{https://www.youtube.com/watch?v=qIhbpchr3gE}
      {https://www.youtube.com/watch?v=qIhbpchr3gE}.
      \label{fig:spinodal_energy_evol}}
  \end{center}
\end{figure}

In Fig.~\ref{fig:spinodal_energy_evol} we demonstrate the evolution of
the energy density for a state initially within the spinodal
region. The exact state chosen is represented as the black dot in
Fig.~\ref{fig:phase_diag}. This initial homogeneous configuration is
slightly perturbed. The evolution of the perturbation has a short
initial regime described well by a linearized analysis around the
homogeneous configuration, where some modes of the perturbation decay
while others grow. After the unstable modes grow large enough, the
evolution enters a non-linear regime where the dynamics become
richer. A thorough discussion of these different regimes can be found
in~\cite{AttBeaCas19}. In short, the growth of the perturbation
creates peaks and valleys in the energy density profile. The initial
separation of these structures depends on the unstable modes that
dominate the first part of the dynamics. They subsequently merge to
eventually form a single low energy domain within a high energy bath,
as shown in the last subfigure of
Fig.~\ref{fig:spinodal_energy_evol}. If the box where the dynamics
takes place is large enough, this state is homogeneous with
temperature~$T=T_c$ and the energy density for the low and high phase
is~$\mathcal{E}_{low}$ and~$\mathcal{E}_{high}$ as shown in
Fig.~\ref{fig:phase_diag}, respectively. This process can produce a GW
spectrum that is distinct from that of a phase transition which takes
place via bubble nucleation and collision, as discussed in detail
in~\cite{BeaCasGia21b}. The conditions under which the spinodal
channel may be favored over the bubble one, are also discussed there.

\subsection{Bubbles}
\label{subsec:bubbles}

The phase transitions we study are expected to mostly take place via
bubble nucleation, expansion and collision. To accurately predict the
GW spectrum of this process, knowledge of several parameters is
required, like critical temperature and strength of the transition
which are thermodynamic in nature and bubble wall velocity that highly
depends on the out-of-equilibrium physics. For the first class of
parameters, holographic calculations have been performed
e.g. in~\cite{AhmBit17, AttBeaCas17, AhmBit18, BeaMat18, AttBeaCas18,
  AttBeaCas19, Ahmadvand:2020fqv, BeaDiaGia21, BigCadCot20,
  BigCadCot21, AreHinHoy20}, whereas for the bubble wall velocity a
holographic calculation from first principles was presented
in~\cite{BeaCasGia21}. This study used~\texttt{SWEC} and so the
bubbles are planar, in the sense that they are invariant along two out
of the three spatial direction of the gauge theory. The focus of this
study is on the interface between the low and high energy density
phases and the velocity of this wall and surface tension is neglected.

In~\cite{BeaCasGia22} this line of research is expanded by allowing
for bubbles with translational invariance only in one spatial
direction, which we call cylindrical. With~\texttt{Jecco} we explore
different types of bubbles: the expanding, collapsing and critical
ones. The reason for this richness is due to the surface tension that
is included in the analysis. In particular, the critical bubble is one
where the inward-pointing surface tension force balances the
outward-pointing coming from the pressure difference between the
inside and outside regions of the bubble. In this study, the free
parameters of the scalar potential are fixed to
\begin{align*}
  \phi_M = 0.85
  \,, \quad
  \phi_Q = 10
  \,.
\end{align*}

\begin{figure}[thp]
 \centering
  \includegraphics[width=0.6\textwidth]{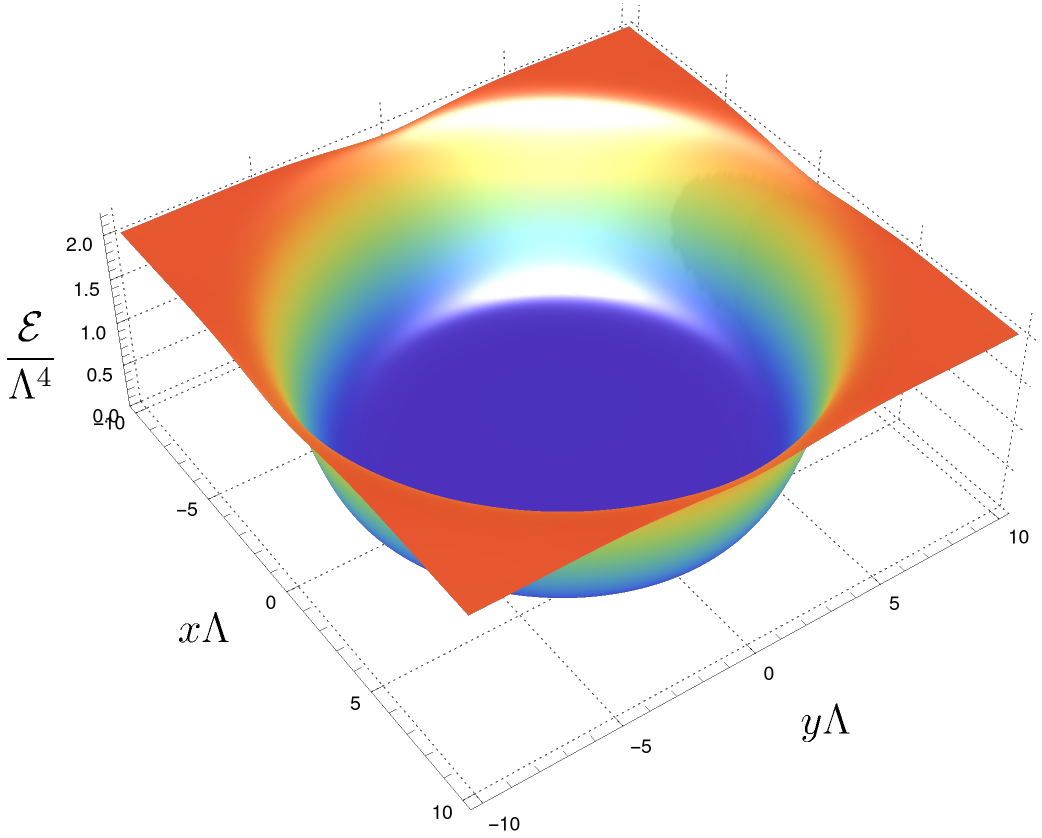} \\
  \includegraphics[width=0.6\textwidth]{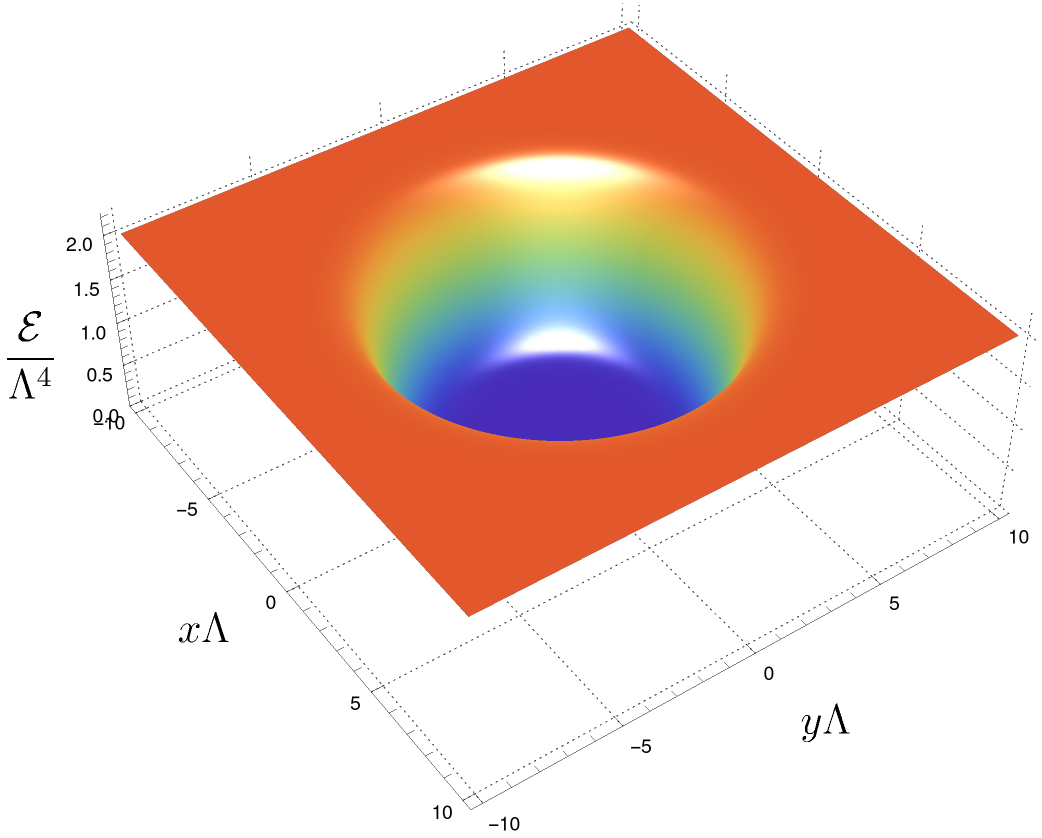} \\
  \includegraphics[width=0.6\textwidth]{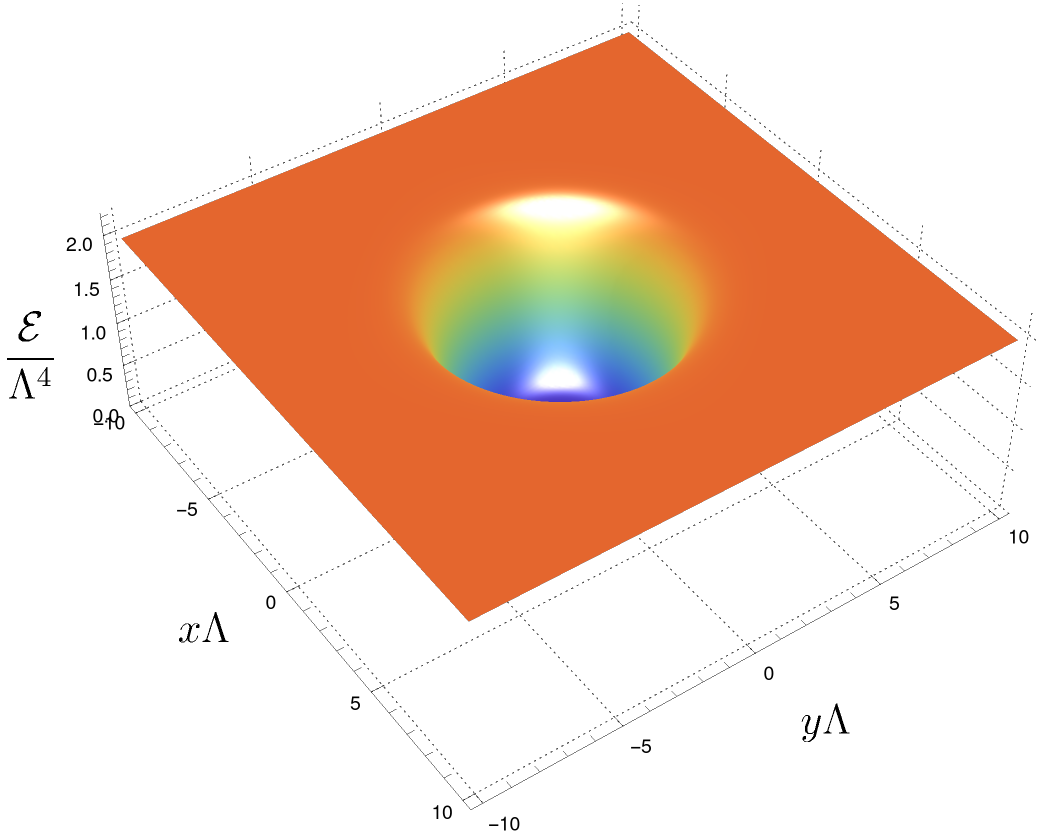}
  \caption[Phase-separated configurations produced with \texttt{Jecco}]
  {
    Phase-separated configurations in a box of size
    $L_x\Lambda=L_y\Lambda=20$ with average energy densities
    $\bar{\mathcal{E}}/\Lambda^4 = 1.0$ (top),
    $\bar{\mathcal{E}}/\Lambda^4=1.6$ (middle) and
    $\bar{\mathcal{E}}/\Lambda^4=1.8$ (bottom).
    \label{Fig:phase_separated}
  }
\end{figure}

In Fig.~\ref{Fig:phase_separated} snapshots of the energy density for
three different cylindrical bubbles constructed and evolved
with~\texttt{Jecco} are shown. To construct them we utilize the
end-state of the spinodal instability and through a series of
manipulations that are described in detail in~\cite{BeaCasGia22} we
build bubbles with different radii. To find the region of criticality
we construct various subcritical and supercritical bubbles and
gradually move towards the radius of criticality. The subcritical
bubbles collapse and the supercritical ones expand. The latter are
important for the GW production scenarios discussed earlier and
in~\cite{BeaCasGia22} we also study their bubble wall velocity, their
profile at late times and the applicability of hydrodynamics in the
description of the phenomenon. A video of an expanding cylindrical
bubble simulated with~\texttt{Jecco} can be found
\href{https://www.youtube.com/watch?v=wFLp0FSeO8Q}{here}.


%% file: sections/final_remarks.tex
\chapter{Final remarks}
\label{chap:final_remarks}

Characteristic formulations of GR are used in a number of cases such
as gravitational waveform modeling, critical collapse and applications
to holography. These formulations are most commonly built upon
Bondi-like gauges.

Despite their extensive use, relatively little attention has been paid
to well-posedness of the resulting PDE problems, which serves as an
obstacle to the construction of rigorous error estimates from
computational work. Motivated by this, we analyzed the EFE in some
popular Bondi-like gauges and demonstrated that the resulting PDE
systems are only weakly hyperbolic. In addition, we showed that this
weak hyperbolicity is caused by the gauge condition~$g^{uA}=0$ common
to all Bondi-like gauges and identified the resulting PDE structure as
a pure gauge effect. To achieve the latter, we had to jump through a
number of technical hoops. We mapped the characteristic free evolution
system to an ADM setup so that the results of~\cite{KhoNov02,HilRic13}
could easily be used. This allowed us to distinguish among the gauge,
constraint, and physical degrees of freedom in the linear, constant
coefficient approximation. Crucially it is known that weakly
hyperbolic pure gauges give rise to weakly hyperbolic formulations. We
were able to show the former in a number of cases.  Specifically, we
have studied three Bondi-like setups: the affine null, the Bondi-Sachs
proper and the double null gauges. All three have the same degenerate
structure rendering the pure gauge subsystem weakly hyperbolic. We
have thus argued that when the EFE are written in a Bondi-like gauge
with at most second derivatives of the metric and there are nontrivial
dynamics in at least two spatial directions, then, due to the weak
hyperbolicity of the pure gauge subsystem, the resulting PDE system is
only WH.

All the hyperbolicity analyses are performed in the linear, frozen
coefficient approximation and we demand that for a system to be
characterized as WH or SH, the definitions of
Sec.~\ref{sec:pde_theory:hyp_degree} are satisfied at each point in
the domain of interest. The latter provides the basis to show
well-posedness for the IVP of variable-coefficient SH systems, as well
as non-linear systems with a SH linearization. Consequently, obtaining
a SH linearization in the frozen coefficient approximation of the
original characteristic systems analyzed here, is the minimum
requirement for the original characteristic system to be SH.

The implication of weak hyperbolicity is that the CIVP and CIBVP of GR
are ill-posed in the natural equivalent of the~$L^2$-norm on these
geometric setups. The obvious approach to circumvent weak
hyperbolicity in characteristic formulations of GR that include up to
second order derivatives of the metric, is to adopt a different
gauge. For applications in CCM this may be necessary, since it is
otherwise not at all clear how a well-posedness result for the
composite PDE problem could be obtained. Building different
characteristic gauges that are strongly hyperbolic could have positive
impact not only in CCM, but also in other dynamical strong gravity
scenarios, as for instance gravitational collapse. So, an interesting
research avenue is to explore different approaches in constructing
strongly hyperbolic characteristic setups, investigate their behavior
in numerical applications and compare their results to known ones from
Bondi-like formulations.

Concerning purely characteristic evolution, symmetric hyperbolic
formulations of GR employing Bondi-like gauges are
known~\cite{Rac13,CabChrTag14,HilValZha19,Rip21}. At first sight this
seems to contradict the claim that any formulation of GR inherits the
pure gauge principal symbol within its own. But these formulations all
promote the curvature to be an evolved variable, so practically they
include higher than second order derivatives of the metric and hence
the results of~\cite{HilRic13} do not apply.  We saw that by taking an
outgoing null derivative of the affine null pure gauge subsystem, we
obtain a strongly hyperbolic PDE. It is thus tempting to revisit the
model of~\cite{HilRic13} to investigate the conjecture that
formulations of GR with evolved curvature can be built that inherit
specific derivatives of the pure gauge subsystem. A deeper
understanding of the relation between the latter and the Bondi-like
formulations analyzed in this thesis could suggest norms in which they
are actually well-posed. Obtaining such a proof would help validate
error estimates for numerical solutions so relevant for applications
in gravitational wave astronomy. Work in this last direction is
ongoing and we reported here some preliminary calculations.

To demonstrate the effect of weak hyperbolicity in practice, we
performed numerical experiments with toy models, as well as in full
GR. In all cases we confirmed that ill-posedness of the continuum PDE
(in the natural equivalent of~$L^2$) for the characteristic problem
serves as an obstruction to convergence of the numerics (in a discrete
approximation to the same norm). For WH toy models that mimic
Bondi-like systems, we found and tested a lopsided norm that is not
equivalent to~$L^2$ in which we recovered well-posedness, but in a
weak form. We found an explicit example for that model where lower
order source terms break this weak well-posedness, as well as a purely
numerically unstable case which highlights that standard numerical
methods are not well developed for WH problems. The tests were
performed with smooth and with noisy given data. For smooth data both
the strongly and weakly hyperbolic systems exhibited good
convergence. But with noisy data only the strongly hyperbolic model
retained this behavior. These findings are compatible with previous
results~\cite{CalHinHus05, CaoHil11}, namely that noisy given data are
essential to reveal weak hyperbolicity in numerical experiments. We
furthermore saw that even with noisy data one might overlook this
behavior if tests are performed in a norm that is not suited to the
particular problem.

Finally, we described in detail~\texttt{Jecco}, a new open source code
that solves the EFE in a characteristic setup. \texttt{Jecco} is a
modular code written in the Julia programming language. At this stage
the code can provide solutions to gravitational setups in
asymptotically AdS spacetimes in five dimensions and with trivial
dynamics along one of the dimensions of the timelike AdS
boundary. These setups are useful to investigate the
out-of-equilibrium dynamics of model strongly coupled plasmas via
holography. We briefly presented examples where we simulated the
evolution of a phase transition for such models. The channels through
which the phase transition can evolve are bubble nucleation or the
spinodal instability. Both could be relevant for primordial GW
production that might be detected by future GW detectors. The PDE
systems currently solved are expected to be only WH since they are
based on Bondi-like gauges. However, through holography we aim to
primarily understand the qualitative behavior of these strongly
coupled systems. To perform rigorous error estimates we would need
alternative SH characteristic formulations adapted to AAdS, which is a
possible research direction. Another possible future task is to enrich
the code with more initial data generating routines, such that we can
simulate e.g. gravitational shockwave collisions that are used to
model heavy ion collisions that take place in terrestrial experiments.


%% file: sections/appendix.tex
\chapter{Appendix for part I}

\section{The divergence theorem}
\label{sec:appA:divergence_thm}

\begin{figure}[h]
  \begin{center}
    \includegraphics[width=0.4\textwidth]{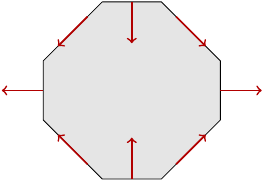}
  \end{center}
  \caption[Divergence theorem: the orientation of the vector normal to
  the boundary]{The orientation of the vector~$N^\mu$ that is normal
    to the boundary~$\p M$: for spacelike~$\p M$, $N^\mu$ points
    inwards, for timelike~$\p M$ it points outwards and for
    null~$\p M$, $N^\mu$ consists of a timelike part that points
    inwards and a spacelike part that points outwards.}
  \label{Fig:div_thm_normals}
\end{figure}

The version of the divergence theorem we apply is based on Sec. 5.1
of~\cite{Nat20}. In brief, given a spacetime~$M$ with boundary~$\p M$
and a spacetime vector~$X^\mu$ the divergence theorem reads
\begin{align*}
  \int_M \nabla_\mu X^\mu \, dV =
  \int_{\p M} X^\mu N_\mu \, d\sigma
  \,,
\end{align*}
where~$dV$ and~$d \sigma$ the volume elements of~$M$ and~$\p M$
respectively and~$N^\mu$ the spacetime vector normal to~$\p M$, with
its orientation illustrated in Fig.~\ref{Fig:div_thm_normals}.

\section{The Gr\"onwall inequality}
\label{sec:appA:Gronwall}

We use the integral version of the \textit{Gr\"onwall inequality}
which states:

Let $I$ denote an interval of the real line of the form $[a, \infty)$,
$[a, b]$ or $[a,b)$, with $a < b$. Let furthermore $\alpha, \beta$ and
$u$ be real-valued functions defined on $I$ and assume that $\beta$
and $u$ are continuous and that the negative part of $\alpha$ is
integrable on every closed and bounded subinterval of $I$. Then
\begin{enumerate}
\item If~$\beta$ is non-negative and~$u$ satisfies the integral inequality
  \begin{align*}
    u(t) \leq \alpha(t) + \int_a^t \beta(s) u(s)ds
    \,, \quad
    \forall t \in I\,,
  \end{align*}
  then
  \begin{align*}
    u(t) \leq \alpha(t) + \int_a^t a(s) \beta(s)
    \textrm{exp}\left(
    \int_s^t \beta(r)dr
    \right)
    ds
    \,,
  \end{align*}
  with~$t \in I$.
\item If in addition~$\alpha(t)$ is non-decreasing, then
  \begin{align*}
    u(t) \leq \alpha(t) 
    \,
    \textrm{exp}\left(
    \int_a^t \beta(s)ds
    \right)
    \,.
  \end{align*}
\end{enumerate}

\section{A symmetric hyperbolic affine null PDE system }
\label{sec:appA:Rip21}

The symmetric hyperbolic characteristic system of~\cite{Rip21} is
reviewed.

\subsection*{Setup and formalism}

The metric ansatz is
\begin{align}
  ds^2 =
  V du + 2 du dr
  -h_{ab} (d\theta^a + W^a du)(d\theta^b + W^b du)
  \,,
  \label{eqn:aff_null_metric}
\end{align}
where~$r$ is an affine parameter for outgoing null geodesics. The
chosen tetrad basis is~$l^\mu,n^\mu,m^\mu,\bar{m}^\mu$, with
\begin{subequations}
  \label{eqn:tetrad_dev_operators}
  \begin{align}
    l^\mu \p_\mu
    & = \p_r
      \equiv D
      \,, \label{eqn:D_op}
    \\
    n^\mu \p_\mu
    &= \p_u + P \p_r + R^a \p_a
      \equiv \Delta
      \,, \label{eqn:Delta_op}
    \\
    m^\mu \p_\mu
    & = Q \p_r + S^a \p_a
      \equiv \delta
      \,,
      \label{eqn:delta_op}
  \end{align}
\end{subequations}
where~$Q$ and~$S^a$ are complex valued functions. Henceforth, the
coordinates on the two-sphere may be labeled
as~$\theta^a = (\theta,\phi)$. The metric and the tetrad are related
via the identities
\begin{subequations}
  \label{eqn:metric_to_tetrad}
  \begin{align}
    g^{\mu\nu}
    & = l^\mu n^\nu + n^\mu l^\nu - m^\mu \bar{m}^\nu - \bar{m}^\mu m^\nu
      \,, \label{eqn:guu_to_tetrad}
    \\
    g_{\mu\nu}
    & = l_\mu n_\nu + n_\mu l_\nu - m_\mu \bar{m}_\nu - \bar{m}_\mu m_\nu
      \,, \label{eqn:gdd_to_tetrad}
  \end{align}
\end{subequations}
which lead to
\begin{subequations}
  \label{eqn:metric_to_tetrad_components}
  \begin{align}
    g^{rr}
    & = -V
      = 2(P - Q \bar{Q})
      \,, \label{eqn:guu_rr_to_tetrad}
    \\
    g^{ra}
    & = -W^a
      = R^a - \bar{S}^a Q - S^a \bar{Q}
      \,, \label{eqn:guu_ra_to_tetrad}
    \\
    g^{ab}
    & = -h^{ab}
      =  - S^a \bar{S}^b -  S^b \bar{S}^a
      \,. \label{eqn:guu_ab_to_tetrad}
  \end{align}
\end{subequations}
The tetrad calculus yields~\cite{NewPen62}
\begin{subequations}
  \label{eqn:tetrad_calculus}
  \begin{align}
    & l_\mu l^\mu = n_\mu n^\mu = m_\mu m^\mu = \bar{m}_\mu \bar{m}^\mu = 0 
      \,, \label{eqn:tetrad_null_condition}
    \\
    & l_\mu n^\mu = -m_\mu \bar{m}^\mu = 1
      \,, \label{eqn:tetrad_unity_condition}
    \\
    & l_\mu m^\mu = l_\mu \bar{m}^\mu = n_\mu m^\mu = n_\mu \bar{m}^\mu = 0
      \,. \label{eqn:tetrad_orthogonal_condition}
  \end{align}
\end{subequations}
In addition to the derivative operators~$D,\Delta,\delta,\bar{\delta}$
defined in~\eqref{eqn:tetrad_dev_operators}, the operators introduced
by Geroch-Held-Penrose (GHP)~\cite{GerHelPen73} may also be used for
convenience:
\begin{subequations}
  \label{eqn:GHP_operators}
  \begin{align}
    &  \th f \equiv (D - p \epsilon - q \bar{\epsilon}) f 
      \,, \quad
      \th^\prime f \equiv (\Delta - p \gamma - q \bar{\gamma}) f
      \,, \label{eqn:th_GHP_op}
      \\
    &  \eth f \equiv (\delta - p \beta - q \bar{\alpha}) f 
      \,, \quad
      \eth^\prime f \equiv (\bar{\delta} - p \alpha - q \bar{\beta}) f
      \,, \label{eqn:eth_GHP_op}
  \end{align}
\end{subequations}
where~$f$ is a scalar of weight~$(p,q)$.

\subsection*{The PDE system}

The system of~\cite{Rip21} consists of the evolution equations for the
metric components, the Ricci rotation coefficients and the Weyl
scalars. The first set is obtained by the tetrad commutation relations
and reads
\begin{subequations}
  \label{eqn:tetrad_eoms}
  \begin{align}
    D P + \gamma + \bar{\gamma} - \bar{\tau} Q - \tau \bar{Q}
    &= 0
      \,, \label{eqn:P_eom}
    \\
    D R^a - \bar{\tau} S^a - \tau \bar{S}^a
    & = 0
      \,, \label{eqn:Ra_eom}
    \\
    D Q + \bar{\alpha} + \beta - \bar{\rho} Q - \rho \bar{Q}
    & = 0
      \,, \label{eqn:Q_eom}
    \\
    D S^a - \bar{\rho} S^a - \rho \bar{S}^a
    & = 0
      \,. \label{eqn:Sa_eom}
  \end{align}
\end{subequations}
The second set comes from the Ricci identities
\begin{subequations}
  \label{eqn:Ricci_eoms}
  \begin{align}
    D \rho - \rho^2 + \sigma \bar{\sigma}
    & = 0
      \,, \label{eqn:rho_eom}
    \\
    D \sigma - \rho \sigma - \bar{\rho} \sigma - \Psi_0
    & = 0
      \,, \label{eqn:sigma_eom}
    \\
    D \tau - \rho \tau - \sigma \bar{\tau} - \Psi_1
    & = 0
      \,, \label{eqn:tau_eom}
    \\
    D \alpha - \rho \alpha - \bar{\sigma} \beta
    & = 0
      \,, \label{eqn:alpha_eom}
    \\
    D \beta - \sigma \alpha - \bar{\rho} \beta - \Psi_1
    & = 0
      \,, \label{eqn:beta_eom}
    \\
    D \gamma - \tau \alpha - \bar{\tau} \beta - \Psi_2
    & = 0
      \,, \label{eqn:gamma_eom}
    \\
    D \lambda - \rho \lambda - \bar{\sigma} \mu
    & = 0
      \,, \label{eqn:lambda_eom}
    \\
    D \mu - \bar{\rho} \mu - \sigma \lambda - \Psi_2
    & = 0
      \,, \label{eqn:mu_eom}
    \\
    D \nu - \bar{\tau} \mu - \tau \lambda - \Psi_3
    & = 0
      \,.
  \end{align}
\end{subequations}
Notice that these two sets involve equations that in the principal
part are advections purely along~$l^\mu$. This is not true for the
Weyl scalars though. Using the Bianchi identities one can obtain
\begin{subequations}
  \label{eqn:Weyl_eoms}
  \begin{align}
    (\th^\prime + \mu) \Psi_0
    - (\eth - 4 \tau) \Psi_1
    - 3 \sigma \Psi_2
    & = 0
      \,, \label{eqn:Psi_0_eom}
    \\
    (D + \th^\prime + 2 \mu - 4 \rho)\Psi_1
    - (\eth - 3 \tau) \Psi_2
    & - \nonumber
    \\
    (\eth^\prime + \nu) \Psi_0
    - 2 \sigma \Psi_3
    & = 0
      \,, \label{eqn:Psi_1_eom}
    \\
    (D + \th^\prime +3 \mu -3 \rho) \Psi_2
    - (\eth -2 \tau) \Psi_3
    & - \nonumber
    \\
    (\eth^\prime + 2 \nu) \Psi_1
    + \lambda \Psi_0 - \sigma \Psi_4
    & = 0
      \,, \label{eqn:Psi_2_eom}
    \\
    (D + \th^\prime + 4 \mu - 2 \rho) \Psi_3
    - (\eth - \tau) \Psi_4
    & - \nonumber
    \\
    (\eth^\prime + 3 \nu) \Psi_2
    + 2 \lambda \Psi_1
    & = 0
      \,, \label{eqn:Psi_3_eom}
    \\
    (D - \rho) \Psi_4
    - \eth^\prime \Psi_3
    + 3 \lambda \Psi_2
    & = 0
      \,. \label{eqn:Psi_4_eom}
  \end{align}
\end{subequations}
The spin and boost weights~$(p,q)$ of the Weyl scalars
are~$(4,0)$, $(2,0)$, $(0,0)$, $(-2,0)$, $(-4,0)$
for $\Psi_0$, $\Psi_1$, $\Psi_2$, $\Psi_3$, $\Psi_4$
respectively~\cite{NewPen62}.

The unknowns of the system can be collected in the state
vector~$\mathbf{v} \equiv
(\mathbf{\Psi},\mathbf{\Gamma},\mathbf{g})^T$ with
\begin{equation}
  \label{eqn:original_state_vector}
  \begin{aligned}
    \mathbf{\Psi}
    & \equiv (\Psi_0, \Psi_1, \Psi_2, \Psi_3, \Psi_4)^T
    \,,
    \\
    \mathbf{\Gamma}
    & \equiv
    (\rho, \sigma, \tau, \alpha, \beta, \gamma, \lambda, \mu, \nu)
    \,,
    \\
    \mathbf{g}
    & \equiv (P, R^\theta, R^\phi, Q, S^\theta, S^\phi)
    \,.
  \end{aligned}
\end{equation}
The source terms involve coupling between the above terms and their
complex conjugates.
%
%
The principal part~$\mathbfcal{A}^\mu \p_\mu \mathbf{v} \simeq 0$ of
the system~\eqref{eqn:Weyl_eoms}, \eqref{eqn:Ricci_eoms},
\eqref{eqn:tetrad_eoms} has the structure
\begin{align}
  \mathbfcal{A}^\mu =
    \begin{pmatrix}
    \mathcal{D}^\mu_{\mathbf{\Psi}}  & 0  & 0 \\
    0 & \mathcal{D}^\mu_{\mathbf{\Gamma}}  & 0 \\
    0  & 0  & \mathcal{D}^\mu_{\mathbf{g}} 
  \end{pmatrix}
              \,,
              \label{eqn:princ_part_full}
\end{align}
with
\begin{subequations}
  \begin{align}
    \mathcal{D}^\mu_{\mathbf{\Psi}}
    & \equiv \begin{pmatrix}
      n^\mu  & -m^\mu  & 0 & 0 & 0 \\
      -\bar{m}^\mu & l^\mu + n^\mu  & - m^\mu & 0 & 0  \\
      0 & -\bar{m}^\mu & l^\mu + n^\mu & -m^\mu & 0 \\
      0 & 0 & -\bar{m}^\mu & l^\mu + n^\mu & - m^\mu \\
      0 & 0 & 0 & -\bar{m}^\mu & l^\mu
    \end{pmatrix}
                                 \,,
                                 \label{eqn:princ_part_Weyl}
    \\
    \mathcal{D}^\mu_{\mathbf{\Gamma}}
    & \equiv id_9 \times l^\mu
      \,,
      \label{eqn:prin_part_Ricci}
    \\
    \mathcal{D}^\mu_{\mathbf{g}}
    & \equiv id_6\times l^\mu
      \,,
      \label{eqn:prin_part_tetrad}
  \end{align}
\end{subequations}
where~$id_n$ denotes the~$n \times n$ identity matrix. The
matrix~$\mathbfcal{A}^\mu$ is Hermitian and positive definite with
respect to the timelike direction~$t^\mu \equiv l^\mu + n^\mu$ i.e.
\begin{align}
  \mathbfcal{A}^t \equiv
  \mathbfcal{A}^\mu t_\mu = \textrm{diag}(1,2,2,2,d_{16})
  \,,
  \label{eqn:At_matrix}
\end{align}
where~$d_n \equiv \textrm{diag}(id_n)$. This system is symmetric
hyperbolic with respect to the timelike
direction~$t^\mu$~\cite{Rip21}.

To understand the propagation speed of the variables let us consider
the spacelike direction~$\rho^\mu \equiv l^\mu - n^\mu$. With respect
to this direction the principal part reads
\begin{align}
  \mathbfcal{A}^\rho \equiv
  \mathbfcal{A}^\mu \rho_\mu = \textrm{diag}(1,0,0,0,-d_{16})
  \,.
  \label{eqn:Arho_matrix}
\end{align}
This yields that~$\Psi_0$ is ingoing at the speed of
light,~$\Psi_1, \Psi_2, \Psi_3$ have vanishing propagation speed
(static) and~$\Psi_4, \mathbf{\Gamma}, \mathbf{g}$ are outgoing at the
speed of light. For completeness
\begin{subequations}
  \begin{align}
    \mathbfcal{A}^l
    &\equiv
      \mathbfcal{A}^\mu l_\mu = \textrm{diag}(d_4, 0 \times d_{16})
      \,,
      \label{eqn:Al_matrix}
    \\
    \mathbfcal{A}^n
    &\equiv
      \mathbfcal{A}^\mu n_\mu = \textrm{diag}(0, d_{19})
      \,,
      \label{eqn:An_matrix}
      \end{align}
\end{subequations}
where the tetrad calculus~\eqref{eqn:tetrad_calculus} has been
applied.

\chapter{Appendix for part II}

\section{Radial equations}
\label{sec:full-eqs}

For completeness, here we list the radial equations obtained from the
metric ansatz~\eqref{eq:metric}.  It is convenient to introduce the
following operators to make the expressions more compact
\begin{equation}
  \begin{aligned}
    f^{\prime}
    & \equiv \p_r f
    \,, \\
    \dot{f}
    & \equiv
    \left( \p_t + \frac{A}{2} \p_r \right) f
    \,,
    \\
    \tilde{f}
    & \equiv
    \left( \p_x - F_x \p_r \right) f
    \,, \\
    \hat{f}
    & \equiv
    \left( \p_y - F_y \p_r \right)
    f \,,
    \\
    \bar{f}
    & \equiv
    \left( \p^2_x - 2 F_x \p_r \p_x + F_x^2 \p^2_r \right) f
    \,,
    \\
    f^{\star}
    & \equiv
    \left( \p^2_y - 2 F_y \p_r \p_y + F_y^2 \p^2_r \right) f
    \,, \\
    f^{\times}
    & \equiv
    \left( \p_x \p_y - F_x \p_r \p_y - F_y \p_r \p_x + F_x F_y \p^2_r
    \right) f
    \,.
  \end{aligned}
  \label{ec:new_der}
\end{equation}

As shown in Table~\ref{tab:system}, by combining Einstein's
equations~\eqref{eq:eom} in a particular way we obtain a nested system
of radial effective ODEs where one can sequentially solve for the
different variables. For this particular case, some of these intrinsic
equations are
coupled.

\begin{align}
  6S^{\prime\prime}
  + S \left(\cosh^2(G)
  \left(B_1^{\prime}\right)^2
  +3 \left(B_2^{\prime}\right)^2
  +\left(G^{\prime}\right)^2
  +\left(\phi^{\prime}\right)^2\right)
  =0
  \,,
  \label{ec:nested_first}
\end{align}

\begin{align}
  &
    2 e^{B_1} S^2 F_x''
    + e^{B_1}
    \left(S^2
    \left(-2
    \left(\cosh ^2(G)
    \left(\tilde{B_1}'-B_1' F_x'\right)
    +B_2' \left(3 \tilde{B_2}-F_x'\right)
    +
    \right. \right. \right.
    \nonumber
  \\
  &
    \left. \left. \left.
    \tilde{G} \left(B_1' \sinh
    (2G)
    + G'\right)
    +\tilde{B_2}'
    + 4 \tilde{\phi } \phi'\right)-2
    \tilde{B_1} B_1' \cosh ^2(G)\right)
    + S \left(-6 \tilde{S}
    \left(B_1' \cosh^2(G)+B_2'\right)
    -
    \right. \right.
    \nonumber
  \\
  &
    \left. \left.
    8 \tilde{S}'+2 S'
    F_x'\right)
    +8 \tilde{S} S'\right)
    +S^2 \left(-2
    G'\left(\hat{B_1}+F_y'\right)
    +\sinh (2 G)
    \left(\hat{B_1}'-B_1'\left(\hat{B_1}+F_y'\right)\right)+
    \right.
    \nonumber
  \\
  &
    \left.
    2 \hat{G}
    B_1' \cosh (2 G)+2 \hat{G}'\right) +3 \hat{S} S \left(B_1' \sinh
    (2 G)+2 G'\right)=0
    \,,
\end{align}

\begin{align}
  &
    2 S^2 F_y''
    + e^{B_1}
    \left(
    S^2 \left(
    2 \left(G'
    \left(\tilde{B_1}-F_x'\right)
    +\tilde{G}'\right)-\sinh (2 G)
    \left(B_1'\left(\tilde{B_1}
    -F_x'\right)+\tilde{B_1}'\right)
    \right. \right.
    -
    \nonumber
  \\
  &
    \left. \left. 
    2 \tilde{G} B_1' \cosh (2 G)\right)
    -3 S \tilde{S} \left(B_1' \sinh
    (2 G)-2 G'\right)
    \right)
    +2 S^2\left(\cosh ^2(G)
    \left(\hat{B_1}'-B_1' F_y'\right)
    \right.
    +
    \nonumber
  \\
  &
    \left.
    B_2' \left(F_y'-3
    \hat{B_2}\right)
    +\hat{G} \left(B_1' \sinh (2
    G)-G'\right)-\hat{B_1} B_1' \cosh^2(G)-\hat{B_2}'
    -4 \hat{\phi }
    \phi '\right)
    +
    \nonumber
  \\
  &
    S \left(6 \hat{S} \left(B_1' \cosh ^2(G)-B_2'\right)
    +2 S' F_y'-8 \hat{S}'\right)+8 \hat{S}S'=0
    \,,
\end{align}

\begin{align}
  &
    12 e^{B_1} S^3 \dot{S}'
    + e^{B_1+B_2}
    \left(S^2
    \left(2 \cosh (G)
    \left(-\hat{G}
    \left(\tilde{B_1}+\tilde{B_2}-F_x'\right)+\tilde{G}
    \left(\hat{B_1}-\hat{B_2}+F_y'\right)
    \right. \right. \right.
    +
    \nonumber
  \\
  &
    \left. \left. \left.
    G'
    \left(\tilde{F_y}+\hat{F_x}\right) -2G^{\times}\right)+2 \sinh
    (G) \left(B_2' \left(\tilde{F_y}+\hat{F_x}\right)+F_y'
    \left(\tilde{B_2}-F_x'\right)
    +\hat{B_2}
    \left(F_x'-4
    \tilde{B_2}\right)
    \right. \right. \right.
    -
    \nonumber
  \\
  &
    \left. \left. \left.
    2{B_2}^{\times}+\tilde{F_y}' -2 \hat{G}
    \tilde{G}-4 \hat{\phi } \tilde{\phi
    }+\hat{F_x}'\right)\right)+S \left(2 \sinh (G) \left(\hat{S}
    \left(F_x'-4 \tilde{B_2}\right)+\tilde{S} \left(F_y'-4
    \hat{B_2}\right)
    \right. \right. \right.
    +
    \nonumber
  \\
  &
    \left. \left. \left.
    4 S'
    \left(\tilde{F_y}+\hat{F_x}\right)-8S^{\times}\right)-8 \cosh
    (G) \left(\hat{S} \tilde{G}+\hat{G} \tilde{S}\right)\right)+8
    \hat{S} \tilde{S} \sinh (G)\right)
    +
    \nonumber
  \\
  &
    e^{2 B_1+B_2} \left(S^2 \left(2
    \sinh (G) \left(\tilde{G} \left(2
    \tilde{B_1}+\tilde{B_2}-F_x'\right)-G'
    \tilde{F_x}+\bar{G}\right)
    +
    \right. \right.
    \nonumber
  \\
  &
    \left. \left.
    \cosh (G) \left(2
    \left(-\left(B_1'+B_2'\right) \tilde{F_x}
    +\bar{B_1}+\bar{B_2}-\tilde{F_x}'+\tilde{G}^2+2 \tilde{\phi
    }^2\right)-2 \left(\tilde{B_1}+\tilde{B_2}\right) F_x'
    +
    \right. \right. \right.
    \nonumber
  \\
  &
    \left. \left. \left.
    2
    \left(\tilde{B_1}{}^2+\tilde{B_2} \tilde{B_1}+2
    \tilde{B_2}{}^2\right)+\left(F_x'\right){}^2\right)\right)
    +S \left(2 \cosh (G) \left(\tilde{S} \left(4
    \left(\tilde{B_1}+\tilde{B_2}\right)-F_x'\right)+4
    \left(\bar{S}-S' \tilde{F_x}\right)\right)
    +
    \right. \right.
    \nonumber
  \\
  &
    \left. \left.
    8 \tilde{G}
    \tilde{S} \sinh (G)\right) -4 \tilde{S}^2 \cosh
    (G)\right)+e^{B_2} \left(S^2 \left(2 \sinh (G) \left(\hat{G}
    \left(-2 \hat{B_1}+\hat{B_2}-F_y'\right)-\hat{F_y}
    G'+G^{\star}\right)
    +
    \right. \right.
    \nonumber
  \\
  &
    \left. \left.
    \cosh (G) \left(2
    \left(\left(B_1'-B_2'\right)
    \hat{F_y}-B_1^{\star}+B_2^{\star}-\hat{F_y}'+\hat{G}^2+2
    \hat{\phi }^2\right)+2 \left(\hat{B_1}-\hat{B_2}\right) F_y'
    +
    \right. \right. \right.
    \nonumber
  \\
  &
    \left. \left. \left.
    2 \left(\hat{B_1}{}^2-\hat{B_2} \hat{B_1}+2
    \hat{B_2}{}^2\right)+\left(F_y'\right){}^2\right)\right)+S
    \left(8 \hat{G} \hat{S} \sinh (G)-2 \cosh (G) \left(\hat{S}
    \left(4 \hat{B_1}-4 \hat{B_2}+F_y'\right)
    +
    \right. \right. \right.
    \nonumber
  \\
  &
    \left. \left. \left.
    4 \hat{F_y} S'-4
    S^{\star}\right)\right)-4 \hat{S}^2 \cosh (G)\right)+e^{B_1}
    \left(8 S^4 V(\phi )+24 \dot{S} S^2 S'\right) = 0
    \,,
\end{align}

\begin{align}
  &
    12 e^{B_1} S^4 \dot{B'_1}+e^{B_1+B_2} \left(6 S^2 \text{sech}(G)
    \left(\hat{G} \left(F_x'-\tilde{B_2}\right)+\tilde{G}
      \left(\hat{B_2}-F_y'\right)+G'
    \left(\tilde{F_y}-\hat{F_x}\right)\right)
    +
    \right.
    \nonumber
  \\
  &
    \left.
    6 S \text{sech}(G)
    \left(\hat{S} \tilde{G}-\hat{G} \tilde{S}\right)\right)+e^{2
    B_1+B_2} \left(-3 S^2 \text{sech}(G) \left(-2 B_2' \tilde{F_x}-2
    \tilde{B_2} F_x'+4 \tilde{B_2}{}^2+2 \bar{B_2}-2 \tilde{F_x}'
    +
    \right. \right.
    \nonumber
  \\
  &
    \left. \left.
    4  \tilde{\phi }^2 +\left(F_x'\right){}^2\right)-6 S \text{sech}(G)
    \left(\tilde{S} \left(\tilde{B_2}+2 F_x'\right)-S'
      \tilde{F_x}+\bar{S}\right)+12 \tilde{S}^2
    \text{sech}(G)\right)
    +
    \nonumber
  \\
  &
    e^{B_2} \left(3 S^2 \text{sech}(G) \left(-2
      B_2' \hat{F_y}-2 \hat{B_2} F_y'+4 \hat{B_2}{}^2+2
      B_2^{\star}+\left(F_y'\right){}^2-2 \hat{F_y}'+4 \hat{\phi
    }^2\right)
    +
    \right.
    \nonumber
  \\
  &
    \left.
    6 S \text{sech}(G) \left(\hat{S} \left(\hat{B_2}+2
        F_y'\right)-\hat{F_y} S'+S^{\star}\right)-12 \hat{S}^2
    \text{sech}(G)\right)+e^{B_1} \left(12 S^4 \tanh (G)
    \left(\dot{B_1} G'+\dot{G} B_1'\right)
    +
    \right.
    \nonumber
  \\
  &
    \left.
    18 S^3 \left(\dot{B_1}
      S'+\dot{S} B_1'\right)\right)=0
 \end{align}

 \begin{align}
   &
     12 e^{B_1} S^4 \dot{G}'+e^{B_1+B_2} \left(6 S^2 \cosh (G)
     \left(B_1' \left(\hat{F_x}-\tilde{F_y}\right)-B_2'
       \left(\tilde{F_y}+\hat{F_x}\right)+\left(\hat{B_1}-F_y'\right)
     \left(\tilde{B_2}-F_x'\right)
     -
     \right. \right.
     \nonumber
   \\
   &
     \left. \left.
     \hat{B_2} \left(\tilde{B_1}-4
         \tilde{B_2}+F_x'\right)+\tilde{B_1}
       F_y'+2B_2^{\times}-\tilde{F_y}'+4 \hat{\phi } \tilde{\phi
       }-\hat{F_x}'\right)+6 S \cosh (G) \left(\hat{S}
     \left(-\tilde{B_1}+\tilde{B_2}+2 F_x'\right)
     +
     \right. \right.
     \nonumber
   \\
   &
     \left. \left.
     \tilde{S}
       \left(\hat{B_1}+\hat{B_2}+2 F_y'\right)-S'
       \left(\tilde{F_y}+\hat{F_x}\right)+2 S^{\times}\right)-24
     \hat{S} \tilde{S} \cosh (G)\right)
     +
     \nonumber
   \\
   &
     e^{2 B_1+B_2} \left(-3 S^2
     \sinh (G) \left(-2 B_2' \tilde{F_x}-2 \tilde{B_2} F_x'+4
       \tilde{B_2}{}^2+2 \bar{B_2}-2 \tilde{F_x}'+4 \tilde{\phi
     }^2+\left(F_x'\right){}^2\right)
     -
     \right.
     \nonumber
   \\
   &
     \left.
     6 S \sinh (G) \left(\tilde{S}
       \left(\tilde{B_2}+2 F_x'\right) -S'
       \tilde{F_x}+\bar{S}\right)+12 \tilde{S}^2 \sinh
     (G)\right)+e^{B_2} \left(-3 S^2 \sinh (G) \left(-2 B_2'
     \hat{F_y}
     -
     \right. \right.
     \nonumber
   \\
   &
     \left. \left.
     2 \hat{B_2} F_y'+4 \hat{B_2}{}^2+2
       B_2^{\star}+\left(F_y'\right){}^2 -2 \hat{F_y}'+4 \hat{\phi
       }^2\right)-6 S \sinh (G) \left(\hat{S} \left(\hat{B_2}+2
     F_y'\right)-\hat{F_y} S'+S^{\star}\right)
     +
     \right.
     \nonumber
   \\
   &
     \left.
     12 \hat{S}^2 \sinh
     (G)\right)+e^{B_1} \left(18 S^3 \left(\dot{S} G'+\dot{G}
       S'\right)-6 \dot{B_1} S^4 B_1' \sinh (2 G)\right)=0
 \end{align}

\begin{align}
  &
    12 e^{B_1} S^4 \dot{B_2}'+e^{B_1+B_2} \left(S^2 \left(2 \cosh (G)
      \left(\hat{G} \left(\tilde{B_1}-2
          \tilde{B_2}-F_x'\right)-\tilde{G} \left(\hat{B_1}+2
    \hat{B_2}+F_y'\right)
    -
    \right. \right. \right.
    \nonumber
  \\
  &
    \left. \left. \left.
    G' \left(\tilde{F_y}+\hat{F_x}\right)
    +2 G^{\times}\right)+2 \sinh (G) \left(2 B_2'
    \left(\tilde{F_y}+\hat{F_x}\right)+F_y' \left(2
    \tilde{B_2}+F_x'\right)+2 \hat{B_2}
    \left(F_x'-\tilde{B_2}\right)
    -
    \right. \right. \right.
    \nonumber
  \\
  &
    \left. \left. \left.
    4 B_2^{\times}-\tilde{F_y}'+2
    \hat{G} \tilde{G} +4 \hat{\phi } \tilde{\phi
    }-\hat{F_x}'\right)\right)+S \left(2 \sinh (G) \left(2
    \left(\hat{S} \left(F_x'-\tilde{B_2}\right)+\tilde{S}
    \left(F_y'-\hat{B_2}\right)+S^{\times}\right)
    -
    \right. \right. \right.
    \nonumber
  \\
  &
    \left. \left. \left.
    S'
    \left(\tilde{F_y}+\hat{F_x}\right)\right) +2 \cosh
    (G)\left(\hat{S} \tilde{G}+\hat{G} \tilde{S}\right)\right)-8
    \hat{S} \tilde{S} \sinh (G)\right)
    +
    \nonumber
  \\
  &
    e^{2 B_1+B_2} \left(S^2 \left(2
    \sinh (G) \left(\tilde{G} \left(-2 \tilde{B_1}+2
    \tilde{B_2}+F_x'\right) +G' \tilde{F_x}-\bar{G}\right)
    -
    \right. \right.
    \nonumber
  \\
  &
    \left. \left.
    \cosh
    (G) \left(2 \left(-\left(B_1'-2 B_2'\right)
    \tilde{F_x}+\bar{B_1}-2 \bar{B_2}-\tilde{F_x}'+\tilde{G}^2+2
    \tilde{\phi }^2\right)-2 \left(\tilde{B_1}-2
    \tilde{B_2}\right) F_x'
    +
    \right. \right. \right.
    \nonumber
  \\
  &
    \left. \left. \left.
    2 \left(\tilde{B_1}{}^2-2
    \tilde{B_2}
    \tilde{B_1}-\tilde{B_2}{}^2\right)+\left(F_x'\right){}^2\right)\right)+S
    \left(-2 \cosh (G) \left(\tilde{S} \left(\tilde{B_1}-2
    \tilde{B_2}+2 F_x'\right)-S' \tilde{F_x}+\bar{S}\right)
    -
    \right. \right.
    \nonumber
  \\
  &
    \left. \left.
    2 \tilde{G} \tilde{S} \sinh (G)\right)+4 \tilde{S}^2 \cosh
    (G)\right)+e^{B_2} \left(S^2 \left(2 \sinh (G) \left(\hat{G}
    \left(2 \left(\hat{B_1}+\hat{B_2}\right)+F_y'\right)+\hat{F_y}
    G'-G^{\star}\right)
    -
    \right. \right.
    \nonumber
  \\
  &
    \left. \left.
    \cosh (G) \left(2 \left(\left(B_1'+2
    B_2'\right) \hat{F_y}-B_1^{\star}-2
    B_2^{\star}-\hat{F_y}'+\hat{G}^2+2 \hat{\phi }^2\right)+2
    \left(\hat{B_1}+2 \hat{B_2}\right) F_y'
    +
    \right. \right. \right.
    \nonumber
  \\
  &
    \left. \left. \left.
    2 \left(\hat{B_1}{}^2+2 \hat{B_2}
    \hat{B_1}-\hat{B_2}{}^2\right)+\left(F_y'\right){}^2\right)\right)+S
    \left(2 \cosh (G) \left(\hat{S} \left(\hat{B_1}+2 \hat{B_2}-2
    F_y'\right)+\hat{F_y} S'-S^{\star}\right)
    -
    \right. \right.
    \nonumber
  \\
  &
    \left. \left.
    2 \hat{G} \hat{S}
    \sinh (G)\right)+4 \hat{S}^2 \cosh (G)\right)+18 e^{B_1} S^3
    \left(\dot{B_2} S'+\dot{S} B_2'\right)=0
\end{align}

\begin{align}
  &
    8 e^{B_1} S^3 \dot{\phi}'+e^{B_1+B_2} \left(S \left(4 \sinh (G)
    \left(\hat{\phi } \left(F_x'-\tilde{B_2}\right)+\tilde{\phi }
    \left(F_y'-\hat{B_2}\right)+\phi '
    \left(\tilde{F_y}+\hat{F_x}\right)-2 \phi ^{\times}\right)
    -
    \right. \right.
    \nonumber
  \\
  &
    \left. \left.
    4 \cosh (G) \left(\hat{\phi } \tilde{G}+\hat{G} \tilde{\phi}\right)\right)
    -4 \sinh (G) \left(\hat{\phi }
    \tilde{S}+\hat{S} \tilde{\phi }\right)\right)+e^{2 B_1+B_2}
    \left(4 S \left(\cosh (G) \left(\tilde{\phi }
    \left(\tilde{B_1}+\tilde{B_2}-F_x'\right)
    -
    \right. \right. \right.
    \nonumber
  \\
  &
    \left. \left. \left.
    \phi '
    \tilde{F_x}+\bar{\phi }\right)+\tilde{G} \tilde{\phi } \sinh
    (G)\right)+4 \tilde{S} \tilde{\phi } \cosh (G)\right)+e^{B_2}
    \left(S \left(4 \hat{G} \hat{\phi } \sinh (G)-4 \cosh (G)
    \left(\hat{\phi }
    \left(\hat{B_1}
    -
    \right. \right. \right. \right. 
    \nonumber
  \\
  &
    \left. \left. \left. \left.
    \hat{B_2}+F_y'\right)
    +
    \hat{F_y} \phi
    '-\phi^{\star}\right)\right)+4 \hat{S} \hat{\phi } \cosh
    (G)\right)+e^{B_1} \left(12 S^2 \left(\dot{\phi } S'+\dot{S} \phi
    '\right) -4 S^3 V'(\phi )\right)=0
\end{align}

\begin{align}
  &
    6 e^{B_1} S^4 A''+e^{B_1+B_2} \left(S^2 \left(6 \cosh (G)
    \left(\left(\hat{B_2}-\hat{B_1}\right) \tilde{G}+\hat{G}
    \left(\tilde{B_1}+\tilde{B_2}\right)-G'
    \left(\tilde{F_y}+\hat{F_x}\right)+2 G^{\times}\right)
    +
    \right. \right.
    \nonumber
  \\
  &
    \left. \left.
    6 \sinh (G) \left(-B_2' \left(\tilde{F_y}+\hat{F_x}\right)+2
    B_2^{\times}+4 \hat{B_2} \tilde{B_2}+2 \hat{G} \tilde{G}+4
    \hat{\phi } \tilde{\phi }-F_x' F_y'\right)\right)+24 S
    \left(\sinh (G) \left(\hat{B_2} \tilde{S}
    +
    \right. \right. \right.
    \nonumber
  \\
  &
    \left. \left. \left.
    \hat{S} \tilde{B_2}-S'
    \left(\tilde{F_y}+\hat{F_x}\right)+2 S^{\times}\right)+\cosh
    (G) \left(\hat{S} \tilde{G}+\hat{G} \tilde{S}\right)\right)-24
    \hat{S} \tilde{S} \sinh (G)\right)
    +
    \nonumber
  \\
  &
    e^{2 B_1+B_2} \left(S^2
    \left(3 \cosh (G) \left(\left(F_x'\right){}^2-2
    \left(-\left(B_1'+B_2'\right) \tilde{F_x}+\tilde{B_1}{}^2+2
    \tilde{B_2}{}^2+\bar{B_1}+\bar{B_2}+\tilde{B_1} \tilde{B_2}
    +\tilde{G}^2
    +
    \right. \right. \right. \right.
    \nonumber
  \\
  &
    \left. \left. \left. \left.
    2 \tilde{\phi }^2\right)\right)-6 \sinh (G)
    \left(\left(2 \tilde{B_1}+\tilde{B_2}\right)\tilde{G}-G'
    \tilde{F_x}+\bar{G}\right)\right)+S \left(-24 \cosh (G)
    \left(\left(\tilde{B_1}+\tilde{B_2}\right) \tilde{S}-S'
    \tilde{F_x}
    +
    \right. \right. \right.
    \nonumber
  \\
  &
    \left. \left. \left.
    \bar{S}\right)-24 \tilde{G} \tilde{S} \sinh
    (G)\right)+12 \tilde{S}^2 \cosh (G)\right) +e^{B_2} \left(S^2
    \left(6 \sinh (G) \left(\left(2 \hat{B_1}-\hat{B_2}\right)
    \hat{G}+\hat{F_y} G'-G^{\star}\right)
    +
    \right. \right.
    \nonumber
  \\
  &
    \left. \left.
    3 \cosh (G)
    \left(\left(F_y'\right){}^2-2 \left(\left(B_1'-B_2'\right)
    \hat{F_y}+\hat{B_1}{}^2+2
    \hat{B_2}{}^2-B_1^{\star}+B_2^{\star}-\hat{B_1}
    \hat{B_2}+\hat{G}^2+2 \hat{\phi }^2\right)\right)\right)
    +
    \right.
    \nonumber
  \\
  &
    \left.
    S \left(24 \cosh (G) \left(\left(\hat{B_1}-\hat{B_2}\right)
    \hat{S}+\hat{F_y} S'-S^{\star}\right)-24 \hat{G} \hat{S} \sinh
    (G)\right)+12 \hat{S}^2 \cosh (G)\right)
    +
    \nonumber
  \\
  &
    e^{B_1} \left(S^4
    \left(6 \left(\dot{B_1} B_1' \cosh ^2(G)+3 \dot{B_2} B_2'+\dot{G}
    G'+4 \dot{\phi } \phi '+4\right)-2 (4 V(\phi )+12)\right)-72
    S^2 \dot{S} S'\right) =0
    \label{ec:nested_last}
\end{align}
 
The following equations (to be solved for $\ddot{S}$ and
$\dot{F}_{x,y}$) are not needed for our evolution scheme, but they are
used in the equation for the gauge condition $\p_t \xi$:
\begin{align}
  &
    6 e^{B_1} \ddot{S} S^3+e^{B_1+B_2} \left(S^2 \left(\sinh (G)
    \left(-A' \left(\tilde{F_y}+\hat{F_x}\right)+\hat{B_2}
    \left(\tilde{A}+2 \dot{F_x}\right)+\tilde{B_2} \left(\hat{A}+2
    \dot{F_y}\right)+2 A^{\times}
    +
    \right. \right. \right.
    \nonumber
  \\
  &
    \left. \left. \left.
    2 \tilde{\dot{F_y}} +2
    \hat{\dot{F_x}}\right)+\cosh (G) \left(\hat{G}
    \left(\tilde{A}+2 \dot{F_x}\right)+\tilde{G} \left(\hat{A}+2
    \dot{F_y}\right)\right)\right)+S \sinh (G) \left(\hat{S}
    \left(\tilde{A}+2 \dot{F_x}\right)
    +
    \right. \right.
    \nonumber
  \\
  &
    \left. \left.
    \tilde{S} \left(\hat{A}+2
    \dot{F_y}\right)\right)\right)+e^{2 B_1+B_2} \left(S^2
    \left(\tilde{G} \sinh (G) \left(-\left(\tilde{A}+2
    \dot{F_x}\right)\right)-\cosh (G) \left(-A' \tilde{F_x}
    +
    \right. \right. \right.
    \nonumber
  \\
  &
    \left. \left. \left.
    \left(\tilde{B_1}+\tilde{B_2}\right) \left(\tilde{A}+2
    \dot{F_x}\right)+\bar{A}+2 \tilde{\dot{F_x}}\right)\right)-S
    \tilde{S} \cosh (G) \left(\tilde{A}+2
    \dot{F_x}\right)\right)-e^{B_1+2 B_2}
    \left(\hat{F_x}-\tilde{F_y}\right)^2
    +
    \nonumber
  \\
  &
    e^{B_2} \left(S^2 \left(\cosh
    (G) \left(A' \hat{F_y}+\left(\hat{B_1}-\hat{B_2}\right)
    \left(\hat{A}+2 \dot{F_y}\right)-A^{\star}-2
    \hat{\dot{F_y}}\right)-\hat{G} \sinh (G) \left(\hat{A}+2
    \dot{F_y}\right)\right)
    -
    \right.
    \nonumber
  \\
  &
    \left.
    S \hat{S} \cosh (G) \left(\hat{A}+2
    \dot{F_y}\right)\right)+e^{B_1} \left(S^4 \left(\dot{B_1}{}^2
    \cosh ^2(G)+3 \dot{B_2}{}^2+\dot{G}^2+4 \dot{\phi }^2\right)-3
    S^3 \dot{S} A'\right)=0
\end{align}

\begin{align}
  &
    4 e^{B_1} S^3 \dot{F_x}'+e^{B_1} \left(2 S^3 \left(\left(\tilde{A}+2
    \dot{F_x}\right) \left(B_1' \cosh ^2(G)+B_2'\right)+2
    \tilde{A}'+2 \tilde{\dot{B_1}} \cosh ^2(G)
    +
    \right. \right.
    \nonumber
  \\
  &
    \left. \left.
    2 \dot{B_1} \left(\tilde{B_1} \cosh ^2(G)+\tilde{G} \sinh (2 G)\right)+2
    \tilde{\dot{B_2}}+6 \dot{B_2} \tilde{B_2}+2 \dot{G} \tilde{G}+8
    \dot{\phi } \tilde{\phi }-A' F_x'\right)
    +
    \right.
    \nonumber
  \\
  &
    \left.
    4 S^2 \left(-S'
    \left(\tilde{A}+2 \dot{F_x}\right)+3 \tilde{S} \left(\dot{B_1}
    \cosh ^2(G)+\dot{B_2}\right)+4 \tilde{\dot{S}}+3 \dot{S}
    F_x'\right)-16 \dot{S} S \tilde{S}\right)
    +
    \nonumber
  \\
  &
    e^{B_1+B_2} \left(4 S
    \sinh (G) \left(\hat{F_x} \left(F_x'-2 \tilde{B_2}\right)+2
    \tilde{B_2} \tilde{F_y}-\tilde{F_x}
    F_y'-F_x^{\times}+\bar{F_y}\right)+4 \tilde{S} \sinh (G)
    \left(\hat{F_x}-\tilde{F_y}\right)\right)
    +
    \nonumber
  \\
  &
    e^{B_2} \left(4 S \cosh
    (G) \left(2 \hat{B_2}
    \left(\hat{F_x}-\tilde{F_y}\right)+\tilde{F_y}
    F_y'-F_y^{\times}-\hat{F_y} F_x'+F_x^{\star}\right)+4 \hat{S}
    \cosh (G) \left(\tilde{F_y}-\hat{F_x}\right)\right)
    +
    \nonumber
  \\
  &
    S^3
    \left(-\left(\hat{A}+2 \dot{F_y}\right) \left(B_1' \sinh (2 G)+2
    G'\right)+4 \hat{B_1} \dot{G}-2 \hat{\dot{B_1}} \sinh (2 G)+2
    \dot{B_1} \left(\hat{B_1} \sinh (2 G)
    -
    \right. \right.
    \nonumber
  \\
  &
    \left. \left.
    2 \hat{G} \cosh (2
    G)\right)-4 \hat{\dot{G}}\right)-6 \hat{S} S^2 \left(\dot{B_1}
    \sinh (2 G)+2 \dot{G}\right)=0
\end{align}

\begin{align}
  &
    4 S^3 \dot{F_y}'+e^{B_1} \left(S^3 \left(\left(\tilde{A}+2
    \dot{F_x}\right) \left(B_1' \sinh (2 G)-2 G'\right)-4 \dot{G}
    \tilde{B_1}+2 \tilde{\dot{B_1}} \sinh (2 G)
    +
    \right. \right.
    \nonumber
  \\
  &
    \left. \left.
    2 \dot{B_1}
    \left(\tilde{B_1} \sinh (2 G) +2 \tilde{G} \cosh (2 G)\right)-4
    \tilde{\dot{G}}\right)+6 S^2 \tilde{S} \left(\dot{B_1} \sinh (2
    G)-2 \dot{G}\right)\right)
    +
    \nonumber
  \\
  &
    e^{B_2} \left(4 S \sinh (G) \left(2
    \hat{B_2} \left(\hat{F_x}-\tilde{F_y}\right) +\tilde{F_y}
    F_y'-F_y^{\times}-\hat{F_y} F_x'+F_x^{\star}\right)+4 \hat{S}
    \sinh (G) \left(\tilde{F_y}-\hat{F_x}\right)\right)
    +
    \nonumber
  \\
  &
    e^{B_1+B_2}
    \left(4 S \cosh (G) \left(\hat{F_x} \left(F_x'-2 \tilde{B_2}\right)
    +2 \tilde{B_2} \tilde{F_y}-\tilde{F_x}
    F_y'-F_x^{\times}+\bar{F_y}\right)+4 \tilde{S} \cosh (G)
    \left(\hat{F_x}-\tilde{F_y}\right)\right)
    +
    \nonumber
  \\
  &
  2 S^3 \left(-A'
    F_y'-\left(\hat{A}+2 \dot{F_y}\right) \left(B_1' \cosh
    ^2(G)-B_2'\right)+2 \hat{A}'-2 \hat{\dot{B_1}} \cosh ^2(G)+2
    \dot{B_1} \left(\hat{B_1} \cosh ^2(G)
    -
    \right. \right.
    \nonumber
  \\
  &
    \left. \left.
    \hat{G} \sinh (2 G)\right)+2
    \hat{\dot{B_2}}+6 \dot{B_2} \hat{B_2} +2 \dot{G} \hat{G}+8
    \dot{\phi } \hat{\phi }\right)+4 S^2 \left(-S' \left(\hat{A}+2
    \dot{F_y}\right)
    +
    \right.
    \nonumber
  \\
  &
    \left.
    3 \hat{S} \left(\dot{B_2}-\dot{B_1} \cosh
    ^2(G)\right)+3 \dot{S} F_y'+4 \hat{\dot{S}}\right)-16 \dot{S}
    \hat{S} S =0
 \end{align}

\section{Apparent horizon finder}
\label{sec:AH}

In order to find the AH we need to compute the expansion of the
outgoing null rays. We can construct the tangent vector to the
outgoing rays using the ingoing null rays, $n$, together with the form
perpendicular to the AH, $s$,
\begin{equation}
  \begin{aligned}
    s
    & =N_s
    \left(
      -\p_t \sigma dt - \p_y \sigma dy - \p_y \sigma dy + dr
    \right)
    \,
    \\
    n & = -N_n \p_r
    \,,
  \end{aligned}
\end{equation}
from where we can compute the vector $s$ by simply raising the
indices. The normalization factors, $N_s$ and $N_n$, can be computed
by imposing $s^2=1$ and $s\cdot n=-1/\sqrt{2}$. Combining these two
vectors we can construct another vector tangent to outgoing
trajectories,
\begin{align}
  l^{\mu}=\sqrt{2}s^{\mu}+n^{\mu}
  \,,
\end{align}
so that it is null, $l^2=0$, and properly normalized, $l\cdot
n=-1$. The expansion of these rays can be computed as
\begin{align}
  \theta_l=h^{\mu\nu}\nabla_{\mu}l_{\nu}
  \,,
\end{align}
where
\begin{align}
  h_{\mu\nu} = g_{\mu\nu}+l_{\mu}n_{\nu}+l_{\nu}n_{\mu}
\end{align}
is the induced metric over hypersurfaces normal to both in- and
out-going null rays. The AH location is given by the condition
$\theta_l=0$. Imposing it at a generic surface, $r=\sigma(x,y)$, we
obtain the following equation:
\begin{align}
  &
    2 e^{B_2} \left(F_y+\partial_y\sigma\right) \left(S \left(e^{B_1}
    \cosh (G) \left(\tilde{G}+G'
    \left(F_x+\partial_x\sigma\right)\right)+e^{B_1} \sinh (G)
    \left(\tilde{B_2}+B_2' \left(F_x+\partial_x\sigma\right)\right)
    +
    \right. \right. 
    \nonumber
  \\
  &
    \left. \left.
    \cosh (G) \left(B_1'
    \left(F_y+\partial_y\sigma\right)+\hat{B_1}\right)-\cosh (G)
    \left(B_2'
    \left(F_y+\partial_y\sigma\right)+\hat{B_2}\right)-\sinh (G)
    \left(G' \left(F_y+\partial_y\sigma
    \right)
    +
    \right. \right. \right.
    \nonumber
  \\
  &
    \left. \left. \left.
    \hat{G}\right)\right) +e^{B_1} \sinh (G)
    \left(\tilde{S}-2 S'
    \left(F_x+\partial_x\sigma\right)\right)-\cosh (G) \left(S'
    \left(F_y+\partial_y\sigma\right)+\hat{S}\right)\right)-2
    e^{B_1+B_2} \left(F_x
    +
    \right.
    \nonumber
  \\
  &
    \left.
    \partial_x\sigma\right) \left(S \left(e^{B_1}
    \left(\cosh (G) \left(\tilde{B_1}+B_1'
    \left(F_x+\partial_x\sigma\right)\right)+\cosh (G)
    \left(\tilde{B_2}+B_2'
    \left(F_x+\partial_x\sigma\right)\right)
    +
    \right. \right. \right.
    \nonumber
  \\
  &
    \left. \left. \left.
    \sinh (G)
    \left(\tilde{G}+G'
    \left(F_x+\partial_x\sigma\right)\right)\right)-\sinh (G)
    \left(B_2'
    \left(F_y+\partial_y\sigma\right)+\hat{B_2}\right)-\cosh (G)
    \left(G' \left(F_y+\partial_y\sigma\right)
    +
    \right. \right. \right.
    \nonumber
  \\
  &
    \left. \left. \left.
    \hat{G}\right)\right)
    +e^{B_1} \cosh (G) \left(\tilde{S}+S' \left(F_x+\partial_x\sigma
    \right)\right)-\sinh (G) \left(S'
    \left(F_y+\partial_y\sigma\right)+\hat{S}\right)\right)
    +
    \nonumber
  \\
  &
    S \left(e^{B_1} \left(2 e^{B_2} \sinh (G) \left(\tilde{F_y}+F_y'
    \left(F_x+\partial_x\sigma\right)+\partial_{xy}\sigma\right)-2
    e^{B_1+B_2} \cosh (G) \left(\tilde{F_x}+F_x'
    \left(F_x+\partial_x\sigma
    \right)
    +
    \right. \right. \right.
    \nonumber
  \\
  &
    \left. \left. \left.
    \partial_{xx}\sigma\right)+6 S \dot{S}\right) +2
    e^{B_1+B_2} \sinh (G) \left(F_x'
    \left(F_y+\partial_y\sigma\right)+\hat{F_x}+\partial_{xy}\sigma\right)-2
    e^{B_2} \cosh (G) \left(F_y'
    \left(F_y
    +
    \right. \right. \right.
    \nonumber
  \\
  &
    \left. \left. \left.
    \partial_y\sigma\right)
    +\hat{F_y}+\partial_{yy}\sigma
    \right)\right) +3 e^{2 B_1+B_2} \cosh (G) S'
    \left(F_x+\partial_x\sigma\right){}^2+3 e^{B_2} \cosh (G) S'
    \left(F_y+\partial_y\sigma\right)^2=0,
\label{eq:expansion-gen}
\end{align}
where every function is evaluated at the $r=\sigma(x,y)$ surface
defining the AH.
When the AH is located surfaces of constant radius, which is what we
impose to find the evolution equation for the gauge function $\xi$,
then Eq.~\eqref{eq:expansion-gen} reduces to
\begin{align}
  \Theta
  &
    \equiv -2 e^{B_1+B_2} F_x \left(S \left(e^{B_1} \left(\cosh (G)
    \left(\tilde{B_1}+B_1' F_x\right)+\cosh (G) \left(\tilde{B_2}+B_2'
    F_x\right)
    +
    \right. \right. \right.
    \nonumber
  \\
  &
    \left. \left. \left.
    \sinh (G) \left(\tilde{G}+F_x G'\right)\right)
    -\sinh (G) \left(B_2' F_y+\hat{B_2}\right)-\cosh (G) \left(F_y
    G'+\hat{G}\right)\right)+e^{B_1} \cosh (G) \left(\tilde{S}
    +
    \right. \right.
    \nonumber
  \\
  &
    \left. \left.
    F_x S'\right)-\sinh
    (G) \left(F_y S'+\hat{S}\right)\right) +2 e^{B_2} F_y \left(S \left(e^{B_1}
    \cosh (G) \left(\tilde{G}+F_x G'\right)+e^{B_1} \sinh (G) \left(\tilde{B_2}
    +
    \right. \right. \right.
    \nonumber
  \\
  &
    \left. \left. \left.
    B_2'
    F_x\right)+\cosh (G) \left(B_1' F_y+\hat{B_1}\right)
    -\cosh (G) \left(B_2' F_y+\hat{B_2}\right)-\sinh (G) \left(F_y
    G'+\hat{G}\right)\right)
    +
    \right.
    \nonumber
  \\
  &
    \left.
    e^{B_1} \sinh (G) \left(\tilde{S}-2 F_x S'\right)-\cosh
    (G) \left(F_y S'+\hat{S}\right)\right) +S \left(e^{B_1} \left(2 e^{B_2} \sinh
    (G) \left(\tilde{F_y}+F_x F_y'\right)
    -
    \right. \right.
    \nonumber
  \\
  &
    \left. \left.
    2 e^{B_1+B_2} \cosh (G)
    \left(\tilde{F_x}+F_x F_x'\right)+6 S \dot{S}\right) +2
    e^{B_1+B_2} \sinh (G) \left(F_y F_x'+\hat{F_x}\right)
    -
    \right.
    \nonumber
  \\
  &
    \left.
    2 e^{B_2} \cosh (G)
    \left(F_y F_y'+\hat{F_y}\right)\right)+3 e^{2 B_1+B_2} F_x^2 \cosh (G) S' +3
    e^{B_2} F_y^2 \cosh (G) S' = 0
    \,.
  \label{ec:expansion}
\end{align}

To start with initial data that satisfies
$\Theta|_{r=\mathrm{const}} = 0$ we first need to find the AH and
adjust $\xi$ accordingly. Solving the Eq.~\eqref{eq:expansion-gen}
provides us the location of the AH at a given time slice $t$. Contrary
to what we have found so far this equation is non-linear, with the
form
\begin{equation}
  \begin{aligned}
    \mathcal{L}
    \left(\sigma,\p\sigma,\p^2\sigma\right)
    & =\alpha_{xx}(t,\sigma,x,y)\p_{xx}\sigma
    +\alpha_{xy}(t,\sigma,x,y)\p_{xy}\sigma
    +\alpha_{yy}(t,\sigma,x,y)\p_{yy}\sigma
    \\
    &\quad
    +\beta_{xx}(t,\sigma,x,y)
    \left(\p_x\sigma\right)^2
    +\beta_{xy}(t,\sigma,x,y)\p_x\sigma\p_y\sigma
    +\beta_{yy}(t,\sigma,x,y)\left(\p_y\sigma\right)^2
    \\
    &\quad
    +\gamma_x(t,\sigma,x,y)\p_x\sigma
    +\gamma_y(t,\sigma,x,y)\p_y\sigma
    +\delta(t,\sigma,x,y)
    =0
    \,,
  \end{aligned}
  \label{eq:AH-nonlin}
\end{equation}
where
\allowdisplaybreaks
\begin{align*}
  \alpha_{xx}
  & =-e^{B_1+B_2} S \cosh (G)
    \,,
  \\
  \alpha_{xy}
  & =2 e^{B_2} S \sinh (G)
    \,,
  \\
  \alpha_{yy}
  &
    =-e^{B_2-B_1} S \cosh (G)
    \,,
  \\
  \beta_{xx}
  &
    =\frac{1}{2} e^{B_1+B_2}
    \left(\cosh (G) S'-2 S
    \left(B_1' \cosh (G)+B_2' \cosh (G)+G' \sinh (G)\right)\right)
    \,,
  \\
  \beta_{xy}
  &
    =e^{B_2} \left(2 S
    \left(B_2' \sinh (G)+G' \cosh (G)\right)-\sinh (G) S'\right)
    \,,
  \\
  \beta_{yy}
  &
    =\frac{1}{2} e^{B_2-B_1}
    \left(2 S \left(B_1' \cosh (G)-B_2' \cosh (G)
    +G' (-\sinh (G))\right)+\cosh (G) S'\right)
    \,,
  \\
  \gamma_x
  &
    =e^{B_2} \left(S
    \left(-e^{B_1} \tilde{G} \sinh (G)
    -e^{B_1} \tilde{B_1} \cosh (G)-e^{B_1} \tilde{B_2} \cosh (G)
    -2 e^{B_1} F_x G' \sinh (G)\right.\right.
  \\
  &\left.\left.-2 e^{B_1} B_1'
    F_x \cosh (G)-2 e^{B_1} B_2' F_x \cosh (G)
    -e^{B_1} \cosh (G) F_x'+2 B_2' F_y \sinh (G)
    +\hat{B_2} \sinh (G)
    \right.\right.
  \\
  &
    \left.\left.
    +2 F_y G' \cosh (G)+\sinh (G) F_y'+\hat{G}
    \cosh (G)\right)-e^{B_1} \tilde{S} \cosh (G)
    +e^{B_1} F_x \cosh (G) S'
    \right.
  \\
  &
    \left.
    -F_y \sinh (G) S'+\hat{S} \sinh (G)\right)
    \,,
  \\
  \gamma_y
  &
    =e^{B_2-B_1} \left(S
    \left(e^{B_1} \tilde{B_2} \sinh (G)
    +e^{B_1} \tilde{G} \cosh (G)+2 e^{B_1} F_x G' \cosh (G)
    +2 e^{B_1} B_2' F_x \sinh (G)
    \right.\right.
  \\
  &
    \left.\left.
    +e^{B_1} \sinh
    (G) F_x'+2 B_1' F_y \cosh (G)-2 B_2' F_y \cosh (G)
    +\hat{B_1} \cosh (G)-\hat{B_2} \cosh (G)
    \right. \right.
  \\
  &
    \left.\left.
    -2 F_y G' \sinh (G)-\cosh (G) F_y'-\hat{G} \sinh
    (G)\right)+e^{B_1} \tilde{S} \sinh (G)-e^{B_1} F_x \sinh (G) S'
    \right.
  \\
  &
    \left.
    +F_y \cosh (G) S'+\hat{S} (-\cosh (G))\right)
    \,,
  \\
  \delta
  &
    =-e^{B_2-B_1} S \left(-F_y
    \left(e^{B_1} \tilde{B_2} \sinh (G)
    +e^{B_1} \tilde{G} \cosh (G)+2 e^{B_1} F_x G' \cosh (G)
    +2 e^{B_1} B_2' F_x \sinh (G)
    \right.\right.
  \\
  &
    \left.\left.
    +e^{B_1}
    \sinh (G) F_x'+\hat{B_1} \cosh (G)-\hat{B_2} \cosh (G)
    -\cosh (G) F_y'-\hat{G} \sinh (G)\right)
    +e^{B_1} F_x \left(e^{B_1} \tilde{G} \sinh (G)
    \right.\right.
  \\
  &
    \left.\left.
    +e^{B_1}
    \tilde{B_1} \cosh (G)+e^{B_1} \tilde{B_2} \cosh (G)
    +e^{B_1} \cosh (G) F_x'-\hat{B_2} \sinh (G)
    -\sinh (G) F_y'-\hat{G} \cosh (G)\right)
    \right.
  \\
  &
    \left.
    +e^{2 B_1}
    \cosh (G) \tilde{F_x}-e^{B_1} \sinh (G) \tilde{F_y}
    +e^{2 B_1} F_x^2 \left(B_1' \cosh (G)+B_2' \cosh (G)
    +G' \sinh (G)\right)
    \right.
  \\
  &
    \left.
    +F_y^2 \left(-B_1' \cosh
    (G)+B_2' \cosh (G)+G' \sinh (G)\right)
    -e^{B_1} \hat{F_x} \sinh (G)+\hat{F_y} \cosh (G)\right)
  \\
  &
    +\frac{1}{2} e^{B_2-B_1} \left(-2 F_y \left(\hat{S} \cosh
    (G)-e^{B_1} \sinh (G) \left(\tilde{S}-F_x S'\right)\right)
    +e^{B_1} F_x \left(2 \hat{S} \sinh (G)
    \right.\right.
  \\
  &
    \left.\left.
    -e^{B_1} \cosh (G) \left(2 \tilde{S}-F_x
    S'\right)\right)+F_y^2 \cosh (G) S'\right)
    +3 \dot{S} S^2
    \,.
\end{align*}

We solve equation~\eqref{eq:AH-nonlin} with the Newton-Kantorovich
method by linearizing the equation around an guessed solution
$\sigma_0(x,y)$. Expanding the operator $\mathcal{L}$ we obtain
\begin{equation}
  \begin{aligned}
    \mathcal{L}
    \left(\sigma,\p\sigma,\p^2\sigma\right)
    & =
    \left(\mathcal{L}
      +\frac{\p\mathcal{L}}{\p\sigma}
      +\frac{\p\mathcal{L}}{\p(\p_x\sigma)}\p_x
      +\frac{\p\mathcal{L}}{\p(\p_y\sigma)}\p_y
      +\frac{\p\mathcal{L}}{\p(\p_{xx}\sigma)}\p_{xx}
      +\frac{\p\mathcal{L}}{\p(\p_{xy}\sigma)}\p_{xy}
    \right.
    \\
    &\qquad
    \left.
      +\frac{\p\mathcal{L}}{\p(\p_{yy}\sigma)}
      \p_{yy}\right)_{\sigma=\sigma_0}\delta\sigma
    +\mathcal{O}\left(\delta\sigma^2\right)
    =0
    \,,
  \end{aligned}
\end{equation}
where $\delta\sigma=\sigma(x,y)-\sigma_0(x,y)$.  The associated linear
problem for the correction $\delta \sigma$ is then
\begin{equation}
  \begin{aligned}
    &
    \left[
      \alpha_{xx}(\sigma_0)\p_{xx}
      +\alpha_{xy}(\sigma_0)\p_{xy}
      +\alpha_{yy}(\sigma_0)\p_{yy}
      +\left(\gamma_x(\sigma_0)
        +2\beta_{xx}(\sigma_0)\p_x\sigma_0
        +\beta_{xy}(\sigma_0)\p_y\sigma_0
      \right)
      \p_x\right.
    \\
    &
    \quad
    \left.
      +\left(\gamma_y(\sigma_0)
        +2\beta_{yy}(\sigma_0)\p_y\sigma_0
        +\beta_{xy}(\sigma_0)\p_x\sigma_0\right)\p_y
      +\p_\sigma\mathcal{L}(\sigma_0)
    \right]
    \delta\sigma
    =
    -\mathcal{L}(\sigma_0,\p\sigma_0,\p^2\sigma_0),
  \end{aligned}
\end{equation}
which has the same functional form as that of
equation~\eqref{ec:xi_evol} and that we solve in the same fashion. For
the purpose of implementing this into the code,what remains is the
rewriting of the coefficients in terms of the outer grid
redefinitions.
